\documentclass[oneside,english]{scrbook}
\usepackage[utf8]{inputenc}
\usepackage{babel}
\usepackage{float}
\usepackage{fancybox}
\usepackage{calc}
\usepackage{bm}
\usepackage{amsmath}
\usepackage{graphicx}
\usepackage{esint}
\usepackage[unicode=true,pdfusetitle,
 bookmarks=true,bookmarksnumbered=false,bookmarksopen=false,
 breaklinks=false,pdfborder={0 0 1},backref=section,colorlinks=false]
 {hyperref}

\makeatletter

\DeclareTextSymbolDefault{\textquotedbl}{T1}
\providecommand{\tabularnewline}{\\}

\newcommand{\lyxaddress}[1]{
	\par {\raggedright #1
	\vspace{1.4em}
	\noindent\par}
}

\usepackage{babel}
\usepackage{cite}

\usepackage{listings}

\makeatother

\usepackage{listings}

\begin{document}
\title{The Quantum and Stochastic Toolbox: xSPDE4.2}
\author{Peter D. Drummond, Run Yan Teh, Manushan Thenabadu\\
Channa Hatharasinghe, Chris McGuigan, Alex Dellios\\
 Ned Goodman, Margaret D. Reid}
\maketitle

\lyxaddress{\textbf{Centre for Quantum Science and Technology Theory, Swinburne
University of Technology, Melbourne, Victoria, Australia.}}

This is the fourth major release of the xSPDE toolbox, which solves
stochastic partial and ordinary differential equations, with applications
in biology, chemistry, engineering, medicine, physics and quantum
technologies. It computes statistical averages, including time-step
and sampling error estimation. xSPDE can provide higher order convergence,
Fourier spectra and probability densities. The toolbox has graphical
output and $\chi^{2}$ statistics, as well as weighted, projected,
or forward-backward equations. It can generate input-output quantum
spectra. The equations can have independent periodic, Dirichlet, and
Neumann or Robin boundary conditions in any dimension, for any vector
component, and at either end of any interval. xSPDE has functions
that can numerically solve both ordinary and partial differential
stochastic equations of any type, obtaining correlations, probabilities
and averages. The toolbox has a core treating stochastic differential
equations, with averages, probability distributions and full error
estimates. There are stochastic extensions treating applications to
partial differential equations, projected equations, quantum stochastic
equations, master equations and quantum phase-space simulations including
Gaussian boson sampling experiments.

\newpage{}

\tableofcontents{}\vspace{10pt}

\noindent \newpage{}

\part{Introduction\label{part:Introduction}}

\textbf{xSPDE is an eXtensible Stochastic Partial Differential Equation
solver}.

There are many equations of this type~\cite{Gardiner2009Stochastic,Drummond2014Quantum,Langevin1908theorie,Karatzas1991brownian,Glasserman2010monte}
in physics, chemistry, engineering, biology, medicine, and finance.
Typical applications are in physics, quantum technology and biostatistics~\cite{Opanchuk2013Quantum,Ng2019Phase,ramesh2022arcsine,peters2022limit,teh2017simulation,teh2018quantum,kiesewetter2022phase,teh2018creation,dechoum2016critical,drummond2017truncated,opanchuk2017one,drummond2017higher,ng2021fate,joseph2021hybrid,drummond2021time,drummond2020initial,ng2019nonlocal,busink2021stochastic,peters2021extremely,peters2022exceptional},
but the code has general applicability. The emphasis in xSPDE is on
combining a simple user interface with a wide range of useful functions,
including the essential features of averaging and global error estimates.
The code enables an efficient use memory and parallelism, which is
vital for large stochastic models, and it is able to be further extended
if needed.

The extensible structure of the code-base permits drop-in replacements
of the algorithms. Different simulations can be carried out sequentially.
This models different stages in an experiment or simulated environment.
It can be used with or without noise terms, and can use a range of
either built-in or user defined integration algorithms. This user
guide describes xSPDE4, which is an improved and extended version
of xSPDE3, and earlier toolboxes~\cite{Kiesewetter2016xSPDE,kiesewetter2023xspde3}.

xSPDE calculates and plots averages and probabilities of arbitrary
functions of any number of complex or real fields, as well as Fourier
transforms in time or space with any given dimensionality. Importantly,
it gives error estimates for both the discretization and sampling
error, but the algorithm, the step-size and the number of samples
used is up to the user to control to obtain the required error levels.

Ordinary and stochastic differential equations of many types can be
treated numerically~\cite{Drummond1991Computer,Kloeden1992numerical},
including stochastic partial differential equations with space dependence~\cite{Werner1997Robust}.
Comparative $\chi^{2}$ statistical tests are available. Additional
libraries exist for projected, forward-backward, and weighted equations.

The algorithms included are designed to be useful and fast in many
practical applications. Higher order convergence is obtained through
order extrapolation. This allows higher-order convergence to be realized
in a uniform way. More complex higher-order algorithms are known \cite{Kloeden1992numerical,burrage2006comment},
which can be included if preferred, as the code is extensible.

The code can be used interactively or in batch mode. All graphs, data,
and input parameters, including default values, can be stored permanently
using standard file-types. It has a fully integrated graphics program,
xGRAPH, which graphs data of any dimensions, including multiple types
of graphical output, error-bars and comparisons.

xSPDE supports parallelism at both vector instruction and multiple
core level using array and parallel loop syntax. This version is Octave/Matlab
based. Matlab is a commercial product, GNU Octave \cite{eaton2012gnu}
is free and open-source. They each have excellent user interfaces
and reliable implementations. Full parallel operation currently requires
the Matlab parallel toolbox.

Part 2 covers SDE theory and numerical solutions. Readers who are
simply interested in how to use the code can go directly to Chapter
\ref{chap:Simulating-an-SDE}, which describes the numerical solution
of SDEs with xSPDE. This includes an explanation of the user interface,
how to input parameters and equations, how to define the output in
terms of functional averages or probabilities, and how to define and
access auxiliary fields and noises. This chapter uses the default
algorithms, and a more detailed explanation is given in chapter \ref{sec:Algorithms}.
Chapter \ref{chap:SDE Theory} has definitions and notations for stochastic
differential equations (SDEs). This is useful for understanding later
chapters. This part includes Ito and Stratonovich calculus, probability
distributions and Fokker-Planck equations. It also explains and defines
the Fourier input-output spectra used in quantum technology.

Part 3 gives the theory and numerical implementation of stochastic
partial differential equations (SPDEs). It includes details of spectral
methods and the interaction picture approach. It has an explanation
of how Fourier transforms and discrete sine or cosine transforms are
implemented. It also explains how boundary conditions can be implemented
using finite differences. In Chapter \ref{sec:Simulating-an-SPDE},
the practical approach to solving stochastic partial differential
equations with xSPDE is explained. The techniques used are an extension
of the ordinary SDE methods, so a thorough understanding of chapter
\ref{chap:Simulating-an-SDE} is strongly recommended.

Part 4 treats phase-space methods, including the Wigner, Q-function,
Glauber P-function and positive P-function. These can be treated simply
using Part 2 or 3, but the purpose of this tool-box is to provide
functions for initializing, propagating and observing quantum systems.
In particular it covers quantum networks such as Gaussian boson sampling
quantum computers, together with scalable verification and photon-counting
algorithms. 

Part 5 treats open quantum system theory and their numerical solutions
using stochastic methods. It treats master equations, phase-space
methods, and stochastic Schrödinger equations. Logic gates are also
included here. Although there are more specialized programs that are
dedicated for this purpose, it is useful to understand how different
types of decoherence can change gate operations. As it has a modular
design, it is possible to include systematic or non-Markovian gate
errors, as well as Markovian noise.

Part 6 is for reference purposes. Chapter \ref{sec:Algorithms} outlines
the integration algorithms used in the manual. It includes a number
of extended integration libraries, applicable to more specialized
problems. This chapter also outlines how integration errors, including
time-step and stochastic errors, can be estimated and displayed. Chapter
\ref{chap:API-reference} provides a reference for the details of
the internals as well as a comprehensive explanation of the input
parameters useful in xSPDE simulations. This explains how to create
projects with separated computation and graphics, as well as workflow
and data storage. It also provides an extensive description of the
visualization aspects of xSPDE, using the integrated xGRAPH function,
which includes an automatic 'cascade' of graphic output where high
dimensional data is reduced to lower dimensional, visualizable data
through projections.

Input parameters related to this are described as well. Data can also
be graphed externally or stored for later analysis if preferred. Both
average and raw trajectory data can be stored. However, the storage
of raw data is generally not recommended, due to the large storage
requirements. Additional examples in Chapter \ref{sec:Examples} demonstrate
how to obtain parametric plots against input parameters. Plots of
one component value against another can be graphed. A function that
analyses convergence rates is also available.

To run xSPDE, an Octave or Matlab environment is needed. A Julia option
will be available in future. The current xSPDE distribution includes
the toolbox: $xspde.mltbx$, or a folder: xspde\_matlab, which includes
the following: 
\begin{itemize}
\item Simulator folder with the core functions
\item Methods folder for the different applications
\item Examples folder that can also be used as templates
\item Graphics folder for the integrated graphics
\item Documentation folder with this user's guide
\item License.txt that contains the BSD license
\end{itemize}
xSPDE can be run interactively as a script, or as a function in batch
mode, either at a local workstation or on a remote cluster. Data can
be either plotted immediately, or saved then plotted later. To simulate
a stochastic equation interactively, first check that the toolbox
or folder is installed.

\textbf{If you have the toolbox file, $xspde.mltbx$, just open it
and click on $install$.} Otherwise the Octave/Matlab path must point
to the xSPDE folder and subfolders. If you have the folders, but not
the toolbox, proceed as follows: 
\begin{itemize}
\item Click on the Octave/Matlab HOME tab (top left), then Set Path 
\item Click on Add with Subfolders 
\item Find the xspde folder in the drop-down menu, select it , then save
the path. 
\end{itemize}
Type $clear$ to clear old data, and enter the inputs and functions
into the command window interactively. For more advanced cases, it
is best to create a function that calls xspde. There are many examples
listed in this manual, and there are more in the Examples folder.
Any of these can be used as templates for building your own simulation
code.

See: \textbf{www.github.com/peterddrummond/xspde\_matlab.} For those
familiar with earlier versions, a list of the main xspde changes since
the documentation of the previously published version (v3.44) \cite{Kiesewetter2016xSPDE,kiesewetter2023xspde3}
is as follows: 
\begin{enumerate}
\item Cell arrays for multiple variables with differing labels and/or spatial
grids
\item Error-checking outputs with both maximum and RMS error estimates
\item Quantum stochastic Schrödinger equations
\item Integration of master equations 
\item Jump algorithms, in addition to Gaussian noise methods 
\item DST and DCT spectral methods for SPDEs with non-zero boundaries
\end{enumerate}
xSPDE is distributed with no guarantee, under an open-source license.
Contributions and bug reports are welcome. An alternative approach
to SPDEs \cite{Collecutt2001xmds,Dennis2013XMDS2} is available in
C++ at http://www.xmds.org/.

\newpage{}

\part{Stochastic differential equations}

\chapter{SDE toolbox\label{chap:Simulating-an-SDE}}

\textbf{This chapter describes how to use the xSPDE numerical toolbox
to solve an SDE to obtain and graph averages, spectra or probability
distributions. For theoretical background, see Chapter (\ref{chap:SDE Theory}).
For detailed examples, see Chapter (\ref{chap:SDE-Examples}) .}

\section{Using xSPDE}

Stochastic equations generally require numerical solutions. To obtain
them, xSPDE has a parameter structure, p, that defines both the equations
and numerical parameters. The equations are defined as user functions
with arguments (fields..,noises.., parameters).

All input parameters are all passed to functions in the structure
$p$. Complete details of the xSPDE input parameters are given in
\ref{sec:Table-of-parameters}. There are default options that allow
one to reduce the required parameter inputs and functions to just
the important ones. The three most essential user-specified functions
are listed below:\\

\begin{tabular}{|c|c|c|}
\hline 
Label & Arguments & Purpose\tabularnewline
\hline 
\hline 
\textbf{initial} & \textbf{$(w,p)$} & \textbf{Function to initialize fields}\tabularnewline
\hline 
\textbf{deriv} & \textbf{$(a,..w,..p)$} & \textbf{Stochastic derivative}\tabularnewline
\hline 
\textbf{observe} & \textbf{$(a,..,p)$} & \textbf{Observable function}\tabularnewline
\hline 
\end{tabular}\\

In the table, $"\ldots"$ indicate optional arguments used if there
are multiple field variables and noise variables. In the simplest
case, the default option is one field and one noise. These can also
be vectors if required. If this is needed, use the fields and noises
input parameters explained below.

xSPDE has two parts, xSIM for the simulations and xGRAPH for automatic
graphic generation. They can be used together in xSPDE, or individually
as a batch job, so that data can be stored and graphed separately.

\subsection{Wiener process}

To use xSPDE to solve for a single trajectory of a simple SDE:
\begin{equation}
\dot{a}=w(t)\,,
\end{equation}
described in more detail in Eq \eqref{eq:Wiener_process} , only two
lines are needed:
\begin{center}
\doublebox{\begin{minipage}[t]{0.75\columnwidth}%
\texttt{p.deriv = @(a,w,p) w;}

\texttt{xspde(p);}%
\end{minipage}} 
\par\end{center}

Here $p.deriv$ defines the time derivative $\dot{a}$ in the input
parameter structure p, while $w$ is a delta-correlated Gaussian noise
generated internally. There are no other input parameters. Default
values are used for initial, observe, ensembles, points, ranges and
olabels. This produces the graph shown in Fig (\ref{fig:The-simplest-case: Wiener}),
for a single trajectory. Results can change with different seeds or
random number generators.
\begin{center}
\begin{figure}
\centering{}\includegraphics[width=0.75\textwidth]{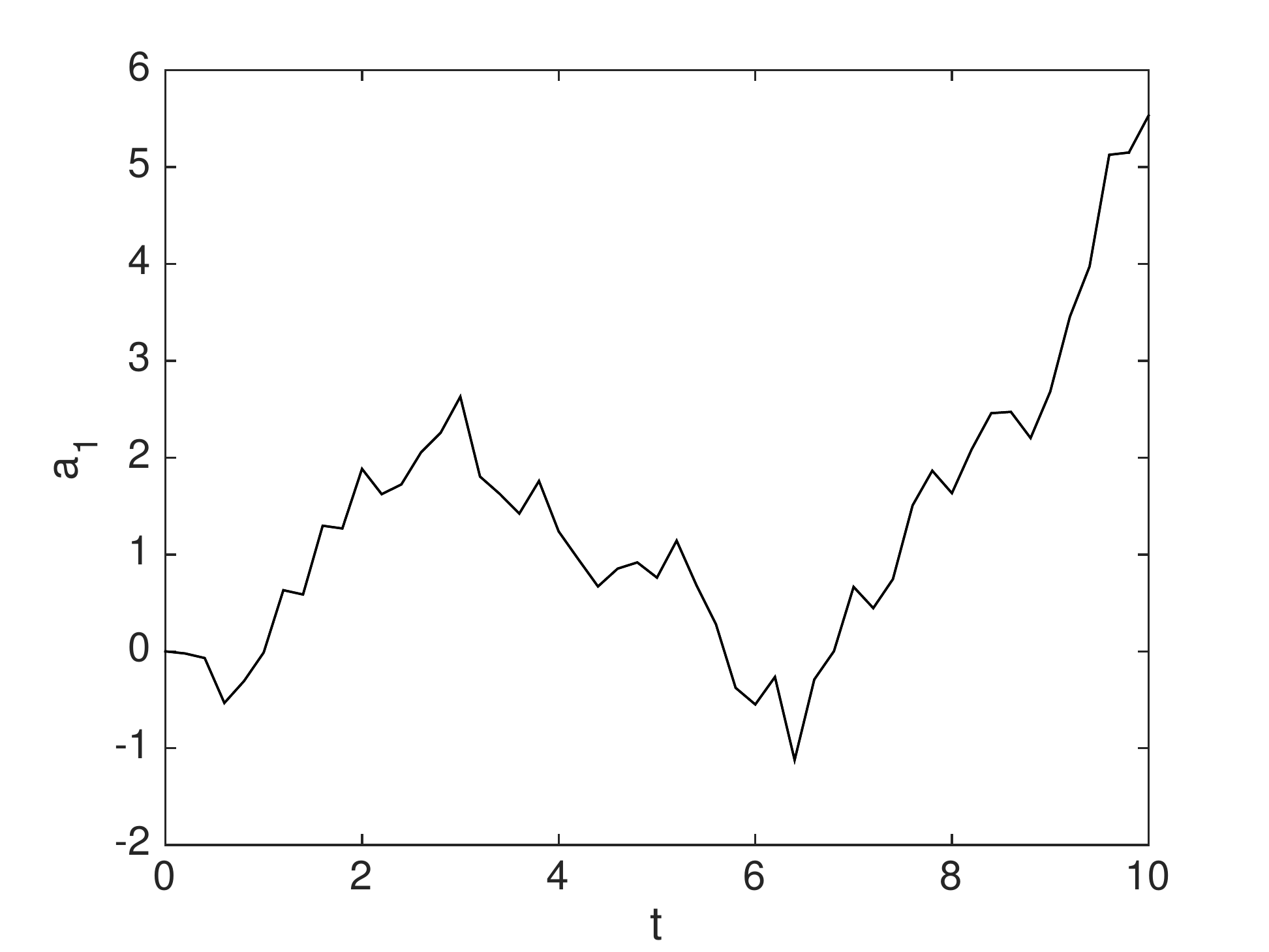}

\caption{\label{fig:The-simplest-case: Wiener}\emph{The simplest example:
a random walk.}}

\vspace{10pt}
\end{figure}
\par\end{center}

At the end of the run, xSPDE reports the RMS errors. There are discretization,
sampling and comparison errors, all normalized by the maximum observable
value, unless compared to a result of zero. In the present simulation,
the discretization or step error is about $10^{-16}$, due to round-off.
This is a single trajectory, but more can be added using the ensembles
input parameter.

\subsection{General derivatives}

All important xSPDE procedures use functions. Functions can be specified
inline, which is the simplest, or externally. The last argument of
more complex xSPDE functions is the parameter structure. An example
already introduced is the derivative function, labeled $p.deriv$.

For example, consider the stochastic differential equation, 
\begin{equation}
\frac{da}{dt}=-ga+w.
\end{equation}
The corresponding derivative code definition is: 
\begin{center}
\doublebox{\begin{minipage}[t]{0.75\columnwidth}%
\texttt{p.deriv = @(a,w,p) - p.G{*}a + w;}%
\end{minipage}} 
\par\end{center}

This code defines the function handle $p.deriv$, which gives the
derivative function, $da/dt$. In this example, it simply returns
the derivative, in terms of the variable $a$, loss parameter $p.G$,
and stochastic noise term $w.$ This user specified inline function
is known internally by the function handle $p.deriv$.

Inside a complete xSPDE simulation input with a parameter values,
it would look like: 
\begin{center}
\doublebox{\begin{minipage}[t]{0.75\columnwidth}%
\texttt{p.G = 0.25;}

\texttt{p.deriv = @(a,w,p) - p.G{*}a + w;}

\texttt{xspde(p);}%
\end{minipage}} 
\par\end{center}

External function handles can also be used. They are useful for complex
functions with more internal logic. A typical script first defines
parameters and function specifications, in a structure, then runs
the simulation code with the parameter structure as an input, as follows: 
\begin{center}
\doublebox{\begin{minipage}[t]{0.75\columnwidth}%
\texttt{p.{[}label1{]} = {[}parameter1{]};}

\texttt{...}

\texttt{p.{[}label2{]} = {[}parameter2{]};}

\texttt{p.deriv = @(a,w,p) {[}derivative{]};}

\texttt{xspde(p);}%
\end{minipage}} 
\par\end{center}

Note the following points to remember: 
\begin{itemize}
\item $\mathtt{p.}[label1]=[parameter1]$ defines a parameter in the structure
$p$. 
\item There are many possible inputs, which all have default values. 
\item You don't have to save the data if you want an immediate plot. 
\item The notation $\mathtt{p.deriv=}\mathtt{@(a,w,p)}\,\,[derivative]$
defines a function, $da/dt$. 
\item In this example, $\mathtt{a}$ is the stochastic variable, $\mathtt{w}$
the random noise, $p$ a structure. 
\item Other labels can be used instead of (a,w,p) if preferred. 
\end{itemize}

\section{SDE parameters\label{sec:Input-parameters}}

All xSPDE simulations use a structure for input data. Most functions
also require a parameter structure, combining the data input with
additional internal parameters. Any naming convention will do for
either structure, as long as you are consistent.

User-defined parameters can be added freely. To ensure that there
is no clash with internal variables, it is best if user defined parameters
start with a capital letter.

The xSPDE inputs have default values, which are used if the input
values are omitted. If you only need the first element of a vector
or array, just input the value required. Parameters can be output
with the verbose switch, \texttt{p.verbose}. This has four levels
of output: $-1$,$0$, $1$ or $2$, with \texttt{p.verbose=0} as
default, giving final error reports. To get more progress details
and individual errors, use \texttt{p.verbose=1.} To eliminate almost
everything, use \texttt{p.verbose=-1.} For maximum information, including
all the internal parameter values, use; 
\begin{center}
\doublebox{\begin{minipage}[t]{0.75\columnwidth}%
\texttt{p.verbose = 2;}%
\end{minipage}} 
\par\end{center}

While this level of detail is not usually needed, it can be useful
to print out all the internal parameters and default values to understand
how the program operates.

\subsection{Parameter table\label{subsec:Simulation-parameters}}

The most common xSPDE simulation parameters used to define the equations
and method of solution, together with their default values are:

\begin{table}[h]
\begin{tabular}{|c|c|c|c|}
\hline 
Label & Type & Default value & Description\tabularnewline
\hline 
\hline 
fields & vector, array or cell & $1$ & Number of stochastic fields\tabularnewline
\hline 
noises & vector, array or cell & fields & Number of noises\tabularnewline
\hline 
inrandoms & vector, array or cell & noises & Number of initial randoms\tabularnewline
\hline 
name & string & ' ' & Simulation name\tabularnewline
\hline 
deriv & function & 0 & The stochastic derivative\tabularnewline
\hline 
initial & function & 0 & Function to initialize variables\tabularnewline
\hline 
method & function & {[}see \ref{sec:Algorithms}{]} & Integration method\tabularnewline
\hline 
ensembles & integer vector & {[}1,1,1{]} & Stochastic ensemble sizes\tabularnewline
\hline 
ranges & real vector & {[}10{]} & Time and space ranges\tabularnewline
\hline 
points & integer vector & {[}51{]} & Output lattice points in {[}t,x,y,z,..{]}\tabularnewline
\hline 
steps & integer & $[1]$ & Intermediate steps per time point\tabularnewline
\hline 
observe\{n\} & function & a & Observable function for averages\tabularnewline
\hline 
compare\{n\} & function & 0 & Comparison function for averages\tabularnewline
\hline 
binranges\{n\}\{m\} & vector & {[}0{]} & Binning ranges for probabilities\tabularnewline
\hline 
\end{tabular}\caption{Table of most common simulation parameters.}
\end{table}

A more detailed explanation of these parameters is found below, and
a complete table is given in section \ref{sec:xSIM-parameters}. Fields
and noises can be cell arrays if there are multiple scalar or vector
fields required.

\subsection{Graphics parameters}

The generated average data can be graphed using any graphics editors,
or else using the internal xGRAPH function defined for this purpose.
An xSPDE simulation can return many different averages. These are
defined in a cell array with indices in braces. The index is used
to address the output data produced.

For each index, one can define parameters that define the quantity
stored, together with corresponding graphics outputs. Some commonly
used options are:\\
~~
\begin{flushleft}
\begin{table}
\begin{centering}
\begin{tabular}{|c|c|c|c|}
\hline 
Label & Type & Default value & Description\tabularnewline
\hline 
\hline 
olabels\{n\} & string & 'a' & Observable label\tabularnewline
\hline 
transverse\{n\} & integer & $0$ & Transverse slices in time\tabularnewline
\hline 
transforms\{n\} & vector & $0$ & Set to 1 for Fourier transforms in time\tabularnewline
\hline 
scatters\{n\} & integer & $0$ & Set to s for s scatter plots in the observable\tabularnewline
\hline 
\end{tabular}
\par\end{centering}
\caption{Table of most common graphics parameters}
\end{table}
~~
\par\end{flushleft}

\begin{flushleft}
The full definition of the options is given in the user guide in sections
\ref{sec:xSIM-parameters} and \ref{sec:Graphics-parameter-table},
although many will be clear from examples.
\par\end{flushleft}

\section{Fields and observables}

Stochastic variables in an SDE are fields, stored in a real or complex
matrix, $a(f,e)$. Here, $f$ is an internal field index, while $e$
is the ensemble index. 
\begin{description}
\item [{fields}] gives the range of the first internal index. This has
a default value of $fields=1$. For multiple labeled stochastic fields,
fields is a cell array that can specify the dimensions of one or more
real or complex arrays, $a(f,e),b(f,e)..$. 
\item [{ensembles}] allows multiple trajectories to be integrated. This
has up to three components. The first component, ensembles(1), gives
a vector of local trajectories, so $e=1,\ldots$ensembles(1). The
second ensemble value species ensembles calculated in series, the
third specifies ensembles calculated in parallel using multiple cores. 
\item [{noises}] are noise dimensions, similar to fields, and used as $w(i,j),$
where the first noise index has $noises$ components. The default
value is $noises=fields$. Like fields, this can be a cell array of
multiple noise dimensions.
\end{description}
In the example above, we could add the fields, dimensions, ensembles
and noises:

\doublebox{\begin{minipage}[t]{0.75\columnwidth}%
\texttt{p.fields = 1;}

\texttt{p.dimensions = 1;}

\texttt{p.noises = 1;}

\texttt{p.ensembles = 1;}%
\end{minipage}}

As these are all default values, this is superfluous in a simple case.
The full definition of ensembles as a vector is given above. If the
third ensemble value is input, it requires the parallel toolbox in
Matlab. For an SDE, $p.dimensions=1$ is the default value, as there
is only a time dimension. This input is only needed for stochastic
partial differential equations, described in Part (\ref{part:Stochastic-partial-differential}).

A more complex valid input could be:

\doublebox{\begin{minipage}[t]{0.75\columnwidth}%
\texttt{p.fields = \{{[}2,3{]},2\};}

\texttt{p.noises = 4;}

\texttt{p.ensembles = {[}10,10{]};}%
\end{minipage}}

This describes an equation with one $2\times3$ array, one $2$-dimensional
vector and one $4$-dimensional noise, integrated with $100$ trajectories
composed of 10 local trajectories repeated 10 times in series.

\subsection{Initial values, points and ranges}

Initial values are required to define any differential equation, and
in a numerical calculation one must also have a defined lattice. 
\begin{description}
\item [{initial}] The initial value is defined by a function\texttt{ $p.initial$}.
This must return either an initial vector of size fields, or else
a random array of size $fields\times ensembles(1)$. The default function
simply returns zero. If there is more than one field variable, initial
is a cell array, which must be specified.
\item [{inrandoms}] are initial random number dimensions, similar to fields,
and used as $v(i,j),$ where the first random dimension has $randoms$
components. The default value is $randoms=noises$. Specifies the
first argument of the function\texttt{ $p.initial(v,p)$} as a real
Gaussian noise vector $v$ with unit variance. The same noise is used
when error-checking, so that changes are from the step-size, not from
random fluctuations. 
\item [{points}] The number of integration points. The default setting
is currently $51$. 
\item [{steps}] The number of integration steps used for each output time-step.
The default is $1$. 
\item [{ranges}] The total integration range in each dimension, the first
element being the maximum integration time $T$. The default setting
is currently $10$. 
\end{description}

\subsection{Observables$ $}
\begin{description}
\item [{observe}] is a cell array of functions of stochastic fields, each
defining an average. xSPDE expects a (named or anonymous) function
that takes two parameters, namely the field matrix $a$ and the input
structure $p$. The function must return a real or complex array,
where the first index is used for a vector observable. xSPDE then
averages over the last index, to calculate the observable. The default
returns all elements of the first cell, as lines.\\
 To plot the variance, for example: 
\end{description}
\doublebox{\begin{minipage}[t]{0.75\columnwidth}%
\texttt{p.observe\{1\} = @(a,p) (a(1,:)-mean(a,2)).\textasciicircum 2;}%
\end{minipage}} 
\begin{description}
\item [{rawdata}] By setting p.rawdata=1 (see section \ref{sec:xSIM-parameters}),
one can also store every trajectory including both fine and coarse
time-step values, but this is very memory-intensive for large simulations. 
\item [{olabels}] is cell array of the output labels associated with each
average, although one can also define additional functional transformation
of the averages to be graphed and label them. 
\end{description}
Observables are computed as a two-dimensional packed array, then unpacked
for storage, giving an array of dimension $(d1,dspacetime,ensembles(1))$.
Here $d1$ is the local observable dimension, so $d1=1$ for a scalar
observable. The space-time dimension is $dspacetime=1$ for an SDE,
otherwise a vector for a SPDE, and $ensembles(1)$ is the size of
the ensemble of trajectories computed in each processor. Once data
is averaged internally over $ensembles(1)$, further transforms of
the averages are available.

\subsection{Using the dot}

All equations entered in xSPDE utilize the Matlab syntax. This is
designed to handle scientific or mathematical matrix and array-based
formulae. It has features to simplify matrix or array equations which
often require a 'dot' or a 'colon'. 
\begin{itemize}
\item Stochastic variables in xSPDE are matrices or arrays, where the last
index is used to treat parallel stochastic trajectories, for greater
efficiency. This requires use of the 'dot' notation to perform multiplication
inside equations. 
\item To multiply vectors, matrices or arrays element-wise, like $a_{ij}=b_{ij}c_{ij}$,
the notation $a=b.*c$ indicates that all the elements are multiplied.
This is used to speed up calculations in parallel. 
\item An equation in xSPDE can apply to many stochastic trajectories in
parallel. Using the dot shortens the equation, and it also means that
a fast parallel arithmetic will be used. The same principle holds
for larger arrays with spatial lattices, treated in in section \ref{sec:Simulating-an-SPDE}. 
\item Broadcasting occurs if one or more dimensions has a unit size. For
example, arrays of size (1,100) and (6,1) can be added or multiplied
to give a (6,100) matrix. 
\item A formula may require addressing the first index - which is the field
component - and treating all the other elements in parallel. To do
this in a compact way, one may use the notation $a(n,:)$, which indicates
that all the subsequent index elements are being addressed as well. 
\item This will ``flatten'' a spatial array into a matrix, in which case
it is better to include space indices.
\end{itemize}
In summary, whenever a formula combines multiplication operations
over spatial lattices or ensembles, \textbf{USE THE DOT}.

\section{Random fields and noises}

\subsection{General structure }

These are specified as cells of scalar, vector or array dimensions,
like the field cells. If there is only a single cell, then just the
dimensions are input. The dimensions should be such that the corresponding
arrays can be added. 

\subsection{Noise fields}

During propagation in time, noises are Gaussian noise fields delta-correlated
in space-time. They are calculated in an analogous way, except with
an additional factor of $1/\sqrt{\Delta t}$ because they are delta
correlated in time. They have a variance of $\sigma^{2}=1/(\Delta t)$. 

During error-checking, which is the default option, the time-step
is halved. The same noise trajectory is used, except that for coarse
steps the two fine step noises are summed, which has the effect of
doubling the variance, as required.

When $unoises\sim=0$, an initial field of uniform random numbers
is generated for jump processes, on the interval $[0,..1/\Delta t]$.
During error-checking the same noise trajectory is used, except that
for coarse steps the minimum of the two fine step noises is used,
so that if there is a jump in at least one of the fine steps, it will
occur in the coarse step.

All noises can all be specified as cells of multiple noise fields,
with arbitrary scalar, vector or array dimensions. They are passed
to the deriv function in the order of {[}noises,unoises{]}.

The maximum number of noise or unoise cells equals the number of field
cells. If more are required, extra field cells should be specified
with zero dimensions. 

\subsection{Initial randoms}

When $inrandoms\sim=0$, initial Gaussian random numbers $\mathbf{v}$
are generated, with unit variance. When $urandoms\sim=0$, an initial
field of uniform random numbers is generated for jump processes. These
can all be specified as cells of multiple random fields, with arbitrary
scalar, vector or array dimensions. 

They are passed to the initial function in the order of {[}randoms,urandoms{]}. 

\subsection{Example}

As an example, consider a matrix SDE where one Gaussian noise is specified
to act on all rows, the other acts on all columns, and initial uniform
randoms act independently on all elements.

\doublebox{\begin{minipage}[t]{0.75\columnwidth}%
\texttt{p.urandoms = {[}2,2{]};}

\texttt{p.fields = {[}2,2{]};}

\texttt{p.noises = \{{[}2,1{]},{[}1,2{]}\};}

\texttt{p.deriv = @(a,u,v,p) -a+u+v;}

\texttt{p.initial= @(w,p) w;}

\texttt{e = xspde(p);}%
\end{minipage}} 

\section{Advanced random walk}

We now return to the random walk, but with some more advanced features:
\begin{equation}
\dot{a}=w(t)\,,
\end{equation}

This is integrated numerically and graphed with $N=points(1)$ points.
The first point stored is the initial value, so there are $N-1$ integration
steps, of length $dt=ranges(1)/(N-1)$. Numerical graphs have discrete
steps, and more detail is obtained if more time steps are used. The
default value is $N=51$, which is predefined in the $xpreferences$
file. This is adjustable by the user. It can also be changed for a
simulation, by inputting a new value of $points$.

\subsection{Simple xSPDE example}

Unless you type clear first, any changes to the input structure are
additive; so in the exercises you should get the combination of all
the previous structure inputs as well as your new input. 
\begin{itemize}
\item \textbf{Run the complete xSPDE script of Example 1 in Matlab.} 
\end{itemize}
It is simple to cut and paste from an electronic file to the command
window. Be careful; pasting can cause subtle changes that may require
correction. Some generated characters may be invalid input characters,
and these will need retyping if this occurs.

You should get the output in Fig (\ref{fig:The-simplest-case: Wiener}). 
\begin{itemize}
\item \textbf{What do you see if you average over $10000$ trajectories
?} 
\end{itemize}
\begin{center}
\doublebox{\begin{minipage}[t]{0.75\columnwidth}%
\texttt{p.ensembles = 10000;}

\texttt{xspde(p);}%
\end{minipage}} 
\par\end{center}
\begin{itemize}
\item \textbf{What do you see if you plot the mean square distance? Note
that variances should increase linearly with $t$.} 
\end{itemize}
\begin{center}
\doublebox{\begin{minipage}[t]{0.75\columnwidth}%
\texttt{p.observe = @(a,p) a.\textasciicircum 2;}

\texttt{p.olabels = '\textless a\textasciicircum 2\textgreater
';}

\texttt{xspde(p);}%
\end{minipage}} 
\par\end{center}
\begin{itemize}
\item \textbf{What if you add a force that takes the particle back to the
origin?} 
\begin{equation}
\dot{a}=-a+w(t)\,,\label{eq: damped_path_with_noise}
\end{equation}
\end{itemize}
\begin{center}
\doublebox{\begin{minipage}[t]{0.75\columnwidth}%
\texttt{p.deriv = @(a,w,p) -a+w;}

\texttt{xspde(p);}%
\end{minipage}} 
\par\end{center}

The corresponding Fokker-Planck equation from Eq \eqref{eq:FPE} is:
\begin{equation}
\frac{\partial P\left(a\right)}{\partial t}=\left[\frac{\partial}{\partial a}+\frac{1}{2}\frac{\partial^{2}}{\partial a^{2}}\right]P\left(a\right).
\end{equation}
It is easy to verify that inserting this dynamical equation into Eq
\eqref{eq:moment_equn} gives the result: 
\begin{equation}
\frac{\partial}{\partial t}\left\langle a^{2}\right\rangle =1-2\left\langle a^{2}\right\rangle 
\end{equation}

\begin{itemize}
\item Solve for $\left\langle a^{2}\left(t\right)\right\rangle $ and use
xSPDE to compare the numerical and analytic solutions. The current
time is accessible as the parameter $p.t$. Can you explain the graph
differences? 
\end{itemize}

\subsection{Discrete Fourier transforms}

While exact in this analytic case, the definition above is impractical
for numerical calculations. In taking measurements and doing simulations,
one has a discrete set of data-points. Assuming the samples are at
fixed intervals, the best one can do in practical cases is a discrete
Fourier transform, with samples $\bar{a}(\bar{t}_{j})$ that are defined
as integrals over each small interval $dt$:

Let $\bar{a}(\bar{t}_{j})$ be the average over a small time interval:
\begin{equation}
\bar{a}(\bar{t}_{j})=\int_{t_{j}}^{t_{j}+dt}a(t)dt\,,\label{eq:field-average}
\end{equation}
then to a good approximation as $dt\rightarrow0$, provided $\omega_{n}$
is not too large, 
\begin{align}
\tilde{a}(\omega_{n}) & =\frac{\Delta t}{\sqrt{2\pi}}\sum_{j=1}^{N}e^{i\omega_{n}\bar{t}_{j}}\bar{a}(\bar{t}_{j})\,\nonumber \\
\bar{a}(\bar{t}_{j}) & =\frac{\Delta\omega}{\sqrt{2\pi}}\sum_{n=1}^{N}e^{-i\omega_{n}\bar{t}_{j}}\tilde{a}(\omega_{n})\,.
\end{align}

These also form an invertible pair provided that $\Delta t\Delta\omega=2\pi/N$.
As well as being more practical, this is very efficient due to the
fast Cooley-Tukey (FFT) algorithm \cite{Cooley1965Algorithm}, allowing
computation on time-scales of $O\left(N\ln N\right)$ rather than
$O\left(N^{2}\right)$ as one might expect.

When taking Fourier transforms in the time-domain, xSPDE does a time-averaging
of all fields over the current time-step, using the available coarse
and fine time-samples. This is done by averaging the field before
and after the stochastic time-step. The methods used for this are
described in greater detail in Section (\ref{sec:Time-domain-spectra}).

\section{Stochastic projections}

It is sometimes necessary to constrain an equation to a sub-manifold
\cite{Joseph2023midpoint}, with an equation of form: 
\begin{equation}
\mathbf{f}\left(\mathbf{a}\right)=0,
\end{equation}
where $\mathbf{f}\left(\mathbf{a}\right)$ is a scalar or vector function
that defines the relevant manifold in Euclidean space. The projected
SDE then has the form of a Stratonovich SDE, where: 
\begin{equation}
\frac{\partial\mathbf{a}}{\partial t}=\mathcal{P}_{\mathbf{a}}^{\parallel}\left[\mathbf{A}\left[\mathbf{a}\right]+\underline{\mathbf{B}}\left[\mathbf{a}\right]\cdot\mathbf{w}(t)\right]\,,
\end{equation}
where $\mathcal{P}_{\mathbf{a}}^{\parallel}$ is a tangential projection
operator at location $\mathbf{a}$ on the sub-manifold, and as usual,
$\mathbf{A}$ is a vector, $\underline{\mathbf{B}}$ a matrix and
$\mathbf{w}$ is a real Gaussian noise vector, delta-correlated in
time. 

Similarly, the general stochastic partial differential equation can
be written in projected form as

\begin{equation}
\frac{\partial\mathbf{a}}{\partial t}=\mathcal{P}_{\mathbf{a}}^{\parallel}\left[\mathbf{A}\left[\mathbf{a}\right]+\underline{\mathbf{B}}\left[\mathbf{a}\right]\cdot\mathbf{w}(t,\mathbf{x})+\underline{\mathbf{L}}\left[\mathbf{\nabla},\mathbf{a}\right]\,\right].
\end{equation}

\newpage{}

\section{Multivariate probabilities}

One can utilize xSPDE to graph probability densities of real observables
instead of averages, if $p.ensembles$ is large. This is achieved
by inputting the observable number and binning range:

\begin{equation}
p.binranges\{n\}=\{oa:ostep:ob\};
\end{equation}

If present, this returns probability density of the $n$-th observable
$o\{n\}$, through binning into ranges of width $ostep$ around the
centers of each bin, starting at $\text{\emph{oa}}$, and ending at
$ob$. The simulation returns a result of $1/ostep$ in the $j-th$
bin if the trajectory is inside the bin, so that $o(j)-ostep/2<o<o(j)+ostep/2$,
and zero otherwise. This gives a probability density on output, plotted
against time. Note that on graphing, an extra dimension is added for
the variable $o$. The probability density at \textit{ntimes} equally
spaced simulation times can be plotted with \textit{p.transverse\{n\}=ntimes}.

The probability can be plotted for any observe function of the stochastic
variable. For these plots, the ordering of axes is: {[}time, space,
observable{]}, where the appropriate axis label can be added to the
graph using the glabels\{n\}\{k\} graphics input, where k identifies
the axis that is being labeled.

The probability density is multivariate for vector observables. This
is possible because the binning ranges are stored in a cell array,
which may contain several bin vectors. If the observable $o\{n\}$
is two-dimensional, then one can input:

\begin{equation}
p.binranges\{n\}=\{oa(1):ostep(1):ob(1),oa(2):ostep(2):ob(2)\};
\end{equation}

On graphing, two extra axes are added for the variable $o$ in this
case. The graphics program xGRAPH will attempt to graph them, but
it is limited by graphical visualization constraints. In general,
an arbitrary observable dimension is possible, but this is also limited
by the sampling and memory, since the number of samples per bin will
decrease rapidly with dimensionality.

The graphics program extracts slices and windows of probabilities
if required. To plot the probabilities of two observables in different
graphs, one for a range of $-5:5$ and the other for 0:25, add the
following inputs before the xsim or xspde command: 
\begin{center}
\doublebox{\begin{minipage}[t]{0.75\columnwidth}%
p.\texttt{binranges\{1\} = \{-5:0.25:5\};}

p.\texttt{binranges\{2\} = \{0:0.5:25\};}%
\end{minipage}} 
\par\end{center}

In the case of a two-dimensional probability density, plotted against
time, there are a total of four graphics dimensions. That is, one
for the probability, one for time, and two for the independent variables
at each time. One can also plot how the probability density changes
in space for the case of a stochastic partial differential equation,
as described in section \ref{sec:Simulating-an-SPDE}.

\subsection{Marginals and labels}

We now treat special cases that often arise.
\begin{itemize}
\item Suppose one has computed two outputs, but only the second one requires
binning? Putting this another way, if the stochastic variable is $[x,y]$,
you can plot $P(x,y)$ easily enough, but what if you want a marginal,
$P(x)$ or $P(y)$? One way is to change the observe function to only
return $x$ or $y$.
\item Sometimes the observe function is written, and one doesn't want to
change it. Then for a marginal, all one has to do is to change the
binning statement, and replace the unwanted binning range by an empty
vector, $[]$. This variable will be omitted, and again there will
be a marginal probability.
\item Probability axis labels can be added using the graphics input ``glabels'',
if you would like axis labels. These correspond to the plotted axes,
so that the label that is integrated over is omitted. In other words,
just enter the labels of variables that are plotted, not those that
are ignored.
\item xSPDE can also integrate matrix or tensor stochastic differential
equations. How does one compute their probabilities? The answer to
this is simple. The observe function can only have a vector output,
so the tensor or matrix has to be expanded into a vector by the observe
function.
\item If there are more binranges than variables, the last ranges that are
input to binranges are ignored. 
\item If there are more variables than binranges data, the last variables
calculated will be ignored.
\end{itemize}

\subsection{Probability summary }

In summary, if p.binranges\{n\} is specified, it takes multidimensional
arrays generated by observe\{n\}, and outputs a probability distribution
instead of the $n$-th average. The first data index or ``line''
dimension gives the independent variables. The last dimension is the
sample index, which is averaged over.

If there are $m$ line index values, and an $m$-dimensional set of
bins, an $m$-dimensional joint probability is computed by adding
up the samples in each bin. Since there are usually more than two
indices in total, xSPDE generates independent probabilities for each
extra index value.

All results really depend on observe\{n\}. If it generates data with
a singleton first index, you get a one-dimensional probability. When
there are more than two indices in the observe data, you get a distinct
probability for each value of the other indices. This gives multiple
probabilities versus the extra indices.

In xSPDE, without binning, and three index values, you get a 2D plot,
not a 3D plot, since the first index indexes the lines, the second
the time axis, and the third is averaged. With binning, binranges
will not ignore the time dimension. It simply generates a new probability
plot for each time.

When there are extra space dimensions as well, there are probabilities
at every space-time point. These are not joint probabilities, since
the curse of dimensionality would make them too high-dimensional and
sparse. One can also reduce the amount of data generated with ``axes''
(see later).

\section{Auxiliary fields and noises}

In some problems, it is useful to access the noise terms, or functions
of the noises and their correlations with the fields at the same time.
This is handled in xSPDE with auxiliary fields or auxfields. These
are fields that are functions of noise terms and the integrated fields.
The number of these is defined in the input structures using the parameter
p.auxfields, which is arbitrary.

Auxiliary fields are calculated using a function p.define, which is
similar to p.deriv, except that it returns the current value of the
auxiliary field, not the derivative. These fields are defined as the
average over the previous step in time of the auxiliary function,
including the noise term. This is essential in calculating spectra,
in order to eliminate systematic errors in Fourier transforms.

More details on this are given in Section (\ref{sec:Time-domain-spectra}).
To access the auxiliary fields, one can compute any observable average
using a p.observe function as usual, or else store the raw trajectories
including auxiliary fields by setting p.rawdata=1. In either case,
the auxiliary fields are appended to the integrated fields by adding
extra cells. 

\subsection{Outputting the noise}

As a simple example, suppose one wishes to calculate the noise terms
and compare them with the field trajectories in a simple Wiener process.
Since there is now an extra cell for the auxiliary field in the define
function, it is passed as an additional field argument to the observe
function. The following code can be used:\\

\doublebox{\begin{minipage}[t]{0.75\columnwidth}%
\texttt{clear}

\texttt{p.auxfields = 1;}

\texttt{p.deriv = @(a,w,p) w;}

\texttt{p.define = @(a,w,p) w;}

\texttt{p.observe = @(a,x,p) {[}a;x{]};}

\texttt{p.olabels = \{'a, w'\};}

\texttt{xspde(p);}%
\end{minipage}} \\

The observe function calculates both rows of the output array, including
the auxiliary field which is defined as the noise term and plotted
as a dashed line. There is no ensemble averaging, and hence no ensemble
error-bars in the example. This is because because no ensembles were
specified in the input parameters. Similarly, there are no time-step
error-bars for this observable, because the fine and coarse noises
are equal to each other after time averaging.

The result that is plotted is therefore the coarse noise, whose correlation
time equals the time step. This is plotted below in Fig ( \ref{fig:The-simplest-case: Wiener-2}),
which plots the same Wiener process as before, except adding the driving
noise term as well. The standard deviation of the noise in a single
step here is $\sqrt{1/dt}$, where $1/dt=50/10=5$ for the default
range of $10$ and default time points of $51.$ Note that noise terms
do not converge at small time-steps for delta-correlated noise, even
when the integrated stochastic process does converge. This is why
it is necessary to choose to plot one or the other, or else to time-average
to obtain a converged result.

If multiple steps are used, only the noise during the last step prior
to the time-point is plotted. 
\begin{center}
\begin{figure}
\centering{}\includegraphics[width=0.75\textwidth]{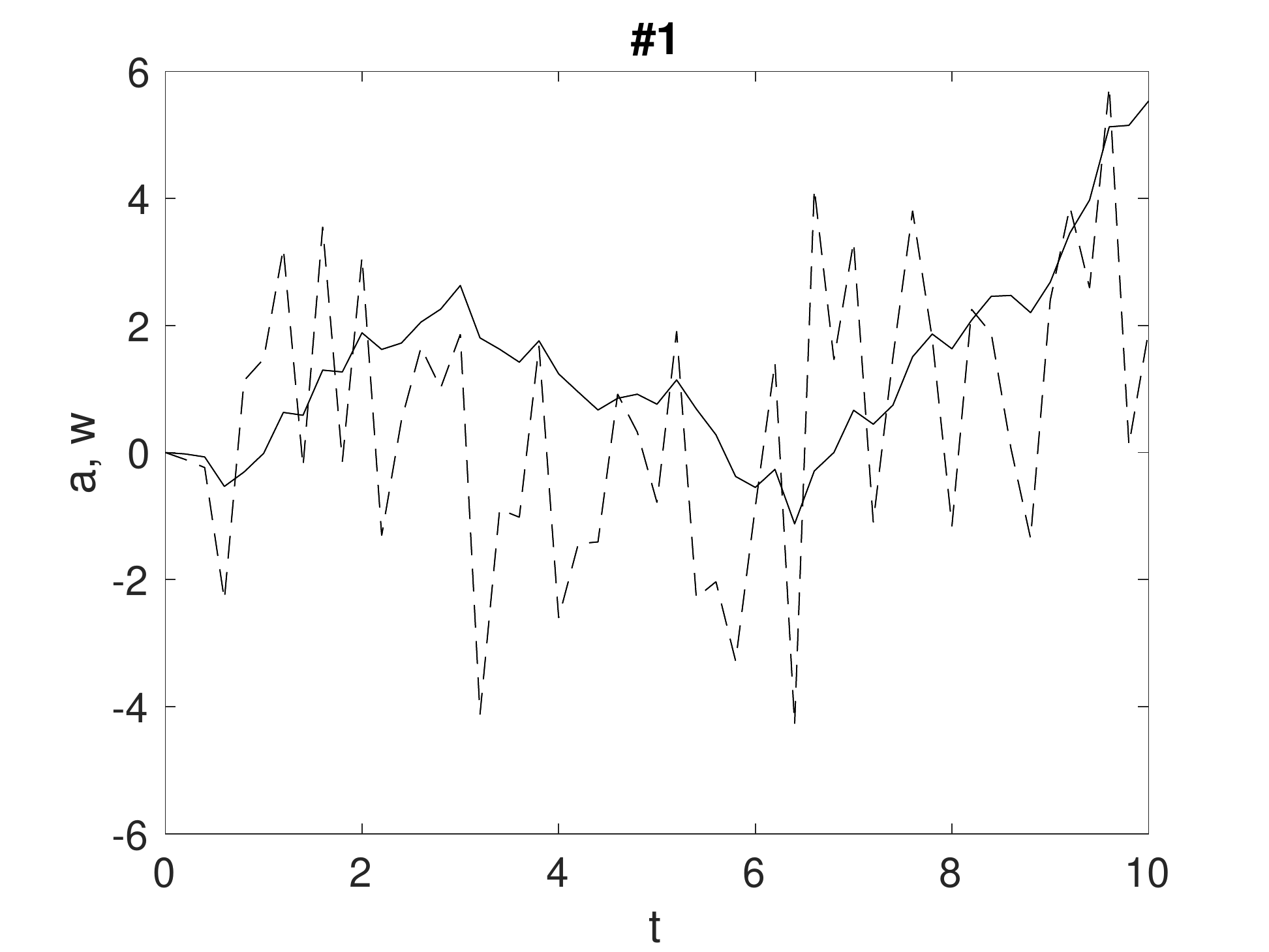}

\caption{\label{fig:The-simplest-case: Wiener-2}\emph{A single trajectory
of a random walk, with the noise terms $w$ graphed using dashed lines,
and the integrated variable $a$ plotted as the solid line.}}
\vspace{10pt}
\end{figure}
\par\end{center}

\section{Time-domain spectra\label{sec:Time-domain-spectra}}

To get an output from a temporally Fourier transformed field, set
$transforms\{n\}=1$ for the observable ($n$) you need to calculate
in transform space. This parameter is a cell array. It can have a
different value for every observable and for every dimension in space-time,
if you have space dimensions as well.

To obtain spectra from Eq \eqref{eq:field-average} with greater accuracy,
all fields are must be averaged internally. The code will use trapezoidal
integration in time over the integration interval, to give the average
midpoint value. This employs the same interval for fine and coarse
integration, to allow comparisons for error-checking. After this,
the resulting step-averaged fields are then Fourier transformed.

In the simplest case of just one internal step, with no error-checking,
this means that the field used to calculate a spectrum is: 
\begin{equation}
\begin{split}\bar{a}_{j}=\left({a}_{j}+{a}_{j+1}\right)/2,\end{split}
\end{equation}
which corresponds to the time in the spectral Fourier transform of:
\begin{equation}
\begin{split}\bar{t}_{j}=\left({t}_{j}+{t}_{j+1}\right)/2.\end{split}
\end{equation}

Note that if any temporal Fourier transform is specified, all the
field variables are time-averaged over a step. This is not strictly
necessary, but it means that there is a reduced code complexity for
cases where there is a Fourier transform for some but not all variables.
As described above, the auxiliary variables are always time-averaged
to allow error-checking, so there is no change for these.

\subsection{Error-checking}

For an error-checking calculation with two internal steps, there are
three successive valuations: $a_{j}$, $a_{j+1/2}$, $a_{j+1}$. In
this case, for spectral calculations one averages according to: 
\begin{equation}
\begin{split}\bar{a}_{j}=\left(a_{j}+2a_{j+1/2}+a_{j+1}\right)/4.\end{split}
\end{equation}

In addition, one must define the noise terms, both for error-checking
and for output, since spectral calculations in quantum input-output
theory include noise terms as well as fields. The noise term used
to calculate a spectrum involving $\bar{a}_{j}$ is $w_{j}$. A coarse
noise term is set equal to the average of two successive fine noise
terms: 
\begin{equation}
\begin{split}\bar{w}_{1}=\frac{1}{2}\left(w_{1}+w_{1/2}\right).\end{split}
\end{equation}
The time integral is carried out numerically as a sum which has $N=points(1)$
time points of interval $dt$. In xSPDE, $dt=T/(N-1)$, where $T=ranges(1)$.
The effective integration time for the Fourier transform time integrals
is 
\begin{equation}
T_{eff}=Ndt=2\pi/d\omega
\end{equation}

When there are larger numbers of steps from using the internal steps
parameter, there are more points to Fourier transform. These additional
frequencies are computed while carrying out the Fourier transform,
but only $N$ low frequency points are saved. The unused high frequency
results are not stored or plotted, to conserve memory.

\section{Scanned parameter plots}

Since xSIM is a function that can be called, plots of results against
simulation parameters are possible. This requires repeated calls to
xSIM with different parameter values, together with data storage in
an xGRAPH compatible form, and a call to xGRAPH. If different random
seeds are required, the seed needs to be reset in each call. The relevant
axes points plotted, labels and the values of scanned parameters also
need to be input.

The simulation function xSIM uses the last data array index, $c$,
to store the data values and up to two corresponding errors. This
takes up three index values. A value of $c=4$ is used to store comparison
data, and its errors if there are any in $c=5,6$. This can be used
for exact results, approximations, or experimental data.

\begin{figure}
\centering{}\includegraphics[width=0.75\textwidth]{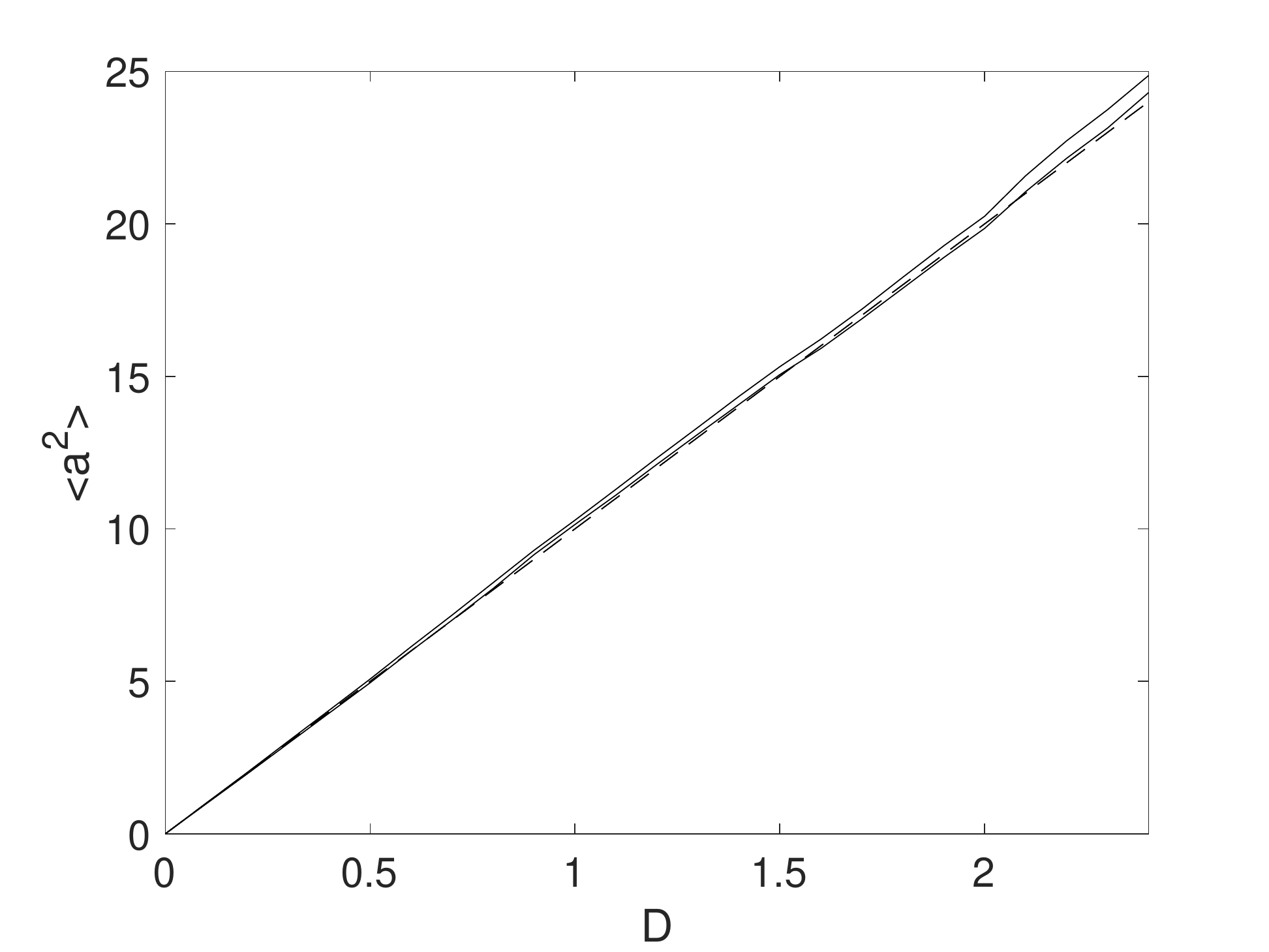}

\caption{\label{fig:Scanned-diffusion-1}\emph{Scanned parameter output with
a variable diffusion, for the case of a pure Wiener process, $\dot{a}=Bw(t)$.
Exact value is the dashed line.}}
\vspace{10pt}
\end{figure}

\subsection{Example: Scanned diffusion}

As an example, consider the simplest possible stochastic equation,
with a scanned diffusion: 
\begin{equation}
\dot{a}=Bw(t)\,.
\end{equation}

The equation is integrated over the interval $t=0:10$, with $a=0$
initially, using $10^{4}$ trajectories to give an expected error
of around $\pm1\%$. The variance of $a$ at $t=10$ is plotted as
a function of $D=B^{2}$, then compared to an exact value. The result
is in Fig (\ref{fig:Scanned-diffusion-1}). The corresponding code
is given as well. 
\begin{center}
\doublebox{\begin{minipage}[t]{0.75\columnwidth}%
\texttt{function e = WienerScan()}

\texttt{p.name = 'Wiener process';}

\texttt{p.ensembles = {[}1000,10{]};}

\texttt{p.points = 12;}

\texttt{p.deriv = @(a,z,p) z{*}p.B;}

\texttt{p.observe = @(a,p) a.\textasciicircum 2;}

\texttt{p.olabels = \{'\textless a\textasciicircum 2\textgreater
'\};}

\texttt{p.glabels\{1\} = \{'D'\};}

\texttt{scanpoints = 25;}

\texttt{data\{1\}\{1\} = zeros(1,scanpoints,4);}

\texttt{for j = 1:scanpoints}

\texttt{\quad{}p.seed = j;}

\texttt{\quad{}p.B = sqrt((j-1){*}0.1);}

\texttt{\quad{}{[}e,data1,input,\textasciitilde{]} = xsim(p);}

\texttt{\quad{}data\{1\}\{1\}(1,j,1:3) = data1\{1\}\{1\}(1,p.points,:);}

\texttt{\quad{}xk\{1\}\{1\}(j) = p.B\textasciicircum 2;}

\texttt{\quad{}D(j) = p.B\textasciicircum 2;}

\texttt{end}

\texttt{data\{1\}\{1\}(1,:,4) = input.ranges(1){*}D(:);}

\texttt{input.xk = xk;}

\texttt{input.axes\{1\}\{1\} = 1:scanpoints;}

\texttt{xgraph(data,input);}

\texttt{end}%
\end{minipage}} 
\par\end{center}

Here $p.deriv$ defines the time derivative function $\dot{a}$, with
$w$ being the delta-correlated Gaussian noise that is generated internally.

\section{Hints}
\begin{itemize}
\item When first using xSPDE, it is a good idea to run the batch test script,
Batchtest. 
\item If the Matlab parallel toolbox is not available, don't use the third
ensemble setting. 
\item To create a project file, it is easiest to start with an existing
example function.
\item Graphics parameters can be included in either xSPDE or xGRAPH inputs. 
\item Comparison functions can be included to compare with analytic results. 
\item Chapter \ref{chap:API-reference} lists the input parameters. 
\end{itemize}
\newpage{}

\chapter{SDE theory\label{chap:SDE Theory}}

\textbf{This chapter describes the basics of stochastic differential
equation (SDE) theory, in order to explain the background to the numerical
methods.}

\section{General form}

A stochastic differential equation (SDE) is an equation with random
noise terms. These were introduced by Langevin to treat small particles
in fluids \cite{Langevin1908theorie}, and extended by Wiener, Ito
and Stratonovich \cite{wiener1930generalized,ito1996diffusion,stratonovich1960theory}.
The theory and its applications to biology, chemistry, engineering,
economics, physics, meteorology and other disciplines are treated
in many texts \cite{Karatzas1991brownian,VanKampen2007stochastic,Gardiner2009Stochastic,klebaner2012introduction,Drummond2014Quantum,Arnold1992stochastic}.

An ordinary stochastic differential equation in one time dimension
is,

\begin{equation}
\frac{\partial\mathbf{a}}{\partial t}=\mathbf{A}\left(\mathbf{a},t\right)+\underline{\mathbf{B}}\left(\mathbf{a},t\right)\cdot\mathbf{\bm{\xi}}(t)\,.\label{eq:SDE}
\end{equation}
Here $\mathbf{a}$ is a real or complex vector, $\mathbf{A}$ is a
vector function, $\underline{\mathbf{B}}$ a matrix function and $\mathbf{w}$
is usually a delta-correlated real Gaussian noise vector such that:
\begin{eqnarray}
\left\langle \xi_{i}\left(t\right)\xi_{j}\left(t'\right)\right\rangle  & = & \delta\left(t-t'\right)\delta_{ij}.\label{eq:noise-correlations}
\end{eqnarray}

One can also have non-Gaussian noise or noise that is not delta-correlated.
Although these are somewhat less commonly treated, these alternatives
are often found in real applications.

These equations can alternatively be written in the equivalent form
that:

\begin{equation}
d\mathbf{a}=\mathbf{A}\left(\mathbf{a},t\right)dt+\underline{\mathbf{B}}\left(\mathbf{a},t\right)\cdot d\bm{w}\,.\label{eq:SDE-4}
\end{equation}
This leads to the relation that:
\begin{equation}
\frac{d\bm{w}}{dt}=\mathbf{\bm{\xi}}(t).
\end{equation}

In a finite time interval $dt$, one can approximately replace the
delta-correlated noise by a fixed noise with variance $\left\langle \xi^{2}\right\rangle =1/dt$,
but if the alternative notation is used, then $\left\langle dw^{2}\right\rangle =dt$.
The first notation is the one mostly used in xSPDE.

\subsection{Observables}

In all cases, there are multiple independent trajectories, and one
is interested in probabilistic averages, where the unweighted average
of an observable $\mathbf{O}\left(\mathbf{a}\right)$, for $N_{s}$
trajectories $\mathbf{a}^{\left(n\right)}$ is: 
\begin{eqnarray}
\left\langle \mathbf{O}\right\rangle _{N_{s}} & = & \frac{1}{N_{s}}\sum_{n}\mathbf{O}\left(\mathbf{a}^{\left(n\right)}\right).\label{eq:averages-1}
\end{eqnarray}
In other types of stochastic equation \cite{graham1977path,Drummond2017forward},
there is a weight $\Omega\left(t\right)$ for each trajectory. This
has an additional equation of motion, where:

\begin{equation}
\frac{\partial\Omega}{\partial t}=A_{\Omega}\left(\mathbf{a},\Omega,t\right)+\underline{B}_{\Omega}\left(\mathbf{a},\Omega,t\right)\cdot\mathbf{\bm{\xi}}(t)\,.\label{eq:SDE-2}
\end{equation}

The results for all mean values are then weighted by the term $\exp\left(\Omega\left(t\right)\right)$,
so that: 
\begin{equation}
\left\langle \mathbf{O}\right\rangle _{\Omega}=\frac{\sum_{n}\mathbf{O}\left(\mathbf{a}^{\left(n\right)}\right)\exp\left(\Omega^{\left(n\right)}\left(t\right)\right)}{\sum_{n}\exp\left(\Omega^{\left(n\right)}\left(t\right)\right)}.\label{eq:Weighted-averages-1}
\end{equation}
This expression reduces to the usual average if the weights are zero,
i.e, $\Omega=0$. Apart from the way that averages are treated, the
weight can simply be regarded as an additional term in the stochastic
differential equations. This simply means that one now has an equation
with an extra random field, so that $\mathbf{a}\rightarrow\left[\mathbf{a},\Omega\right]$,
together with a modified expression for the averages. This, in fact,
is how these equations are solved.

For reasons of efficiency, it is best to use ``breeding'' algorithms
to treat these numerically. This replicates highly weighted trajectories
with $\Omega^{\left(n\right)}\left(t\right)\gg0$ and removes trajectories
with $\Omega^{\left(n\right)}\ll0$, that have negligible weight.
The numerical method is described in section \ref{sec:Algorithms}.
The remainder of this chapter will focus on the most commonly treated
case of unweighted, Gaussian, delta-correlated noise.

\section{Stochastic calculus}

In the case of delta-correlated noise, the trajectories are not differentiable.
As a result, there are two main variants of stochastic calculus used
to define the derivatives, called Ito or Stratonovich \cite{Gardiner2009Stochastic,Arnold1992stochastic},
and xSPDE can be used for either type. The default algorithms are
designed for Stratonovich cases, since this is just ordinary calculus.
Ito calculus can be treated also, either using the directly applicable
Euler method, or else by appropriate transformations to a Stratonovich
form. One can also have a time-reversed or implicit Ito calculus \cite{Drummond1991Computer},
which is directly solved using an implicit Ito-Euler method.

A single step in time of duration $\Delta t$ uses finite noises $\mathbf{\bm{\xi}}$
which are defined to be delta-correlated in the small time-step limit,
so that $\left\langle \xi_{i}\xi_{j}\right\rangle =\delta_{ij}/\Delta t.$

\subsection{Types of stochastic calculus}

The limits as $\Delta t\rightarrow0$ are taken differently for the
different types of stochastic calculus. Let $\mathbf{a}_{0}=\mathbf{a}\left(t_{0}\right)$,
$t_{1}=t_{0}+\Delta t$, $\mathbf{a}_{1}=\mathbf{a}\left(t_{1}\right)$,
$\bar{\mathbf{a}}=\left(\mathbf{a}_{1}+\mathbf{a}_{0}\right)/2$,
and $\bar{t}=t+\Delta t/2$, then the next step in time is: 
\begin{itemize}
\item Ito calculus - uses \textbf{initial-time} derivative evaluations 
\end{itemize}
\begin{equation}
\mathbf{a}_{1}=\mathbf{a}_{0}+\left[\mathbf{A}^{(I)}\left(\mathbf{a}_{0},t_{0}\right)+\underline{\mathbf{B}}\left(\mathbf{a}_{0},t_{0}\right)\cdot\mathbf{\mathbf{\bm{\xi}}}\right]\,\Delta t\,\,.\label{eq:SDE-1}
\end{equation}

\begin{itemize}
\item Stratonovich calculus - uses \textbf{midpoint} derivative evaluations 
\end{itemize}
\begin{equation}
\mathbf{a}_{1}=\mathbf{a}_{0}+\left[\mathbf{A}\left(\bar{\mathbf{a}},\bar{t}\right)+\underline{\mathbf{B}}\left(\bar{\mathbf{a}},\bar{t}\right)\cdot\mathbf{\mathbf{\bm{\xi}}}\right]\,\Delta t\,.\label{eq:SDE-1-1}
\end{equation}

\begin{itemize}
\item Backward Ito calculus - uses \textbf{final-time} derivative evaluations 
\end{itemize}
\begin{equation}
\mathbf{a}_{1}=\mathbf{a}_{0}+\left[\mathbf{A}^{(I+)}\left(\mathbf{a}_{1},t_{1}\right)+\underline{\mathbf{B}}\left(\mathbf{a}_{1},t_{1}\right)\cdot\mathbf{\mathbf{\bm{\xi}}}\right]\,\Delta t\,\,.\label{eq:SDE-1-2}
\end{equation}
The drift term $\mathbf{A}$ is changed in Ito or implicit Ito calculus,
if the noise coefficient $B$ depends on the stochastic variable.
Defining $\partial_{n}\equiv\partial/\partial a_{n}$ and using the
Einstein convention of summing over repeated indices, one has the
following relationships: 
\begin{align}
A_{i}^{(I)} & =A_{i}+\frac{1}{2}B_{jk}\partial_{j}B_{ik},\nonumber \\
A_{i}^{(I+)} & =A_{i}-\frac{1}{2}B_{jk}\partial_{j}B_{ik}.\label{eq:ItovsStratonovich}
\end{align}

Methods used for solving stochastic equations depend on the type of
stochastic calculus. The default methods used in xSPDE are for Stratonovich
calculus. Other methods are available as well, for both forward and
backward Ito calculus. Alternatively, one can use the conversion formulae
to change the equation.

\section{Example: random walk}

The first example of an SDE is the simplest possible stochastic equation
or Wiener process: 
\begin{equation}
\dot{a}=w(t)\,.\label{eq:Wiener_process}
\end{equation}

This has the solution that 
\begin{equation}
a\left(t\right)=a\left(0\right)+\int_{0}^{t}w\left(\tau\right)d\tau,
\end{equation}
which means that the initial mean value does not change in time: 
\begin{equation}
\left\langle a\left(t\right)\right\rangle =\left\langle a\left(0\right)\right\rangle .\label{eq:Wiener_mean}
\end{equation}

\subsection{Variance solution}

The noise correlation is non-vanishing from Eq \eqref{eq:noise-correlations},
so the variance must increase with time: 
\begin{align}
\left\langle a^{2}\left(t\right)\right\rangle  & =\left\langle a^{2}\left(0\right)\right\rangle +\int_{0}^{t}\int_{0}^{t}\left\langle w\left(\tau\right)w\left(\tau'\right)\right\rangle d\tau d\tau'\nonumber \\
 & =\left\langle a^{2}\left(0\right)\right\rangle +\int_{0}^{t}\int_{0}^{t}\delta\left(\tau-\tau'\right)d\tau d\tau'.
\end{align}

Integrating the delta function gives unity, which means that the second
moment and the variance both increase linearly with time:

\begin{align}
\left\langle a^{2}\left(t\right)\right\rangle  & =\left\langle a^{2}\left(0\right)\right\rangle +\int_{0}^{t}d\tau\nonumber \\
 & =\left\langle a^{2}\left(0\right)\right\rangle +t.\label{eq:Wiener_mean_square}
\end{align}

The probability follows an elementary diffusion equation: 
\begin{equation}
\frac{\partial P}{\partial t}=\frac{1}{2}\frac{\partial^{2}P}{\partial a^{2}}\,,\label{eq:FPE-1}
\end{equation}
which is an example of Eq \eqref{eq:FPE}. From this equation and
using Eq \eqref{eq:moment_equn}, the first two corresponding moment
equations in this case are 
\begin{align}
\frac{\partial}{\partial t}\left\langle a\right\rangle = & \left\langle \frac{1}{2}\frac{\partial^{2}}{\partial a^{2}}a\,\right\rangle =0\nonumber \\
\frac{\partial}{\partial t}\left\langle a^{2}\right\rangle = & \left\langle \frac{1}{2}\frac{\partial^{2}}{\partial a^{2}}a^{2}\,\right\rangle =1.
\end{align}

These differential equations are satisfied by the solutions obtained
directly from the stochastic equations, namely Eq \eqref{eq:Wiener_mean}
and Eq \eqref{eq:Wiener_mean_square}.

\section{Interaction picture}

The interaction picture allows one to eliminate linear terms in the
time derivatives. It is especially useful for stochastic partial differential
equations, but it is applicable to stochastic equations as well. Suppose
there are linear terms $\underline{\mathbf{L}}$, so that $\mathbf{A}\left(\mathbf{a},t\right)=\mathbf{A}_{1}\left(\mathbf{a},t\right)+\underline{\mathbf{L}}\cdot\mathbf{a}\,$,
where $\underline{\mathbf{L}}$ is a constant matrix. The interaction
picture defines local variables $\tilde{\mathbf{a}}$ for the fields
$\mathbf{a}$.

It is convenient to introduce an abbreviated notation as: 
\begin{equation}
\begin{split}\begin{aligned}D\left(\mathbf{a}\right)=\mathbf{A}_{1}\left(\mathbf{a},t\right)+\underline{\mathbf{B}}\left(\mathbf{a},t\right)\cdot\mathbf{w}(t)\end{aligned}
\end{split}
,\label{eq:deriv_without_linear_term-1}
\end{equation}
so that one can write the differential equation as: 
\begin{equation}
\begin{split}\frac{\partial\mathbf{a}}{\partial t}=D\left(\mathbf{a}\right)+\underline{\mathbf{L}}\cdot\mathbf{a}.\end{split}
\end{equation}

\subsection{Linear propagator}

Next, we define a linear propagator. This is given formally by: 
\begin{equation}
\begin{split}\underline{\mathbf{P}}\left(\Delta t\right)=\exp\left(\Delta t\underline{\mathbf{L}}\right)\end{split}
.
\end{equation}
where $\Delta t=t-\bar{t}$, and $\bar{t}$ is the interaction picture
origin. Transforming the field $\mathbf{a}$ to an interaction picture
is achieved on defining: 
\begin{equation}
\tilde{\mathbf{a}}=\underline{\mathbf{P}}^{-1}\left(\Delta t\right)\mathbf{a}.
\end{equation}
As a result, the equation of motion is: 
\begin{equation}
\begin{split}\frac{\partial\tilde{\mathbf{a}}}{\partial t}=D\left(\underline{\mathbf{P}}\left(\Delta t\right)\tilde{\mathbf{a}}\right).\end{split}
\end{equation}

This removes linear terms, which can cause stiffness in the equations,
increasing the discretization error. Given the case of a completely
linear ODE or SDE, the trajectory solutions will be exact up to round-off
errors.

\section{Stochastic equations with jumps}

Many stochastic equations involve a discrete Poisson or jump process,
which xSPDE can also solve. These are common in many fields, from
financial modeling to open quantum systems. The fundamental noise
is then a discrete jump or Poisson process, $dN$, which in our applications
has the integer values $0$ or $1$. 

Including this, a combined jump-diffusion Ito SDE can be written \cite{higham2005numerical}:
\begin{equation}
\Delta\bm{a}=\left[\mathbf{A}\left(\mathbf{a},t\right)+\underline{\mathbf{B}}\left(\mathbf{a},t\right)\cdot\bm{\xi}(t)\right]\Delta t+\underline{\bm{C}}\left(\mathbf{a},t\right)\cdot\Delta\bm{N}_{\lambda}(t),\label{eq:SDE-jump}
\end{equation}
where the $i-th$ jump process intensity is $\lambda_{i}\left(\mathbf{a},t\right)$.
This is defined such that:
\begin{equation}
\lambda_{i}\left(\mathbf{a},t\right)=\lim_{\Delta t\rightarrow0}\frac{1}{\Delta t}P(\Delta N_{i\lambda}(t)=1).
\end{equation}

Defining this general equation requires two additional parameters,
$\underline{\bm{C}}\left(\mathbf{a},t\right)$ and $\bm{\lambda}\left(\mathbf{a},t\right)$,
to specify the jump rate and its effect on the independent variable
$\mathbf{a}$. Such equations are treated in the mathematics and numerical
literature, but quantum physics applications require a variable jump
rate, which is sometimes ignored.

Such equations are often specified in the Ito picture in the usual
mathematical literature. The fact that $\underline{\bm{C}}\left(\mathbf{a},t\right)$
and $\bm{\lambda}\left(\mathbf{a},t\right)$ can depend on the field
$\mathbf{a}$ means that the equations do not follow standard calculus,
just as with continuous SDE equations.

\section{Probability distributions}

Stochastic equations generate trajectories distributed with a probability
density $P\left(\mathbf{a}\right)$. These can be defined as an average
and hence can be evaluated stochastically, since: 
\begin{equation}
P\left(\mathbf{a}'\right)=\left\langle \delta\left(\mathbf{a}'-\mathbf{a}\right)\right\rangle .
\end{equation}
Here $\left\langle ..\right\rangle \equiv\left\langle ..\right\rangle _{\infty}$
is the infinite ensemble limit of the average over many trajectories.
The probability can be shown to follow a Fokker-Planck equation (FPE)
with positive semi-definite diffusion matrix, \cite{Risken1996,Gardiner2009Stochastic}:
\begin{equation}
\frac{\partial P}{\partial t}=\mathcal{L}P=\left[-\partial_{n}A_{n}^{(I)}+\frac{1}{2}\partial_{n}\partial_{m}B_{nk}B_{mk}\right]P\,,\label{eq:FPE}
\end{equation}
where the differential operators act on all terms to their right.

\subsection{Distribution averages}

The average of any observable $\mathbf{O}\left(\mathbf{a}\right)$
is obtained either by averaging over the stochastic trajectories numerically,
or by analytic calculations, using: 
\begin{eqnarray}
\left\langle \mathbf{O}\right\rangle  & = & \int\mathbf{O}\left(\mathbf{a}\right)P\left(\mathbf{a}\right)d\mathbf{a}.\label{eq:averages}
\end{eqnarray}

The dynamics of an observable or moment follows an adjoint equation,
where $\tilde{\mathcal{L}}$ is the adjoint of $\mathcal{L}$: 
\begin{equation}
\left\langle \frac{\partial\mathbf{O}}{\partial t}\right\rangle =\left\langle \tilde{\mathcal{L}}\mathbf{O}\right\rangle ,\label{eq:moment_equn}
\end{equation}
where: 
\begin{equation}
\left\langle \tilde{\mathcal{L}}\mathbf{O}\right\rangle =\left\langle \left[A_{n}^{(I)}\partial_{n}+\frac{1}{2}B_{nk}B_{mk}\partial_{n}\partial_{m}\right]\mathbf{O}\right\rangle .
\end{equation}
This equation allows the time-evolution of averages to be calculated
analytically in simple cases, given an initial distribution. However,
in more complex cases, a numerical simulation of the stochastic equations
is more practical, and this can be carried out with xSPDE or other
software.

\section{Probability of a Wiener process}

The Wiener process with an arbitrary noise strength has the stochastic
equation: 
\begin{equation}
\dot{a}=bw\left(t\right).
\end{equation}

The probability density satisfies the Fokker-Planck equation for diffusion,
\begin{equation}
\frac{\partial P}{\partial t}=\frac{b^{2}}{2}\frac{\partial^{2}}{\partial a^{2}}P\,.
\end{equation}

Then, if $x$ initially is Gaussian distributed, this has a Gaussian
distribution at time $t$ with: 
\begin{equation}
P\left(a\right)=\frac{1}{\sqrt{2\pi\sigma^{2}\left(t\right)}}\exp\left[-\frac{\left(a-\bar{a}\left(t\right)\right)^{2}}{2\sigma^{2}\left(t\right)}\right].
\end{equation}

Here: 
\begin{align}
\bar{a}\left(t\right) & =\bar{a}\left(0\right)\\
\sigma^{2}\left(t\right) & =\sigma^{2}\left(0\right)+b^{2}t.\nonumber 
\end{align}

\subsection{Distributions of functions}

Any function of the stochastic variables has a corresponding probability
density. For example, the distribution of $a^{2}$ has a $\chi^{2}$
distribution with a single degree of freedom, such that if $y=\left(a-\bar{a}\left(t\right)\right)^{2}/\sigma^{2}\left(t\right)$,
then:

\begin{equation}
P\left(y\right)=\frac{1}{\sqrt{2\pi y}}\exp\left[-\frac{y}{2}\right].
\end{equation}
Hence: 
\begin{equation}
P\left(a^{2}\right)=\frac{1}{\left|a-\bar{a}\left(t\right)\right|\sqrt{2\pi\sigma^{2}\left(t\right)}}\exp\left[-\frac{\left(a-\bar{a}\left(t\right)\right)^{2}}{2\sigma^{2}\left(t\right)}\right].
\end{equation}

More generally, it is often not known what the exact analytic solutions
are, and a numerical solution is employed. This can either use the
stochastic equation directly, or the Fokker-Planck equation, although
it is generally difficult to scale this to many variables or to partial
differential equations,

That is why we focus on the stochastic equation approach here, which
can be used to numerically calculate either the mean values or the
probability distributions in general cases.

\section{Fourier transforms}

Frequency spectra have many uses, especially for understanding the
steady-state fluctuations of any physical system in the presence of
noise, typically either thermal or quantum-mechanical, although the
noise could have other sources.

The time-domain spectral definition used here is: 
\begin{align}
\tilde{a}(\omega) & =\frac{1}{\sqrt{2\pi}}\int e^{i\omega t}a(t)dt\,\nonumber \\
a(t) & =\frac{1}{\sqrt{2\pi}}\int e^{-i\omega t}\tilde{a}(\omega)d\omega.\,
\end{align}

As a simple example, a sinusoidal oscillation in the form 
\begin{equation}
a(t)=\cos\left(\omega_{0}t\right).
\end{equation}
between $t=-T/2$ and $t=T/2$ has a Fourier transform given by: 
\begin{align}
\tilde{a}(\omega) & =\frac{1}{2\sqrt{2\pi}}\int_{-T/2}^{T/2}\left[e^{i\left(\omega-\omega_{0}\right)t}+e^{i\left(\omega+\omega_{0}\right)t}\right]dt\,\\
 & =\frac{T}{2\sqrt{2\pi}}\left[sinc\left(\left(\omega-\omega_{0}\right)\frac{T}{2}\right)+sinc\left(\left(\omega+\omega_{0}\right)\frac{T}{2}\right)\right].\nonumber 
\end{align}

\chapter{SDE Examples\label{chap:SDE-Examples}}

\section{Complex damped spectrum}

Consider the spectrum of Eq \eqref{eq: damped_path_with_noise}, with
a complex noise, 
\begin{equation}
\left\langle w\left(t\right)w^{*}\left(t'\right)\right\rangle =2\delta\left(t-t'\right),
\end{equation}

Th script below solves an SDE with a complex Gaussian initial condition
having $\left\langle \left|a\left(0\right)\right|^{2}\right\rangle =1$,
so it is in the steady-state initially:

\begin{equation}
\frac{\partial a}{\partial t}=-a+w_{1}(t)+iw_{2}(t)\,.\label{eq:SDE-3-1}
\end{equation}

The equation is such that the initial distribution is also the equilibrium
probability distribution so the numerical simulation uses a random
initial equation near the equilibrium value, and a range of $t=100$,
with $640$ points. Here there are two real noises.

The input parameters are given below. There are parallel operations
here, for ensemble averaging, so we \textbf{USE THE DOT}. 
\begin{center}
\doublebox{\begin{minipage}[t]{0.75\columnwidth}%
\texttt{clear}

\texttt{p.points = 640;}

\texttt{p.ranges = 100;}

\texttt{p.noises = 2;}

\texttt{p.ensembles = 10000;}

\texttt{p.initial = @(v,p) (v(1,:)+1i{*}v(2,:))/sqrt(2);}

\texttt{p.deriv = @(a,w,p) -a + w(1,:)+1i{*}w(2,:);}

\texttt{p.observe = @(a,p) a.{*}conj(a);}

\texttt{p.transforms = 1;}

\texttt{p.olabels = '\textbar a(\textbackslash omega)\textbar\textasciicircum 2';}

\texttt{xspde(p);}%
\end{minipage}} 
\par\end{center}

Note that p.\texttt{transforms = 1} tells xSPDE to Fourier transform
the field over the time coordinate before averaging, to give a spectrum.
Both observe and transforms could be cell arrays, but the this is
not needed with a single observable. The first argument $v$ of the
initial function is a random field, used to initialize the stochastic
variable.

\begin{figure}[H]
\centering{} \includegraphics[width=0.75\textwidth]{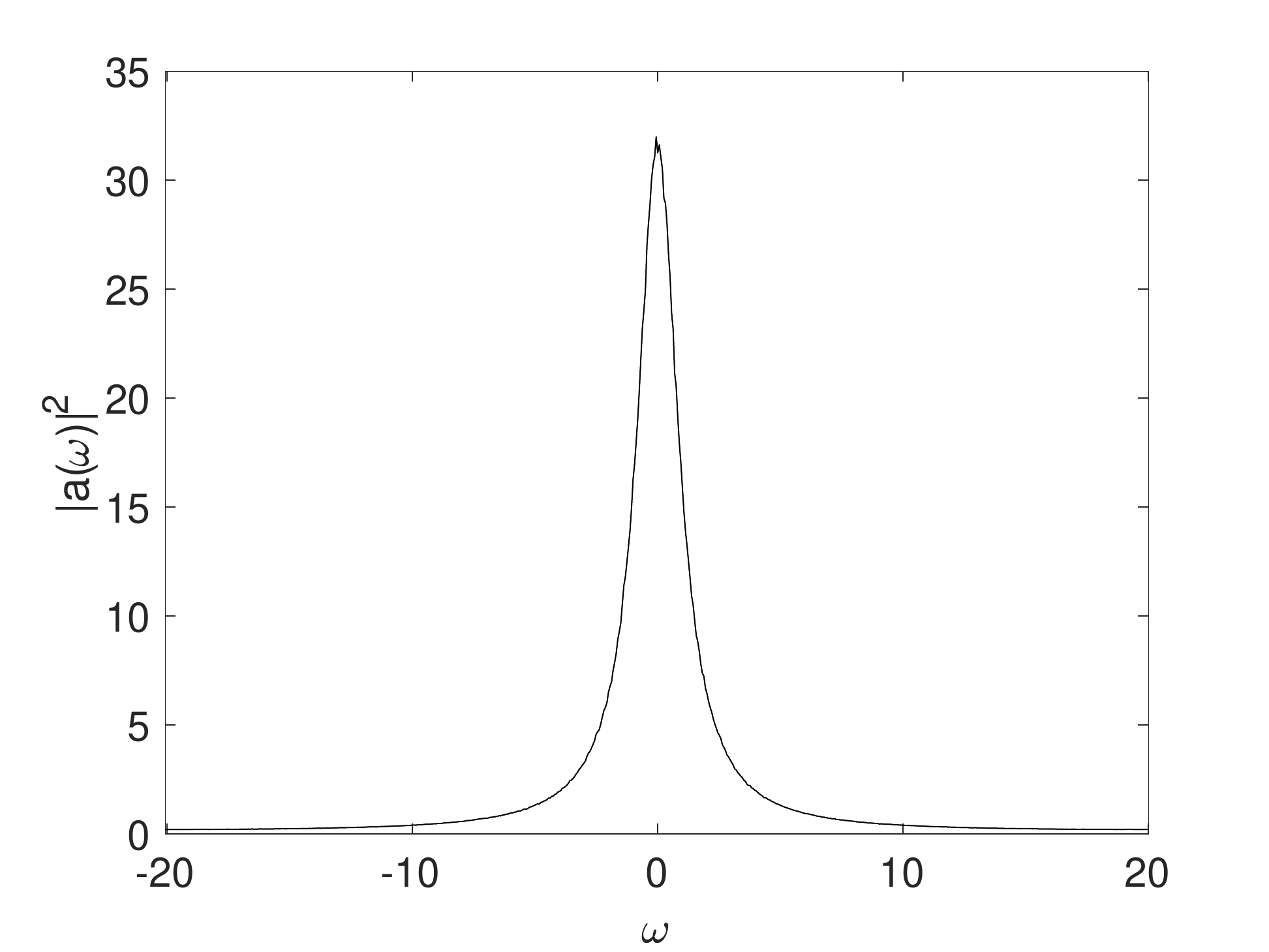}

\caption{\emph{Complex damped spectrum obtained from the interactive script.}}
\vspace{10pt}
\end{figure}

To define as many observables as you like, use a cell array of observe
function handles: 
\begin{center}
\doublebox{\begin{minipage}[t]{0.75\columnwidth}%
\texttt{p.observe\{1\} = ..;}

\texttt{p.observe\{2\} = ..;}%
\end{minipage}} 
\par\end{center}

To learn more, try the following: 
\begin{itemize}
\item \textbf{Simulate over a range of $t=200$. What changes do you see?
Why?} 
\item \textbf{Change the equation to the laser noise equations introduced
in the next section (Laser quantum noise). Why is the spectrum much
narrower?} 
\end{itemize}

\subsection{Reducing the frequency cut-off}

If the number of time-points is reduced, the maximum spectral frequency
is reduced. In the example below, a comparison is included, to compare
with the exact result, and both the time domain and freqency domain
outputs are plotted.

The computed ordinary and spectral variances are compared with exact
solutions and graphed, where 
\begin{align}
\lim_{t\rightarrow\infty}\left\langle \left|a\left(t\right)\right|^{2}\right\rangle  & =1.\nonumber \\
\left\langle \left|a\left(\omega\right)\right|^{2}\right\rangle  & =\frac{T}{\pi\left(1+\omega^{2}\right)}.
\end{align}

\begin{center}
\doublebox{\begin{minipage}[t]{0.9\columnwidth}%
\texttt{function {[}e{]} = Equilibrium()}

\texttt{p.name = 'Equilibrium spectrum';}

\texttt{p.points = 50;}

\texttt{p.steps = 4;}

\texttt{p.ranges = 50;}

\texttt{p.seed = 241;}

\texttt{p.noises = 2;}

\texttt{p.ensembles = {[}100,50{]};}

\texttt{p.initial = @(w,\textasciitilde ) (w(1,:)+1i{*}w(2,:))/sqrt(2);}

\texttt{p.deriv = @(a,w,\textasciitilde ) -a + w(1,:)+1i{*}w(2,:);}

\texttt{p.observe\{1\} = @(a,\textasciitilde ) a.{*}conj(a);}

\texttt{p.observe\{2\} = @(a,\textasciitilde ) a.{*}conj(a);}

\texttt{p.transforms = \{0,1\};}

\texttt{p.olabels = \{'\textbar a(t)\textbar\textasciicircum 2','\textbar a(\textbackslash omega)\textbar\textasciicircum 2'\};}

\texttt{p.compare = \{@(p) 1, @(p)p.ranges(1)./(pi{*}(1+p.w.\textasciicircum 2))\};}

\texttt{e = xspde(p);}

\texttt{end}%
\end{minipage}} 
\par\end{center}

\paragraph{Notes}
\begin{itemize}
\item A fixed random seed is input using the p.seed parameter. 
\item The p.transforms cell array gives a Fourier transform for p.observe\{2\}
only. 
\item A small number of ensembles and time-steps is used to improve error
visibility. 
\end{itemize}
\begin{figure}[H]
\centering{}\includegraphics[width=0.75\textwidth]{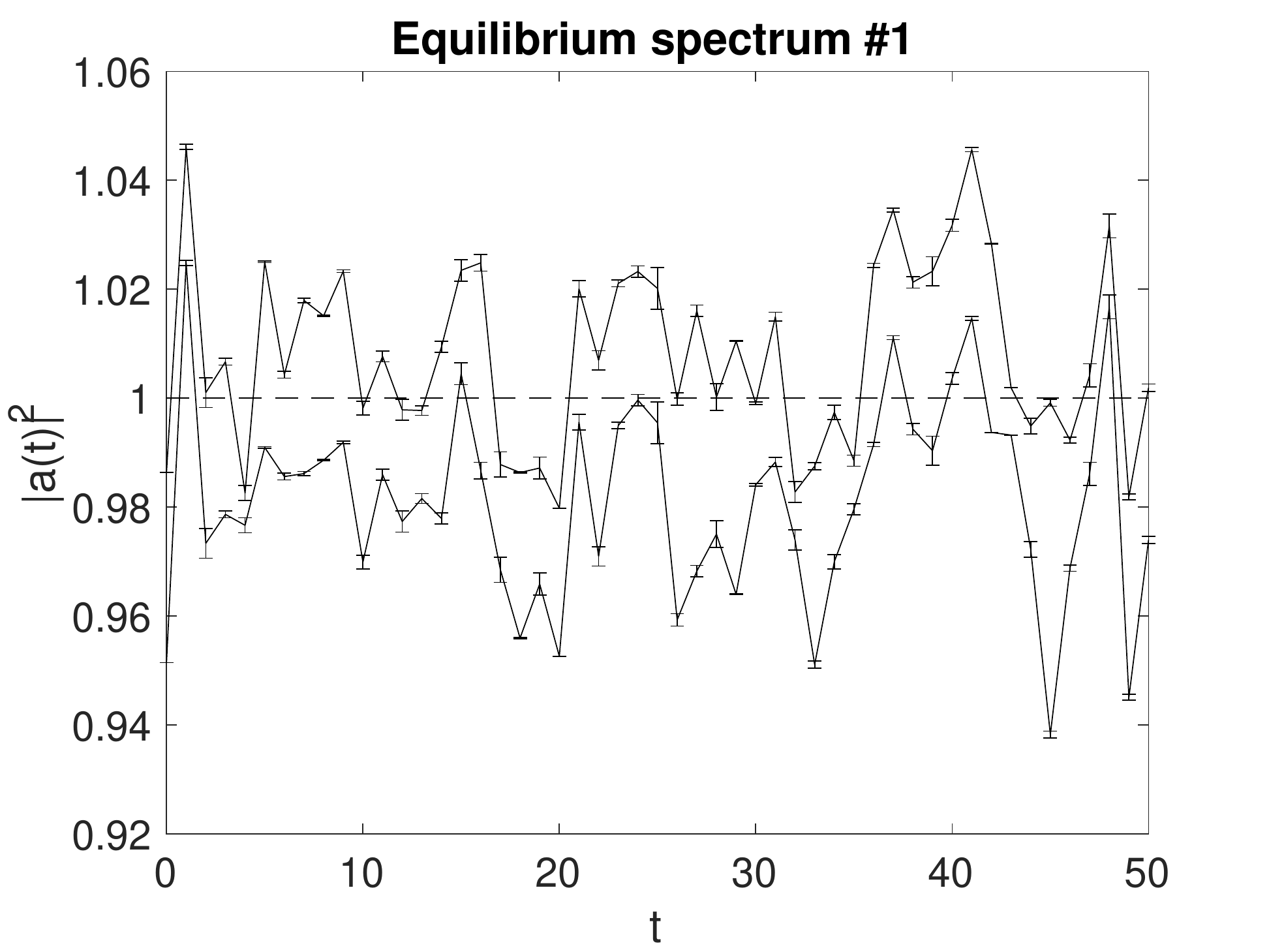}\\
 \includegraphics[width=0.75\textwidth]{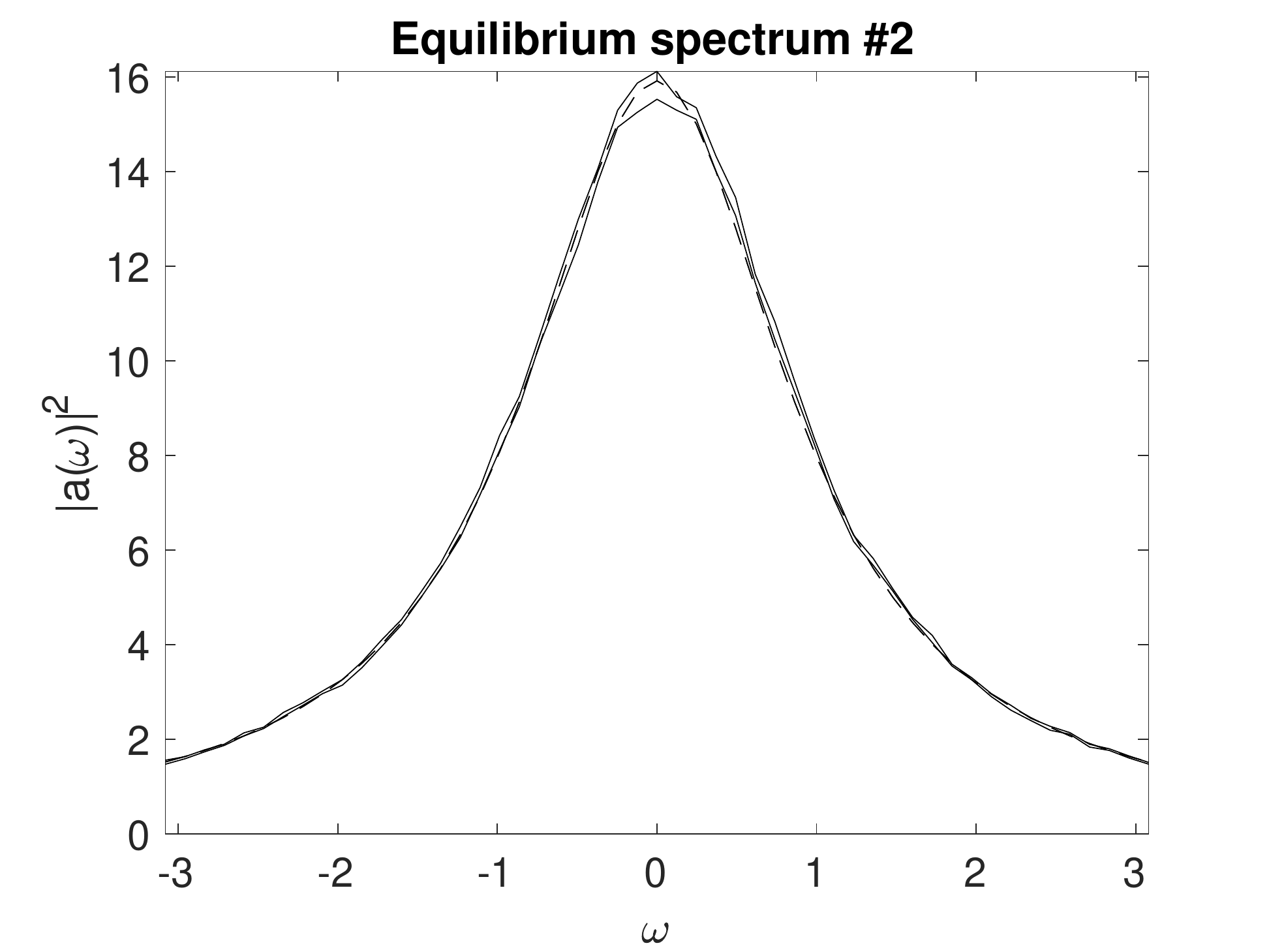}

\caption{\emph{Top figure: Mean amplitude squared, showing invariant behavior
with time, apart from sampling errors. Bottom figure: Mean spectrum
as a function of frequency. The dashed lines are exact results, solid
lines are upper and lower sampling error bounds $(\pm\sigma)$, from
sampling the stochastic equations, the error-bars are errors due to
the step-size. Error bars are less than the minimum size for graphics
display in the bottom figure.}}
\vspace{10pt}
\end{figure}

\pagebreak{}

\section{The Black-Scholes equation}

A well-known Ito-type stochastic equation is called the Black-Scholes
equation \cite{black1976pricing}, used to price financial options.
It describes the fluctuations in a stock or commodity value: 
\begin{equation}
da=\mu a\,dt+a\sigma\,dw,
\end{equation}
where $\left\langle dw^{2}\right\rangle =dt$. As the noise is multiplicative,
the equation is different in Ito and Stratonovich calculus. The corresponding
Stratonovich equation, as used in xSPDE for the standard default integration
routine is: 
\begin{equation}
\dot{a}=\left(\mu-\sigma^{2}/2\right)a+a\sigma w(t).
\end{equation}

\begin{figure}
\centering{}\includegraphics[width=0.75\textwidth]{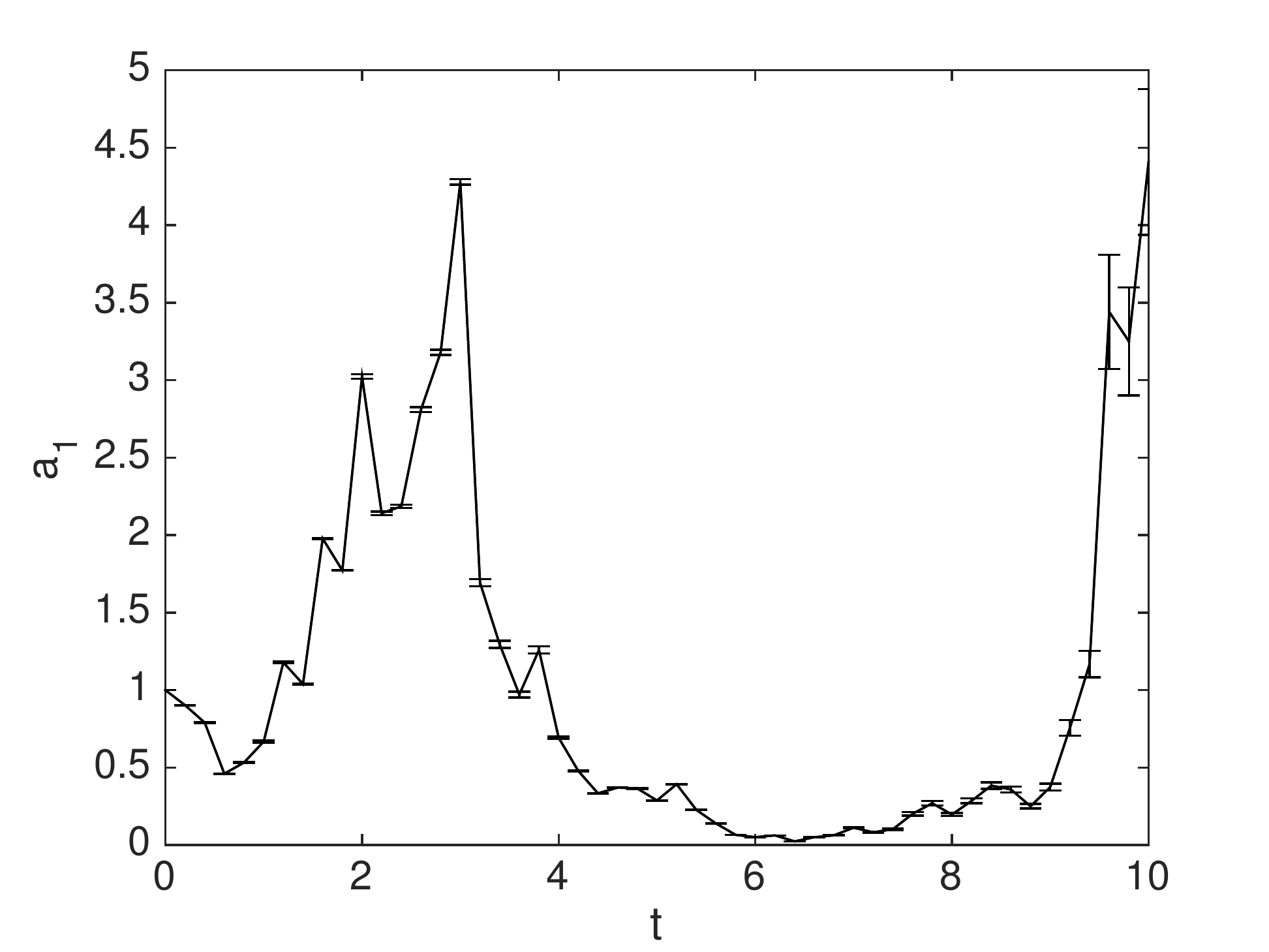}

\caption{\label{fig:The-Black-Scholes}\emph{Simulation of the Black-Scholes
equation describing stock prices.}}
\vspace{10pt}
\end{figure}

An interactive xSPDE script in Matlab is given below with an output
graph in Fig (\ref{fig:The-Black-Scholes}). This is for a startup
with a volatile stock having $\mu=0.1,\,\sigma=1$. The spiky behavior
is typical of multiplicative noise, and also of the more risky stocks
in the small capitalization portions of the stock market. 
\begin{center}
\doublebox{\begin{minipage}[t]{0.75\columnwidth}%
\texttt{clear}

\texttt{p.initial = @(v,p) 1;}

\texttt{p.deriv = @(a,w,p) -0.4{*}a+a.{*}w;}

\texttt{xspde(p);}%
\end{minipage}} 
\par\end{center}

Here $p.initial$ describes the initialization function. The first
argument of $@(v,p)$ is $v$, an initial random variable with unit
variance. The error-bars are estimates of step-size error. Errors
can be reduced by using more time-steps.

To learn more, try the following: 
\begin{itemize}
\item \textbf{Solve for a more mature stock, with less volatility, having
$\mu=0.1,\,\sigma=0.1$.} 
\end{itemize}

\section{Kubo oscillator}

The Kubo oscillator is widely used to model environmental noise in
solid-state environments. The function below solves the relevant multiplicative
SDE with initial condition $a\left(0\right)=1$ and:

\begin{equation}
\frac{\partial a}{\partial t}=iaw(t)\,.\label{eq:SDE-3}
\end{equation}

The function employs the RK4 method, although other algorithms can
be used instead. It has both vector and series ensembles, then stores
the computed averages with a comparison of the variance and an exact
solution, 
\begin{equation}
\left\langle a^{n}\right\rangle =e^{-tn^{2}/2}.
\end{equation}

\begin{center}
\doublebox{\begin{minipage}[t]{0.9\columnwidth}%
\texttt{function {[}e{]} = Kubo()}

\texttt{p.name = 'Kubo oscillator';}

\texttt{p.ensembles = {[}1000,8{]};}

\texttt{p.method = @RK4;}

\texttt{p.initial = @(w,p) 1;}

\texttt{p.deriv = @(a,w,p) 1i{*}w.{*}a(1,:) ;}

\texttt{p.file = 'Kubo.mat';}

\texttt{p.observe\{1\} = @(a,p) a;}

\texttt{p.olabels\{1\} = \{'\textless{} a \textgreater '\};}

\texttt{p.observe\{2\} = @(a,p) a.\textasciicircum 2;}

\texttt{p.olabels\{2\} = \{'\textless{} a\textasciicircum 2\textgreater '\};}

\texttt{p.compare = \{@(p) exp(-p.t/2),@(p) exp(-2{*}p.t)\};}

\texttt{e = xsim(p);}

\texttt{p2.name = 'Kubo oscillator edited title';}

\texttt{xgraph(p.file,p2);}

\texttt{end}%
\end{minipage}} 
\par\end{center}

\paragraph{Notes}
\begin{itemize}
\item The algorithm is changed from the default to RK4. 
\item The data is stored to 'Kubo.mat'. 
\item This is re-read and edited using a second parameter structure, p2. 
\end{itemize}
\begin{figure}[H]
\centering{}\includegraphics[width=0.75\textwidth]{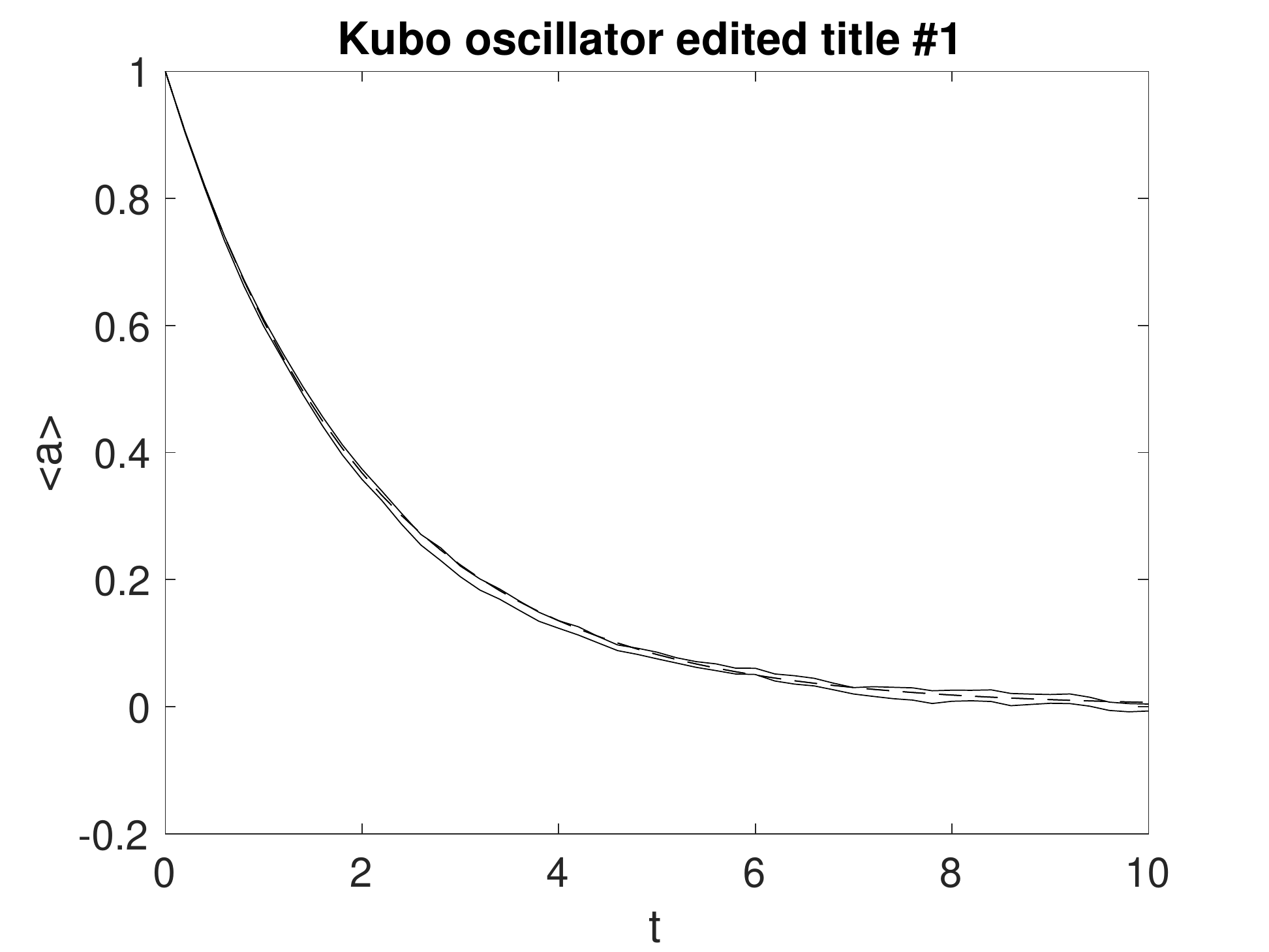}\\
 \includegraphics[width=0.75\textwidth]{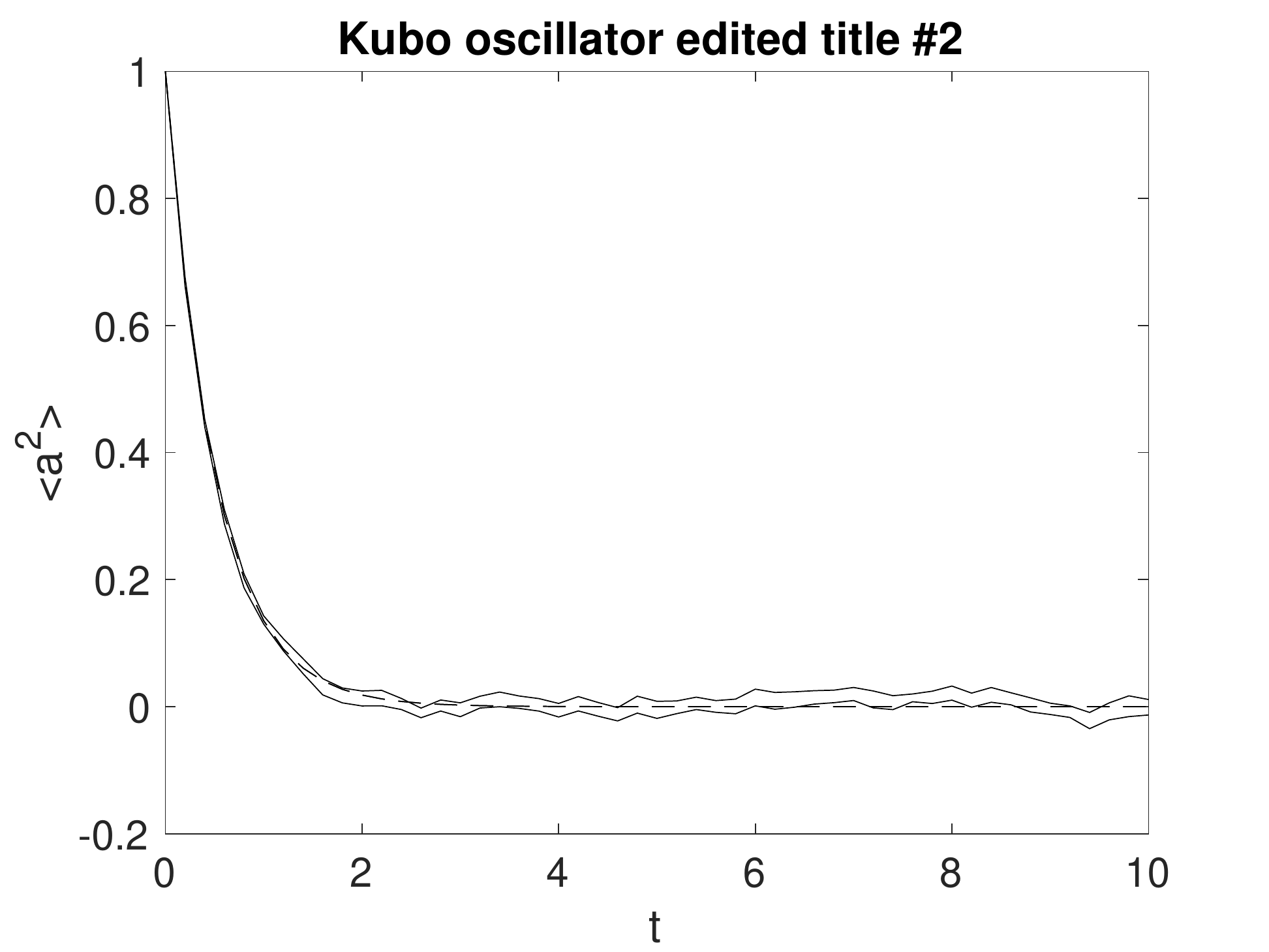}

\caption{\emph{Example: Kubo oscillator. The graph shows the sampling error-bars
as two parallel lines. The discretization error-bars are less than
the minimum, and are not shown.}}
\vspace{10pt}
\end{figure}

\subsection*{Exercises}
\begin{itemize}
\item Simulate the Kubo oscillator with three ensemble levels to allow parallel
computation.
\item Increase the first ensemble size to check how it modifies the sampling
errors. 
\item Try a higher order moment as an observable.
\item This function generates a data file, \texttt{kubo.mat}. If you run
this twice, note that the new data overwrites the old data. 
\item Try including modified graphics parameters when running \texttt{xGRAPH,}
if the first graphs you generate need changes.
\end{itemize}
\newpage{}

\section{Loss and gain with noise}

This solves an SDE with a complex Gaussian distributed initial condition
having $\left\langle \left|a\left(0\right)\right|^{2}\right\rangle =1$
and a sequence of SDE equations, such that 
\begin{equation}
\frac{\partial a}{\partial t}=\begin{cases}
-a+w_{1}(t)+iw_{2}(t) & 0<t<4\\
a+w_{1}(t)+iw_{2}(t) & 4<t<8
\end{cases}\,.\label{eq:SDE-3-1-1}
\end{equation}

The computed variance is compared with an exact solution, 
\begin{equation}
\left\langle a^{2}\right\rangle =\begin{cases}
1 & 0<t<4\\
2e^{2\left(t-4\right)t}-1 & 4<t<8
\end{cases}.
\end{equation}

. 
\begin{center}
\doublebox{\begin{minipage}[t]{0.9\columnwidth}%
\texttt{function {[}e{]} = Gain()}

\texttt{p.name = 'Loss with noise';}

\texttt{p.ranges = 4;}

\texttt{p.noises = 2;}

\texttt{p.ensembles = {[}10000,1,10{]};}

\texttt{p.initial = @(w,\textasciitilde ) (w(1,:)+1i{*}w(2,:))/sqrt(2);}

\texttt{p.deriv = @(a,w,p) -a + w(1,:)+1i{*}w(2,:);}

\texttt{p.observe = @(a,\textasciitilde ) a.{*}conj(a);}

\texttt{p.olabels = '\textbar a\textbar\textasciicircum 2';}

\texttt{p.compare = @(p) 1;}

\texttt{p2 = p;}

\texttt{p2.steps = 2;}

\texttt{p2.name = 'Gain with noise';}

\texttt{p2.deriv = @(a,w,\textasciitilde ) a + w(1,:)+1i{*}w(2,:);}

\texttt{p2.compare = @(p) 2{*}exp(2{*}(p.t-4))-1;}

\texttt{e = xspde(p,p2);}

\texttt{end}%
\end{minipage}} 
\par\end{center}

\paragraph{Notes}
\begin{itemize}
\item Low and high level parallel ensembles optimize use of multi-core vector
hardware. 
\item Two distinct simulations are run in series, with a change in the equation. 
\item The simulation name is changed in sequence 2, to distinguish the graphical
outputs 
\end{itemize}
\begin{figure}[H]
\centering{}\includegraphics[width=0.75\textwidth]{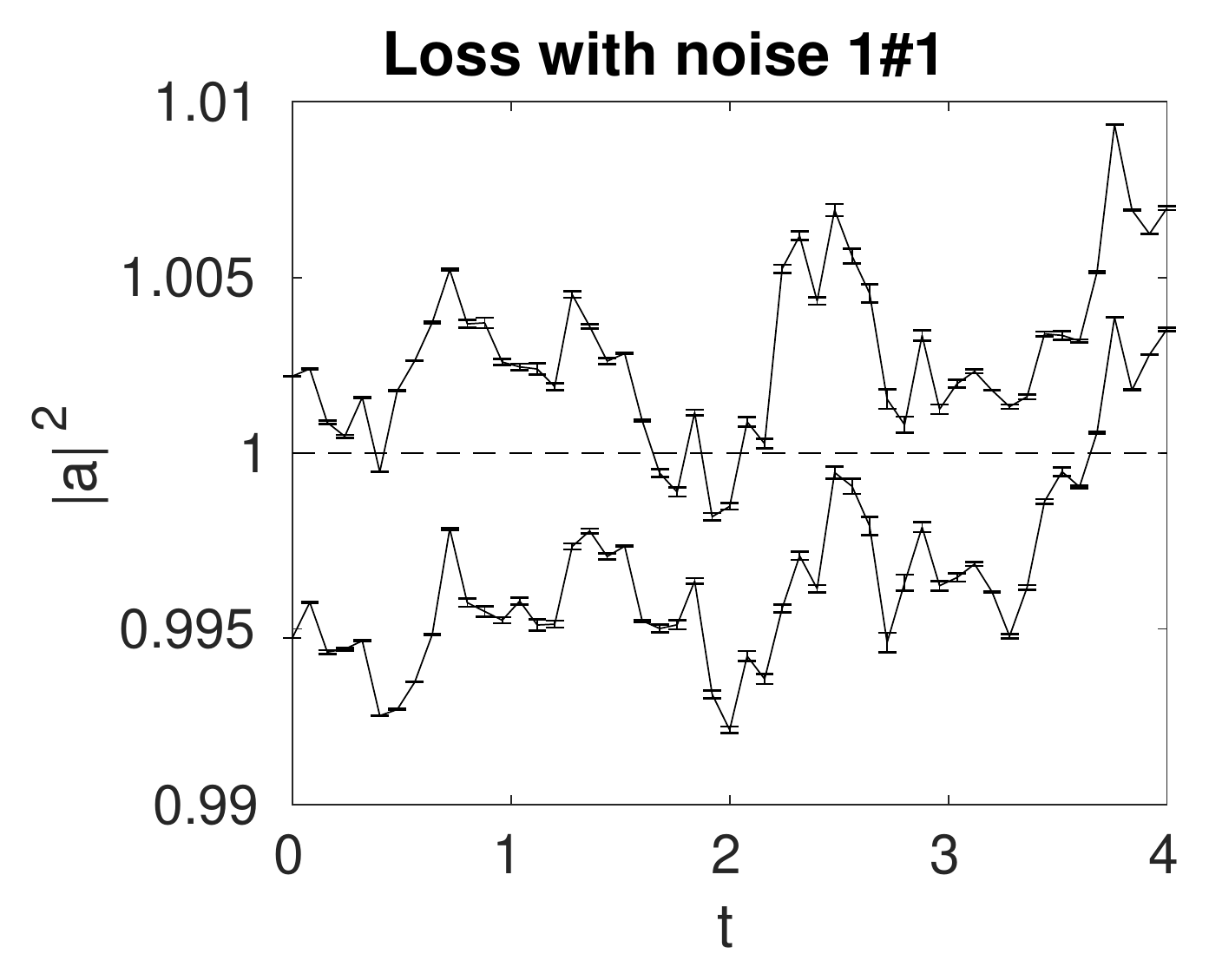}\\
 \includegraphics[width=0.75\textwidth]{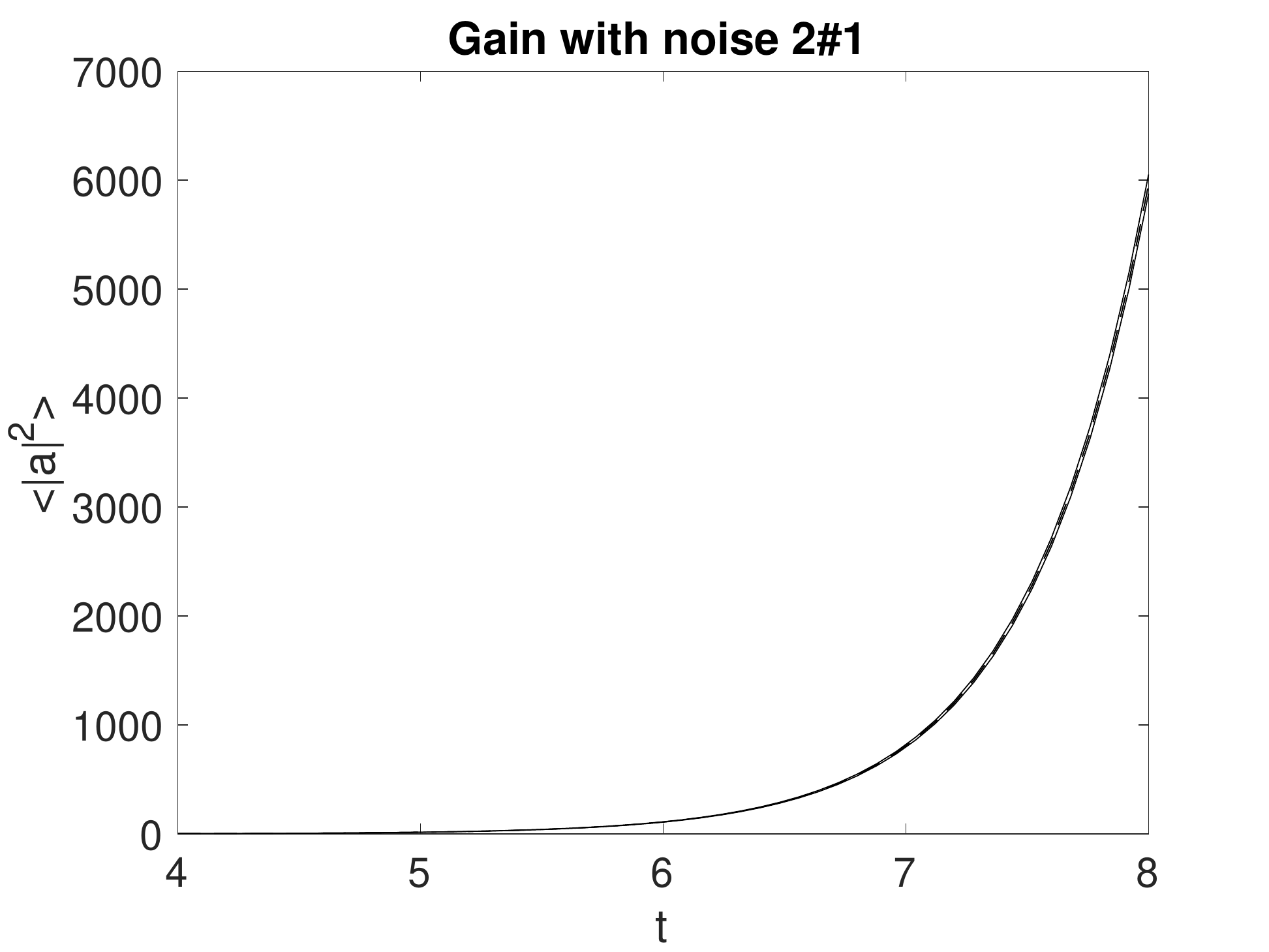}

\caption{\emph{Top figure: amplitude squared with loss balanced by noise. Bottom
figure, amplitude squared with gain. Graphs show excellent agreement
with theory up to the sampling errors of less than $\pm0.005$ in
the initial phase, shown by the parallel lines, with step errors of
order $\pm0.001$ indicated by error-bars.}}
\vspace{10pt}
\end{figure}

\pagebreak{}

\section{Probability of a Wiener process}

The script below solves an SDE with an initial condition $\left\langle a\left(0\right)\right\rangle ^{2}=\frac{1}{4}$
and

\begin{equation}
\dot{a}=w(t)\,.
\end{equation}

It saves the probability density and compares this with an exact solution:

\begin{eqnarray}
P\left(x,t\right) & = & \frac{1}{\sqrt{2\pi\sigma^{2}\left(t\right)}}e^{-\frac{x^{2}}{2\sigma^{2}\left(t\right)}}\nonumber \\
\sigma^{2}\left(t\right) & = & \frac{1}{4}+t\,.
\end{eqnarray}

\paragraph{Notes}
\begin{itemize}
\item The script outputs a 3D plot of $P\left(x,t\right)$, together with
the time evolution of $P\left(0,t\right)$ 
\item There are 5 ``transverse'' plots of transient probabilities at intermediate
times. 
\item Legends are plotted to identify the simulated and the analytic comparison
lines. 
\end{itemize}
\begin{center}
\doublebox{\begin{minipage}[t]{0.9\columnwidth}%
\texttt{function e = Wienerprob()}

\texttt{p.name = 'Wiener SDE distribution';}

\texttt{p.noises = 1;}

\texttt{p.points = 10;}

\texttt{p.ensembles = {[}10000,10{]};}

\texttt{p.initial = @(v,p) v/2;}

\texttt{p.sig = @(p) .25 + p.r\{1\};}

\texttt{p.deriv = @(a,w,p) w;}

\texttt{p.observe\{1\} = @(a,p) a;}

\texttt{p.compare\{1\} = @gaussprob;}

\texttt{p.transverse\{1\} = 5;}

\texttt{p.olabels\{1\} = 'P(x)';}

\texttt{p.binranges\{1\} = \{-5:0.25:5\};}

\texttt{p.legends\{1\} = \{'Sampled P(x,\textbackslash tau) \textbackslash pm
\textbackslash sigma',...}

\texttt{'Exact P(x,\textbackslash tau)'\};}

\texttt{p.xlabels = \{'\textbackslash tau','x'\};}

\texttt{e = xspde(p);}

\texttt{end}

\texttt{\%}

\texttt{function p = gaussprob(p)}

\texttt{p = exp(-(p.r\{2\}.\textasciicircum 2)./(2{*}p.sig(p)))./sqrt(2{*}pi{*}p.sig(p));}

\texttt{end}%
\end{minipage}} 
\par\end{center}

\begin{figure}[H]
\begin{centering}
\includegraphics[width=0.75\textwidth]{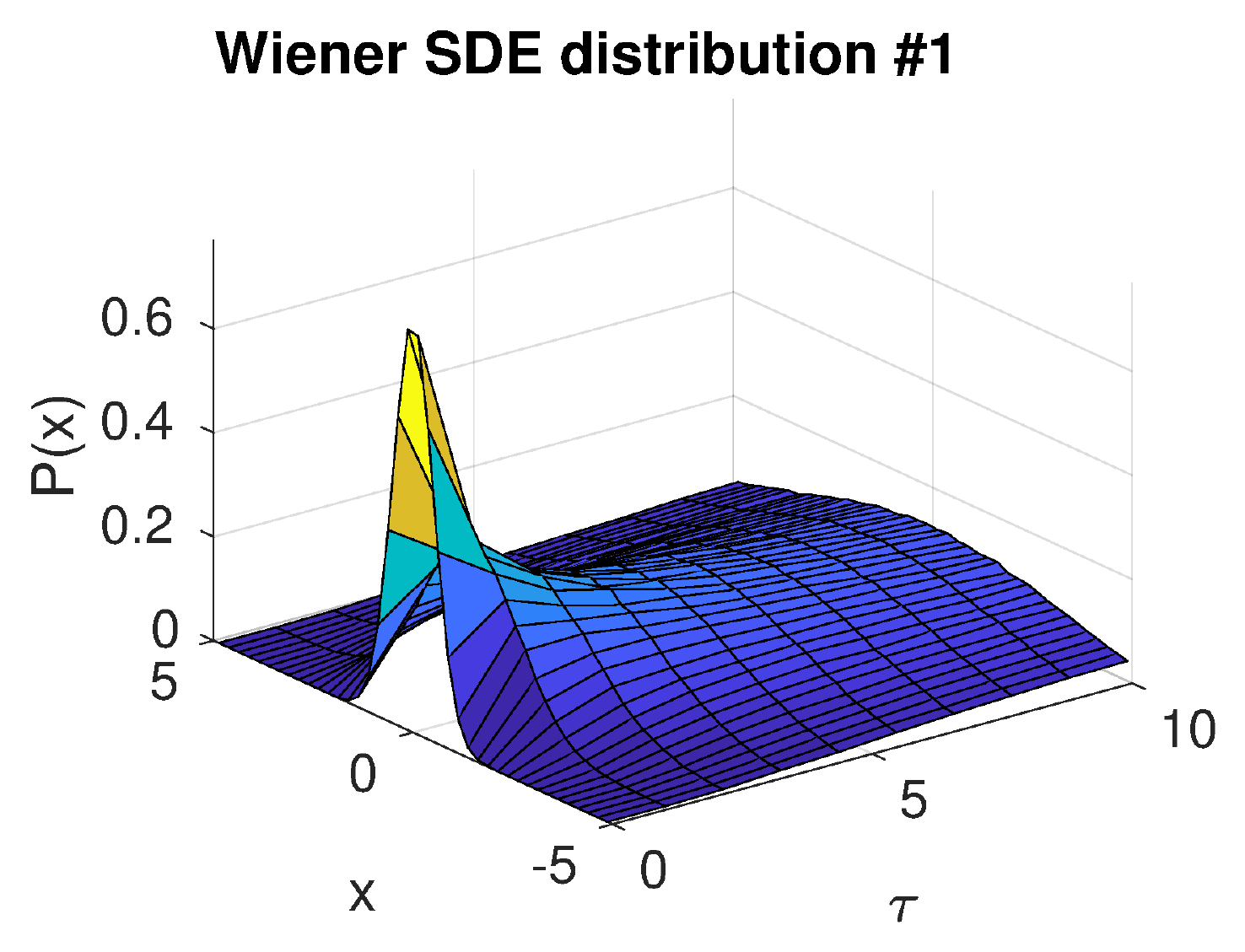}
\par\end{centering}
\begin{centering}
\includegraphics[width=0.75\textwidth]{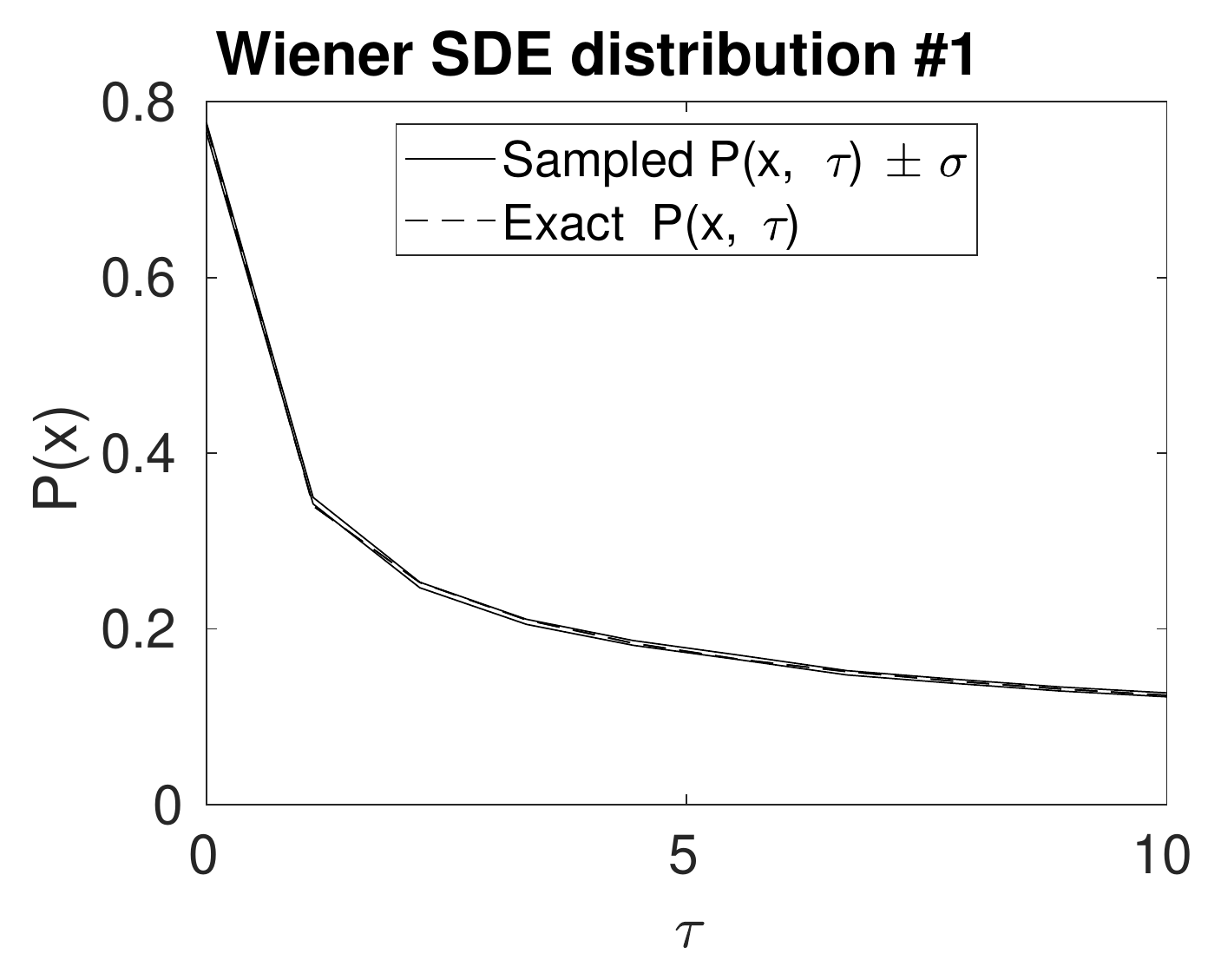}
\par\end{centering}
\centering{}

\caption{\emph{Top figure: 3D plot of the computed probability density of the
simulated Wiener process as a function of time ($\tau$) and \textquotedblleft position\textquotedblright{}
($x$). Bottom figure: Time evolution of the computed probability
density for $x=0$. The solid lines indicate upper and lower sampling
error bounds, while the dashed line indicates theoretical predictions.}}
\vspace{10pt}
\end{figure}

\begin{figure}[H]
\begin{centering}
\includegraphics[width=0.75\textwidth]{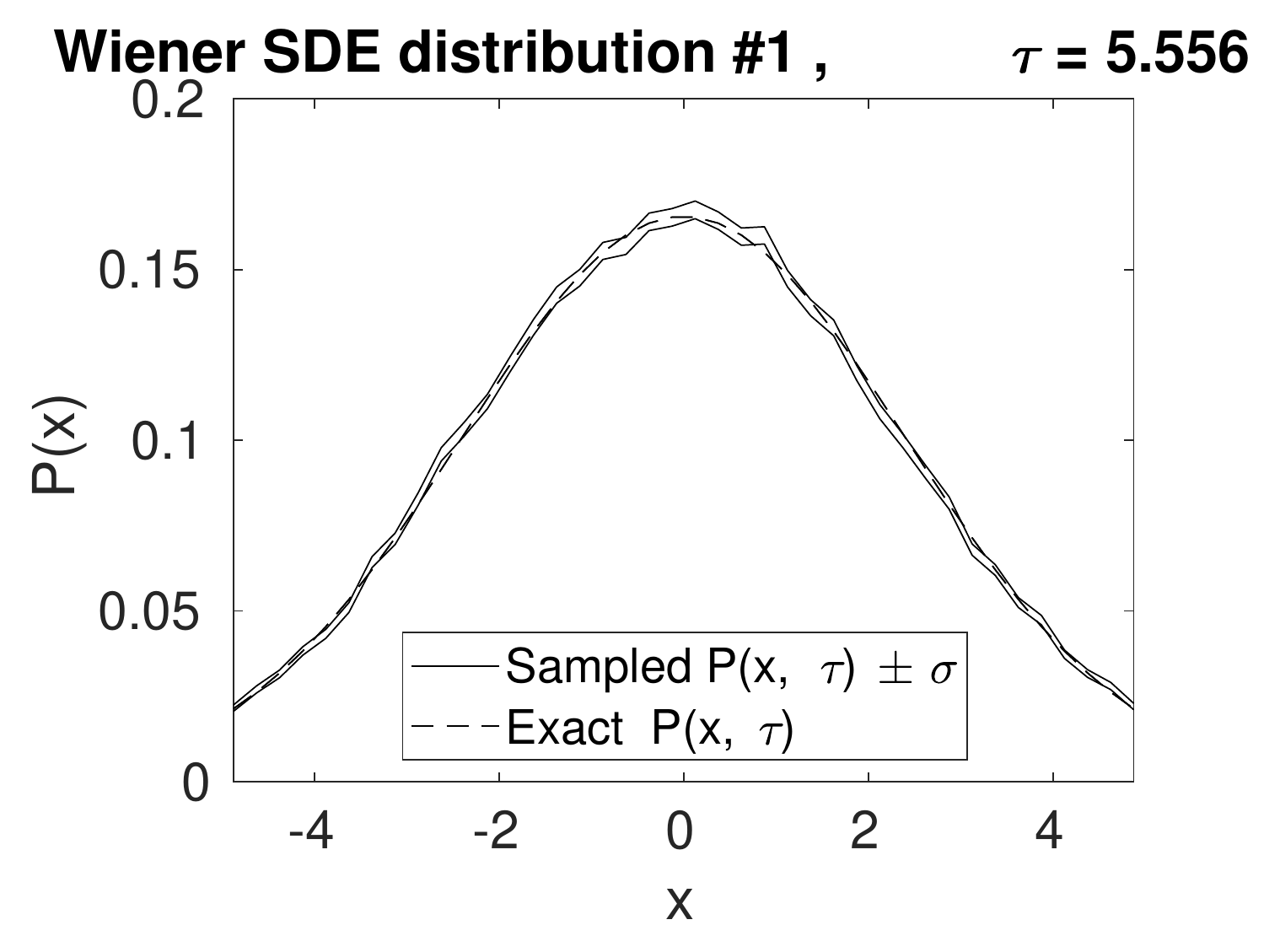}
\par\end{centering}
\begin{centering}
\includegraphics[width=0.75\textwidth]{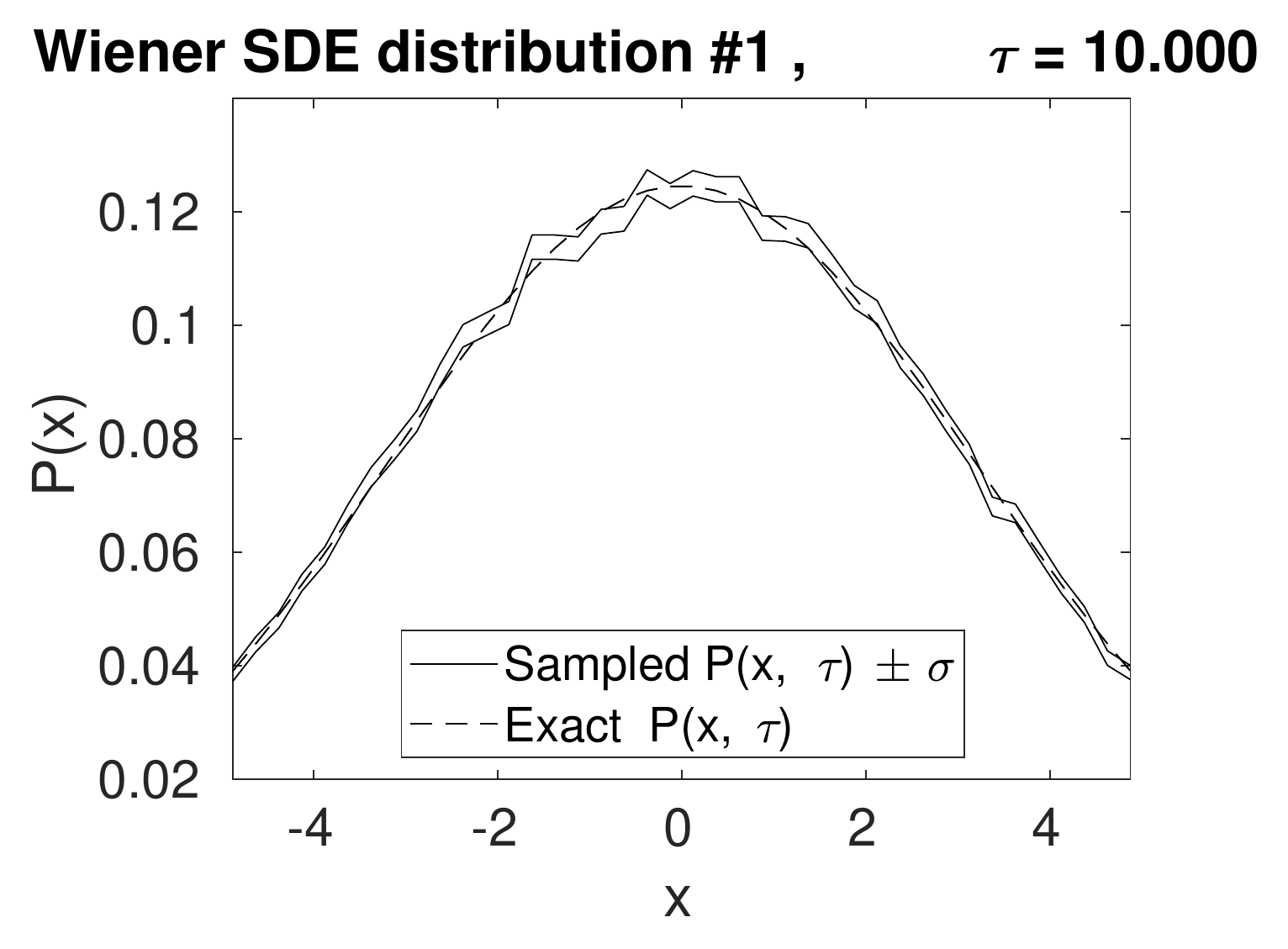}
\par\end{centering}
\centering{}

\caption{\emph{Top and bottom figure: Computed probability densities of the
simulated Wiener process at $\tau=5.556$ and $\tau=10$, respectively.
In total, 5 of these transverse plots are generated, however, only
2 are presented here.}}
\vspace{10pt}
\end{figure}

\pagebreak{}

\section{Projected SDE on a catenoid}

This solves an SDE with 3 field variables $\mathbf{a}=\left(a_{1},a_{2},a_{3}\right)^{T}$.
The Stratonovich diffusion equation is 
\begin{eqnarray}
\frac{\partial\mathbf{a}}{\partial t} & = & \mathcal{P}_{\mathbf{a}}^{\parallel}\left[\mathbf{w}\right]\,,
\end{eqnarray}
where $\mathcal{P}_{\mathbf{a}}^{\parallel}\left[\cdot\right]$ indicates
a projected onto the surface of a catenoid manifold defined by 
\begin{eqnarray}
f & =x_{1}^{2}+x_{2}^{2}-\sinh^{2}\left(x_{3}\right)-1 & =0\,.
\end{eqnarray}
The initial condition is given by $\mathbf{a}\left(0\right)=\left(1,0,0\right)^{T}$.
Here $\mathbf{w}=\left(w_{1},w_{2},w_{3}\right)^{T}$ consists of
3 independent noise variables

\paragraph{Notes}
\begin{itemize}
\item This is a projected sde case 
\item The Euclidean distance from the initial point is computed 
\item This is compared with the predicted analytic value $\left\langle R^{2}\right\rangle =2t$. 
\end{itemize}
\begin{center}
\doublebox{\begin{minipage}[t]{0.9\columnwidth}%
\texttt{function {[}e{]} = Catenoid}

\texttt{p.name = '3D Catenoid diffusion';}

\texttt{p.iterproj = 3;}

\texttt{p.X0 = {[}1,0,0{]}';}

\texttt{p.fields = 3;}

\texttt{p.ranges = 5;}

\texttt{p.points = 51;}

\texttt{p.ensembles = {[}400, 10{]};}

\texttt{p.compare\{2\} = @(p) 2{*}p.t;}

\texttt{p.deriv = @(a, w, p) w;}

\texttt{p.initial = @(w, p) p.X0;}

\texttt{p.observe\{2\} = @(a, p) sum((p.X0-a).\textasciicircum 2,1);}

\texttt{p.diffplot\{2\} = 1;}

\texttt{p.function\{1\} = @(o, p) o\{2\}.\textasciicircum 2;}

\texttt{p.olabels = \{'\textbackslash langle R\textasciicircum 2
\textbackslash rangle\textasciicircum 2','\textbackslash langle
R\textasciicircum 2 \textbackslash rangle'\};}

\texttt{p.project = @Catproj;}

\texttt{p.method = @MPnproj;}

\texttt{e = xspde(p);}

\texttt{end}%
\end{minipage}} 
\par\end{center}

\begin{center}
\begin{figure}[H]
\centering{}\includegraphics[width=0.75\textwidth]{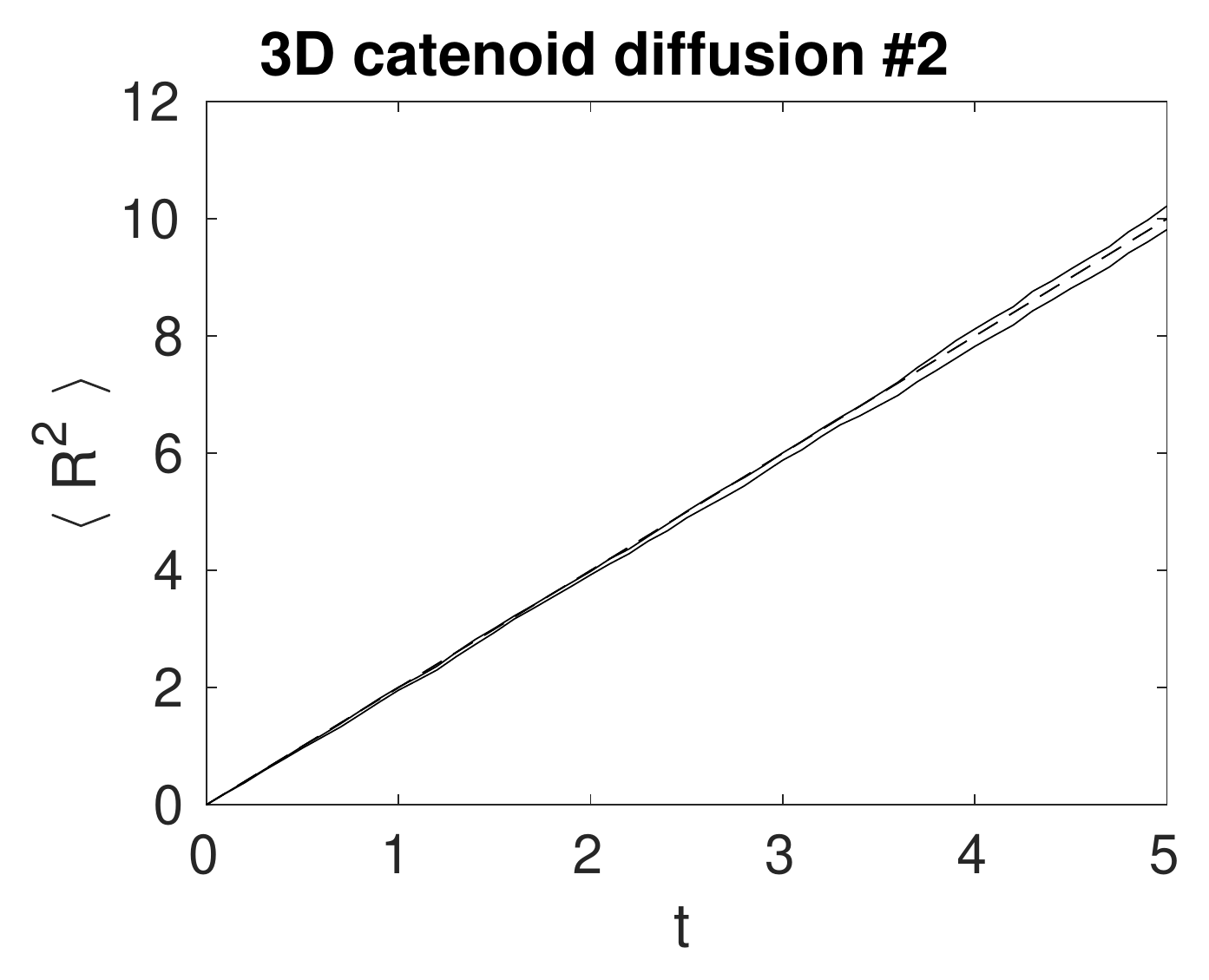}

{}\includegraphics[width=0.75\textwidth]{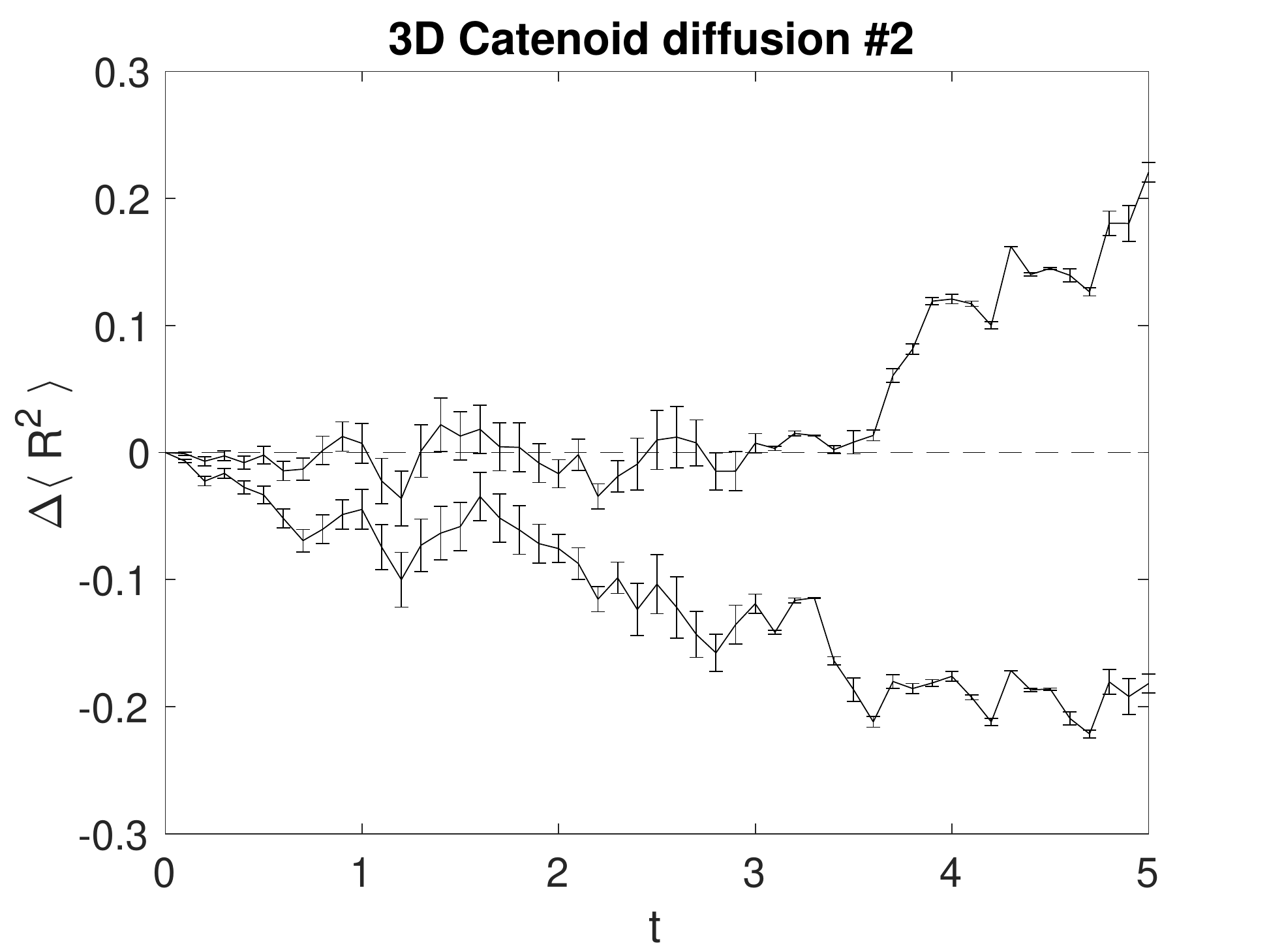}

\caption{\emph{(Top) Tme evolution of the catenoid squared Euclidean diffusion
distance $\left|\mathbf{x}_{0}-\mathbf{x}\left(t\right)\right|^{2}$,
where $\mathbf{x}_{0}=\left(1,0,0\right)^{T}$, as a function of time.
The solid lines are stochastic error bounds. The dashed line is the
theoretical prediction.}\protect \\
\emph{(Bottom) Differences between the distance $\left|\mathbf{x}_{0}-\mathbf{x}\left(t\right)\right|^{2}$
and the exact result. }}
\vspace{10pt}
\end{figure}
\par\end{center}

\pagebreak{}

\section{Cell array coupled SDE}

This solves an SDE with two different variables, one a vector and
another a scalar. The equation is
\begin{eqnarray}
\frac{\partial\mathbf{a}}{\partial t} & = & \left[\begin{array}{c}
2\\
1
\end{array}\right]-a+\left[\begin{array}{cc}
1 & 0\\
0 & 0.5
\end{array}\right]\times\left[\begin{array}{c}
w(1)\\
w(2)
\end{array}\right]\nonumber \\
\frac{\partial b}{\partial t} & = & -b+a\left(1\right)\,.
\end{eqnarray}
The initial condition is given by $\mathbf{a}\left(0\right)=\left(0,2\right)^{T}$,
$b\left(0\right)=2$. Here $\mathbf{w}=\left(w_{1},w_{2}\right)^{T}$
has 2 independent noise variables. The output consists of one graph
of the mean. and a second of the variance of all three variables.
The $b$ variable is effectively driven by colored noise, consisting
of the finite bandwidth $a(1$) variable. The $\bm{a}$ vector has
two white noise inputs of different size.

One can prove the following are the solutions to the stochastic equations:
\begin{align*}
a_{i}\left(t\right) & =e^{-t}\left(a_{i}\left(0\right)+\int_{0}^{t}e^{\tau}\left[e_{i}+w_{i}(\tau)\right]d\tau\right)\\
 & =e^{-t}\left(a_{i}\left(0\right)+e_{i}\left(e^{t}-1\right)+\int_{0}^{t}e^{\tau}w_{i}(\tau)d\tau\right)\\
b\left(t\right) & =e^{-t}\left(2+\int_{0}^{t}\left[2\left(e^{t}-1\right)+\int_{0}^{\tau}e^{\tau'}w_{1}(\tau')d\tau'\right]d\tau\right)\\
 & =e^{-t}\left(2\left(e^{t}-t\right)+\int_{0}^{t}\int_{0}^{\tau}e^{\tau'}w_{1}(\tau')d\tau d\tau'\right)
\end{align*}
Here, $\bm{e}=[2,1]^{T}$ are the two driving terms, and the resulting
mean values are:
\begin{align*}
\bar{a}_{1}\left(t\right) & =2(1-e^{-t})\\
\bar{a}_{2}\left(t\right) & =1+e^{-t}\\
\bar{b}\left(t\right) & =2\left(1-te^{-t}\right)
\end{align*}

Defining the diffusion terms as $\bm{D}=[1,0.25]^{T}$, the two stochastic
equation variances are:
\begin{align*}
\Delta a_{i}^{2}\left(t\right) & =e^{-2t}\int_{0}^{t}\int_{0}^{t}e^{\tau+\tau'}\left\langle w_{i}(\tau)w_{i}(\tau')\right\rangle d\tau d\tau'\\
 & =e^{-2t}\int_{0}^{t}e^{2\tau}d_{i}d\tau=\frac{D_{i}}{2}\left(1-e^{-2t}\right)
\end{align*}

The additional equation variance is more complex. One must solve for
the $a_{1}$ variable, and use this as an external colored noise term
driving the last equation. The result for the variance is as follows:

\begin{align*}
\Delta b^{2}\left(t\right) & =e^{-2t}\int_{0}^{t}\int_{0}^{\tau_{1}}\int_{0}^{t}\int_{0}^{\tau_{2}}e^{\tau_{3}+\tau_{4}}\left\langle w_{1}(\tau_{3})w_{1}(\tau_{4})\right\rangle d\tau_{1}..d\tau_{4}\\
 & =e^{-2t}\int_{0}^{t}d\tau_{1}\int_{0}^{t}d\tau_{2}\left(\int_{0}^{\min\left(\tau_{1},\tau_{2}\right)}e^{2\tau_{3}}d\tau_{3}\right)\\
 & =\frac{1}{2}e^{-2t}\int_{0}^{t}d\tau\left(\left(e^{2\tau}-1\right)\left(t+\frac{1}{2}\right)-e^{2\tau}\tau\right)\\
 & =\frac{1}{4}\left[1-\left(1+2t+2t^{2}\right)e^{-2t}\right]
\end{align*}

~~
\begin{center}
\noindent\doublebox{\begin{minipage}[t]{1\columnwidth - 2\fboxsep - 7.5\fboxrule - 1pt}%
\texttt{function e = Cellarraysde()}

\texttt{p.fields = \{2,1\}; }

\texttt{p.noises = 2;}

\texttt{p.initial = \{@(u,v,p) {[}0;2{]},@(u,v,p) 2;\};}

\texttt{p.ensembles = {[}100,100{]};}

\texttt{p.deriv\{1\} = @(a,b,w,p) {[}2;1{]} - a + {[}1,0;0,0.5{]}{*}w;}

\texttt{p.deriv\{2\} = @(a,b,w,p) - b + a(1,:);}

\texttt{p.observe = \{@(a,b,p) {[}a;b{]},@(a,b,p) {[}a.\textasciicircum 2;b.\textasciicircum 2{]}\};}

\texttt{p.output\{2\} = @(o,p) o\{2\} - o\{1\}.\textasciicircum 2;}

\texttt{p.compare\{1\} = @(p) {[}2{*}(1-exp(-p.t));1+exp(-p.t);2{*}(1-p.t.{*}exp(-p.t)){]};}

\texttt{p.compare\{2\} = @(p) {[}0.5{*}(1-exp(-2{*}p.t));...}

\texttt{0.125{*}(1-exp(-2{*}p.t));0.25{*}(1(1+2{*}p.t+2{*}p.t.\textasciicircum 2).{*}exp(-2{*}p.t)){]};}

\texttt{p.olabels\{1\} = '\textless a(i)\textgreater , \textless b\textgreater ';}

\texttt{p.olabels\{2\} = '\textless{[}\textbackslash Delta a(i){]}\textasciicircum 2\textgreater ,
\textless{[}\textbackslash Delta b{]}\textasciicircum 2\textgreater ';}

\texttt{e = xspde(p);}%
\end{minipage}} 
\par\end{center}

~~

\paragraph{Notes}
\begin{itemize}
\item This uses cell arrays to define vector and scalar variables
\item Cell arrays of initial and deriv functions are also required
\item The output is first an observe average, then an output function.
\item Two vector compare functions are used to check results
\item The RMS errors in the observables are reported as a normalised error,
compared to the maximum value of the corresponding output. See Section
\ref{sec:Error-outputs} for explanations. 
\item RMS errors are: Step=0.000518 Samp=0.00645 Diff=0.00896 ~
\end{itemize}
~
\begin{center}
\begin{figure}[H]
\centering{}\includegraphics[width=0.75\textwidth]{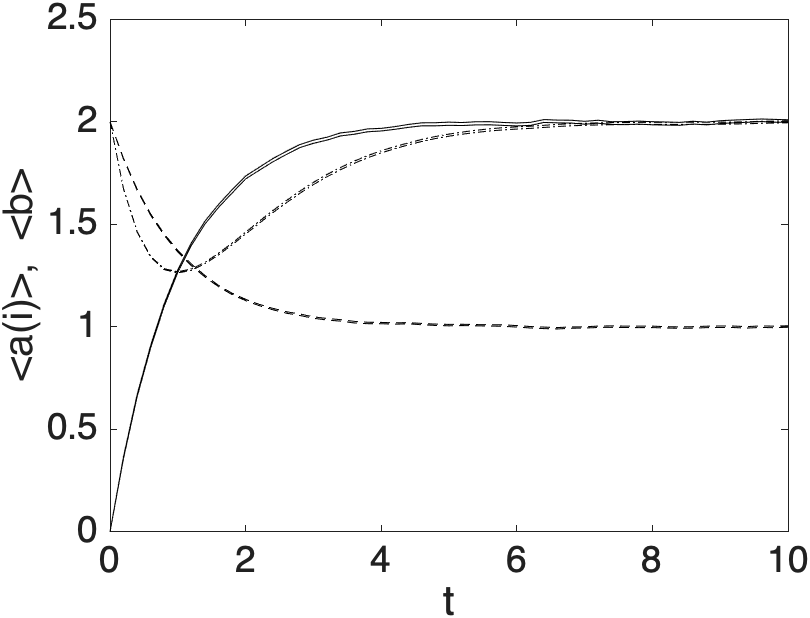}

~~

\includegraphics[width=0.75\textwidth]{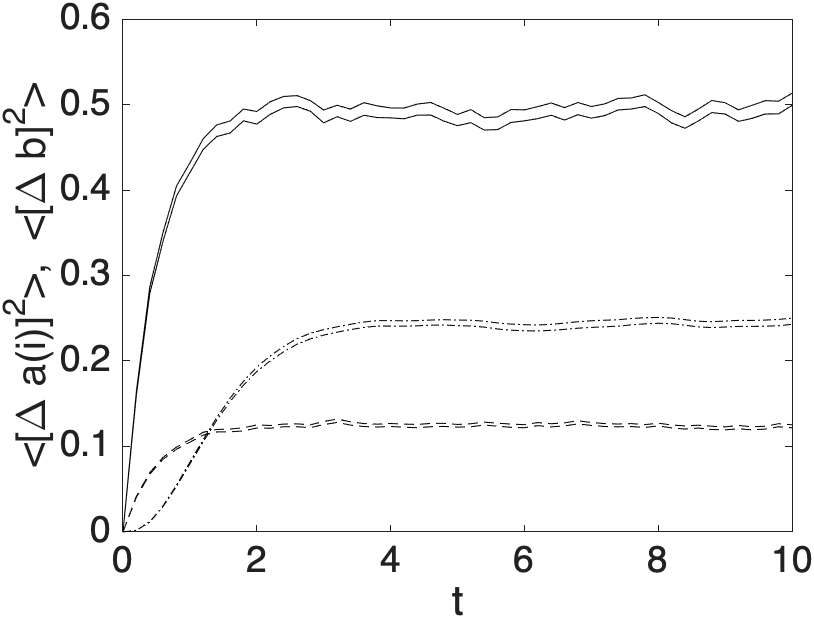}

\caption{\emph{Time evolution of the mean (top) and variance (bottom) of each
variable. The two lines indicate the stochastic error bounds. The
solid lines are $a(1)$, the dashed lines are $a(2)$, the dot-dash
lines are $b$. For clarity, the comparison lines are not graphed.
\label{fig:CellarraySDE-means}}}
\vspace{10pt}
\end{figure}
\par\end{center}

\newpage{}

\part{Stochastic partial differential equations\label{part:Stochastic-partial-differential}}

\chapter{SPDE toolbox \label{sec:Simulating-an-SPDE}}

\textbf{This chapter describes how to simulate a PDE or SPDE, including
choosing spectral or finite difference methods and specifying boundary
conditions. For theoretical background, see Chapter (\ref{chap:SPDE Theory}).
For detailed examples, see Chapter (\ref{chap:SPDE-examples}).}

\section{SPDE parameters}

A stochastic partial differential equation or $SPDE$ for a complex
vector field is defined in both time $t$ and space dimension(s) $\mathbf{x}$.
The total dimensions includes both time and space. To solve a stochastic
partial differential equation xSPDE involves a similar procedure to
the case of the SDE, covered in section \ref{chap:Simulating-an-SDE}.

The numerical solutions require additional parameters to define the
spatial grid, and to define the linear transformations in an interaction
picture, if spectral methods are used. The SPDE input parameters extend
those already introduced in (\ref{subsec:Simulation-parameters}).
Some new and extended parameters are listed in the table below: 
\begin{center}
\begin{tabular}{|c|c|c|c|}
\hline 
Label & Type & Typical value & Description\tabularnewline
\hline 
\hline 
dimensions & integer & $2$ & Space-time dimensions\tabularnewline
\hline 
linear\{c\} & function & @(p) p.Dx & Linear interaction picture function\tabularnewline
\hline 
ranges & real vector & {[}10,10,...{]} & Ranges in time and space\tabularnewline
\hline 
transforms\{c,d\} & integer vector & $[1,0,1,..]$ & Space-time transform switch\tabularnewline
\hline 
points\{c\} & integer vector & {[}51,35,..{]} & Output lattice points in {[}t,x,y,z,..{]}\tabularnewline
\hline 
\textit{origins} & real vector & {[}0,-5,..{]} & Space-time integration origin\tabularnewline
\hline 
boundaries\{c,d\} & integer array & $[0,0;0,0;..]$ & Boundary type per field index\tabularnewline
\hline 
boundval\{c,d\} & cell array & $\{0,0;0,0;..\}$ & Boundary value per field index\tabularnewline
\hline 
boundfun & function & @(a,c,d,p) ... & Boundary value function\tabularnewline
\hline 
\end{tabular}
\par\end{center}

Setting $dimensions>1$ defines an (S)PDE as opposed to an ordinary
(S)DE. In the xSPDE implementation, the space-time dimensions are
unlimited, but large space-time dimensions become memory-intensive
and slow. There is a practical limit of less than ten space-time dimensions
with current digital computers, owing to exponential growth of memory
and corresponding CPU time requirements at large space dimensionality.

The cell index $c$ can be omitted in cell arguments like boundval\{c,d\}
if there is only one field cell. Using boundval will only specify
boundary values that are static in time. These can be any combinations
of Dirichlet and/or Neumann/Robin. Using boundfun allows boundary
values that can vary in time or are dynamic functions of the field
cells. Definitions of boundfun have four arguments, with the first
one a field cell array: see (\ref{sec:xSIM-parameters}). 

\section{Multidimensional Wiener process}

To solve for a single four-dimensional trajectory with three space
dimensions, as in Eq \eqref{eq:Wiener_process-1} , just type in: 
\begin{center}
\doublebox{\begin{minipage}[t]{0.75\columnwidth}%
\texttt{p.dimensions = 4;}

\texttt{p.deriv = @(a,w,p) w;}

\texttt{xspde(p);}%
\end{minipage}} 
\par\end{center}

Here $p.deriv$ defines the time derivative $\dot{a}$ in the input
parameter structure p, while $w$ is a delta-correlated Gaussian noise
generated internally. Apart from the dimensions, there are no other
parameters, so default values are used. This produces the graph shown
in Fig (\ref{fig:The-simplest-case: Wiener-1}), which gives a single
trajectory using the default lattice settings. 
\begin{center}
\begin{figure}[H]
\centering{}\includegraphics[width=0.75\textwidth]{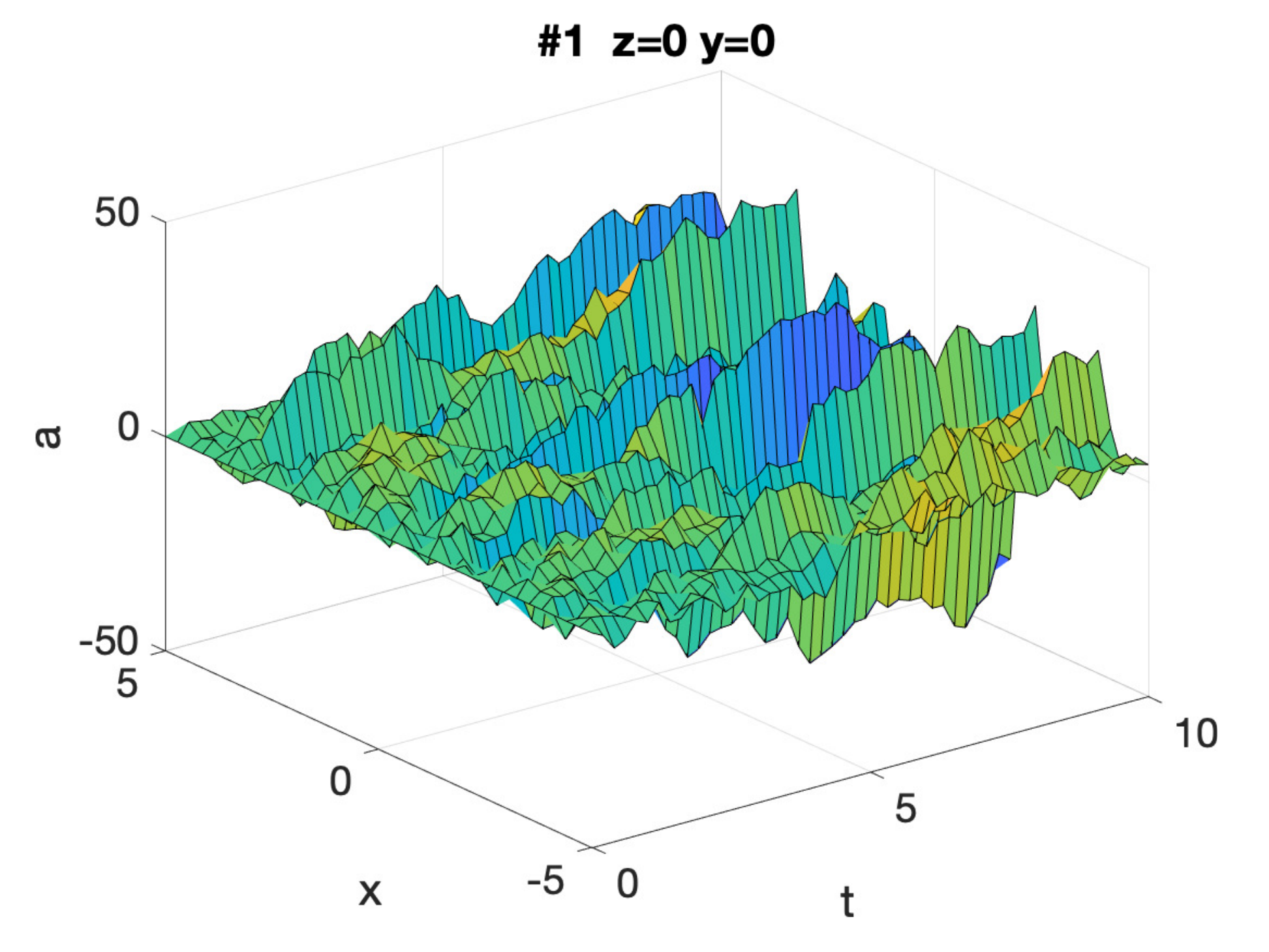}

\caption{\label{fig:The-simplest-case: Wiener-1}\emph{A multidimensional random
walk of a three-dimensional field projected onto $y=z=0$.}}
\vspace{10pt}
\end{figure}
\par\end{center}

For more interesting problems than this, more parameters are needed,
as explained next.

\subsection{Initial conditions}

Initial conditions in the previous example have their default value
of zero. For other values, they are set at the initial time of $t=O_{1}$
with a user-defined function p.initial, so that: 
\begin{equation}
a(O_{1})=initial(v,p)
\end{equation}
The initial function includes initial random fields $v=\left[v^{x},v^{k}\right]$.
Their correlations are either delta correlated or spatially correlated.
To allow this, the input parameter $randoms$ is a vector such that:
$randoms(1)$ is the number of delta-correlated random fields, $v^{x}$,
and $randoms(2)$ is the number of correlated random fields, $v^{k}$.
All random fields in the initial function, even if correlated using
filters in momentum space, are transformed to position space before
use. If there is no filtering, $v^{x}$ and $v^{k}$ have the same
correlations.

\subsection{Additional damping term}

As another very simple example, consider the SPDE

\begin{eqnarray}
\frac{\partial a}{\partial t} & = & -\frac{1}{4}a+x\cdot w\label{eq:simple_spde_example}
\end{eqnarray}

The system has one spatial dimension, or $d=2$ space-time dimensions,
one field and one noise variable. We suppose that the initial noise
variance is Gaussian, with: 
\begin{equation}
a(0,x)=10v(x).
\end{equation}
We want to consider $10,000$ stochastic trajectories per sub-ensemble
with$10$ sub-ensembles. We will set the origin for $x$ to $0$.
The variable $a$ will be initialized as delta-correlated in space
with a gaussian standard deviation on the lattice of $\sigma=10/\sqrt{\Delta V}$.
As our observable, we consider the second moment of $a$.

This is simulated through the following xSPDE code: 
\begin{center}
\doublebox{\begin{minipage}[t]{0.75\columnwidth}%
\texttt{clear;}

\texttt{p.name = 'simple SPDE';}

\texttt{p.dimensions = 2;}

\texttt{p.ensembles = {[}10000,10{]};}

\texttt{p.origins = {[}0,0{]};}

\texttt{p.noises = 1;}

\texttt{p.initial = @(v,p) 10{*}v;}

\texttt{p.observe = @(a,\textasciitilde ) a.\textasciicircum 2;}

\texttt{p.olabels = '\textless a\textasciicircum 2\textgreater
';}

\texttt{p.deriv = @(a,w,p) -0.25{*}a + p.x .{*} w;}

\texttt{xspde(p);}%
\end{minipage}} 
\par\end{center}

With this input, Matlab produces two output graphs:

\begin{figure}
\includegraphics[width=0.5\textwidth]{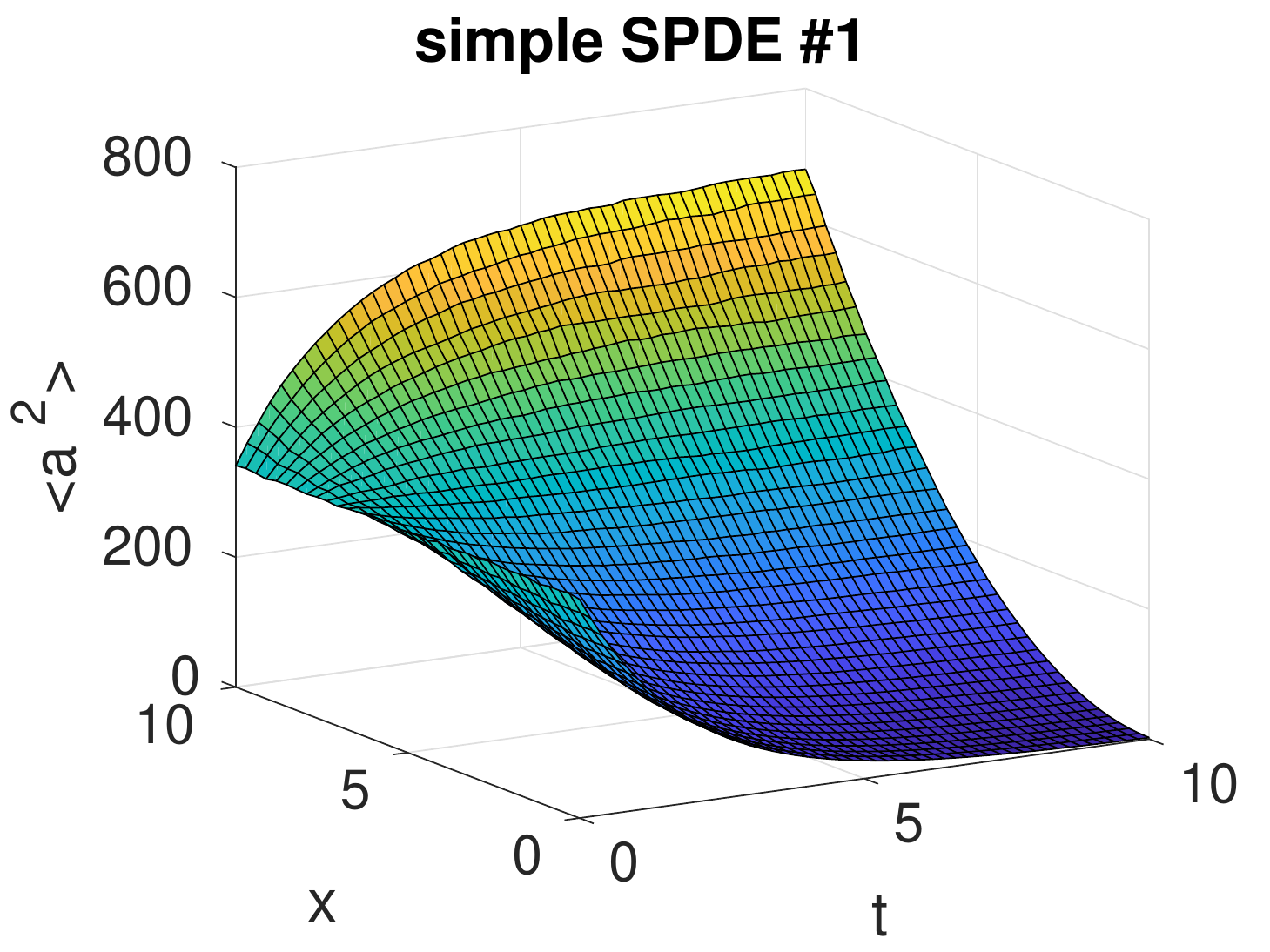}\includegraphics[width=0.5\textwidth]{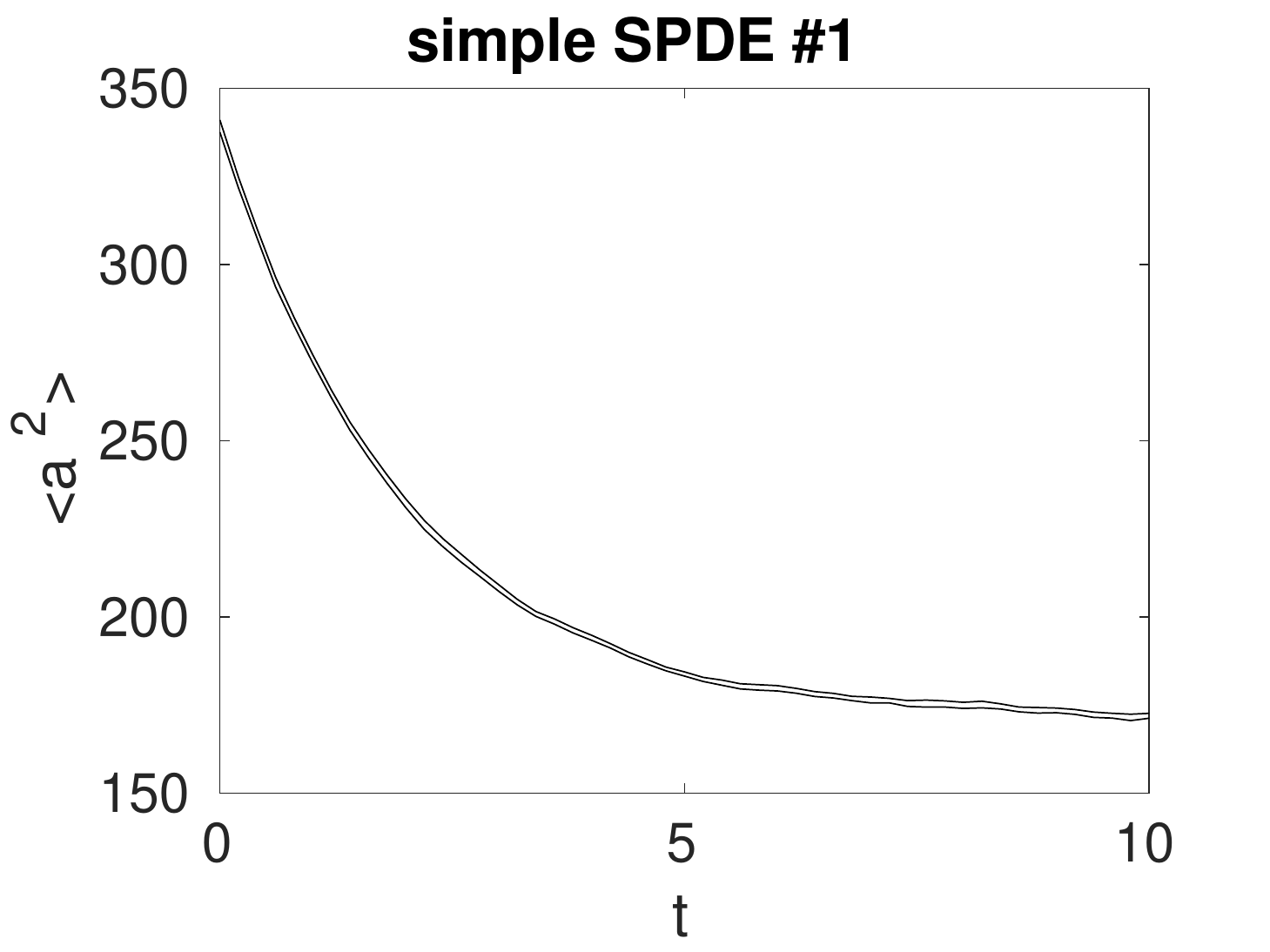}

\caption{\emph{Example: simple SPDE output graphs.}}
\vspace{10pt}
\end{figure}

The second graph shows the time evolution for $x$ at the mid-point,
$x=5$.\textbf{{} }The variances are larger than they would be in
the SDE case, where one might expect an initial variance of $\left\langle a^{2}(0)\right\rangle =100$.
The reason for this is that the initial and propagating random noise
fields are replaced by a lattice of noise terms with a variance of
$1/\Delta V$. This causes an increase in each local noise.

\section{Differential operators }

xSPDE has finite difference and spectral methods for direct differentiation.
These derivatives are obtained through function calls $D1$ and $D2$
respectively for first and second derivatives, which use a fixed grid
spacing. As elsewhere, they can be replaced by user-written functions
if preferred. Generally they require smaller steps in time than spectral
methods, when used to define the derivative.

\subsection{Finite difference first derivatives}

The code to take a first order spatial derivative with finite difference
methods is carried out using the xSPDE function D1() with arguments
(o, {[}d, c, ind ,{]} p).

This takes a scalar or vector o and returns a first derivative in
an axis direction d. Set d = 2 for an x-derivative, d = 3 for a y-derivative,
and so on. Time derivatives are ignored at present. Derivatives are
returned at all lattice locations.

If the direction $d$ is omitted, an x-derivative is returned. The
next optional input is $c$, the cell index, which is needed to identify
the boundary conditions. If the cell index is omitted, $c=1$ is assumed.
Finally ind, which is a vector of one or more field indices can be
input. If omitted, all indices are differentiated. 

These derivatives can be used both in calculating propagation and
in calculating observables. The boundary condition is set by the boundaries
input. Any boundary of any dimension, cell or index can be made periodic,
which is the default, or Neumann, or Robin/Dirichlet. Boundary values
and/or a boundary function can also be input, as described in the
next subsection.

\subsection{Finite difference second derivatives}

The code to take a second order spatial derivative with finite difference
methods is carried out using the xSPDE D2 function with arguments
(o, {[}d, c, ind ,{]} p).

This takes a scalar or vector o and returns the second derivative
in axis direction d. Set d = 2 for an x-derivative, d = 3 for a y-derivative
and so on. All other properties are exactly the same as D1.

Without using the interaction picture, the stochastic equation of
Eq \eqref{eq:Nonlinear-SPDE-example} is specified in xSPDE using
finite differences as 
\begin{center}
\doublebox{\begin{minipage}[t]{0.75\columnwidth}%
\texttt{p.dimensions = 3;}

\texttt{p.steps = 50;}

\texttt{p.deriv = @(a,w,p) D2(a,2,p)+D2(a,3,p)+a - a.\textasciicircum 3
+...}

\texttt{w/10;}

\texttt{xspde(p);}%
\end{minipage}} 
\par\end{center}

\begin{figure}
\includegraphics[width=0.5\textwidth]{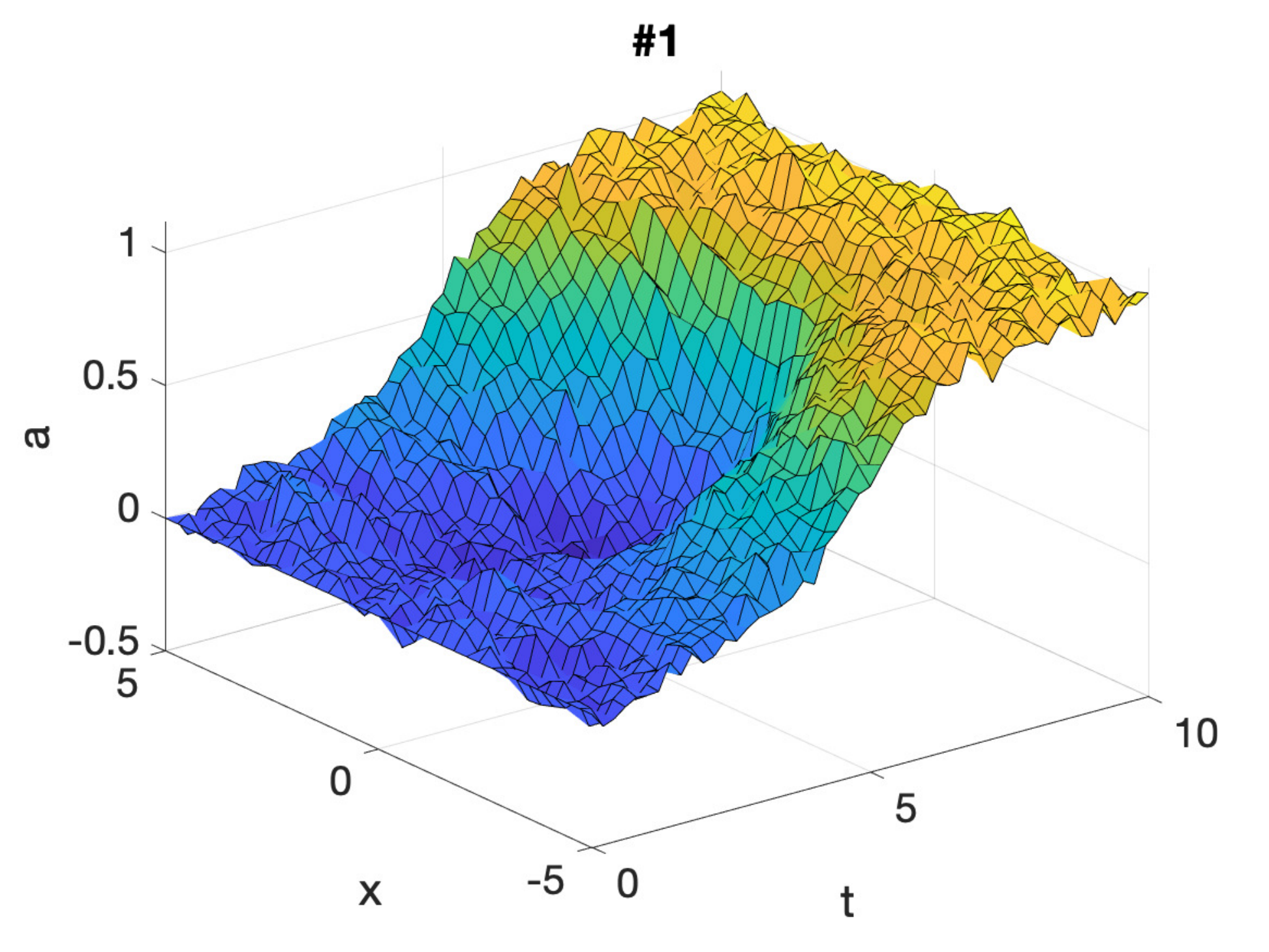}\includegraphics[width=0.5\textwidth]{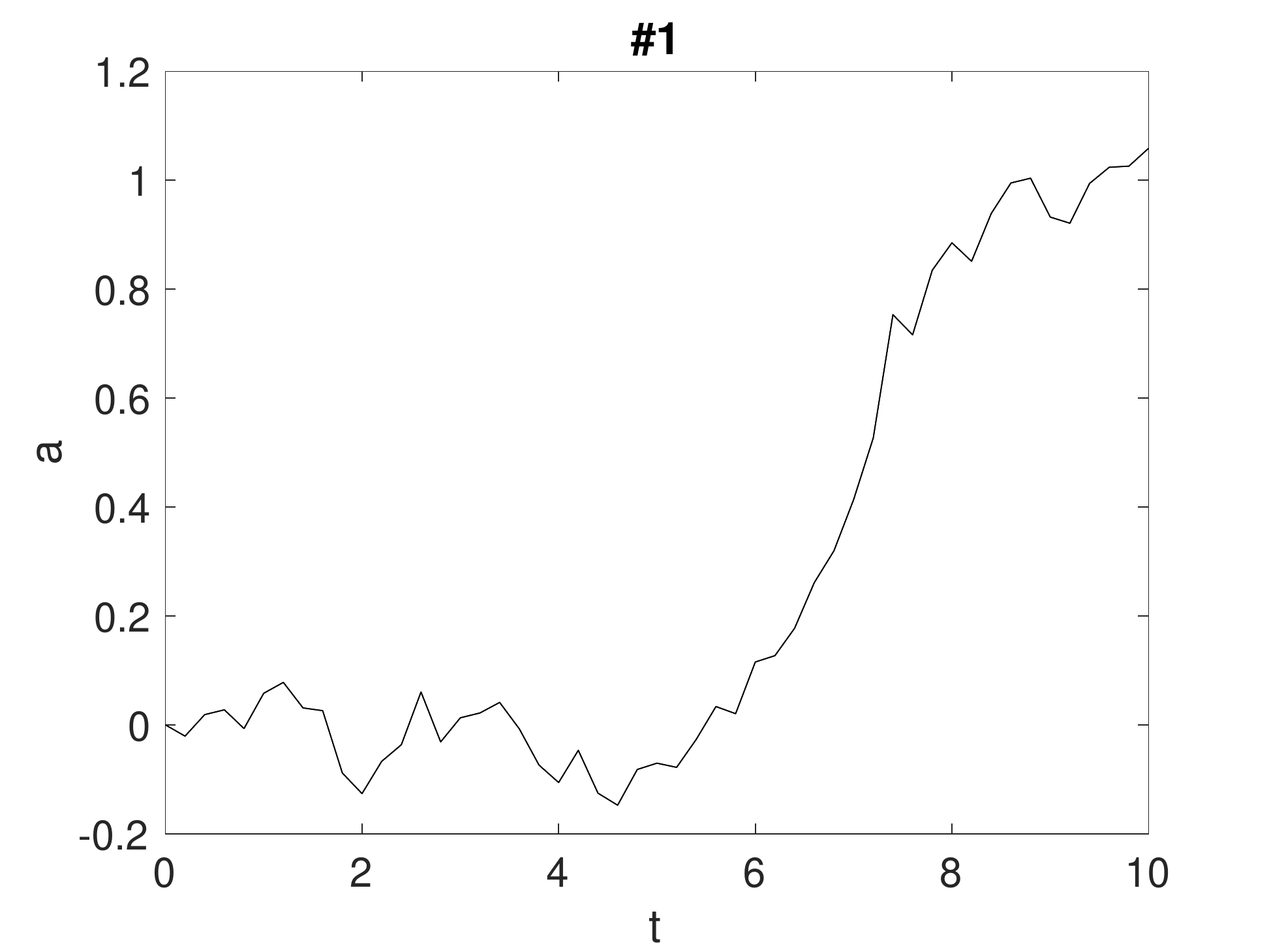}

\caption{\emph{Two space-dimensional example graphs, direct differentiation}.\label{fig:Two-space-dimensional-example-direct-diff}}
\vspace{10pt}
\end{figure}

This gives the same result as with the linear propagator, although
requiring smaller step-sizes for numerical stability, with an output
graph shown in Fig (\ref{fig:Two-space-dimensional-example-direct-diff}).
Note that the parameters and noises are slightly different!

\subsection{Spectral derivatives}

The code to take n-th order spatial derivative with spectral methods
is carried out using the xSPDE DS function with arguments DS(o, {[}n,
d, c, ind ,{]} p).

This takes a scalar or vector o and returns the n-th derivative in
direction d. If $n$ is omitted, a first spectral derivative is assumed.
Set d = 2 for an x-derivative, d = 3 for a y-derivative and so on.
All other properties are the same as D1,D2. These operators can only
be used for propagation, so that they act on a field.

The spectral methods used in xSPDE use Fourier and trigonometric methods,
as described next.

\section{Spectral propagators}

Using a linear spectral propagator in an SPDE can give gives better
accuracy, and allows use of the interaction picture. This is included
for all built-in xSPDE algorithms, provided the linear function is
defined in the parameter structure. Variables $p.D\{i\}$ (with placeholders
$p.Dx,p.Dy,p.Dz$ for the first 3 spatial dimensions) provide access
to the operator. Higher-order derivatives are found through potentiating
$p.Dx$ accordingly.

For example, the $2$-dimensional Laplacian operator 
\begin{equation}
\nabla^{2}=\frac{\partial^{2}}{\partial x^{2}}+\frac{\partial^{2}}{\partial y^{2}}
\end{equation}
corresponds to a linear differential operator specified as: 
\begin{equation}
p.linear=@(p)\,\,\,p.Dx.^{2}+p.Dy^{2};
\end{equation}
For a comprehensive list of variables accessible through the $p$-structure,
refer to sec. \ref{sec:Table-of-parameters}.

As explained in section \ref{sec:IP-implementation}, the general
equation solved can be written in differential form as

\begin{equation}
\frac{\partial\mathbf{a}}{\partial t}=\mathbf{A}\left[\mathbf{a}\right]+\underline{\mathbf{B}}\left[\mathbf{a}\right]\cdot\mathbf{w}(t)+\underline{\mathbf{L}}\left[\mathbf{\nabla},\mathbf{a}\right]\,.
\end{equation}

The linear function $L$ can be input either inside the derivative
function using finite difference operators described below, or as
a separate linear function, to allow for an interaction picture in
which case: 
\begin{equation}
\underline{\mathbf{L}}\left[\mathbf{\nabla},\mathbf{a}\right]=\,\underline{\mathbf{L}}\left[\mathbf{\nabla}\right]\mathbf{a}\,.
\end{equation}
This depends on momentum space coordinates, which involves Fourier
transforms so that no space dependence is allowed. It is also possible
to use finite differences, in which case the derivative terms are
included as part of the derivative function deriv.

The usual FFT spectral methods require periodicity, and may have either
even or odd linear derivatives. The four other boundary combinations
must be used with an interaction picture derivative that only has
even powers of linear derivatives. Odd derivatives and nonlinear derivative
terms can also be included by using finite difference derivatives
in the deriv functions.

The field $p.x$ is provided by the parameter structure, and corresponds
to the variable $x$ in Eq \eqref{eq:simple_spde_example}. All parameters
are preceded by the structure label, p. For two or three space dimensional
problems, $x,y,z$ are placeholders for $r\{2\},r\{3\},r\{4\}$, and
spatial variables of even higher dimensional problems can be accessed
through $r\{n\}$.

Note that where numerical labels or indices are used, the convention
is that time is the first dimension.

\subsection{One space-dimensional example}

A famous partial differential equation is an exactly soluble equation
for a soliton, the nonlinear Schrödinger equation (NLSE): 
\begin{equation}
\frac{da}{dt}=\frac{i}{2}\left[\nabla^{2}a-a\right]+ia\left|a\right|^{2}.
\end{equation}

Together with the initial condition that $a(0,x)=sech(x)$, this has
a soliton, an exact solution that doesn't change in time: 
\begin{eqnarray}
a(t,x) & = & sech(x).
\end{eqnarray}
The spatial integral is simply: 
\begin{eqnarray}
\int sech(x)dx & = & \pi.
\end{eqnarray}

An xSPDE code that solves this using periodic boundary conditions
is given below, together with code that compares the numerical solution
with the exact solutions for the soliton and the integral: 
\begin{center}
\doublebox{\begin{minipage}[t]{0.75\columnwidth}%
\texttt{p.name = 'NLS soliton';}

\texttt{p.dimensions = 2;}

\texttt{p.initial = @(v,p) sech(p.x);}

\texttt{p.deriv = @(a,\textasciitilde ,p) 1i{*}a.{*}(conj(a).{*}a);}

\texttt{p.linear = @(p) 0.5{*}1i{*}(p.Dx.\textasciicircum 2-1.0);}

\texttt{p.olabels = \{'a(x)','\textbackslash int a(x) dx'\};}

\texttt{p.observe\{2\} = @(a,p) Int(a, p);}

\texttt{p.compare\{1\} = @(p) sech(p.x);}

\texttt{p.compare\{2\} = @(p) pi;}

\texttt{e = xspde(p);}%
\end{minipage}} 
\par\end{center}

Due to finite boundaries and discrete spatial lattice, the agreement
is not perfect. The errors can be reduced by increasing the range
of the integration domain and improving the resolution with more points.

\subsection{Two space-dimensional example}

As another example, consider the two-dimensional nonlinear stochastic
equation, with periodic boundary conditions:

\begin{eqnarray}
\frac{\partial a}{\partial t} & = & \nabla^{2}a\left(\mathbf{x},t\right)+a\left(\mathbf{x},t\right)-a\left(\mathbf{x},t\right)^{3}+\eta\left(\mathbf{x},t\right).\label{eq:Nonlinear-SPDE-example}
\end{eqnarray}

Using the interaction picture allows for the absorption of both the
Laplacian and the first-order term by the \textit{p.linear} parameter,
which results in 
\begin{center}
\doublebox{\begin{minipage}[t]{0.75\columnwidth}%
\texttt{...}

\texttt{p.linear = @(p) (p.Dx.\textasciicircum 2+p.Dy.\textasciicircum 2)
+ 1;}

\texttt{p.deriv = @(a,w,\textasciitilde ) -a.\textasciicircum 3
+ w;}

\texttt{xspde(p);}%
\end{minipage}} 
\par\end{center}

With this input, Matlab produces two output graphs as shown in Fig
(\ref{fig:Two-space-dimensional-example}):

\begin{figure}
\includegraphics[width=0.5\textwidth]{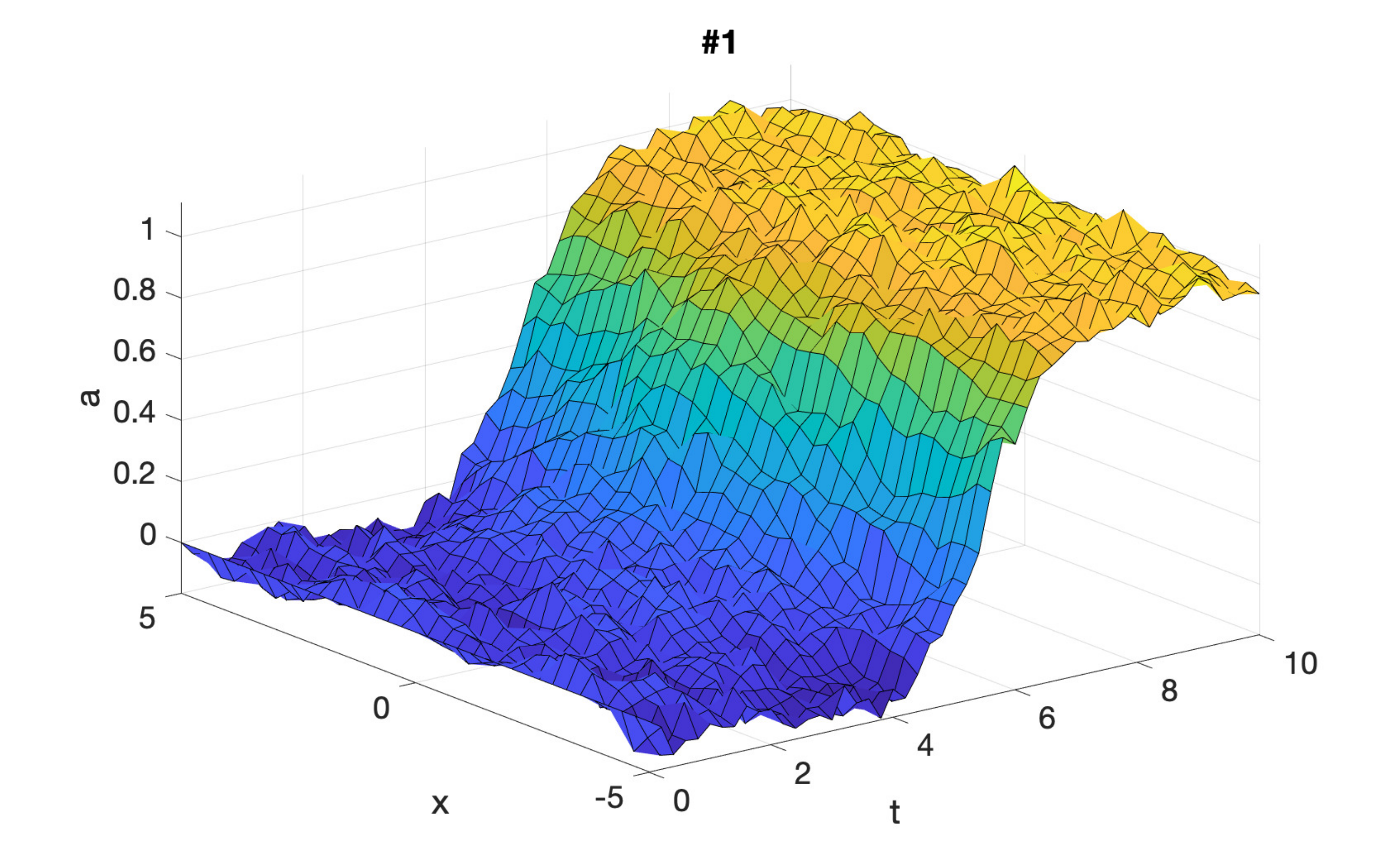}\includegraphics[width=0.5\textwidth]{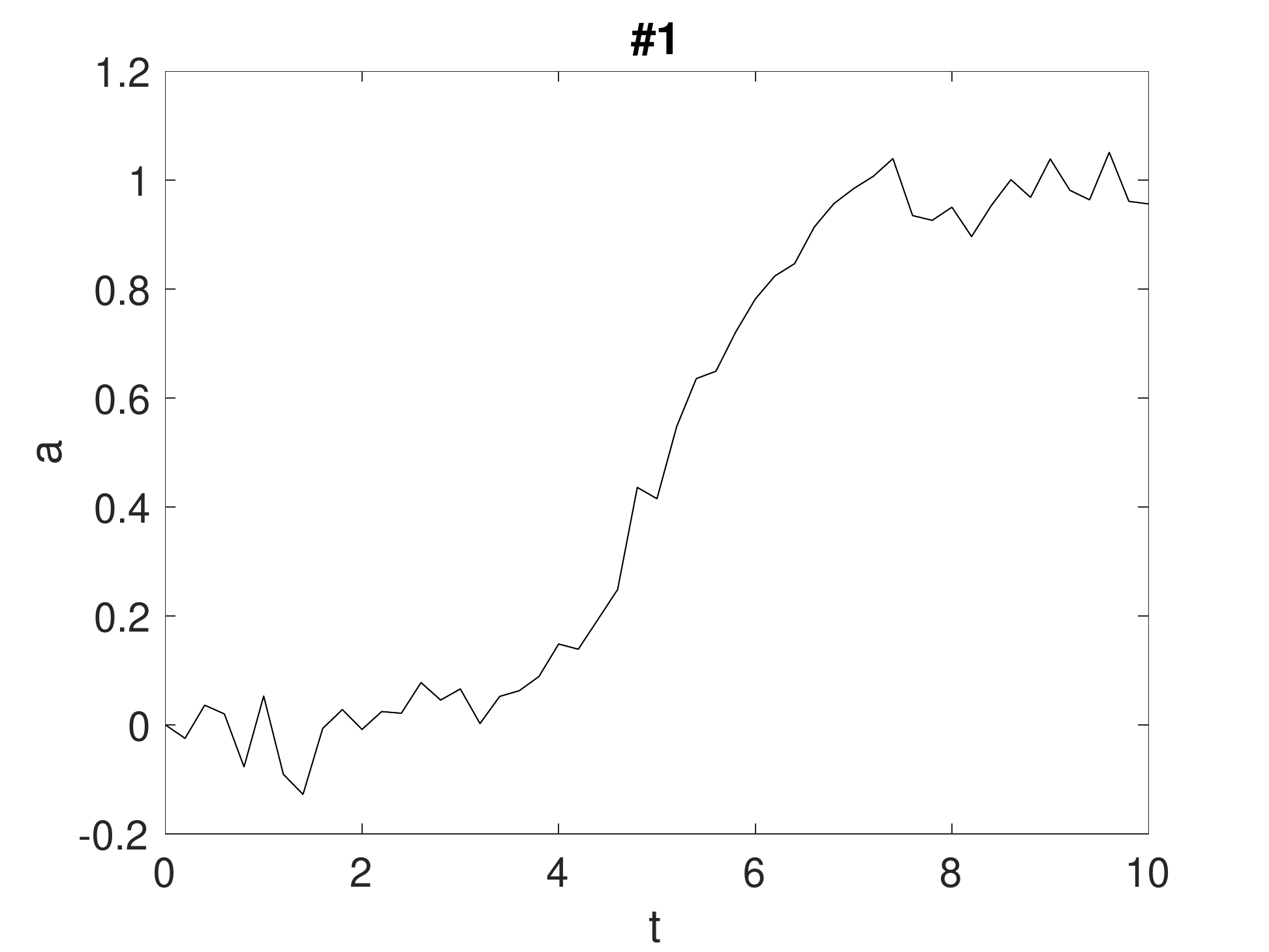}

\caption{\emph{Two space-dimensional example graphs.\label{fig:Two-space-dimensional-example}}}
\vspace{10pt}
\end{figure}

\section{Transverse lattice}

\subsection{SPDE spatial lattice}

Stochastic fields in an SPDE are stored in a cell array that can include
one or more real or complex arrays, $a(f,\mathbf{i},e)$. Here $f$
is the internal field index, $\bm{i}$ is a $d-1$ dimensional spatial
lattice index for d space-time dimensions, and $e$ is the ensemble
index. For a cell array of multiple fields, $a,b,c$ .., these must
each have either the same number of points or one point per dimension.
When specifying the spatial lattice, one must define: 
\begin{description}
\item [{dimensions}] The dimensionality in time and space. The default
is an SDE: $d=1$. 
\item [{points}] The number of integration points. The first cell default
points is $\mathbf{N}=[51,35,35..]$. This can have a cell index if
there are multiple field cells, provided the indices are compatible
for broadcasting, eg $points=\{[35,25],[35,1]\}$, to allow different
dimensions to be input. After the first or base cell dimension, the
default is the previous cell dimension, with ones in the higher dimensions
unless a value is already present in the previous cell.
\item [{steps}] The number of intermediate steps per plotted point. The
default is $\mathbf{N}=[1,1,1..]$. This is the same for all cells,
and can modify any space-time dimension. 
\item [{ranges}] The integration ranges in each dimension. The default
is $\mathbf{R}=[10,10,10..]$. This is the same for all cells.
\item [{origins}] The origins of the space-time integration domains. By
default, the origin is $O\left(1\right)=0$ for the time coordinate
and $\mathbf{O}=-\mathbf{R}/2$ for the space coordinates ($\mathbf{R}$
is the $ranges$ variable) such that the spatial grid is symmetric
around $\mathbf{r}=0$. 
\end{description}
There is a restriction, which is that the first cell must have the
maximum number of points in each dimension. 

The spatial points in the n-th dimension are at $r\{n\}=O(n),O(n)+dx(n),\ldots O(n)+R(n)$,
so the spacing for $N(n)$ points is $dx(n)=R(n)/\left(N(n)-1\right)$.
For periodic boundaries, the boundaries where fields and derivatives
are equal are at $O(n)-dx(n)/2$ and $O(n)+R(n)+dx(n)/2$ . For other
boundary types, the boundary values are defined at the first and last
points.

\subsection{Field indices}

In the functions $deriv$, $initial$ and $observe$, the field and
noise variables $a$ and $w$ have extended dimensionality compared
to the $1$-dimensional case, to index the transverse lattice. The
indices are $a\left(f,\mathbf{i},e\right)$, where the: 
\begin{description}
\item [{field}] index $f$ corresponds to the field index for $a$ and
the noise index for $w$. 
\item [{intermediate}] indices $\mathbf{i},$ which are absent in the $1$-dimensional
case, correspond to the spatial grid and have the same structure.
For example, in the case with dimensions = 3, indicating one time
index and two spatial dimension, $\mathbf{i}$ corresponds to the
two space indices. 
\item [{last}] index $e$ corresponds to the stochastic trajectory. 
\end{description}
For storing space coordinates like $p.x$, the first and last index
are $f=e=1$. Where Fourier transforms are used internally, the momentum
arrays have zero momentum as the first index to follow standard discrete
Fourier transform conventions. This is changed to a symmetric convention
in all stored graphics data outputs that are functions of momentum
space.

\subsection{Integrals and averages}

\label{xsim:averages} There are functions available in xSPDE for
spatial grid averages and integrals, to handle the spatial grid. These
are \textbf{Ave} and \textbf{Int,} which are used to calculate observables
for plotting. They operate in parallel over the lattice dimensions,
by taking a vector or scalar quantity, for example a single field
component, and returning an average or a space integral. In each case
the first argument is the field, the second argument is a vector defining
the type of operation, and the last argument is the parameter structure.
If there are two arguments, the operation vector is replaced by its
default value.

\paragraph*{Int(o, {[}dx, {]} p)}

Integrates over the spatial grid to allow calculation of global quantities.
To take an integral over the spatial grid, use the xSPDE function
Int with arguments (o, {[}dx, {]} p). 

This function takes a scalar or vector quantity o, and returns a trapezoidal
space integral over selected dimensions with vector measure dx. If
$dx(j)>0$ an integral is taken over dimension j. Dimensions are labelled
from j = 1,2,3 ... as in all xSPDE standards. Time integrals are ignored
at present. To integrate over an entire lattice, set dx = p.dx, which
is the default value if dx is omitted, otherwise set dx(j) = p.dx(j)
for selected dimensions j.

If momentum-space integrals are needed, first use the transforms switch
to make sure that the field is Fourier transformed before being averaged,
and \textbf{always} input dk instead of dx. 

\textbf{Warning: if dk is omitted, xSPDE assumes that you want to
use the default of dx! This will generate incorrect results for momentum
integrals.}

\paragraph*{Ave(o, {[}av, {]} p)}

Spatial grid averages can be used to obtain stochastic results with
reduced sampling errors if the overall grid is homogeneous. An average
is carried out using the built in xSPDE function Ave() with arguments
(o, {[}av, {]} p).

This takes a vector or scalar field or observable, defined on the
lattice, and returns an average over the spatial lattice. The input
is a field a or observable o, and an optional averaging switch av.
If $av(j)>0$, an average is taken over dimension j. Space dimensions
are labelled from j = 2,3... as elsewhere. If the av vector is omitted,
the average is taken over all space directions.

\section{Boundary conditions}

\subsection{Transverse boundary types}

Transverse boundary conditions must be given for all partial differential
equations. Common transverse boundary types are of three types: Neumann
(specified derivative), periodic, or Dirichlet (specified field).
These are obtained using $boundaries\{c,d\}=-1,0,1$, which is specified
for each cell $c$, space dimension $d>1$, field index $i$ and boundary
$j$.

If boundaries are omitted for any dimension the default is $0$, which
gives periodic boundaries in that dimension for all field indices,
and permits the use of Fourier transforms and an interaction picture
as described above.

The value of $boundaries\{c,d\}$ is a matrix whose column index $(i)$
is the field index, and whose row index (j) is given by $j=1,2$ for
the lower and upper boundary type respectively.

Spatial derivatives or other functions linking different spatial points
can be specified either in the functionals $\boldsymbol{A}\left[\mathbf{a},\mathbf{r}\right]$,
$\underline{\mathbf{B}}\left[\mathbf{a},\mathbf{r}\right]$ or else
in the $linear$ function, provided the derivative terms are linear
functions of the fields. Use of the $linear$ function allows an interaction
picture algorithm, with increased efficiency. The $linear$ function
is available with all boundary conditions, but works best in periodic
cases.

The default boundary conditions are periodic. The implicit setting
of this is that periodicity is enforced such that $a\left(o_{i}-dx_{i}/2\right)=a\left(o_{i}+r_{i}+dx_{i}/2\right)$
, which is the usual discrete Fourier transform requirement.

Otherwise, the differential equation boundaries are specified at $a\left(o_{i}\right)$,
$a\left(o_{i}+r_{i}\right)$, using the cell-array input $boundaries\{c,d\}(i,j)$,
which is defined per field cell, space dimension ($d=2,3..$), field
index ($i=1,2..$) and boundary $j=(1,2)$. Here $d>1$ is the transverse
dimension, not including time, which only has an initial condition.

In summary the available boundary types are: 
\begin{description}
\item [{Neumann:}] For specified derivative boundaries, $boundaries\{c,d\}(i,j)=-1$ 
\item [{Periodic:}] For periodic boundaries, $boundaries\{c,d\}(i,j)=0$ 
\item [{Dirichlet:}] For specified field boundaries, $boundaries\{c,d\}(i,j)=1$ 
\end{description}
These are specified in a cell array: $boundaries\{c,d\}(i,1)$ sets
the lower boundary type in dimension d for the i-th field component,
while $boundaries\{c,d\}(i,2)$ gives the upper boundary type. Each
space dimension, variable and boundary is set independently. In xSPDE,
the equations are always initial value problems in time, so the time
dimension boundary specification for $d=1$ is not included.

\paragraph{Example: boundary types in a 2-dimensional PDE}

Suppose there is one field cell with two field indices, and mixed
boundaries in space: Dirichlet for a lower boundary at $x=0$, and
and Neumann for an upper boundary at $x=1$, for the first field $a(1,:)$,
with the opposite combination in the second field component, $a(2,:)$,
hence 
\begin{center}
\doublebox{\begin{minipage}[t]{0.75\columnwidth}%
\texttt{p.boundaries\{2\} = {[}1,-1;-1,1{]};}%
\end{minipage}} 
\par\end{center}

The field cell index is $c=1$, which can be omitted here.

\subsection{Transverse boundary values}

For non-vanishing, specified boundary conditions, the boundary values
can be entered either using $boundval\{c,d\}(a,p)$, if they are constant
in time, \textbackslash or else, if they are dynamical, the function
$boundfun(a,c,d,p)$ is specified. This returns the boundary values
used for the fields or derivatives in a particular cell $c$ and dimension
$d>1$ as an array of dimension $b(\mathbf{j},e))$, where $\mathbf{j}=i,\mathbf{k}$.

Here $a$ is the current field cell array, $i=j_{1}$ is the field
index, and $\mathbf{k}$ is the space index, where $j_{d}$ is the
index of the dimension whose boundary values are specified. For this
dimension, only two values are needed: $j_{d}=1,2$ for the lower
and upper boundary values, which could either be field values or their
derivatives. An ensemble index $e$ is also needed if the boundary
values are stochastic.

Boundary values can be a function of both the fields ($a$) and internal
variables like the current time ($t$). These may have stochastic
initial values at $t=0$ which are calculated only once. In such cases
the boundary values must first be initialized, so the routine $boundfun(a,c,d,p)$
is first internally initialized with time $t<origin(1)$, and with
random Gaussian values in the input field $a$. These are delta-correlated
in space, i.e., with the same definition as ``inrandoms''. The xSDPE
program stores the returned values $b$ for the boundaries in an internal
cell array, $boundval\{c,d\}$, for later use if required.

The default boundary value is zero, if not specified.

\subsection{Example: boundaries in a 2-dimensional PDE}

Suppose there are two fields, and we wish to set boundary values.

We take boundary values as Dirichlet for $x=0$ and Neumann for $x=1$
in field variable 1, and Neumann for $x=0$ and Dirichlet for $x=1$
in field variable 2. Suppose the boundary values are different from
the default values of $a=0$, $\partial_{x}a=0$, so that: 
\begin{align}
a_{1}\left(x=0\right) & =1,\nonumber \\
\partial_{x}a_{1}\left(x=1\right) & =a_{1}\left(x=1\right).\nonumber \\
\partial_{x}a_{2}\left(x=0\right) & =-a_{2}\left(x=0\right)\nonumber \\
a_{2}\left(x=1\right) & =-1.
\end{align}

These are set in the following code: 
\begin{center}
\doublebox{\begin{minipage}[t]{0.75\columnwidth}%
\texttt{p.boundfun = @mybfun}

\texttt{p.boundaries\{2\} = {[}1,-1;-1,1{]};}

\texttt{...}

\texttt{function b = mybfun(a,\textasciitilde ,\textasciitilde ,p)}

\texttt{\% b = mybfun(a,c,d,p) calculates boundary values}

\texttt{b(1,2,:) = a(1,end,:);}

\texttt{b(2,1,:) = -a(2,end,:);}

\texttt{b(1,1,:) = 1+0{*}a(1,end,:);}

\texttt{b(2,2,:) = -1+0{*}a(1,end,:);}

\texttt{end}%
\end{minipage}} 
\par\end{center}

\subsection{Transverse plots}

A number of plots at equally spaced points in time can be generated
through. For example, adding the line below creates 3 time-sliced
plots at $t=0,5,10$: 
\begin{center}
\doublebox{\begin{minipage}[t]{0.75\columnwidth}%
\texttt{p.transverse\{1\} = 3;}%
\end{minipage}} 
\par\end{center}

\section{Output transforms}

For graphical output, Fourier transforms involve a sum over the lattice
points using a discrete Fourier transform at the lattice points $x_{i}$,
so that:

\begin{equation}
\tilde{a}(\omega_{i},\mathbf{k}_{i})=\frac{dtd\mathbf{x}}{\left[2\pi\right]^{d/2}}\sum_{j_{1}\ldots j_{d}}\exp\left[i\left(\omega_{i_{1}}t_{j_{1}}-\mathbf{k}_{\mathbf{i}}\cdot\mathbf{x}_{\mathbf{j}}\right)\right]a(t_{j_{1}},\mathbf{x}_{\mathbf{j}})\,
\end{equation}
The momenta $k_{i}$ have an interval of 
\begin{equation}
dk_{i}=\frac{2\pi}{n_{i}dx_{i}}
\end{equation}
with $k_{i}$ values given for even n by: 
\begin{equation}
k_{i}=\left(1-\frac{n_{i}}{2}\right)dk_{i},\ldots\frac{n_{i}}{2}dk_{i}
\end{equation}
and for odd n by: 
\begin{equation}
k_{i}=\frac{1-n_{i}}{2}dk_{i},\ldots\frac{n_{i}-1}{2}dk_{i}
\end{equation}

Once Fourier transformed, the $observe$ function can be used to take
any further functions or combinations of Fourier transformed fields
prior to averaging. Important points to keep in mind are as follows: 
\begin{itemize}
\item Fourier transforms are specified for the k-th observe function independently
of all other functions, by specifying $transforms\{k\}=\left[\ell_{1,}\ldots\ell_{d,}\right]$. 
\item Here $\ell_{j}=0,1$ is a logical switch, set to $\ell_{j}=1$ if
the $j-th$ dimension requires a Fourier transform, and to $\ell_{j}=0$
if there is no Fourier transform. 
\item The internal fields $p.k\{1\},\ldots p.k\{d\}$ are available for
use in making functions of momentum for use with observations. 
\item In propagation calculations, the momentum lattice values start with
$k=0,\ldots$, following standard Matlab and FFT conventions. 
\item For storing and graphing, momentum lattice values are reordered to
start with $k=-k_{max},\ldots$, following standard graphics and mathematical
conventions. 
\end{itemize}

\section{Initial random fields}

When $randoms\sim=0$, an initial Gaussian random field $\mathbf{v}^{x}$
is generated with delta-correlations in $x$-space. This corresponds
to a variance of $1/\Delta V$ for a lattice volume of $\Delta V=\Delta x\Delta y\Delta z$
for three space dimensions, and similarly for other cases. 

When $krandoms\sim=0$, an initial random field $\tilde{\mathbf{v}}^{k}$
is generated with delta-correlations in $k$-space. This can be filtered
with a user-specified filter function to give $\tilde{\mathbf{v}}^{kf}$,
then inverse Fourier transformed to give $v^{k}$. Both random fields
are passed to the $initial$ function as an extended vector $\left[v^{x},v^{k}\right]$,
for field initialization in space. 

When $urandoms\sim=0$, an initial field of uniform random numbers
is generated for jump processes.

These can all be specified as cells of multiple random fields, and
passed to the initial function in the order of {[}randoms,krandoms,urandoms{]}.

For multiple fields, one may combine random fields of different dimensionality
using different cell-array indices. The ``missing'' dimension has
only one point. This has a corresponding reduced lattice volume of
$\Delta V_{r}=\Delta x\Delta y$, if the third dimension has only
a single point for this field. If all space dimensions are missing,
then $\Delta V_{r}=1$.

There is a user specified filter function available, to modify random
fields $\tilde{v}^{k}$, that are delta-correlated in momentum space
using a filter function, 'rfilter' so that $v_{i}^{kf}\left(\mathbf{k}\right)=f_{i}^{(r)}\left(\mathbf{v}^{k}\left(\mathbf{k}\right)\right)$,
before being used. The corresponding correlations are: 
\begin{eqnarray}
\left\langle v_{i}^{x}\left(\mathbf{x}\right)v_{j}^{x}\left(\mathbf{x}'\right)\right\rangle  & = & \delta\left(\mathbf{x}-\mathbf{x}'\right)\delta_{ij}\sim\frac{1}{\Delta V}\delta_{\mathbf{x},\mathbf{x}'}\delta_{ij}\nonumber \\
\left\langle \tilde{v}_{i}^{k}\left(\mathbf{k}\right)\tilde{v}_{j}^{k}\left(\mathbf{k}'\right)\right\rangle  & = & \delta\left(\mathbf{k}-\mathbf{k}'\right)\delta_{ij}\sim\frac{1}{\Delta K}\delta_{\mathbf{k},\mathbf{k}'}\delta_{ij}\nonumber \\
\left\langle \tilde{v}_{i}^{kf}\left(\mathbf{k}\right)\tilde{v}_{j}^{kf}\left(\mathbf{k}'\right)\right\rangle  & = & \left\langle f_{i}^{(r)}\left(\tilde{\mathbf{v}}^{k}\left(\mathbf{k}\right)\right)f_{j}^{(r)}\left(\tilde{\mathbf{v}}^{k}\left(\mathbf{k}'\right)\right)\right\rangle .
\end{eqnarray}

On a lattice, we replace the Dirac continuous delta-function by a
discrete Kronecker delta function scaled by an inverse volume element
either in space ($\Delta V$) or momentum ($\Delta K$) . The xSPDE
Fourier transforms are given by a symmetric Fourier transform, so
that if we inverse Fourier-transform the $k-$space inrandoms, without
filtering, then: 
\begin{equation}
v^{k}(\mathbf{x})=\frac{1}{\left[2\pi\right]^{(d-1)/2}}\int e^{i\mathbf{k}\cdot\mathbf{x}}\tilde{v}^{k}(\mathbf{k})d\mathbf{k}\,
\end{equation}

These have random initial values that are real and delta-correlated
in space, so that: 
\begin{equation}
\left\langle v^{x}\left(\mathbf{x}\right)v^{x}\left(\mathbf{x}'\right)\right\rangle =\delta\left(\mathbf{x}-\mathbf{x}'\right).
\end{equation}
The corresponding noises in position space are correlated according
to:

\begin{align}
\left\langle v^{k}\left(\mathbf{x}\right)\left(v^{k}\left(\mathbf{x}'\right)\right)^{*}\right\rangle  & =\frac{1}{\left[2\pi\right]^{(d-1)}}\int e^{i(\mathbf{k}\cdot\mathbf{x}-\mathbf{k}'\cdot\mathbf{x}')}\left\langle \tilde{v}^{k}\left(\mathbf{k}\right)\tilde{v}^{k}\left(\mathbf{k}'\right)\right\rangle d\mathbf{k}d\mathbf{k}'\nonumber \\
 & =\frac{1}{\left[2\pi\right]^{(d-1)}}\int e^{i(\mathbf{x}-\mathbf{x}')\cdot\mathbf{k}}d\mathbf{k}\nonumber \\
 & =\delta\left(\mathbf{x}-\mathbf{x}'\right).
\end{align}
Similarly, if we don't conjugate the k-noise, then: 
\begin{equation}
\left\langle v^{k}\left(\mathbf{x}\right)v^{k}\left(\mathbf{x}'\right)\right\rangle =\delta\left(\mathbf{x}+\mathbf{x}'\right).
\end{equation}

However, if we define $\tilde{v}^{c}\left(\mathbf{k}\right)=\left[\tilde{v}_{1}^{k}\left(\mathbf{k}\right)+i\tilde{v}_{2}^{k}\left(\mathbf{k}\right)\right]/\sqrt{2}$
, then we obtain complex noise that is only delta correlated when
conjugated. 
\begin{align}
\left\langle v^{c}\left(\mathbf{x}\right)\left(v^{c}\left(\mathbf{x}'\right)\right)^{*}\right\rangle  & =\delta\left(\mathbf{x}-\mathbf{x}'\right)\nonumber \\
\left\langle v^{c}\left(\mathbf{x}\right)v^{c}\left(\mathbf{x}'\right)\right\rangle  & =0.
\end{align}
This is obtainable with the x-space noise as well, but the utility
of the k-space noise is that it can be filtered to have nonlocal correlations
in space if required.

\section{Noise fields}

During propagation in time, noises are Gaussian noise fields delta-correlated
in space-time. They are calculated in an analogous way, except with
an additional factor of $1/\sqrt{\Delta t}$ because they are delta
correlated in time. They have a variance of $\sigma^{2}=1/(\Delta t\Delta V)$.
Reduced dimension cells with volume $V_{r}$ have a noise variance
$\sigma^{2}=1/(\Delta t\Delta V_{r})$. 

There is a user specified scaling function available, to take random
knoises $w^{k}$ in momentum space that are then scaled using a filter
function, 'nfilter' so that $w_{i}^{kf}\left(\mathbf{k}\right)=f_{i}^{(n)}\left(\mathbf{w}^{k}\left(\mathbf{k}\right)\right)$,
before being used:

\begin{eqnarray}
\left\langle w_{i}^{x}\left(t,\mathbf{x}\right)w_{j}^{x}\left(t,\mathbf{x}'\right)\right\rangle  & = & \delta\left(\mathbf{x}-\mathbf{x}'\right)\delta\left(t-t'\right)\delta_{ij}\nonumber \\
\left\langle \tilde{w}_{i}^{k}\left(t,\mathbf{k}\right)\tilde{w}_{j}^{k}\left(t,\mathbf{k}'\right)\right\rangle  & = & \delta\left(\mathbf{k}-\mathbf{k}'\right)\delta\left(t-t'\right)\delta_{ij}\nonumber \\
\left\langle \tilde{w}_{i}^{kf}\left(t,\mathbf{k}\right)\tilde{w}_{j}^{kf}\left(t',\mathbf{k}'\right)\right\rangle  & = & \left\langle f_{i}^{(n)}\left(\tilde{\mathbf{w}}^{k}\left(t,\mathbf{k}\right)\right)f_{j}^{(n)}\left(\tilde{\mathbf{w}}^{k}\left(t',\mathbf{k}'\right)\right)\right\rangle .
\end{eqnarray}

When $unoises\sim=0$, an initial field of uniform random numbers
is generated for jump processes.

All noises can all be specified as cells of multiple noise fields,
and are passed to the deriv function in the order of {[}noises,knoises,unoises{]}.

\newpage{}

\chapter{SPDE theory\label{chap:SPDE Theory}}

\textbf{This chapter describes the basics of stochastic partial differential
equation (SPDE) theory, in order to explain the background to the
numerical methods.}

\section{SPDE definitions}

A stochastic partial differential equation or SPDE is defined in both
time $t$ and one or more space dimensions $\mathbf{x}$. We suppose
there are $d$ total space-time dimensions. The space-time coordinate
is denoted as $\mathbf{r}=\left(r^{1},\ldots r^{d}\right)=\left(t,\mathbf{x}\right)=\left(t,x,y,z,...\right)$.

The stochastic partial differential equation solved is written in
differential form as

\begin{equation}
\frac{\partial\mathbf{a}}{\partial t}=\mathbf{A}\left[\mathbf{\nabla},\mathbf{a},\mathbf{r}\right]+\underline{\mathbf{B}}\left[\mathbf{\nabla},\mathbf{a},\mathbf{r}\right]\cdot\mathbf{w}(\mathbf{r})+\mathbf{L}\left[\mathbf{\nabla},\mathbf{a},\mathbf{r}\right]\cdot\mathbf{a}.\label{eq:spde}
\end{equation}
Here, $\mathbf{a}=\left[a_{1},\dots a_{f}\right]$ is a real or complex
vector field, $\mathbf{A}$ is a vector function of fields and space
and $\underline{\mathbf{B}}$ a matrix function. The new feature is
that terms can now include the operator $\nabla$, which is a differential
term in a real space $\mathbf{x}$. The exact structure of these terms
is important, and not all such equations have well-behaved solutions
\cite{quastel2015one,lam1998improved}.

In many common cases, the noise term $\mathbf{w}$ is delta-correlated
in time and space: 
\begin{eqnarray}
\left\langle w_{i}\left(\mathbf{r}\right)w_{j}\left(\mathbf{r}'\right)\right\rangle  & = & \delta\left(t-t'\right)\delta\left(\mathbf{x}-\mathbf{x}'\right)\delta_{ij}.
\end{eqnarray}
One can also have noise with a finite correlation length defined by
a noise correlation function $N_{ij}\left(\mathbf{x}-\mathbf{x}'\right)$
in space so that: 
\begin{eqnarray}
\left\langle w_{i}\left(\mathbf{r}\right)w_{j}\left(\mathbf{r}'\right)\right\rangle  & = & \delta\left(t-t'\right)N_{ij}\left(\mathbf{x}-\mathbf{x}'\right).
\end{eqnarray}
It is even possible to have noise with a finite correlation time.
Currently, these are not directly treated in xSPDE, although user
definitions of this are possible by adding a customized noise function.

Additionally, the initial field has a probability distribution. In
most examples, we suppose that this initial random field distribution
can be generated as a function of Gaussian distributed initial random
fields $\mathbf{v}\left(\mathbf{x}\right)$, where: 
\begin{equation}
\left\langle v_{i}\left(\mathbf{x}\right)v_{j}\left(\mathbf{x}'\right)\right\rangle =\delta\left(\mathbf{x}-\mathbf{x}'\right)\delta_{ij}.
\end{equation}

However, it is also possible that the initial random fields are also
not delta-correlated, so that

\begin{equation}
\left\langle v_{i}\left(\mathbf{x}\right)v_{j}\left(\mathbf{x}'\right)\right\rangle =R_{ij}\left(\mathbf{x}-\mathbf{x}'\right).
\end{equation}
Both finite correlation length and delta-correlated noise and random
terms can be used in xSPDE simulations, with finite correlation lengths
defined through a Fourier transform method.

\section{Boundary conditions\label{subsec:Boundary-conditions}}

There are three types of boundaries that are commonly used. They are
specified independently for each space dimension $j=2,\ldots d$,
field component $i=1,\dots f,$ and lower or upper location $\ell=1,2$.
Each has a specific boundary type. These are described with a numerical
code $bt$, as: 
\begin{description}
\item [{Dirichlet}] (specified value, $bt=1$): $a_{i}\left(r^{1},r^{2},\dots\hat{r}_{\ell}^{j},\dots\right)=f_{ij\ell}\left(\mathbf{r},\mathbf{a}\right)$
. 
\item [{Periodic}] ($bt=0$): $a_{i}\left(r^{1},r^{2},\dots\hat{r}_{\ell}^{j},\dots\right)=a_{i}\left(r^{1},r^{2},\dots\hat{r}_{3-\ell}^{j},\dots\right)$
. 
\item [{Robin/Neumann}] (specified derivative, $bt=-1$): $\frac{\partial}{\partial r^{j}}a_{i}\left(r^{1},r^{2},\dots\hat{r}_{\ell}^{j},\dots\right)=g_{ij\ell}\left(\mathbf{r},\mathbf{a}\right)$. 
\end{description}
The coordinates $\hat{r}_{\ell}^{j}=\left(r_{1}^{j},r_{2}^{j}\right)$
are locations where boundary conditions are enforced. There are five
types of boundary combinations of these for each dimension and field
variable. Note that the boundary type can change the error stability
properties of an equation. 

Periodic boundaries can't be combined with other types, as this defines
both boundaries: 
\begin{description}
\item [{a)}] periodic-periodic- P-P: \textquotedbl 0,0\textquotedbl{} 
\item [{b)}] Dirichlet-Dirichlet- D-D: \textquotedbl 1,1\textquotedbl{} 
\item [{c)}] Robin-Robin- R-R: \textquotedbl -1,-1\textquotedbl{} 
\item [{d)}] Robin-Dirichlet- R-D: \textquotedbl -1,1\textquotedbl{} 
\item [{e)}] Dirichlet-Robin- D-R: \textquotedbl 1,-1\textquotedbl{} 
\end{description}
Just as with the derivative term, each of these types can change with
dimension and field component. Specified field or derivative values
can be any user-defined functions of space, time, and field amplitude
or simply have fixed values. Currently, all combinations of boundaries
can be treated in xSPDE.

\section{Spatial grid and boundaries}

The location of the boundary at $\hat{r}_{\ell}^{j}$ is important
in solving (S)PDEs, especially if high accuracy is required, or if
field values at the boundary are needed.

Suppose the spatial grid spacing is $\Delta x$ and the number of
grid points in a particular dimension $d$ is $points(d)=N$, then
the maximum range from the first to last computed point is: 
\begin{equation}
\begin{split}R=(N-1)\Delta x=ranges(d).\end{split}
\end{equation}
Noting that $\mathbf{r}=\left(t,\mathbf{x}\right)$, and $\Delta\mathbf{r}=\left(\Delta t,\Delta\mathbf{x}\right),$this
means that the space-time points for an origin vector $\bm{O}$are
at: 
\begin{equation}
r_{i}=O_{i}+(i-1)\Delta r_{i}.
\end{equation}
There are two slightly different spatial boundary locations used in
xSPDE, depending on the type of boundary conditions specified, as
follows:

\subsection{Periodic boundary}

For the default case of a periodic boundary, the logical boundary
location is arbitrary. The indices are arranged as though on a circle
from $1:N$. It is useful to suppose the boundary is simultaneously
at $\hat{r}_{1}^{j}=r_{1}^{j}-\Delta r^{j}/2$ and at $\hat{r}_{2}^{j}=r_{N_{j}}^{j}+\Delta r^{j}/2$.
Neither upper or lower logical 'boundary' is at a grid point. The
effective range of the domain is $R^{j}+\Delta r^{j}$, due to this
displaced boundary. 

Only the values at $N$ points are computed, and one must regard the
point where the periodicity is enforced as interpolating between the
last and first point.

\subsection{Non-periodic boundary}

For the case of a non-periodic boundary, including Dirichlet, Robin
and Neumann boundary conditions, the indices are in a line from $1:N$.
The lower and upper lower boundaries are at $\hat{r}_{1}^{j}=r_{1}^{j}$
and at $\hat{r}_{2}^{j}=r_{N_{j}}^{j}$. In some PDE methods the logical
boundaries are outside the grid boundaries, but that is not the case
here. Unlike the periodic case, boundaries are enforced at the first
and last point. 

This is different to what is found in most trigonometric transform
software, but this approach allows for a unified treatment of multiple
types of algorithm. For finite difference derivatives at the boundaries,
this leads to the usual result that the central difference approximation
to the second derivative is of first order (in $\Delta r$) at the
boundaries, while it is of second order elsewhere.

With spectral methods, the derivative boundaries are obtained with
a combination of an interaction picture transform and additional polynomial
terms. For simplicity, only linear, even order space derivatives are
included in the linear propagator for the interaction picture (see
\ref{sec:Interaction-picture}), which means that other space derivatives
must be included using finite differences.

\section{Multidimensional walk}

The simplest example of an SPDE is the multidimensional Wiener process:
\begin{equation}
\dot{a}=w(t,\mathbf{x})\,.\label{eq:Wiener_process-1}
\end{equation}

This has a solution that is identical in appearance to an SDE: 
\begin{equation}
a\left(t,\mathbf{x}\right)=a\left(0,\mathbf{x}\right)+\int_{0}^{t}w\left(\tau,\mathbf{x}\right)d\tau.
\end{equation}
Just as for an SDE, this means that the initial mean value does not
change in time: 
\begin{equation}
\left\langle a\left(t,\mathbf{x}\right)\right\rangle =\left\langle a\left(0,\mathbf{x}\right)\right\rangle .\label{eq:Wiener_mean-1}
\end{equation}

Since there are no spatial derivatives here, boundary values are not
important. One can regard this as having periodic boundaries, which
by the xSPDE conventions means that no boundary conditions are enforced
- since periodic boundaries do not alter computed values when there
are no derivatives.

\subsection{Variance solution}

The noise correlation is non-vanishing from Eq \eqref{eq:noise-correlations},
so the variance must increase with time: 
\begin{align}
\left\langle a^{2}\left(t,\mathbf{x}\right)\right\rangle  & =\left\langle a^{2}\left(0,\mathbf{x}\right)\right\rangle +\int_{0}^{t}\int_{0}^{t}\left\langle w\left(\tau,\mathbf{x}\right)w\left(\tau',\mathbf{x}\right)\right\rangle d\tau d\tau'\nonumber \\
 & =\left\langle a^{2}\left(0,\mathbf{x}\right)\right\rangle +\delta^{d-1}\left(0\right)\int_{0}^{t}\int_{0}^{t}\delta\left(\tau-\tau'\right)d\tau d\tau'.
\end{align}

Integrating the temporal delta function gives unity. The spatial delta-function
is replaced by $1/\Delta V$ in a discretized lattice calculation
at points $\mathbf{x}_{j}$ with cell volume $\Delta V=\prod\Delta x_{j}$,
which means that the second moment and the variance both increase
linearly with time:

\begin{align}
\left\langle a^{2}\left(t,\mathbf{x}_{j}\right)\right\rangle  & =\left\langle a^{2}\left(0,\mathbf{x}_{j}\right)\right\rangle +t/\Delta V.\label{eq:Wiener_mean_square-1}
\end{align}

The probability on the lattice for observing lattice field values
$a_{j}$ follows an elementary diffusion equation: 
\begin{equation}
\frac{\partial P}{\partial t}=\frac{1}{2\Delta V}\sum_{j}\frac{\partial^{2}P}{\partial a_{j}^{2}}\,,\label{eq:FPE-1-2}
\end{equation}
which is an example of Eq \eqref{eq:FPE}. From this equation and
using Eq \eqref{eq:moment_equn}, the first two corresponding moment
equations in this case are 
\begin{align}
\frac{\partial}{\partial t}\left\langle a_{j}\right\rangle = & \left\langle \frac{1}{2}\frac{\partial^{2}}{\partial a_{j}^{2}}a_{j}\,\right\rangle =0\nonumber \\
\frac{\partial}{\partial t}\left\langle a_{j}^{2}\right\rangle = & \left\langle \frac{1}{2\Delta V}\frac{\partial^{2}}{\partial a_{j}^{2}}a_{j}^{2}\,\right\rangle =\frac{1}{\Delta V}.
\end{align}

These differential equations are satisfied by the solutions obtained
directly from the stochastic equations, but as one can see, the coupling
between the lattice points provides more interesting behavior. This
requires derivative terms such as Laplacians.

\section{Interaction picture\label{sec:Interaction-picture}}

To treat Laplacians, spectral or interaction-picture methods can be
very efficient, with much lower errors and much faster run-times.
This is because they do not have the stability problems of finite
difference methods when treating higher-order derivatives, which allows
much larger time-steps to be used.

To explain the interaction picture algorithm, SPDEs often contain
terms which are linear in the field variables $\mathbf{a}$, including
derivative operators acting on $\mathbf{a}$. This can be treated
exactly using an \textit{interaction picture}, which leads to dramatically
reduced time-step errors and higher stability \cite{Drummond1993Simulation,Werner1997Robust},
by using a spectral method to compute derivatives. These methods are
also very useful in non-stochastic PDEs.

In summary, the interaction picture provides a means to solve for
linear space-derivative terms in the propagation in an efficient way.
This is based on introducing local variables $\tilde{\mathbf{a}}$
for the field variables $\mathbf{a}$. It is convenient for the purposes
of describing such interaction picture methods to introduce an abbreviated
notation as: 
\begin{equation}
\begin{split}\begin{aligned}\mathcal{D}\left[\mathbf{a},\mathbf{r}\right]=\mathbf{A}\left[\mathbf{\nabla},\mathbf{a},\mathbf{r}\right]+\underline{\mathbf{B}}\left[\mathbf{\nabla},\mathbf{a},\mathbf{r}\right]\cdot\mathbf{w}(\mathbf{r})\end{aligned}
\end{split}
\label{eq:deriv_without_linear_term}
\end{equation}
Hence, we can write the differential equation as: 
\begin{equation}
\begin{split}\frac{\partial\boldsymbol{a}}{\partial t}=\mathcal{D}\left[\mathbf{a},\mathbf{r}\right]+\underline{\mathbf{L}}\left[\boldsymbol{\nabla}\right]\cdot\boldsymbol{a}.\end{split}
\end{equation}

Here $\underline{\mathbf{L}}\left[\boldsymbol{\nabla}\right]$ should
include the highest order derivatives, as these have the largest eigenvalues,
but lower-order derivative terms may occur in the other terms.

\subsection{Linear propagator}

Next, we define a linear propagator. This is given formally by: 
\begin{equation}
\begin{split}\mathcal{P}\left(t,\bar{t}\right)=\exp\left(\Delta t\underline{\mathbf{L}}\left[\boldsymbol{\nabla}\right]\right)\end{split}
.
\end{equation}
where $\Delta t=t-\bar{t}$, $\bar{t}$ is the interaction picture
origin, and the notation includes setting boundary values. Transforming
the field $\mathbf{a}$ to an interaction picture is achieved on defining:
\begin{equation}
\tilde{\mathbf{a}}=\mathcal{P}^{-1}\left(t,\bar{t}\right)\mathbf{a}.
\end{equation}
As a result, the equation of motion is: 
\begin{equation}
\begin{split}\frac{\partial\tilde{\mathbf{a}}}{\partial t}=\mathcal{D}\left[\mathcal{P}\left(t,\bar{t}\right)\tilde{\mathbf{a}},t\right].\end{split}
\end{equation}

This allows an SPDE to be treated with transformations using Fourier
or discrete sine/cosine transforms. Our implementation uses a diagonal
linear operator L without space-dependence. The linear operator can
have any derivative in the periodic case, but only even order derivatives
in the Dirichlet and Neumann case.

As well as the linear term, derivatives and nonlinear functions that
are not tractable with spectral methods can appear in the residual
term $\mathcal{D}\left[\mathbf{a},\mathbf{r}\right]$, where they
are treated using finite difference techniques. As a result, while
the interaction picture does not handle all possible derivative terms,
it also does not restrict them from being used elsewhere in the equations.

Other methods exist in the literature. Improved convergence properties
are obtained for some problems in a spectral picture using an exact
solution of a linear part of the drift term \cite{bao2005fourth,Blakie2005Projected},
or stochastic noise terms \cite{jentzen2009numerical}, as well as
the Laplacian terms. The xSPDE code has user-definable functions that
can be adapted to include these.

\section{Fourier transforms}

It is often useful to transform a field to implement the interaction
picture, or to extract nonlocal correlation properties in space. The
Fourier transforms or spectrum definitions used in xSPDE are given
by the symmetric Fourier transform definition: 
\begin{align}
\tilde{a}(\mathbf{k}) & =\mathcal{F}\left(a(\mathbf{x})\right)\nonumber \\
 & =\frac{1}{\left[2\pi\right]^{\left(d-1\right)/2}}\int e^{-i\mathbf{k}\cdot\mathbf{x}}a(\mathbf{x})d\mathbf{x}\,.
\end{align}

The inverse Fourier transform is the function: 
\begin{align}
a(\mathbf{x}) & =\mathcal{F}^{-1}\left(\tilde{a}\right)\nonumber \\
 & =\frac{1}{\left[2\pi\right]^{\left(D-1\right)/2}}\int e^{i\mathbf{k}\cdot\mathbf{x}}\tilde{a}(\mathbf{k})d\mathbf{k}\,.
\end{align}

In simulations, this is not combined with any time (or space) averaging
as in the temporal Fourier transforms. The reason for this is that
the interaction picture transformations must be invertible, which
is the case for a point-based discrete Fourier transform.

\subsection{Normalization}

During propagation, we define temporary internal fields $A\left(\mathbf{k}_{\mathbf{n}}\right)$,
that are normalized using FFT conventions: 
\begin{align}
A\left(\mathbf{k}_{\mathbf{n}}\right) & =\sum_{j_{2}=1}^{N_{2}}\ldots\sum_{j_{d}=1}^{N_{d}}e^{-i\mathbf{k}_{n}\cdot\mathbf{x}_{\mathbf{j}}}a\left(\mathbf{x}_{\mathbf{j}}\right)\,\nonumber \\
a\left(\mathbf{x}_{\mathbf{j}}\right) & =\frac{1}{\prod_{k=2}^{D}N_{k}}\sum_{n_{2}=1}^{N_{2}}\ldots\sum_{n_{D}=1}^{N_{D}}e^{i\mathbf{k}_{\mathbf{n}}\cdot\mathbf{x}_{\mathbf{j}}}A\left(\mathbf{k}_{\mathbf{n}}\right)\,.
\end{align}
Otherwise, for graphical and output averages, we define Fourier transforms
using physics and mathematics conventions:

\begin{align}
\tilde{a}\left(\mathbf{k}_{\mathbf{n}}\right) & =\prod_{d=2}^{D}\left[\frac{\Delta x_{d}}{\sqrt{2\pi}}\right]\sum_{j_{2}=1}^{N_{2}}\ldots\sum_{j_{d}=1}^{N_{D}}e^{-i\mathbf{k}_{n}\cdot\mathbf{x}_{\mathbf{j}}}a\left(\mathbf{x}_{\mathbf{j}}\right)\,\nonumber \\
a\left(\mathbf{x}_{\mathbf{j}}\right) & =\prod_{d=2}^{D}\left[\frac{\Delta k_{d}}{\sqrt{2\pi}}\right]\sum_{n_{2}=1}^{N_{2}}\ldots\sum_{n_{d}=1}^{N_{D}}e^{i\mathbf{k}_{\mathbf{n}}\cdot\mathbf{x}_{\mathbf{j}}}\tilde{a}\left(\mathbf{k}_{\mathbf{n}}\right)\,.
\end{align}

Note that this rescaling is consistent, because 
\begin{equation}
\Delta x_{d}\Delta k_{d}=\frac{2\pi}{N_{d}}.
\end{equation}

\section{Trigonometric transforms}

Taking the interaction picture approach, we now consider other types
of boundary conditions, which we initially assume here are either
a zero field (Dirichlet) or a zero derivative (Neumann). We will only
treat cases of even order derivatives, which do not change the trigonometric
function. Any odd order derivatives are taken to be included in the
finite difference ($\mathcal{D}$) term.

\subsection{Zero boundary cases}

In the spectral transform method in one space dimension, with zero
boundaries, one uses a trigonometric function, $T\left(kx\right)=T_{1}\sin\left(kx\right)+T_{2}\cos\left(kx\right)$
to expand as: 
\begin{align}
a_{i}\left(t,x\right) & =\sum_{n}a_{i,n}(t)T(k_{i,n}x),
\end{align}
The discrete inverse transform allows evaluation at sample points
$x_{j}$, in order to satisfy the boundary conditions: 
\begin{equation}
a_{i,n}(t)=\sum_{j}a_{i}(t,x_{j})\tilde{T}(k_{n}x_{j}),
\end{equation}
The trigonometrical function is defined such that:

\begin{equation}
\partial_{x}^{2p}T(kx)=\left(-k^{2}\right)^{p}T(kx).
\end{equation}

The propagated equation is exactly soluble for the sampled points,
since for each component 
\begin{align}
\mathcal{L}\cdot a(t,x_{j}) & =\sum_{ijn}\mathcal{L}a_{n}(t)T(k_{n}x_{j}),\nonumber \\
 & =-\sum_{ijnp}L_{p}\left(-k_{n}^{2}\right)^{p}a_{n}(t)T(k_{n}x_{j})).
\end{align}

Hence, 
\begin{equation}
a_{n}(t)=\exp\left(\sum L_{p}\left(-k_{n}^{2}t\right)^{p}t\right)a_{n}(0).
\end{equation}

This is an exact solution, provided the initial condition has the
given expansion. This of course is usually an approximation itself,
which should be checked by changing the grid. There are no other approximations
made on the transverse derivative. Provided the $k$ values are the
same, this propagator is identical for all types of trigonometric
and Fourier transforms.

As explained in (\ref{subsec:Boundary-conditions}), there are five
boundary combinations that are possible in each dimension and field
component. Each has a corresponding xSPDE boundary type and spectral
integrator. Each boundary type is specified to depend on the space
dimension and the field component, as well as having boundary values
depending on time and any field value.

Currently, all can be treated in xSPDE using finite differences, and
each type of boundary also has a spectral method that preserves the
boundary requirement. In principle one can define the trigonometric
transforms to correspond to whole symmetries whose boundary is at
a grid point, as used in xSPDE, or half symmetries which are half-way
between two grid points. 

All spectral methods used in xSPDE make use of boundaries at a grid
point, in order to compute the relevant terms, which means that there
is greater compatibility with the finite difference methods, when
the boundaries are at the grid points. Differential equations can
also have first order terms, which currently require using either
finite differences or periodic boundaries. 

It is possible to compute first-order derivatives with spectral methods,
but these turn sine transforms into cosine transforms. This is not
compatible with trigonometric interaction picture transformations
used in XSPDE. As a result, any odd-order derivative terms must be
computed using finite differences in all cases , except for the periodic
case, where either method can be used.

In summary, spectral transforms can all be implemented using fast
FTT, discrete sine (DST) or cosine (DCT) transforms. The spectral
method used is specific to the boundary type. The definitions used
here correspond to the standard definitions \cite{Frigo1998FFTW,Frigo2005Design},
except for one-based indexing, normalization, and extra points at
the boundaries, explained below.

\subsection{Finite boundary values}

For the case of finite boundaries, a combination of trigonometric
and polynomial functions are used to expand the fields, so that: 
\begin{align}
a_{i}\left(t,x\right) & =b_{i}\left(t,x\right)+u_{i}(t,x)\nonumber \\
 & =b_{i}\left(t,x\right)+\sum_{n}u_{i,n}(t)T(k_{i,n}x),
\end{align}

Here, the functions $b_{i}\left(t,x\right)$ are inhomogeneous polynomial
terms specified to satisfy the non-vanishing boundary conditions such
that: 
\begin{align}
\dot{b} & =\mathcal{L}\cdot b,
\end{align}
while the trigonometric expansion simply has to satisfy the equation
with zero boundaries.

\section{\label{sec:Spectral-transforms-and}Transforms and boundaries}

For Dirichlet or Neumann/Robin boundaries, the following expansion
can be employed in each dimension. We only describe one space dimension
for simplicity with: 
\begin{equation}
u=\sum_{n=1}^{\infty}\left[S_{n}\sin\left(k_{n}x\right)+C_{n}\cos\left(k_{n}x\right)\right]e^{\sum L_{p}\left(-k_{n}^{2}\right)^{p}t},
\end{equation}
where $k_{n},C_{n},S_{n}$ are chosen to satisfy the initial and boundary
conditions. Boundaries are taken, for the purposes of explanation,
as being from $x=0$ to $x=R$. This is not the case in the actual
code, which can treat arbitrary boundary locations due to the use
of the optional origins input to change the origin.

Unlike fast Fourier transform (FFT) definitions, there are multiple
distinct trigonometric transforms . These are generally labeled DST-(n)
and DCT-(n), where $n=I..IV$. They correspond to distinct boundary
combinations, as explained below.

Suppose there are $N$ computational grid-points. For the spatial
grid (1-based), this corresponds to $x_{n}=\left(n-1\right)\Delta x$,
$n=1,...,N$ with $\Delta x=\frac{R}{N-1}$ , so we have $x_{1}=0$
and $x_{N}=R$, as elsewhere in the manual.

In carrying out a discrete transform on $N_{T}$ points, with standard
trigonometric transform definitions of $N_{T}$, there are \textbf{less}
transform grid points required if some boundary values are defined
due to Dirichlet boundaries, hence $N_{T}<N$. This is because xSPDE
stores the full computational range, $N$, with boundary values. 

Sometimes one may wish to refer to the corresponding periodic Fourier
transform size, $N_{FT}$. This is $N_{FT}=2N_{T}=2\left(N-1\right)$,
except for DST-I , when it is $N_{FT}=2(N_{T}+1)=2\left(N-1\right)$. 

An unnormalized inverse gives the original array multiplied by $N_{FT}/4=\left(N-1\right)/2$,
where $N_{FT}=2\left(N-1\right)$ is the periodic size, so our definitions
include a normalization of $\sqrt{2/\left(N-1\right)}$. Here $N_{T}$,
the number of points in the standard DST/DCT definitions, differs
from \textbf{both} the xSPDE computation grid size $N$ that includes
both boundaries, and also from the periodic size, which always includes
one (periodic) boundary.

Our notation is based on standard discrete sine and cosine transform
definitions. Here we use $1-$based indices throughout. For all coordinates,
including these examples of discrete Fourier transforms, with an origin
at $\bm{r}=0$ and an integration range of $\bm{R}$, we define:\textbf{
\begin{align}
r_{n}^{d} & =\left(n-1\right)\Delta r^{d}.\nonumber \\
\Delta r^{d} & =R^{d}/(N^{d}-1).
\end{align}
}

If we regard the transforms as having arguments of form $k_{j}\cdot r_{n}$,
the momentum spacings given below are such that: 
\begin{align}
\Delta k & =\frac{\pi}{R}\nonumber \\
\Delta x\Delta k & =\frac{\pi}{N-1}.
\end{align}
The internal momentum definitions used in the propagator calculations
are therefore different to those used in external graphs and in periodic
boundary cases.

\textbf{The following lists the trigonometric transforms required
to obtain the transform $\tilde{u}_{k}$ from $u_{n}$, and vice-versa,
for the four non-periodic boundary types in each dimension and field
index. }

\subsection{D-D case: Discrete map (DST-I)}

Let $u(0)=u_{1}=0$, and $u(R)=u_{N}=0$. The discrete representation
of $u$ is: 

\paragraph{Forward transform: DST-I}

\begin{align}
\tilde{u}_{k} & =\sqrt{\frac{2}{N-1}}\sum_{n=2}^{N-1}u_{n}\sin\left(\pi\frac{\left(k-1\right)\left(n-1\right)}{N-1}\right).
\end{align}

\paragraph{Inverse transform: DST-I
\begin{align}
u_{n} & =\sqrt{\frac{2}{N-1}}\sum_{k=2}^{N-1}\tilde{u}_{k}\sin\left(\pi\frac{\left(k-1\right)\left(n-1\right)}{N-1}\right).
\end{align}
}

The forward transform does \textbf{not} require the values at the
end-points of $n=1$ and $n=N$, which are set to zero in this case.
This is implicit in the sine expansion, since $\text{\ensuremath{\sin}(n\ensuremath{\pi)=0}.}$
Second derivatives are proportional to $\left(k-1\right)^{2}$.

\subsection{R-R case: Discrete map (DCT-I)}

Let $u'(0)=0$, and $u'(R)=0$. The discrete representation of $u$
is: 

\paragraph{Forward transform: DCT-I}

\begin{align}
\tilde{u}_{k} & =\sqrt{\frac{2}{N-1}}\left(\frac{1}{2}\left(u_{1}+(-1)^{n-1}u_{N}\right)+\sum_{n=2}^{N-1}u_{n}\cos\left(\pi\frac{\left(k-1\right)\left(n-1\right)}{N-1}\right)\right).
\end{align}

\paragraph{Inverse transform: DCT-I}

\begin{align}
u_{n} & =\sqrt{\frac{2}{N-1}}\left(\frac{1}{2}\left(\tilde{u}_{1}+(-1)^{n-1}\tilde{u}_{N}\right)+\sum_{k=2}^{N-1}\tilde{u}_{k}\cos\left(\pi\frac{\left(k-1\right)\left(n-1\right)}{N-1}\right)\right).
\end{align}

The forward transform requires the values at the end-points of $n=1$
and $n=N$, which are not zero in this case. It is equal (up to a
factor) to a discrete Fourier transform of $2\left(N-1\right)$ real
numbers $u_{n}$ with even symmetry about $n=1$ and $n=N$. As a
result, the equivalent discrete derivatives at both the end-points
are zero. Second derivatives are proportional to $\left(k-1\right)^{2}$.

\subsection{D-R case: Discrete map (DST-II/III)}

Let $u(0)=0$, and $u'(R)=0$. The discrete representation of $u$
is: 

\paragraph{Forward transform: DST-III}

\begin{align}
\tilde{u}_{k}\left(t\right) & =\sqrt{\frac{2}{N-1}}\left((-1)^{\left(n-1\right)}u_{N}/2+\sum_{n=2}^{N-1}u_{n}\sin\left[\frac{\pi}{N-1}\left(k-\frac{1}{2}\right)(n-1)\right]\right).
\end{align}

\paragraph{Inverse transform: DST-II}

\begin{align}
u_{n} & =\sqrt{\frac{2}{N-1}}\left(\sum_{k=1}^{N-1}\tilde{u}_{k}\left(t\right)\sin\left[\frac{\pi}{N-1}\left(k-\frac{1}{2}\right)n\right]\right).
\end{align}

The forward transform does \textbf{not} require the value at $n=1$,
which is zero in this case. This is implicit in the sine expansion,
since $\text{\ensuremath{\sin}(n\ensuremath{\pi)=0}.}$ This transform
implies a boundary condition that is odd around $n=1$, and even around
$n=N$. Second derivatives are proportional to $\left(k-1/2\right)^{2}$.

\subsection{R-D case Discrete map (DCT-II/III)}

Take $u'(0)=u(R)=0$. The discrete representation of $u$ is:

\paragraph{Forward transform: DCT-III}

\begin{align}
\tilde{u}_{k} & =\sqrt{\frac{2}{N-1}}\left(u_{1}/2+\sum_{n=2}^{N-1}u_{n}\cos\left[\frac{\pi}{N-1}\left(k-\frac{1}{2}\right)(n-1)\right]\right).
\end{align}

\paragraph{Inverse transform: DCT-II}

\begin{align}
u_{n} & =\sqrt{\frac{2}{N-1}}\sum_{k=1}^{N}\tilde{u}_{k}\cos\left[\frac{\pi}{N-1}\left(k-\frac{1}{2}\right)\left(n-1\right)\right].
\end{align}

The forward transform does not require the value at $n=N$, which
is zero in this case. This transform implies a boundary condition
that is even around $n=1$, and odd around $n=N$. Second derivatives
are proportional to $\left(k-1/2\right)^{2}$.

\section{Frequency or momentum grid \label{sec:IP-implementation}}

The frequency or momentum grid spacing is defined for all output graphs
and periodic Fourier transforms as 
\begin{equation}
\begin{split}\Delta k=\frac{2\pi}{N\Delta x}\end{split}
.
\end{equation}

The internal momentum grid spacing used can differ from this, depending
on the transforms used in the interaction picture. As explained above
in Section (\ref{sec:Spectral-transforms-and}), the internal momenta
for trigonometric transforms are:

\begin{equation}
\begin{split}\Delta k=\frac{\pi}{\left(N-1\right)\Delta x}\end{split}
.
\end{equation}
This is because the xSPDE algorithms allow the use of a sequence of
interaction pictures. Each successive interaction picture is referenced
to $t=t_{n}$, for the n-th step starting at $t=t_{n}$, so $\boldsymbol{a}_{I}(t_{n})=\boldsymbol{a}(t_{n})\equiv\boldsymbol{a}_{n}$.
It is also possible to solve stochastic partial differential equations
in xSPDE using explicit derivatives, but this is less efficient.

A discrete Fourier transform (DFT) using a fast Fourier transform
method is employed for the interaction picture (IP) transforms used
with periodic boundaries. This is normalized differently to the graphed
Fourier transforms, but the difference is not computationally significant.
However, the $\Delta k$ used internally changes with the precise
type of trigonometric transform used in other cases.

In one dimension, the DFT is usually defined by a sum over indices
starting with zero, rather than the Matlab convention of one. Hence,
if $\tilde{m}=m-1$: 
\begin{equation}
\begin{split}A_{\tilde{n}}=\mathcal{F}\left(a\right)=\sum_{\tilde{m}=0}^{N-1}a_{\tilde{m}}\exp\left[-2\pi i\tilde{m}\tilde{n}/N\right]\end{split}
.
\end{equation}
For periodic boundaries, the IP Fourier transform can be written in
terms of an FFT as 
\begin{equation}
\begin{split}\boldsymbol{A}\left(\boldsymbol{k}_{\boldsymbol{n}}\right)=\prod_{j}\left[\sum_{\tilde{m}_{j}}\exp\left[-i\left(dk_{j}dx_{j}\right)\tilde{m}_{j}\tilde{n}_{j}\right]\right]\end{split}
.
\end{equation}
The inverse FFT Fourier transforms divide by the correct factors of
$\prod_{j}N_{j}$ to ensure invertibility. Due to the periodicity
of the exponential function, negative momenta are obtained if we consider
an ordered lattice such that: 
\begin{equation}
\begin{split}\begin{aligned}k_{j} & =(j-1)\Delta k\,\,\,(j\le N/2)\\
k_{j} & =(j-1-N)\Delta k\,\,(j>N/2)
\end{aligned}
.\end{split}
\end{equation}
This Fourier transform is then multiplied by the appropriate factor
to propagate in the interaction picture, then an inverse Fourier transform
is applied. While it is not scaled for interaction picture transforms,
an additional scaling factor is applied to obtain transformed fields
in any averages for output plots.

In other words, in the averages 
\begin{equation}
\begin{split}\tilde{a}_{n}=\frac{\Delta x}{\sqrt{2\pi}}A_{\tilde{n}'}.\end{split}
\end{equation}
where the indexing change indicates that graphed momenta are stored
from negative to positive values. For plotted frequency spectra a
\textbf{positive} sign is used in the frequency exponent of the transform
to frequency space, to agree with common physics conventions.

\section{Derivatives}

\subsection{Spectral derivatives}

For spectral derivatives in the interaction picture, we define $D_{x}\left(k\right)$
to obtain a derivative. To explain, one integrates by parts: 
\begin{equation}
\begin{split}D_{x}^{p}\tilde{\boldsymbol{a}}\left(\boldsymbol{k}\right)=\left[ik_{x}\right]^{p}\tilde{\boldsymbol{a}}\left(\boldsymbol{k}\right)=\frac{1}{\left(2\pi\right)^{d/2}}\int d\boldsymbol{x}e^{-i\boldsymbol{k}\cdot\boldsymbol{x}}\left[\frac{\partial}{\partial x}\right]^{p}\boldsymbol{a}\left(\boldsymbol{x}\right).\end{split}
\end{equation}
This means, for example, that to calculate a one dimensional space
derivative in a Fourier interaction picture routine, one uses:

\begin{equation}
\nabla_{x}\rightarrow D_{x}.
\end{equation}

Here Dx is an array of momenta in cyclic order in dimension $d$ as
defined above, suitable for an FFT calculation. The imaginary $i$
is not needed to give the correct sign, as it is included in the derivative
array. In two dimensions, a full two-dimensional Laplacian is:

\begin{equation}
\boldsymbol{\nabla}^{2}=\nabla_{x}^{2}+\nabla_{y}^{2}\rightarrow D_{x}^{2}+D_{y}^{2}.
\end{equation}

Then, on inverting the transform 
\begin{equation}
\left[\frac{\partial}{\partial x}\right]^{p}\boldsymbol{a}\left(\boldsymbol{x}\right)=\frac{1}{\left(2\pi\right)^{d/2}}\int d\boldsymbol{x}e^{i\boldsymbol{k}\cdot\boldsymbol{x}}\left[D_{x}\left(\boldsymbol{k}\right)\right]^{p}\tilde{\boldsymbol{a}}\left(\boldsymbol{k}\right).
\end{equation}

\subsection{Finite difference derivatives}

For calculating derivatives using finite differences, the following
central differencing method is used, away from the boundaries: 
\[
\nabla_{x}a\left(x_{i}\right)\rightarrow\frac{1}{2\Delta x}\left[a\left(x_{i+1}\right)-a\left(x_{i-1}\right)\right]
\]

\begin{equation}
\nabla_{x}^{2}a\left(x_{i}\right)\rightarrow\frac{1}{\Delta x^{2}}\left[a\left(x_{i+1}\right)-2a\left(x_{i}\right)+a\left(x_{i-1}\right)\right].
\end{equation}

This raises the question of how to calculate derivatives at the boundary,
for example at the lower boundary $x_{1}$, where $a\left(x_{0}\right)$
is not known, and similarly at the upper boundary. The answer depends
on the boundary type \cite{crank1947practical}, and is obtained by
extending the boundary to additional points $a\left(x_{0}\right)$
and $a\left(x_{N+1}\right)$ that are assumed to extend the boundary
condition:

\paragraph{Periodic:~$a\left(x_{0}\right)=a\left(x_{N}\right)$ }

\[
\nabla_{x}a\left(x_{1}\right)\rightarrow\frac{1}{2\Delta x}\left[a\left(x_{2}\right)-a\left(x_{N}\right)\right]
\]

\begin{equation}
\nabla_{x}^{2}a\left(x_{1}\right)\rightarrow\frac{1}{\Delta x^{2}}\left[a\left(x_{2}\right)-2a\left(x_{2}\right)+a\left(x_{N}\right)\right].
\end{equation}

\paragraph{Dirichlet: $\tilde{a}\left(x_{1}\right)$ specified: ~$a\left(x_{0}\right)=\tilde{a}\left(x_{1}\right)$ }

\[
\nabla_{x}a\left(x_{1}\right)\rightarrow\frac{1}{2\Delta x}\left[a\left(x_{2}\right)-\tilde{a}\left(x_{1}\right)\right]
\]

\begin{equation}
\nabla_{x}^{2}a\left(x_{1}\right)\rightarrow\frac{1}{\Delta x^{2}}\left[a\left(x_{2}\right)-\tilde{a}\left(x_{1}\right)\right].
\end{equation}

\paragraph{Robin/Neumann: $\tilde{a}'\left(x_{1}\right)$ specified: ~$a\left(x_{0}\right)=a\left(x_{2}\right)-2\tilde{a}'\left(x_{1}\right)\Delta x$ }

\[
\nabla_{x}a\left(x_{1}\right)\rightarrow\tilde{a}'\left(x_{1}\right)
\]

\begin{equation}
\nabla_{x}^{2}a\left(x_{1}\right)\rightarrow\frac{2}{\Delta x^{2}}\left[a\left(x_{2}\right)-a\left(x_{1}\right)-\tilde{a}'\left(x_{1}\right)\Delta x\right].
\end{equation}

In all cases the boundary value is evaluated as part of the derivative
evaluation, so it can be a nonlinear function of $\mathbf{a}$.

\newpage{}

\chapter{SPDE examples\label{chap:SPDE-examples}}

\section{Gaussian diffraction}

Free diffraction and absorption of a Gaussian wave-function in $d-1=s$
space dimensions, is given by the partial differential equation (PDE):
\begin{equation}
\frac{da}{dt}=-\frac{\gamma}{2}a+\frac{i}{2}D\nabla^{2}a.
\end{equation}

The corresponding stochastic partial differential equation (SPDE)
includes additional noise, so that:

\begin{equation}
\frac{da}{dt}=-\frac{\gamma}{2}a+\frac{i}{2}D\nabla^{2}a+bw(t,x).
\end{equation}

The xSPDE spectral definition in space is: 
\begin{equation}
\tilde{a}(t,\mathbf{k})=\frac{1}{\left[2\pi\right]^{s/2}}\int e^{i\mathbf{k}\cdot\mathbf{x}}a(t,\mathbf{x})d\mathbf{x}\,.
\end{equation}

Together with the initial condition that $a(0,x)=exp(-\left|\mathbf{x}\right|^{2}/2)$,
this has an exact solution for the diffracted intensity with $b=0$,
in either ordinary space or momentum space: 
\begin{eqnarray}
\left|a\left(t,\mathbf{x}\right)\right|^{2} & = & \frac{1}{\left(1+\left(Dt\right)^{2}\right)^{s/2}}exp\left(-\left|\mathbf{x}\right|^{2}/\left(1+\left(Dt\right)^{2}\right)-\gamma t\right)\nonumber \\
\left|\tilde{a}\left(t,\mathbf{k}\right)\right|^{2} & = & exp\left(-\left|\mathbf{k}\right|^{2}-\gamma t\right).
\end{eqnarray}

\subsection*{Exercises}
\begin{itemize}
\item Simulate Gaussian diffraction in three dimensions using an xSPDE function 
\item Check your results against the exact solution 
\item The example below stores data in a standard Matlab file. 
\end{itemize}
\begin{center}
\doublebox{\begin{minipage}[t]{0.75\columnwidth}%
\texttt{function {[}e{]} = Gaussian()}

\texttt{p.dimensions = 4;}

\texttt{p.name = 'Gaussian diffraction';}

\texttt{p.initial = @(v,p) exp(-0.5{*}(p.x.\textasciicircum 2+p.y.\textasciicircum 2+p.z.\textasciicircum 2));}

\texttt{p.linear = @(p) 1i{*}0.05{*}(p.Dx.\textasciicircum 2+p.Dy.\textasciicircum 2+p.Dz.\textasciicircum 2);}

\texttt{p.observe = @(a,p) a.{*}conj(a);}

\texttt{p.olabels = '\textbar a(t,x)\textbar\textasciicircum 2';}

\texttt{p.file = 'Gaussian.mat';}

\texttt{p.images = 4;}

\texttt{e = xsim(p);}

\texttt{xgraph(p.file);}

\texttt{end}%
\end{minipage}} 
\par\end{center}
\begin{itemize}
\item \textbf{Add an additive complex noise of $0.01(w_{1}+iw_{2}$) to
the Gaussian differential equation, then replot with an average over
$100$ samples.} 
\item Work out the exact solution and repeat the comparisons. 
\end{itemize}
Note that for this, you'll need to add: $p.deriv=@(a,w,p)\,\,..+0.01*(w(1,:)+i*w(2,:))$

\section{Stochastic Ginzburg-Landau equation}

Including two space dimensions, or space-time dimensions of $d=3$,
an example of a SPDE is the stochastic Ginzburg-Landau equation. This
describes symmetry breaking. The system develops a spontaneous phase
which varies spatially as well. The model is used to describe lasers,
magnetism, superconductivity, superfluidity and particle physics:
\begin{equation}
\dot{a}=\left(1-\left|a\right|^{2}\right)a+bw(t)+c\nabla^{2}a
\end{equation}
where 
\begin{equation}
\left\langle w(x)w^{*}(x')\right\rangle =2\delta\left(t-t'\right)\delta\left(x-x'\right).
\end{equation}

The following new ideas are introduced for this problem: 
\begin{enumerate}
\item \textbf{$\mathtt{dimensions}$ is the space-time dimension.} 
\item \textbf{The} 'dot' \textbf{notation used for parallel operations over
lattices}. 
\item \textbf{$\mathtt{linear}$ is the linear operator - a Laplacian in
these cases.} 
\item \textbf{$\mathtt{images}$ produces movie-style images at discrete
time slices.} 
\item \textbf{$\mathtt{Dx}$ indicates a derivative operation, $\partial/\partial x$.} 
\item \textbf{$-5<x<5$ is the default xSPDE coordinate range in space.} 
\end{enumerate}

\subsection*{Exercises}
\begin{figure}
\centering{}\includegraphics[width=0.75\textwidth]{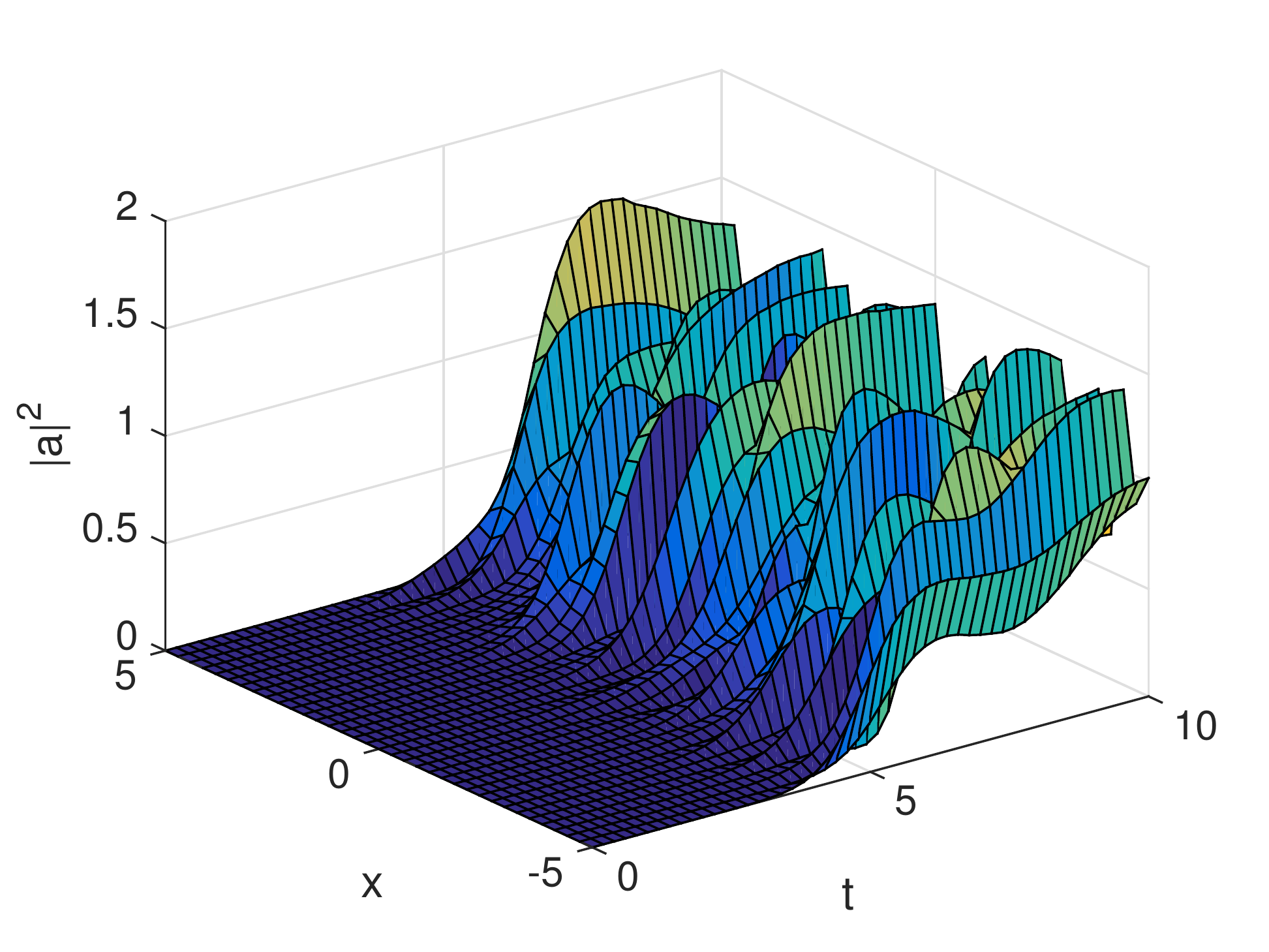}

\caption{\label{fig:Symmetry-breaking}\emph{Simulation of the stochastic equation
describing symmetry breaking in two dimensions. Spatial fluctuations
are caused by the different phase-domains that interfere. The graph
obtained here is projected onto the $y=0$ plane.}}
\vspace{10pt}
\end{figure}

\begin{enumerate}
\item \textbf{Solve the stochastic G-L equation for $b=0.001$ and $c=0.01i$.} 
\item \textbf{Change to a real diffusion so that $c=0.1$.} 
\end{enumerate}
In the first case, you should get the output graphed in Fig (\ref{fig:Symmetry-breaking})
. 
\begin{center}
\doublebox{\begin{minipage}[t]{0.75\columnwidth}%
\texttt{clear;}

\texttt{p.name = 'Extended laser gain equation';}

\texttt{p.noises = 2;}

\texttt{p.dimensions = 3;}

\texttt{p.steps = 10;}

\texttt{p.linear = @(p) 1i{*}0.01{*}(p.Dx.\textasciicircum 2+p.Dy.\textasciicircum 2);}

\texttt{p.observe = @(a,\textasciitilde ) abs(a).\textasciicircum 2;}

\texttt{p.images = 6;}

\texttt{p.olabels = '\textbar a\textbar\textasciicircum 2';}

\texttt{p.deriv = @(a,w,\textasciitilde ) (1-abs(a(1,:).\textasciicircum 2)).{*}a(1,:)+0.001{*}(w(1,:)+1i{*}w(2,:));}

\texttt{xspde(p)}%
\end{minipage}} 
\par\end{center}

Here the notation $a(1,:)$ means that the operation is repeated over
all values of the subsequent indices, which are the two spatial lattice
indices in this case.

\newpage{}

\section{NLS soliton}

The famous nonlinear Schrödinger equation (NLSE) is: 
\begin{equation}
\frac{da}{dt}=\frac{i}{2}\left[\nabla^{2}a-a\right]+ia\left|a\right|^{2}.
\end{equation}

Together with the initial condition that $a(0,x)=sech(x)$, this has
a soliton \cite{Scott1973soliton}, an exact solution that doesn't
change in time: 
\begin{eqnarray}
a(t,x) & = & sech(x).
\end{eqnarray}
The Fourier transform at $k=0$ is simply: 
\begin{eqnarray}
\tilde{a}(t,0) & = & \frac{1}{\sqrt{2\pi}}\int sech(x)dx=\sqrt{\frac{\pi}{2}}.
\end{eqnarray}

\subsection*{Exercises}
\begin{itemize}
\item \textbf{Solve the NLSE for a soliton using a function instead of a
script, then include an additive complex noise of $0.01(w_{1}+iw_{2}$)
to the differential equation, and plot again with an average over
$1000$ samples.} 
\end{itemize}
\newpage{}

\section{Planar noise}

The next example is growth of thermal noise of a two-component complex
field in a plane, given by the equation 
\begin{equation}
\frac{d\mathbf{a}}{dt}=\frac{i}{2}\nabla^{2}\mathbf{a}+\mathbf{w}(t,x).
\end{equation}
where $\mathbf{\zeta}$ is a delta-correlated complex noise vector
field: 
\begin{equation}
w_{j}(t,\mathbf{x})=\left[w_{j}^{re}(t,\mathbf{x})+i\zeta_{j}^{im}(t,\mathbf{x})\right]/\sqrt{2},
\end{equation}
with the initial condition that the initial noise is delta-correlated
in position space 
\begin{equation}
a(0,\mathbf{x})=\mathbf{\zeta}^{(in)}(\mathbf{x})
\end{equation}
where: 
\begin{equation}
\mathbf{\zeta}^{(in)}(\mathbf{x})=\left[\mathbf{\zeta}^{re(in)}(\mathbf{x})+i\mathbf{\zeta}^{im(in)}(\mathbf{x})\right]/\sqrt{2}
\end{equation}

This has an exact solution for the noise intensity in either ordinary
space or momentum space: 
\begin{eqnarray}
\left\langle \left|a_{j}\left(t,\mathbf{x}\right)\right|^{2}\right\rangle  & = & (1+t)/dV\nonumber \\
\left\langle \left|\tilde{a}_{j}\left(t,\mathbf{k}\right)\right|^{2}\right\rangle  & = & (1+t)/dV_{k}\nonumber \\
\left\langle \tilde{a}_{1}\left(t,\mathbf{k}\right)\tilde{a}_{2}^{*}\left(t,\mathbf{k}\right)\right\rangle  & = & 0.
\end{eqnarray}

Here, the noise is delta-correlated, and $dV$, $dV_{k}$ are the
cartesian space and momentum space lattice cell volumes, respectively.
Suppose that $N_{s}=N_{x}N_{y}$ is the total number of spatial points,
and there are $N_{x(y)}$ points in the x(y)-direction, so then: 
\begin{eqnarray}
dV & = & dxdy\\
dV_{k} & = & dk_{x}dk_{y}=\frac{(2\pi)^{2}}{V}.\nonumber 
\end{eqnarray}

In the simulations, two planar noise fields are propagated, one using
delta-correlated noise, the other with noise transformed to momentum
space to allow filtering. This allows use of finite correlation lengths
when needed, by including a frequency filter function that is used
to multiply the noise in Fourier-space. The Fourier-space noise variance
is the square of the filter function.

The parameter $p.noises$ indicates how many noise fields are generated,
while $p.knoises$ indicates how many of these are spatially correlated,
via Fourier transform, filter and inverse Fourier transform. These
appear as additional noise cells. The filtered noises have a finite
correlation length in general, but in this example are delta-correlated.

\subsection*{Exercises}
\begin{itemize}
\item \textbf{Solve the planar noise growth equation} 
\end{itemize}
\begin{center}
\doublebox{\begin{minipage}[t]{0.75\columnwidth}%
\texttt{function {[}e{]} = PlanarExample()}

\texttt{p.name = 'Planar noise growth';}

\texttt{p.dimensions = 3;}

\texttt{p.fields = 2;}

\texttt{p.ranges = {[}1,5,5{]};}

\texttt{p.steps = 2;}

\texttt{p.noises = 2;}

\texttt{p.knoises = 2;}

\texttt{p.inrandoms = 2; }

\texttt{p.krandoms = 2; }

\texttt{p.ensembles = {[}10,4,4{]};}

\texttt{p.initial = @Initial;}

\texttt{p.deriv = @Da;}

\texttt{p.linear = @(p) 0.5{*}1i{*}(p.Dx.\textasciicircum 2+p.Dy.\textasciicircum 2);}

\texttt{p.observe = @(a,p) a(1,:).{*}conj(a(1,:));}

\texttt{p.olabels = '\textless\textbar a\_1(x)\textbar\textasciicircum 2\textgreater ';}

\texttt{p.compare = @(p) (1+p.t)/p.dv;}

\texttt{p.images = 4;}

\texttt{e = xspde(p);}

\texttt{end }

\texttt{function a0 = Initial(u,v,\textasciitilde ) }

\texttt{a0(1,:,:) = (u(1,:,:)+1i{*}u(2,:,:))/sqrt(2); }

\texttt{a0(2,:,:) = (v(1,:,:)+1i{*}v(2,:,:))/sqrt(2); }

\texttt{end}

\texttt{function da = Da(\textasciitilde ,w,z,\textasciitilde ) }

\texttt{da(1,:) = (w(1,:)+1i{*}w(2,:))/sqrt(2); }

\texttt{da(2,:) = (z(1,:)+1i{*}z(2,:))/sqrt(2); }

\texttt{end }%
\end{minipage}} 
\par\end{center}
\begin{itemize}
\item \textbf{Add a decay rate of $-a$ to the differential equation, then
plot again} 
\item \textbf{Add growth and nonlinear saturation terms} 
\end{itemize}

\section{Gross-Pitaevskii equation}

The next example is a stochastic Gross-Pitaevskii (GP) equation \cite{Gardiner2003Stochastic}
in two dimensions, 
\begin{equation}
\frac{da}{dt}=\frac{i}{2}\nabla^{2}a-ia(V(r)-i\kappa(r)+\left|a\right|^{2})+\epsilon\eta\label{eq:Stochastic GPE}
\end{equation}
where $\eta$ is a correlated complex noise vector field: 
\begin{equation}
\eta(t,\mathbf{x})=w_{1}(t,\mathbf{x})+iw_{2}(t,\mathbf{x}),
\end{equation}
with the initial condition that the initial random field and the noise
are both filtered in momentum space 
\begin{equation}
a(0,\mathbf{x})=a_{0}(\mathbf{x})+\epsilon\zeta^{(in)}(\mathbf{x})
\end{equation}
where: 
\begin{equation}
\zeta^{(in)}(\mathbf{x})=v_{1}(\mathbf{x})+iv_{2}(\mathbf{x})
\end{equation}

We add a Gaussian filter in momentum space for both the initial random
field and noise so that, if $\tilde{w}\left(\mathbf{k}\right)$ is
a delta-correlated noise in momentum space: 
\begin{align}
w\left(\mathbf{k}\right) & =\tilde{w}\left(\mathbf{k}\right)\exp\left(-\left|\mathbf{k}\right|^{2}\right)\nonumber \\
v\left(\mathbf{k}\right) & =\tilde{v}\left(\mathbf{k}\right)\exp\left(-\left|\mathbf{k}\right|^{2}\right)
\end{align}

This allows use of finite correlation lengths when needed, by including
a frequency filter function that is used to multiply the noise in
Fourier-space. The Fourier-space noise variance is the square of the
filter function.

The first noise index, $p.noises(1)$, indicates how many noise fields
are generated that are delta-correlated in $x$, while $p.noises(2)$
indicates how many of these are spatially correlated, via Fourier
transform, filter and inverse Fourier transform. These appear to the
user as additional noises, so the total is $p.noises(1)+p.noises(2)$.
The filtered noises have a finite correlation length.

\subsection*{Exercises}
\begin{itemize}
\item \textbf{Solve the stochastic GP equation \eqref{eq:Stochastic GPE},
with a noise coefficient of $b=0.1$, $V=0.01\left|\mathbf{x}\right|^{2},$
$\kappa=0.001\left|\mathbf{x}\right|^{4}$, and a stored output data
file.} 
\end{itemize}
\begin{center}
\doublebox{\begin{minipage}[t]{0.75\columnwidth}%
\texttt{function {[}e{]} = GPE()}

\texttt{p.name = 'GPE';}

\texttt{p.dimensions = 3;}

\texttt{p.points = {[}101,64,64{]};}

\texttt{p.ranges = {[}1,20,20{]};}

\texttt{p.noises = 0;}

\texttt{p.knoises = 2;}

\texttt{p.inrandoms = 0;}

\texttt{p.krandoms = 2;}

\texttt{p.rfilter = @(w,p) w.{*}exp(-p.kx.\textasciicircum 2-p.ky.\textasciicircum 2);}

\texttt{p.nfilter = @(v,p) v.{*}exp(-p.kx.\textasciicircum 2-p.ky.\textasciicircum 2);}

\texttt{b = @(xi) .1{*}(xi(1,:,:)+1i{*}xi(2,:,:));}

\texttt{p.initial = @(u,v,p) (p.x+1i{*}p.y)./(1+10{*}(p.x.\textasciicircum 2
+p.y.\textasciicircum 2))+b(v);}

\texttt{V = @(p) 0.01{*}(p.x.\textasciicircum 2 + p.y.\textasciicircum 2)-0.001{*}1i{*}(p.x.\textasciicircum 2
+p.y.\textasciicircum 2).\textasciicircum 2;}

\texttt{p.deriv = @(a,v,w,p) -1i{*}a.{*}(V(p)+conj(a).{*}a)+b(w);}

\texttt{p.linear = @(p) 0.5{*}1i{*}(p.Dx.\textasciicircum 2+p.Dy.\textasciicircum 2);}

\texttt{p.observe\{1\} = @(a,p) a.{*}conj(a);}

\texttt{p.images = \{2\};}

\texttt{p.imagetype = \{2\};}

\texttt{p.olabels = \{'\textbar a\textbar\textasciicircum 2'\};}

\texttt{p.file = 'GPE.mat';}

\texttt{e = xsim(p);}

\texttt{xgraph(p.file,p);}

\texttt{end}%
\end{minipage}} 
\par\end{center}

\newpage{}

\section{Characteristic equation}

The next example is the characteristic equation for a traveling wave
at constant velocity \cite{courant2008methods}. It is included to
illustrate what happens at periodic boundaries, when Fourier-transform
methods are used for propagation. There are a number of methods known
to prevent this effect, including addition of absorbers - called apodization
- at the boundaries. The equation is: 
\begin{equation}
\frac{da}{dt}+\frac{da}{dx}=0.
\end{equation}

Together with the initial condition that $a(0,x)=sech(2x+5)$, this
has an exact solution that propagates at a constant velocity: 
\begin{eqnarray}
a(t,x) & = & sech(2(x-t)+5).
\end{eqnarray}
The time evolution at $x=0$ is simply: 
\begin{eqnarray}
a(t,0) & = & sech(2(t-5/2)).
\end{eqnarray}

\subsection*{Exercises}
\begin{itemize}
\item \textbf{Solve the characteristic equation given above, noting the
effects of periodic boundaries.} 
\end{itemize}
\begin{center}
\doublebox{\begin{minipage}[t]{0.75\columnwidth}%
\texttt{function {[}e{]} = Characteristic()}

\texttt{p.name = 'Characteristic';}

\texttt{p.dimensions = 2;}

\texttt{p.initial = @(v,p) sech(2.{*}(p.x+2.5));}

\texttt{p.deriv = @(a,z,p) 0{*}a;}

\texttt{p.linear = @(p) -p.Dx;}

\texttt{p.olabels = \{'a\_1(x)'\};}

\texttt{p.compare = @(p) sech(2.{*}(p.t-2.5));}

\texttt{e = xspde(p);}

\texttt{end}%
\end{minipage}} 
\par\end{center}
\begin{itemize}
\item \textbf{Recalculate with the opposite velocity, and a new exact solution.} 
\end{itemize}

\section{Nonlinear Anderson localization}

A random potential prevents normal wave-packet spreading in quantum-mechanics.
This is Anderson localization \cite{Anderson1958Absence}: a famous
property of quantum mechanics in a random potential. A typical experimental
method is to confine an ultra-cold Bose-Einstein condensate (BEC)
in a trap, then release the BEC in a random external potential produced
by a laser \cite{billy2008direct}. The expansion rate of the BEC
is reduced by the Anderson localization due to the random potential.
Physically, the observable quantity is the particle density $n=\left|\psi\right|^{2}$,
but there is a complication, which is that there are nonlinearities
from atomic scattering \cite{pikovsky2008destruction}.

This can be treated either using a Schrödinger equation with a random
potential, at low density, or using the Gross-Pitaevskii (GP) equation
to include atom-atom interactions at the mean field level. In this
example of a problem where strong localization occurs, the general
equations are:

\begin{equation}
\frac{\partial\psi}{\partial t}=\frac{1}{i\hbar}\left[-\frac{\hbar^{2}}{2m}\nabla^{2}+V\left(\mathbf{r}\right)+g\left|\psi\right|^{2}\right]\psi.
\end{equation}

In calculations, it is best to use a dimensionless form by rescaling
coordinates and fields. A simple way to simulate this with xSPDE is
to treat $\psi$ as a scaled field $a(1),$ and to assume the random
potential field $V\left(\mathbf{r}\right)$ as caused by interactions
with second random field $\left|a(2)\right|^{2}$. This has the advantage
that it is similar to the actual experiment and allows one to treat
time-dependent potentials as well, if desired.

With the rescaling, this simplifies to: 
\begin{equation}
\frac{\partial a_{1}}{\partial\tau}=i\left[\frac{\partial}{\partial\zeta^{2}}^{2}-\left|a_{2}\right|^{2}-\left|a_{1}\right|^{2}\right]a_{1}.
\end{equation}

A convenient initial condition is to use: 
\begin{eqnarray}
a_{1} & = & a_{0}\exp(-\zeta^{2})\nonumber \\
\left\langle a_{2}(\zeta)a_{2}(\zeta')\right\rangle  & = & v\delta\left(\zeta-\zeta'\right).
\end{eqnarray}

\subsection*{Exercise}
\begin{itemize}
\item \textbf{Solve Schrödinger's equation without a random potential, to
observe expansion.} 
\item \textbf{Include a random potential $v$, to observe localization.} 
\item \textbf{Experiment with nonlinear terms and higher dimensions.} 
\end{itemize}
The GP equation is a mean field approximation; this is still not a
full solution of the many-body problem! Also, the experiments are
more complicated than this, and actually observe the momentum distribution.

\section{Nonlinear Schrödinger equation with Neumann boundary conditions\label{subsec:Nonlinear-Schrodinger-equation}}

This solves a (1+1)-dimensional PSDE with an initial condition of
$a\left(t=0,x\right)=sech\left(x\right)$ and

\begin{eqnarray}
\frac{\partial a}{\partial t} & = & i\cdot\left(a\cdot\left(\left|a\right|^{2}-\frac{1}{2}\right)+\frac{1}{2}\frac{\partial^{2}a}{\partial x^{2}}\right)\,.
\end{eqnarray}
The solution is subject to Neumann boundary conditions with boundary
values at zero

\begin{eqnarray}
\frac{\partial a}{\partial x}\left(t,\pm x_{m}\right) & = & 0\,.
\end{eqnarray}

The equation is a deterministic nonlinear Schrödinger equation, which
applies to nonlinear optics, Bose-Einstein condensates and plasma
physics. The observables are $o_{1}\equiv\left|a\right|^{2}$ and
$o_{2}\equiv\int_{-x_{m}}^{x_{m}}\left|\frac{\partial}{\partial x}a\right|^{2}dx$.

\paragraph{Notes}
\begin{itemize}
\item The boundary conditions are specified with p.boundaries\{2\}, which
is the x-dimension. 
\item The integration differential $dx$ does not have to be entered, as
this is the default. 
\item Three transverse graphs were specified, but they aren't reproduced
here. 
\item As there is only one field, which is the default, this does not need
to be given. 
\item Since there is no noise, the default integration method was RK4. 
\end{itemize}
\begin{center}
\doublebox{\begin{minipage}[t]{0.9\columnwidth}%
\texttt{function {[}e{]} = SolitonDerivN()}

\texttt{p.dimensions = 2;}

\texttt{p.points = {[}101,101{]};}

\texttt{p.ranges = {[}10,15{]};}

\texttt{p.initial = @(v,p) sech(p.x);}

\texttt{p.observe\{1\} = @(a,p) a.{*}conj(a);}

\texttt{p.observe\{2\} = @(a,p) Int(abs(D1(a,2,p)).\textasciicircum 2,p);}

\texttt{p.olabels = \{'\textbar a\textbar\textasciicircum 2','\textbackslash int
\textbar da/dx\textbar\textasciicircum 2 dx'\};}

\texttt{p.name = 'NLS soliton:spectral method + Neumann';}

\texttt{p.boundaries\{2\} = {[}-1,-1{]};}

\texttt{p.transverse = \{3\};}

\texttt{p.deriv = @(a,\textasciitilde ,p) 1i{*}a.{*}(conj(a).{*}a);}

\texttt{p.linear = @(p) 0.5{*}1i{*}(p.Dx.\textasciicircum 2-1);}

\texttt{e = xspde(p);}

\texttt{end}%
\end{minipage}} 
\par\end{center}

\newpage{}

\begin{figure}[H]
\begin{centering}
\includegraphics[width=0.75\textwidth]{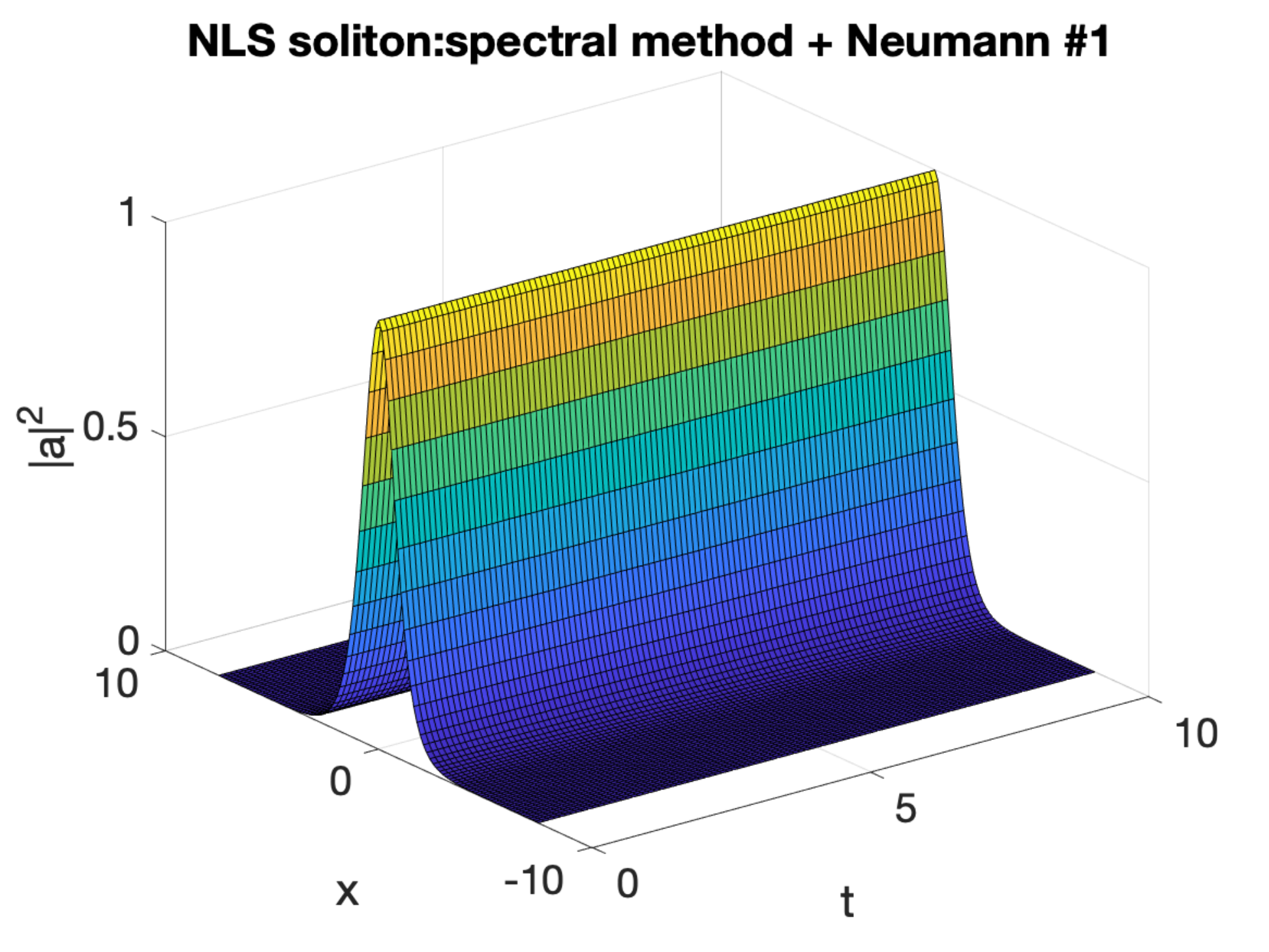}
\par\end{centering}
\centering{}\includegraphics[width=0.75\textwidth]{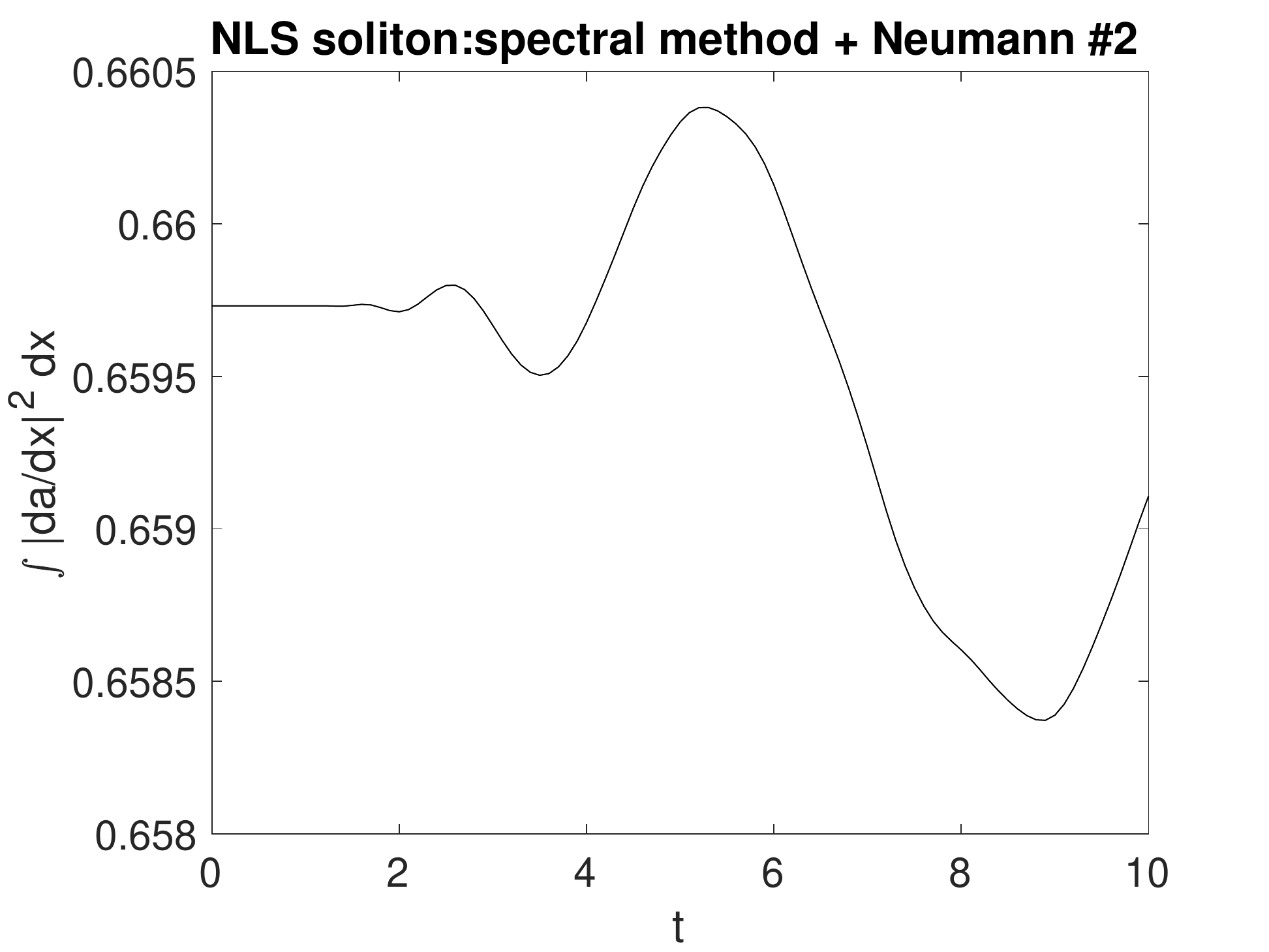}

\caption{\emph{Top figure: Evolution of the field modulus squared of an NLS
soliton with Neumann boundaries. }\protect \protect \\
 \emph{Bottom figure: Evolution of the integrated modulus squared
of the gradient for an NLS soliton with Neumann boundaries, showing
how the reflected fields at the boundaries change the result even
though this is not readily visible above.}}
\vspace{10pt}
\end{figure}

\pagebreak{}

\section{Planar noise growth}

This solves a (1+2)-dimensional PSDE describing the growth of noise
in a planar vector field with a diffraction term giving rise to noise
dispersion. The equation is:

\begin{eqnarray}
\frac{\partial\mathbf{a}}{\partial t} & = & \frac{i}{2}\left(\frac{\partial^{2}}{\partial x^{2}}+\frac{\partial^{2}}{\partial x^{2}}\right)\mathbf{a}+\mathbf{\eta}\left(t,x\right)\,.
\end{eqnarray}
The initial conditions are that $\mathbf{a}=\left(\mathbf{v}_{x}+i\mathbf{v}_{y}\right)/\sqrt{2}$,
where: 
\begin{equation}
\left\langle v_{i}\left(\mathbf{x}\right)v_{j}\left(\mathbf{x}'\right)\right\rangle =\delta\left(\mathbf{x}-\mathbf{x}'\right)\delta_{ij}
\end{equation}
the noise correlations are that $\mathbf{\eta}=\left(\mathbf{w}_{x}+i\mathbf{w}_{y}\right)/\sqrt{2}$,
where:

\begin{eqnarray}
\left\langle w_{i}\left(\mathbf{r}\right)w_{j}\left(\mathbf{r}'\right)\right\rangle  & = & \delta\left(t-t'\right)\delta_{ij}\left(\mathbf{x}-\mathbf{x}'\right)
\end{eqnarray}

The solution has periodic boundary conditions. The noise correlations
for the second field are specified in momentum space. As there are
no filters, the noise terms are delta-correlated in both momentum
($\mathbf{k}$) and in space ($x$). The exact results for comparison
within each field are similar in position and momentum space: 
\begin{align}
\left\langle \left|a_{i}\left(t,\mathbf{x}\right)\right|^{2}\right\rangle  & =\left(1+t\right)/\Delta A_{x}.\nonumber \\
\left\langle \left|a_{i}\left(t,\mathbf{k}\right)\right|^{2}\right\rangle  & =\left(1+t\right)/\Delta A_{k}.
\end{align}

Here, $\Delta A_{x,k}$ is the area of a lattice cell in space or
momentum space. This is $\Delta A_{x}=1/49$ for the parameters used.
The correlations are proportional to $N_{s}$, the number of points
in the spatial lattice, which is $35^{2}=1225$ for the spatial lattice
used:

\begin{align}
\int\left\langle \left|a_{i}\left(t,\mathbf{x}\right)\right|^{2}\right\rangle d\mathbf{x}=\int\left\langle \left|a_{i}\left(t,\mathbf{k}\right)\right|^{2}\right\rangle d\mathbf{k} & =N_{s}\left(1+t\right).
\end{align}

\paragraph{Notes}
\begin{itemize}
\item All three types of ensemble are used 
\item The much lower sampling error after integration is evident in the
graphs 
\item Spatially resolved graphs show larger sampling errors 
\item The integration method is mid-point, as it is stochastic. 
\item Two k-space noises are specified, but they aren't filtered. 
\item Under these conditions, x-space and k-space noise are identical. 
\end{itemize}
\begin{center}
\doublebox{\begin{minipage}[t]{0.9\columnwidth}%
\texttt{function {[}e{]} = Planar()}

\texttt{p.name = 'Planar noise growth';}

\texttt{p.dimensions = 3;}

\texttt{p.fields = 2;}

\texttt{p.ranges = {[}1,5,5{]};}

\texttt{p.points = 10;}

\texttt{p.noises = 2;}

\texttt{p.knoises = 2;}

\texttt{p.inrandoms = 2;}

\texttt{p.krandoms = 2;}

\texttt{p.ensembles = {[}10,2,12{]};}

\texttt{p.initial = @Initial;}

\texttt{p.deriv = @D\_planar;}

\texttt{p.linear = @(p) 1i{*}0.5{*}(p.Dx.\textasciicircum 2+p.Dy.\textasciicircum 2);}

\texttt{p.observe\{1\} = @(a,p) Int(a(1,:).{*}conj(a(1,:)),p);}

\texttt{p.observe\{2\} = @(a,p) Int(a(2,:).{*}conj(a(2,:)),p.dk,p);}

\texttt{p.observe\{3\} = @(a,p) real(Ave(a(1,:).{*}conj(a(2,:)),p));}

\texttt{p.observe\{4\} = @(a,p) a(2,:).{*}conj(a(2,:));}

\texttt{p.transforms = \{{[}0,0,0{]},{[}0,1,1{]},{[}0,1,1{]}\};}

\texttt{p.olabels\{1\} = '\textless\textbackslash int\textbar{}
a\_1(x)\textbar\textasciicircum 2 d\textasciicircum 2x\textgreater{}
';}

\texttt{p.olabels\{2\} = '\textless\textbackslash int\textbar{}
a\_2(k)\textbar\textasciicircum 2 d\textasciicircum 2k\textgreater{}
';}

\texttt{p.olabels\{3\} = '\textless\textless{} a\_1(k) a\textasciicircum{*}\_2(k)\textgreater\textgreater ';}

\texttt{p.olabels\{4\} = '\textless{} \textbar a\_2(x)\textbar\textasciicircum 2\textgreater ';}

\texttt{p.compare\{1\} = @(p) (1+p.t){*}p.nspace;}

\texttt{p.compare\{2\} = @(p) (1+p.t){*}p.nspace;}

\texttt{p.compare\{3\} = @(p) 0.0;}

\texttt{e = xspde(p);}

\texttt{end}\\

\texttt{function a0 = Initial(u,v,\textasciitilde )}

\texttt{a0(1,:) = (u(1,:)+1i{*}u(2,:))/sqrt(2);}

\texttt{a0(2,:) = (v(1,:)+1i{*}v(2,:))/sqrt(2);}

\texttt{end}\\

\texttt{function da = D\_planar(\textasciitilde ,u,w,\textasciitilde )
\%\%Derivatives}

\texttt{da(1,:) = (u(1,:)+1i{*}u(2,:))/sqrt(2);}

\texttt{da(2,:) = (w(1,:)+1i{*}w(2,:))/sqrt(2);}

\texttt{end}%
\end{minipage}} 
\par\end{center}

\begin{figure}[H]
\begin{centering}
\includegraphics[width=0.75\textwidth]{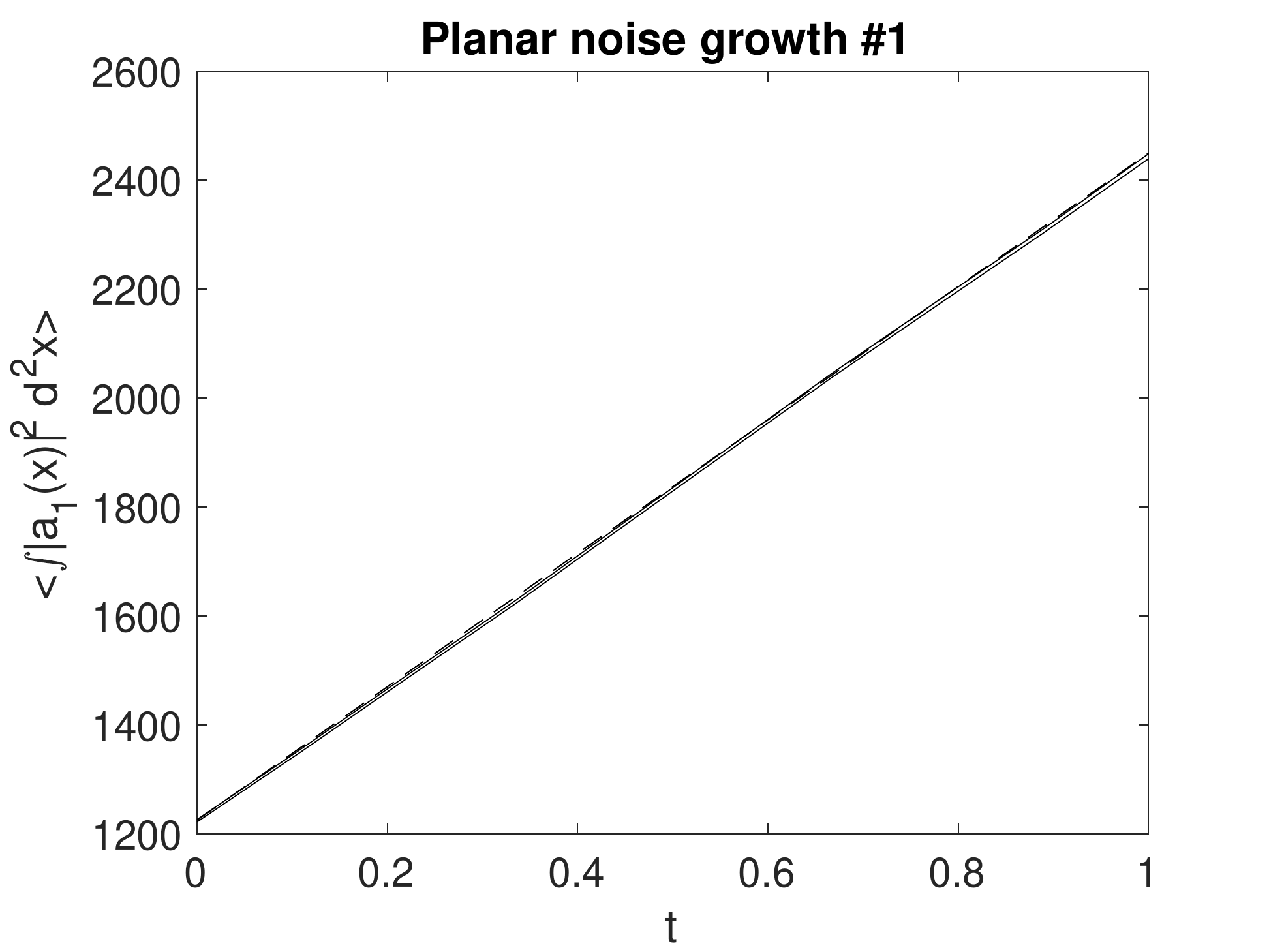}
\par\end{centering}
\centering{}\includegraphics[width=0.75\textwidth]{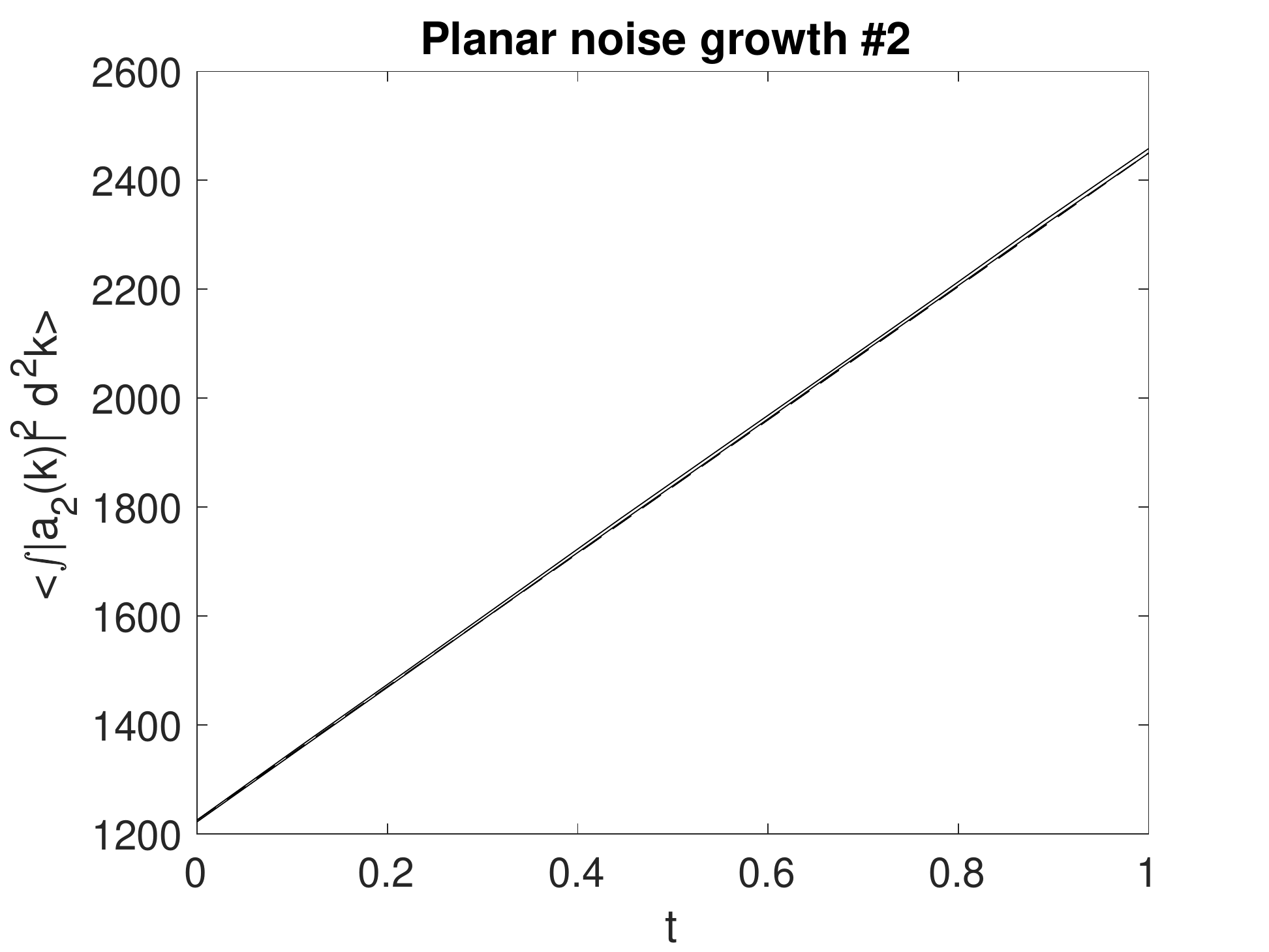}

\caption{\emph{Top and bottom figure: Time evolution of the integrated modulus
square of the first and second field, respectively. The solid lines
indicate upper and lower bounds of the stochastic error, which the
dashed lines indicate theoretical predictions.}}
\vspace{10pt}
\end{figure}

\begin{figure}[H]
\begin{centering}
\includegraphics[width=0.75\textwidth]{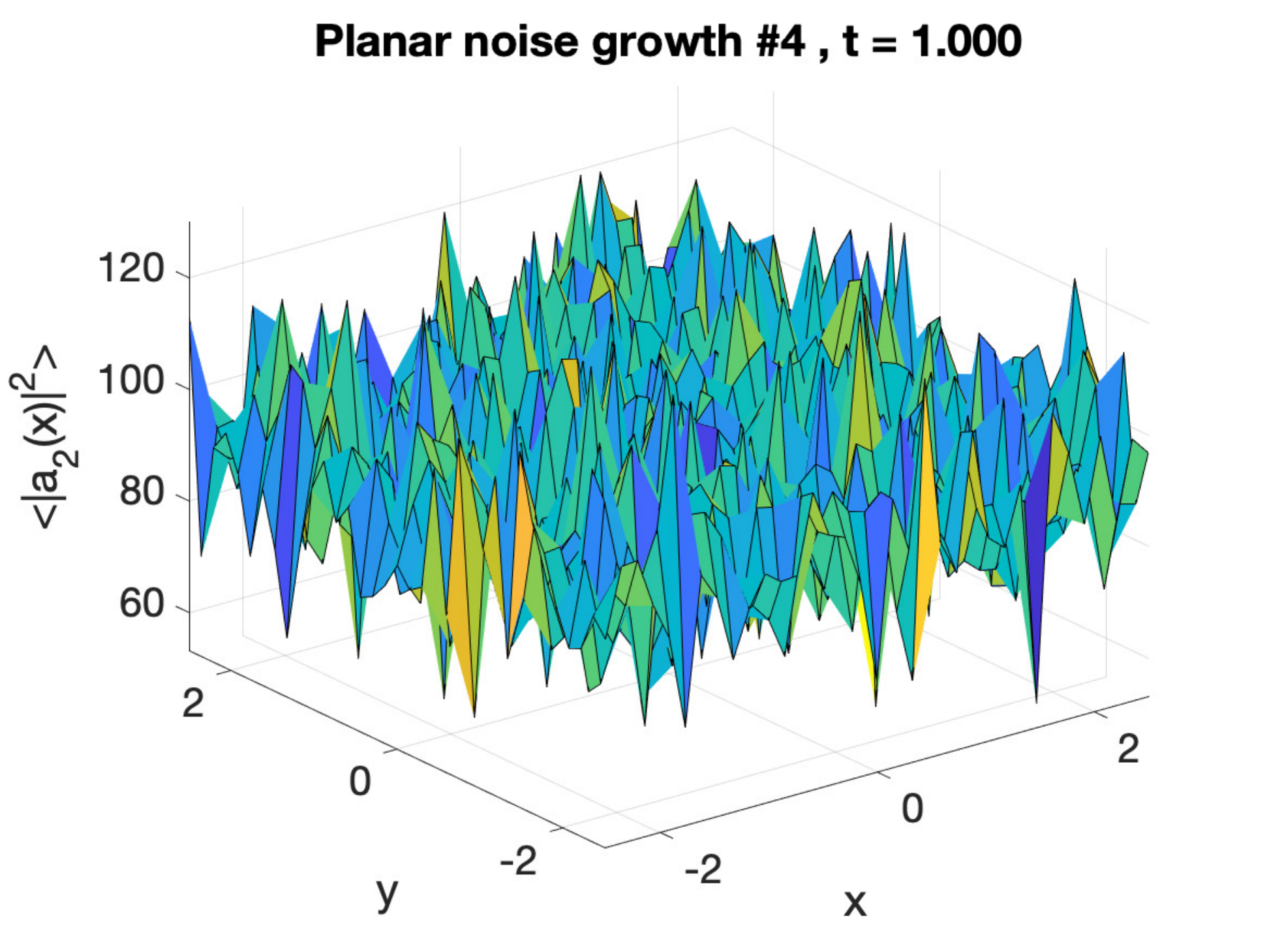}
\par\end{centering}
\centering{}\includegraphics[width=0.75\textwidth]{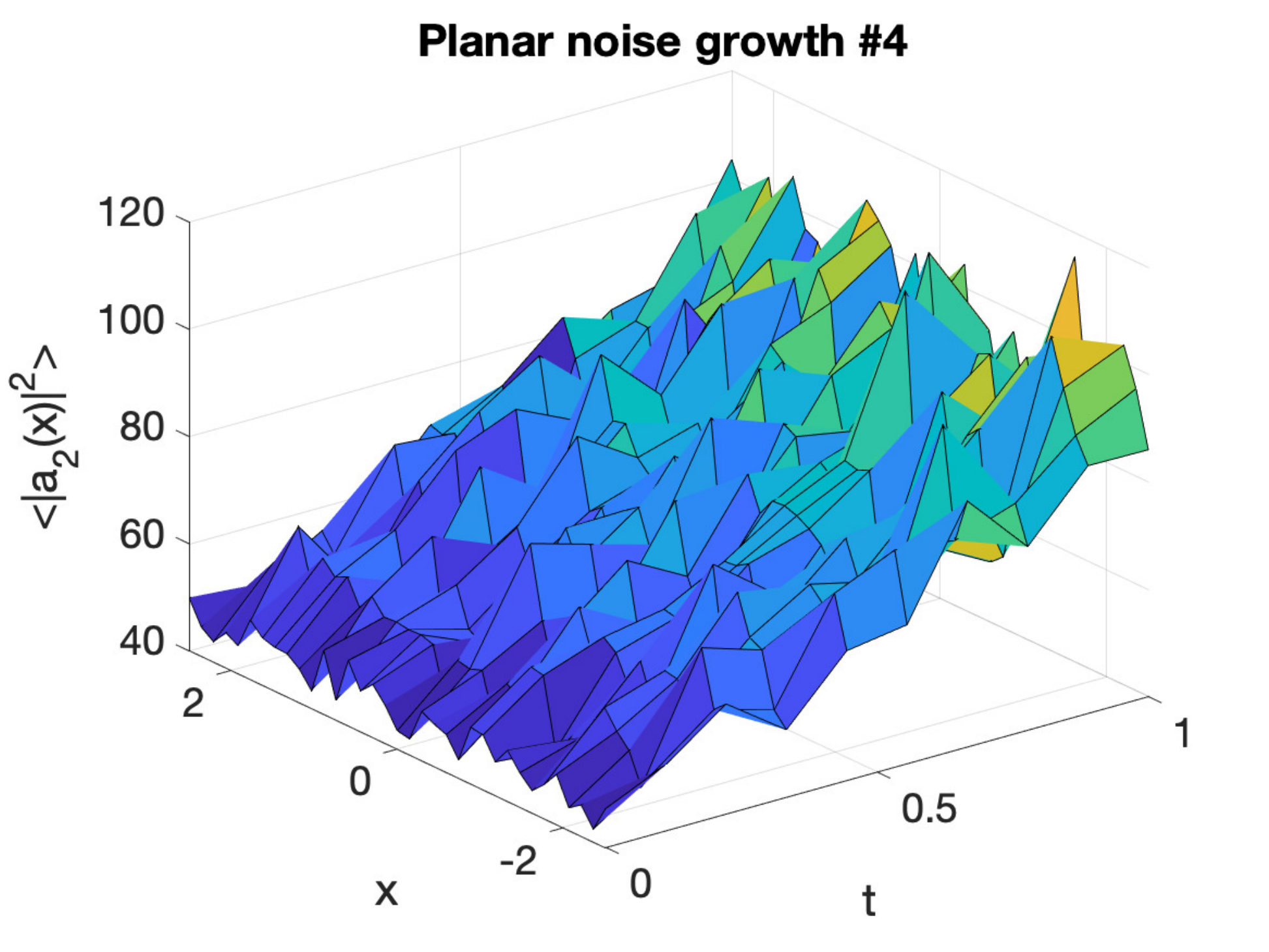}

\caption{\emph{Top figure: 3D plot of the modulus square of $a_{2}$ at $t=1$
as a function of $x$ and $y$. Bottom figure: 3D plot of the modulus
square of $a_{2}$ for $y=0$ as a function of $x$ and $t$.}}
\vspace{10pt}
\end{figure}

\pagebreak{}

\section{Gross-Pitaevskii equation with vortex formation}

This solves a (1+2)-dimensional PDE called the Gross-Pitaevskii equation.
In addition to the standard GPE terms, it includes the vortex forming
term $\left(\mathbf{x}\times\nabla\right)a$. There is just one ensemble
member, to demonstrate how a single trajectory can be imaged. The
equation is:

\begin{eqnarray}
\frac{\partial a}{\partial t} & = & \left(\frac{1}{2}\nabla^{2}a-\left\Vert \left(\left(V\left(\mathbf{x}\right)+200\left|a\right|^{2}\right)+0.6i\cdot\left(\mathbf{x}\times\nabla\right)\right)a\right\Vert \right)\nonumber \\
V\left(\mathbf{x}\right) & = & 0.35\left(x^{2}+y^{2}\right)\nonumber \\
\left\Vert b\left(\mathbf{x}\right)\right\Vert  & = & \frac{b\left(\mathbf{x}\right)}{\int\left|b\right|^{2}d\mathbf{x}}\,.
\end{eqnarray}

Here,$\left\Vert \cdot\right\Vert $ is the normalized derivative
and $\times$ indicates the two-dimensional cross-product. The system
is initialized as

\begin{eqnarray}
a\left(t=0,\mathbf{x}\right) & = & 0.1\cdot\exp\left(-V\left(\mathbf{x}\right)\right)\,.
\end{eqnarray}

\paragraph{Notes}
\begin{itemize}
\item This is a deterministic partial differential equation case 
\item The $15$ intermediate steps used are necessary to reduce integration
errors 
\item The trap potential is an inline function, and is not a parameter 
\item Normalization is used because otherwise particle number is not conserved 
\item The output includes transverse images to show how the vortices develop 
\item Different imagetypes are used to show different 3D features 
\end{itemize}
\begin{center}
\doublebox{\begin{minipage}[t]{0.9\columnwidth}%
\texttt{function {[}e{]} = GPEvortex2D()}

\texttt{p.name = 'GPEvortex2D';}

\texttt{p.dimensions = 3;}

\texttt{p.fields = 1;}

\texttt{p.points = {[}50,40,40{]};}

\texttt{p.ranges = {[}15,16,16{]};}

\texttt{p.steps = 15;}

\texttt{g = 200;}

\texttt{om = 0.6;}

\texttt{L = @(a,p) 1i{*}(p.x.{*}D1(a,3,p)-p.y.{*}D1(a,2,p));}

\texttt{V = @(p) 0.35{*}(p.x.\textasciicircum 2+p.y.\textasciicircum 2);}

\texttt{p.initial = @(v,p) 0.1{*}exp(-V(p));}

\texttt{rho = @(a) g{*}conj(a).{*}a;}

\texttt{p.deriv = @normda;}

\texttt{p.da1 = @(a,w,p) -a.{*}(V(p)+rho(a))+om{*}L(a,p);}

\texttt{p.linear = @(p) 0.5{*}(p.Dx.\textasciicircum 2+p.Dy.\textasciicircum 2);}

\texttt{p.observe\{1\} = @(a,p) a(1,:).{*}conj(a(1,:));}

\texttt{p.observe\{2\} = @(a,p) a(1,:).{*}conj(a(1,:));}

\texttt{p.images = \{2,2\};}

\texttt{p.imagetype = \{1,2\};}

\texttt{p.olabels = \{'\textbar a\textbar\textasciicircum 2','\textbar a\textbar\textasciicircum 2'\};}

\texttt{e = xspde(p);}

\ 

\texttt{function b = normda(a,w,p)}

\texttt{\% b = NORMDA(a,z,p) is a normalized derivative}

\texttt{\% Takes a derivative and returns a normalized step}

\texttt{b = a+p.da1(a,w,p){*}p.dtr;}

\texttt{norm = sqrt(Int(abs(b).\textasciicircum 2,p.dx,p));}

\texttt{b = (b./norm-a)/p.dtr;}

\texttt{end}

\texttt{end}%
\end{minipage}} 
\par\end{center}

\begin{figure}[H]
\begin{centering}
\includegraphics[width=0.75\textwidth]{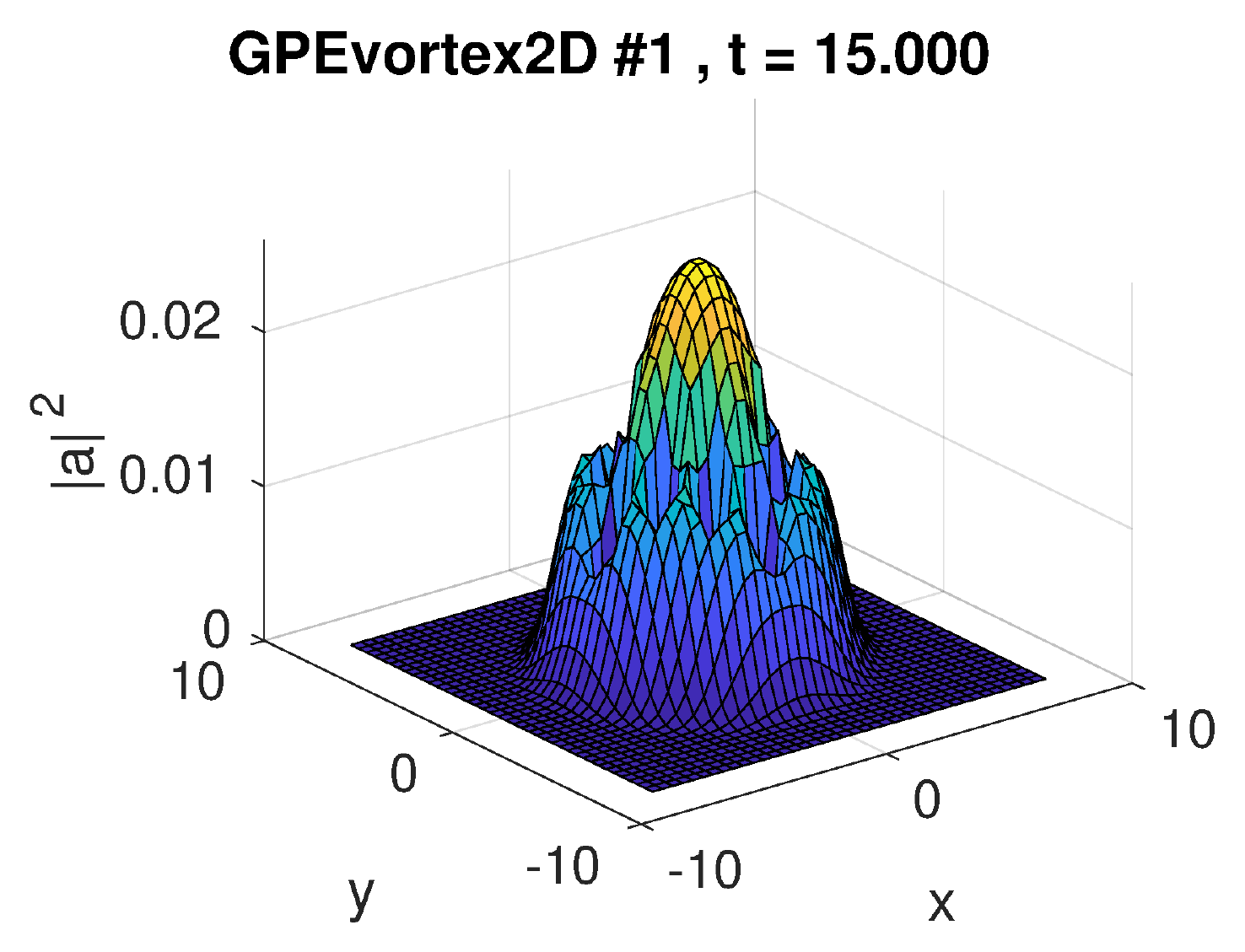}
\par\end{centering}
\centering{}\includegraphics[width=0.75\textwidth]{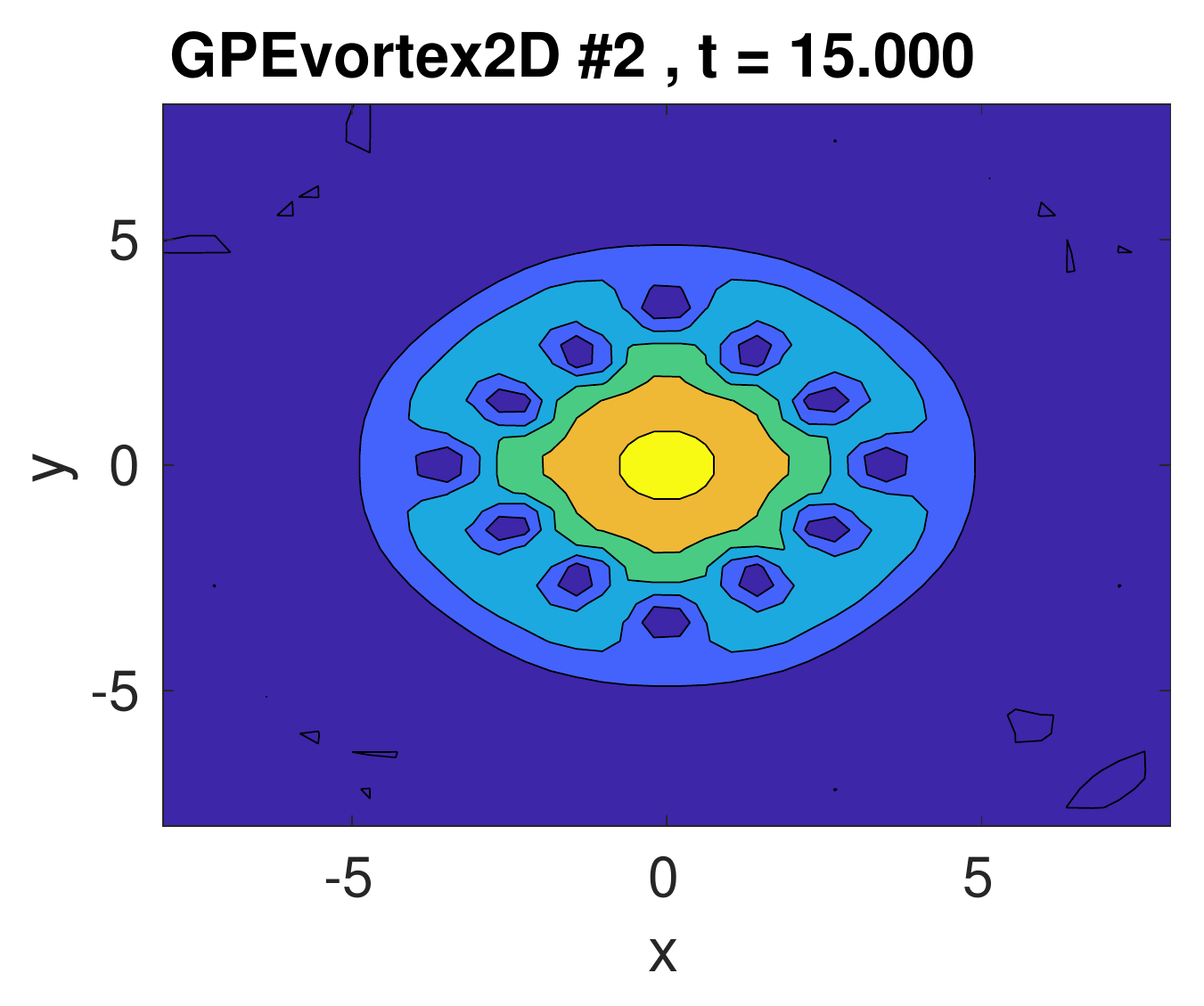}

\caption{\emph{Top and bottom figure: The computed solution for $\left|a\right|^{2}$
at $t=15$ as a function of $x,y$ as a 3D plot (top) and as a color
map (bottom).}}
\vspace{10pt}
\end{figure}

\newpage{}

\section{Heat equation with non-periodic boundaries \label{subsec:Vector_heat}}

This example solves a (1+1)-dimensional PDE with an initial condition
of $\mathbf{a}\left(t=0,x\right)=\mathbf{f}\left(x\right)$ and

\begin{eqnarray}
\frac{\partial\mathbf{a}}{\partial t} & = & \frac{\partial^{2}\mathbf{a}}{\partial x^{2}}\,.
\end{eqnarray}
The solution is subject to either periodic boundary conditions or
Dirichlet and/or Neumann with boundary values of zero at $x_{\pm}=[0,\pi]$
so that $a\left(t,x_{\pm}\right)=0\,$ or $\partial a/\partial x\left(t,x_{\pm}\right)=0$.
Each component has different combinations of boundary types. Using
spectral methods the solutions here are exact, up to round-off errors
of order $10^{-15}$, and are also much faster than with finite differences,
which is demonstrated in the example.

In all cases the grid range is from $x=0$ to $x=\pi,$ and the time
duration is from $t=0$ to $t=4$. In the examples, the spectral propagation
error is reduced by more than $10^{10}$ and the time is reduced by
a factor of $20$ compared to the finite-difference methods. The periodic
method has boundaries just outside the grid.

\paragraph*{Dirichlet-Dirichlet}

With $a(0)=a(\pi)=0$, the exact solution has the form: 
\begin{align}
a & =\sum_{n=1}^{2}S_{n}\sin\left(nx\right)e^{-n^{2}t}.
\end{align}
Suppose that 
\begin{equation}
a(x,0)=1-2x/\pi+4\sin\left(x\right)+\sin\left(2x\right),
\end{equation}
and 
\begin{align*}
a_{x}(-\pi/2) & =2-4e^{-t}-e^{-4t}\\
a_{x}(\pi/2) & =+4e^{-t}+e^{-4t}
\end{align*}

For this case: 
\begin{equation}
a(x,t)=1-2x/\pi+4\sin\left(x\right)e^{-t}+\sin\left(2x\right)e^{-4t}.
\end{equation}

\paragraph*{Neumann-Neumann}

With $\partial_{x}a(0)=\partial_{x}a(\pi)=0$, the exact solution
has the form: 
\begin{align}
a & =\sum_{n=0}^{\infty}C_{n}\cos\left(nx\right)e^{-n^{2}t}.
\end{align}
Suppose that 
\begin{equation}
a(x,0)=5+4\cos\left(x\right)+\cos\left(2x\right),
\end{equation}

For this case: 
\begin{equation}
a(x,t)=5+4\cos\left(x\right)e^{-t}+\cos\left(2x\right)e^{-4t}.
\end{equation}

\paragraph*{Dirichlet-Neumann}

Here $a(0)=\partial_{x}a(\pi)=0$, the exact solution has the form:
\begin{align}
a & =\sum_{n=1}^{\infty}S_{n}\sin\left((2n-1)x/2\right)e^{-(2n-1)^{2}t/4}.
\end{align}
Suppose that 
\begin{equation}
a(x,0)=4\sin\left(x/2\right)+\sin\left(3x/2\right),
\end{equation}

For this case: 
\begin{equation}
u(x,0)=4\sin\left(x/2\right)e^{-t/4}+\sin\left(3x/2\right)e^{-9t/4}.
\end{equation}

\paragraph*{Neumann-Dirichlet}

Here $\partial_{x}a(0)=a(\pi)=0$, the general solution has the form:
\begin{align}
a & =\sum_{n=1}^{\infty}C_{n}\cos\left((2n-1)x/2\right)e^{-(2n-1)^{2}t/4}.
\end{align}
Suppose that 
\begin{equation}
a(x,0)=4\cos\left(x/2\right)+\cos\left(3x/2\right).
\end{equation}

For this case: 
\begin{equation}
a(x,t)=4\cos\left(x/2\right)e^{-t/4}+\cos\left(3x/2\right)e^{-9t/4}.
\end{equation}

\paragraph*{Periodic}

Here $a(0)=a(\epsilon\pi)$, where $\epsilon=N/\left(N-1\right)$
accounts for the periodic boundaries being outside the grid range,
so the general solution has the form: 
\begin{align}
a & =\sum_{n=1}^{\infty}S_{n}\sin\left(2nx/\epsilon\right)e^{-4n^{2}t/\epsilon^{2}}\nonumber \\
 & +\sum_{n=0}^{\infty}C_{n}\cos\left(2nx/\epsilon\right)e^{-4n^{2}t/\epsilon^{2}}.
\end{align}
Suppose that 
\begin{equation}
a(x,0)=2+\cos\left(2x/\epsilon\right)+\sin\left(4x/\epsilon\right).
\end{equation}

For this case: 
\begin{equation}
u(x,0)=2+2\cos\left(2x/\epsilon\right)e^{-4t/\epsilon^{2}}+\sin\left(4x/\epsilon\right)e^{-16t/\epsilon^{2}}.
\end{equation}

\paragraph{Notes}
\begin{itemize}
\item This is a deterministic partial differential equation, although noise
can be added 
\item Different boundary conditions apply to each component 
\item Sequential integration is used, but the initial condition is just
recycled. 
\item In p1, the $80$ intermediate steps are necessary to reduce finite-difference
errors 
\end{itemize}
\doublebox{\begin{minipage}[t]{0.9\columnwidth}%
\texttt{function {[}e{]} = Boundaries()}

\texttt{p.dimensions = 2;}

\texttt{p.points = {[}51,51{]};}

\texttt{p.order = 0;}

\texttt{p.verbose = 1;}

\texttt{p.method = @MP;}

\texttt{p.fields = 5;}

\texttt{p.ranges = {[}4,pi{]};}

\texttt{p.origins = {[}0,0{]};}

\texttt{p.initial = @heat\_in;}

\texttt{p.observe = \{@(a,p) a(1,:),@(a,p) a(2,:),@(a,p) a(3,:)...}

\texttt{@(a,p) a(4,:),@(a,p) a(5,:)\};}

\texttt{p.compare = \{@heat\_1,@heat\_2,@heat\_3,@heat\_4,@heat\_5\};}

\texttt{p.diffplot = \{1,1,1,1,1\};}

\texttt{p.olabels = \{'a, DD','a, NN','a, DN','a, ND','a, PP'\};}

\texttt{p.name = 'Heat test, spectral';}

\texttt{p.boundaries\{2\}= {[}1,1;-1,-1;1,-1;-1,1;0,0{]};}

\texttt{p1 = p;}

\texttt{p.linear = @(p) p.Dx.\textasciicircum 2;}

\texttt{p1.deriv = @(a,w,p) D2(a,2,p);}

\texttt{p1.steps = 40;}

\texttt{p1.transfer = @(\textasciitilde ,\textasciitilde ,p) heat\_in(0,p);}

\texttt{p1.name = 'Heat test, finite diffs';}

\texttt{e = xspde(\{p,p1\});}

\texttt{end}

\ 

\texttt{function a = heat\_in(\textasciitilde ,p)}

\texttt{a(1,:) = 4{*}sin(p.x)+sin(2{*}p.x);}

\texttt{a(2,:) = 5+4{*}cos(p.x)+cos(2{*}p.x);}

\texttt{a(3,:) = 4{*}sin(p.x/2)+sin(3{*}p.x/2);}

\texttt{a(4,:) = 4{*}cos(p.x/2)+cos(3{*}p.x/2);}

\texttt{a(5,:) = 2+cos(2{*}p.x/1.02)+sin(4{*}p.x/1.02);}

\texttt{end}

\ 

\texttt{function o = heat\_1(p)}

\texttt{o = 4{*}sin(p.x).{*}exp(-p.t)+sin(2{*}p.x).{*}exp(-4{*}p.t);}

\texttt{end}

\texttt{function o = heat\_2(p)}

\texttt{o = 5+4{*}cos(p.x).{*}exp(-p.t)+cos(2{*}p.x).{*}exp(-4{*}p.t);}

\texttt{end}

\texttt{function o = heat\_3(p)}

\texttt{o = 4{*}sin(p.x/2).{*}exp(-p.t/4)+sin(3{*}p.x/ 2).{*}exp(-9{*}p.t/4);}

\texttt{end}

\texttt{function o = heat\_4(p)}

\texttt{o = 4{*}cos(p.x/2).{*}exp(-p.t/4)+cos(3{*}p.x/2).{*}exp(-9{*}p.t/4);}

\texttt{end}

\texttt{function o = heat\_5(p)}

\texttt{o = 2+cos(2{*}p.x/1.02).{*}exp(-4{*}p.t/1.02\textasciicircum 2)+...}

\texttt{sin(4{*}p.x/1.02).{*}exp(-16{*}p.t/1.02\textasciicircum 2);}

\texttt{end}%
\end{minipage}}

\newpage{}

\begin{figure}[H]
\centering{}\includegraphics[width=0.75\textwidth]{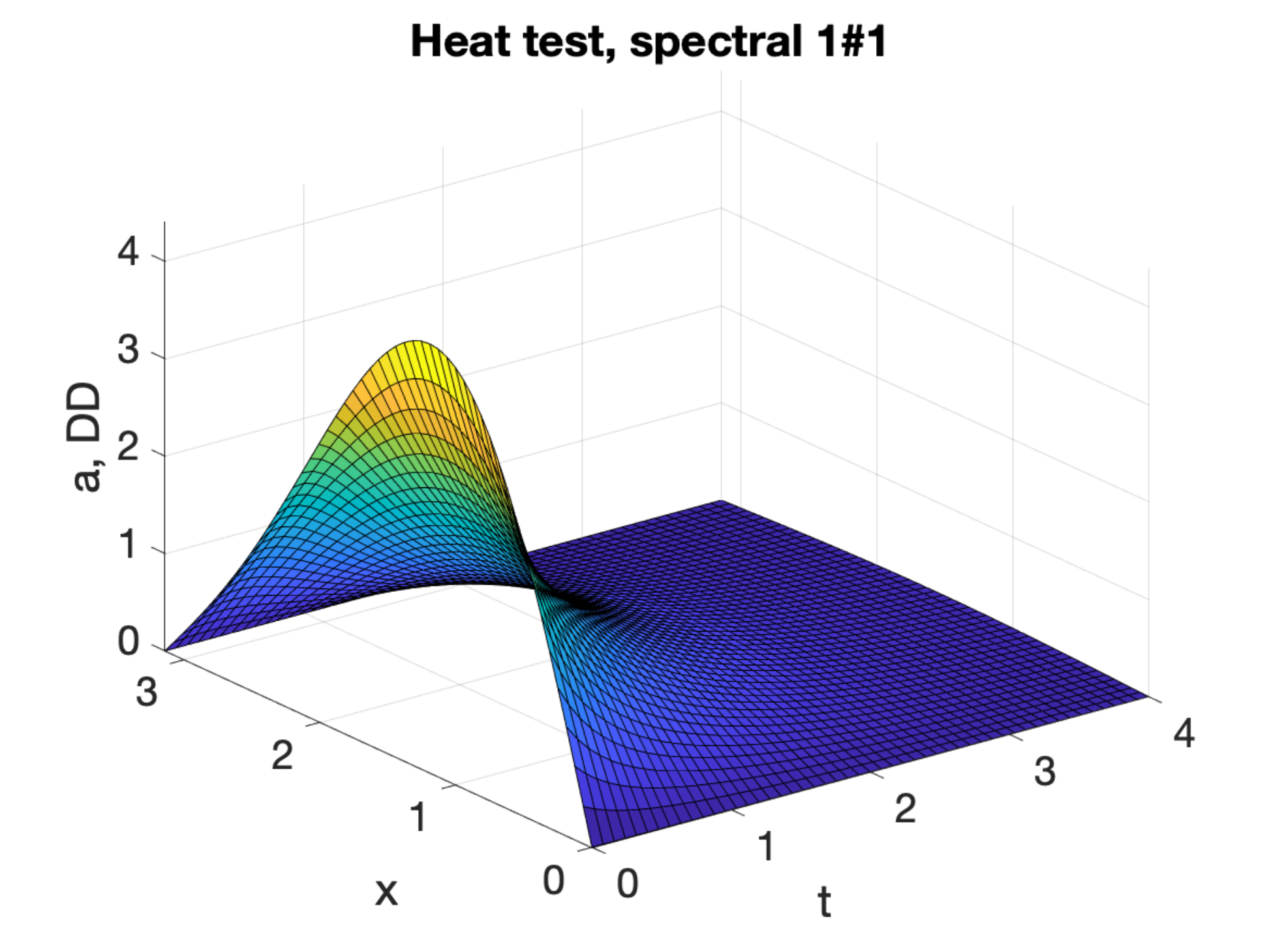}

\includegraphics[width=0.75\textwidth]{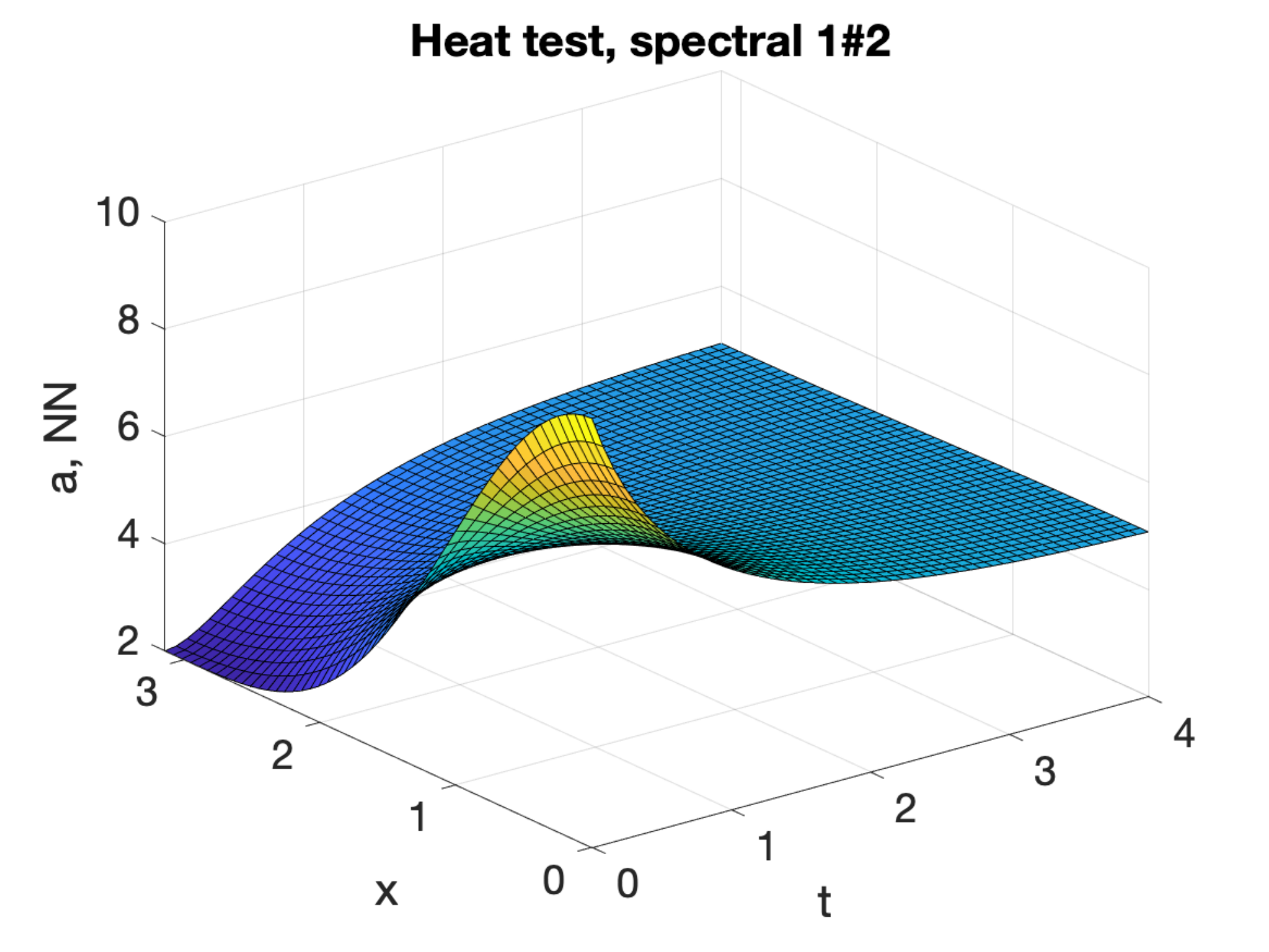}

\caption{\emph{Top figure: Spectral solution for $a$ as a function of time
and position with Dirichlet-Dirichlet boundaries. Bottom figure: Plot
of the solution with Neumann-Neumann boundaries.}}
\vspace{10pt}
\end{figure}

\newpage{}

\section{Peregrine solitary wave with arbitrary boundary conditions\label{subsec:Nonlinear-Schrodinger-equation-1}}

Peregrine solitary waves are models for isolated large ocean waves.
They are modeled as solutions to a (1+1)-dimensional PDE,

\begin{eqnarray}
\frac{\partial a}{\partial t} & = & i\cdot\left(a\cdot\left|a\right|^{2}+\frac{1}{2}\frac{\partial^{2}a}{\partial x^{2}}\right)\,.
\end{eqnarray}
The Peregrine solution on an infinite domain is:

\begin{eqnarray}
a & = & e^{it}\left(\frac{4\left(1+2it\right)}{1+4\left(t^{2}+x^{2}\right)}-1\right)\,.
\end{eqnarray}

In the example, this is solved using finite boundary conditions, with
initial values and boundary values that correspond to the exact solution.

\paragraph{Notes}
\begin{itemize}
\item The boundary conditions are specified with p.boundaries\{2\}, which
is the x-dimension. 
\item Four different boundary conditions are specified for a four component
field.
\item Spectral methods are used for accuracy
\item No noise is included in the example below, but it can be added
\item The boundary values are time-dependent, and are specified with p.boundfun
\item For improved stability, the integration method is the semi-implicit
\textbf{MP} method.
\end{itemize}

\paragraph{Errors\protect \\
}

The reported errors in this example are reduced using second order
extrapolation, by specifying p.order =2. This gives the following
RMS average errors for the output intensity, averaged over all four
cases:
\begin{itemize}
\item Step=0.000621 
\item Diff=0.00019 
\end{itemize}
The 'Step' error is from subtracting the most accurate results from
lower accuracy results. With extrapolation specified, the most accurate
results are the extrapolated results. The less accurate ones are at
half the specified time-step. This is averaged over all space-time
points, and normalized by the maximum intensity of $\left|a\right|^{2}=9$. 

The 'Diff' error is from comparing the most accurate results with
the analytic solution. This demonstrates a typical case where the
time-step error is an upper bound to the true error. The maximum error
occurs at large times, and is greater than the RMS error by about
$10\times$in this case. All four boundary types used give similar
results and errors.

\newpage{}
\begin{center}
\doublebox{\begin{minipage}[t]{0.9\columnwidth}%
\texttt{function e = Peregrine()}

\texttt{\% e = Peregrine() tests xSPDE for a nonlinear Schrodinger
equn.}

\texttt{\% Using NN,DD,DN,ND boundary values with a spectral method}

\texttt{\% Uses time dependent boundary values for a peregrine solution}

\texttt{p.dimensions = 2;}

\texttt{p.noises = 1;}

\texttt{p.fields = 4;}

\texttt{p.order = 2;}

\texttt{p.ranges = {[}10,10{]};}

\texttt{p.origins = {[}-5,-5{]};}

\texttt{p.points = {[}51,161{]};}

\texttt{p.method = @MP;}

\texttt{p.olabels = \{'\textbar a\textbar\textasciicircum 2 ,
DD','\textbar a\textbar\textasciicircum 2 , NN','\textbar a\textbar\textasciicircum 2
, DN','\textbar a\textbar\textasciicircum 2 , ND'\};}

\texttt{p.boundaries\{2\} = {[}1,1;-1,-1;1,-1;-1,1{]};}

\texttt{p.boundfun = @boundval;}

\texttt{sol = @(p) abs(per(p.x,p.t).\textasciicircum 2);}

\texttt{p.initial = @(\textasciitilde ,p) per(p.x,p.origins(1))+zeros(4,1,1);}

\texttt{p.compare = \{@(p) sol(p),@(p) sol(p),@(p) sol(p),@(p) sol(p)\};}

\texttt{p.observe = \{@(a,p) a(1,:),@(a,p) a(2,:),...}

\texttt{@(a,p) a(3,:),@(a,p) a(4,:)\};}

\texttt{p.output = \{@(o,p) abs(o\{1\}).\textasciicircum 2,@(o,p)
abs(o\{2\}).\textasciicircum 2,...}

\texttt{@(o,p) abs(o\{3\}).\textasciicircum 2,@(o,p) abs(o\{4\}).\textasciicircum 2\};}

\texttt{p.name = 'Peregrine solution';}

\texttt{p.steps = 20;}

\texttt{p.deriv = @(a,w,p) 1i{*}a.{*}((conj(a).{*}a));}

\texttt{p.linear = @(p) 0.5{*}1i{*}p.Dx.\textasciicircum 2;}

\texttt{e = xspde(p);}

\texttt{end}~\\

\texttt{function {[}p,varargout{]} = per(x,t)}

\texttt{\% Generates peregrine solutions with alpha = 1/2, beta =
A0 = 1}~\\

\texttt{p = exp(1i{*}t).{*}(4{*}(1+2{*}1i{*}t)./(1+4.{*}(t.\textasciicircum 2+x.\textasciicircum 2))-1);}

\texttt{if nargout == 2}

\texttt{dp = -8{*}x.{*}exp(1i{*}t).{*}(4{*}(1+2{*}1i{*}t)./(1+4.{*}(t.\textasciicircum 2+x.\textasciicircum 2)).\textasciicircum 2);}

\texttt{varargout\{1\} = dp;}

\texttt{end}~\\

\texttt{end}

\texttt{function bound = boundval(\textasciitilde ,\textasciitilde ,\textasciitilde ,p)}

\texttt{\% Generates nonzero, time dependent boundary values}

\texttt{{[}p,dp{]} = per(p.origins(2),p.t);}

\texttt{bound = \{p,p;dp,-dp;p,-dp;dp,p\};}

\texttt{end}%
\end{minipage}} 
\par\end{center}

\newpage{}

\begin{figure}[H]
\begin{centering}
\includegraphics[width=0.75\textwidth]{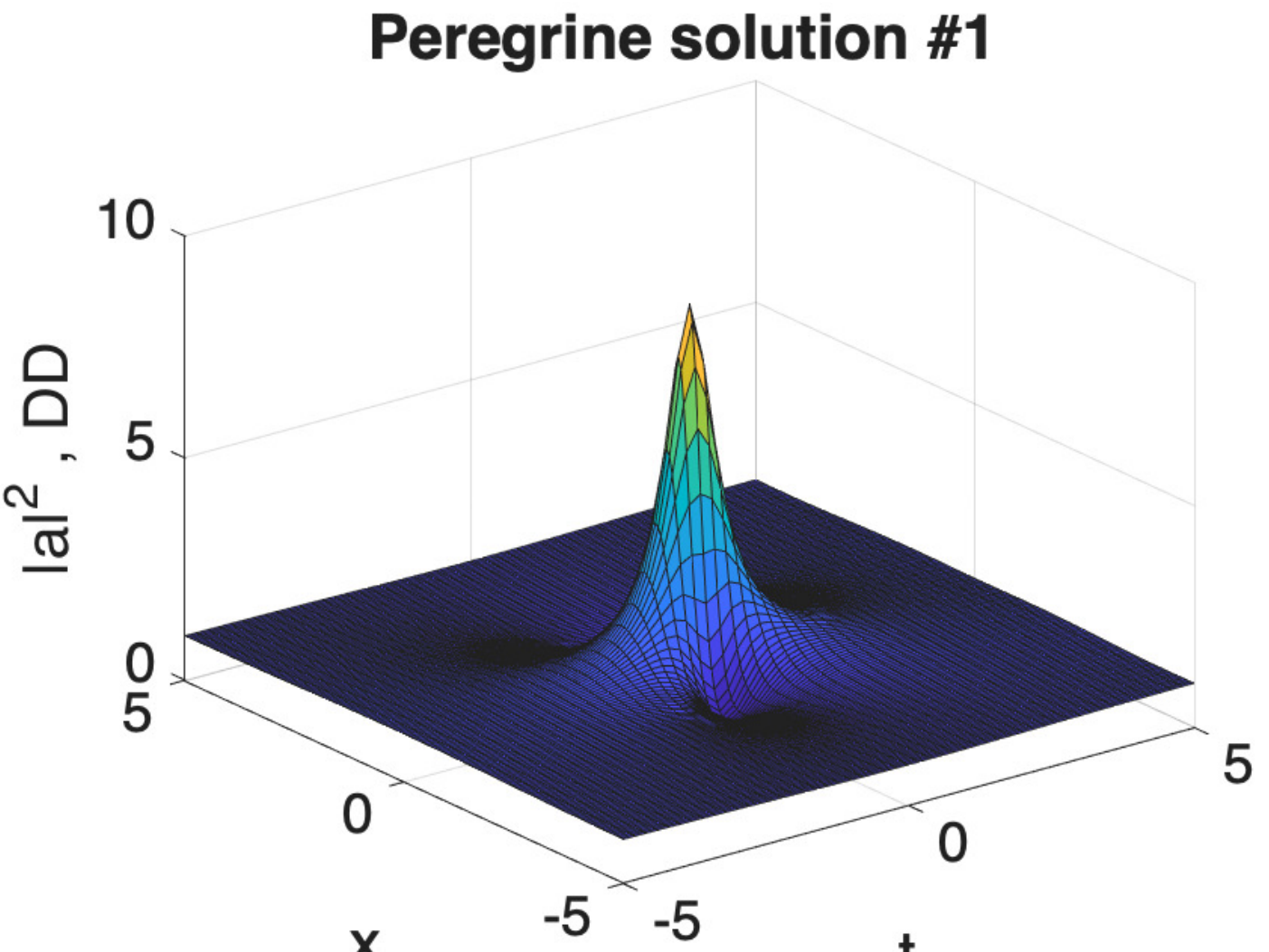}
\par\end{centering}
\begin{centering}
\centering{}\includegraphics[width=0.75\textwidth]{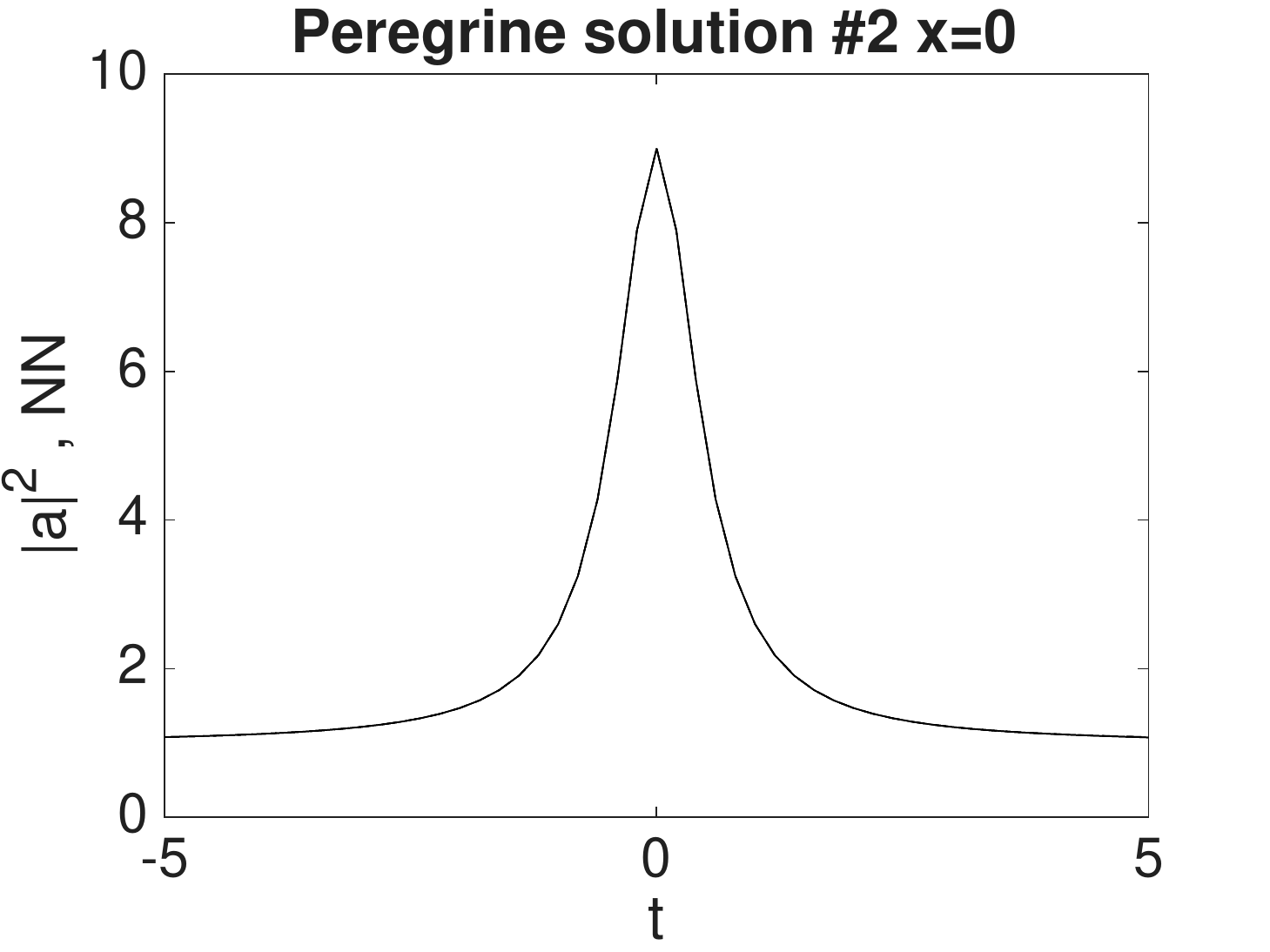}
\par\end{centering}
\caption{\emph{Top figure: Peregrine solution with Dirichlet-Dirichlet boundaries
in space-time. }\protect \protect \\
 \emph{Bottom figure: Peregrine solution with Neumann-Neumann boundaries
at $x=0.$}}
\vspace{10pt}
\end{figure}

\newpage{}

\part{Quantum phase-space}

\chapter{Phase-space toolbox \label{chap:Quantum-phase-space-toolbox} }

\textbf{This chapter describes how to use the xSPDE numerical toolbox
to solve network and quantum dynamical problems in phase-space. For
theoretical background, see Chapter (\ref{chap:Quantum-phase-space-theory}).
For detailed examples, see Chapter (\ref{chap:Phase-space-examples}).}

\section{Quantum phase-space}

Initially developed in the first half of 20th century, phase-space
representations of quantum mechanics have been extensively developed,
and utilized to simulate large bosonic Hilbert spaces, as well as
small ones. Multiple such methods exist, including the classical Wigner
\cite{Wigner1932Quantum}, Husimi Q and Glauber-Sudarshan \cite{Glauber1963Coherent,Sudarshan_1963_P-Rep}
P representations, as well as the non-classical positive-P \cite{Drummond1980Generalised}
and gauge-P \cite{Deuar2002Gauge} methods. If the reader is unfamiliar
with these methods, they are referred to the theory sections in Chapter
\ref{chap:Quantum-phase-space-theory} and the original literature
to obtain further explanations of these methods.

These different approaches have areas of applicability that depend
on the Hilbert space dimension, as explained below.

\subsection{Phase-space methods}

This is used for phase-space mappings. Details are: 
\begin{description}
\item [{p.phase~=~1}] - for a normally-ordered P or positive-P representation. 
\item [{p.phase~=~2}] - for a symmetrically-ordered Wigner-representation. 
\item [{p.phase~=~3}] - for an anti-normally-ordered Q-representation. 
\end{description}

\section{Laser amplification noise}

Laser quantum noise is commonly modeled \cite{Louisell1973Quantum,Carmichael2002Statistical,gardiner2004quantum}
using SDEs in a normally ordered quantum phase-space representation.
Consider a model for the quantum noise of a single mode laser as it
turns on, near threshold:

\begin{equation}
\dot{a}=ga+bw(t)
\end{equation}
where the noise is complex, $w=\left(w_{1}+iw_{2}\right)$, so that:
\begin{equation}
\left\langle w(t)w^{*}(t')\right\rangle =2\delta\left(t-t'\right)\,.
\end{equation}
Here the coefficient $b$ describes the quantum noise of the laser,
and is inversely proportional to the equilibrium photon number.

As an example, try the following: 
\begin{itemize}
\item \textbf{Solve for the case of $g=0.25$, $b=0.01$} 
\end{itemize}
using xSPDE input is:\\

\texttt{}%
\doublebox{\begin{minipage}[t]{0.75\columnwidth}%
\texttt{clear}

\texttt{p.noises = 2;}

\texttt{p.observe = @(a,p) abs(a).\textasciicircum 2;}

\texttt{p.olabels = '\textbar a\textbar\textasciicircum 2';}

\texttt{p.deriv = @(a,w,p) 0.25{*}a + 0.01{*}(w(1)+1i{*}w(2));}

\texttt{xspde(p);}%
\end{minipage}}\texttt{}~\\

This input script can either be copied into a new script, or simply
pasted into the Command Window, noting that one should usually type
clear first when starting new interactive simulations. Most lasers
have many photons and hence much less noise than this. At this small
gain, numerical errors are negligible, so the program reports only
small RMS average errors:
\begin{itemize}
\item Errors: Step=0.000621 Samp=0 Diff=0 Chisq/k=0
\end{itemize}
For larger gain, error-bars will display on the graph. These are calculated
from the difference between using steps of size $dt$ and steps of
size $dt/2$. They only appear if greater than a minimum relative
size, typically $1\%$ of the graph size, which can be set by the
user. Here the time-step errors are too small to be graphed, there
is no sampling error because only one trajectory is requested, and
there is neither difference nor $\chi^{2}$ error, since there are
no comparisons specified. 

Note that specifying \texttt{p.phase = 1 }is not required, even though
the phase-space is normally-ordered, which is the default phase-space
method. This is because there is no vacuum noise in this classical
noise case, and only the simple Glauber-Sudarshan P-representation
is needed. Non-classical states require a positive P-representation. 

\begin{figure}
\centering{}\includegraphics[width=0.75\textwidth]{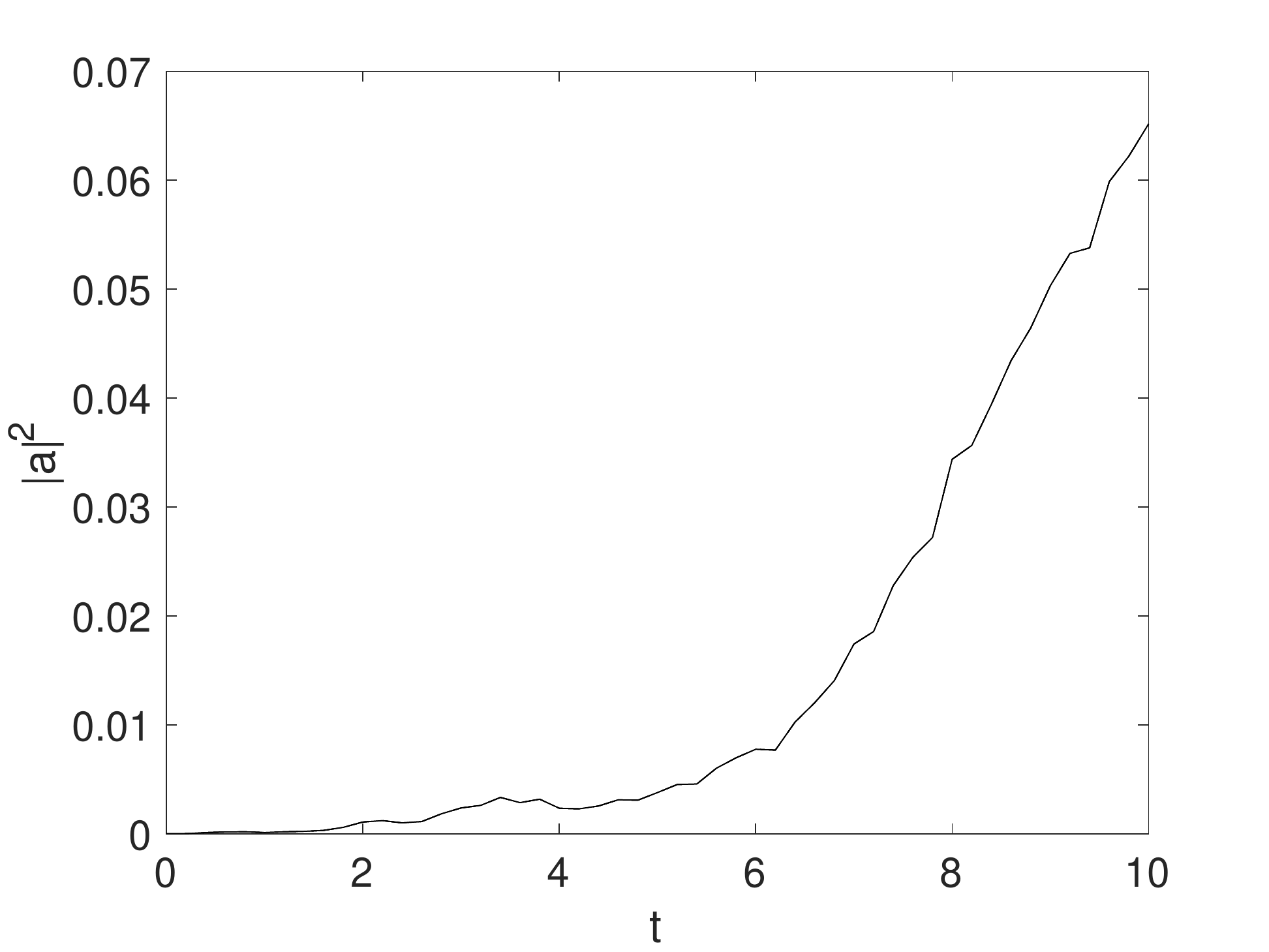}

\caption{\label{fig:Unsaturatedlaser-1}\emph{Simulation of the stochastic
equation describing a laser turning on.}}
\vspace{10pt}
\end{figure}

\section{Input parameters}

More complex input parameters are stored in a structure which is input
to the program. This is a superset of the parameters already defined.
In the definitions below, the structure name is omitted, but we use
$p$ in the examples.

While the quantum monte-carlo and phase-space toolboxes have some
common parameters, they are distinct toolboxes that correspond to
different Hilbert space bases. One must choose one or the other by
either setting $quantum>0$ or $phase>0$. The defaults used are given
in the phasepreferences function.

Common parameters with the standard SDE and SPDE simulations include
ensembles, noises, inrandoms, points, steps, ranges. All have default
values, but other choices can be input to over-ride the default values.

\subsection{\label{subsec:Phase-space-parameters}Phase-space parameters }

Phase-space simulations can be initiated with a set of parameters
that generate a default initial function. This is suitable for network
simulations with gaussian inputs of squeezed and thermalized states. 
\begin{description}
\item [{dimension}] is for the type of calculation. For network transformations,
dimension = 0.
\item [{sqz}] is the initial squeezing parameter vector, $\bm{r}$, per
mode. 
\item [{alpha}] is the initial coherent amplitude vector per mode.
\item [{matrix}] is a matrix transformation function used to create network
transformations. 
\item [{tr}] is an amplitude transmission vector used to attenuate the
initial state. 
\item [{thermal}] is a thermal fraction vector used to initialize a thermalized
squeezed state. 
\item [{initial}] functions initialise the equations. If omitted, the
initial conditions are defined using the parameters above.
\end{description}

\section{Gaussian boson sampling outputs}

If one is interested in computing output probabilities for Gaussian
boson sampling (GBS) photonic quantum networks, observe functions
are available for comparisons with either experimental data, or exactly
known test cases. 

As every experimental group treats its data differently, xSPDE4 does
not attempt to provide universal code for extracting and binning experimental
data. This will be provided in a seperate codepackage. Instead, xSPDE4
focuses on simulating such networks in phase-space. 

If p.dimension=0, where the mode index can be large, the mode index
is treated as a space index rather than a line index. This aids in
graphical rendering with xGRAPH. 

\subsection{Observe functions}

The main observe functions correspond to binned photon-counting probability
distributions and marginal moments of various operator observables.
A summary, and brief description of the current list is given below:\\

\begin{center}
\begin{tabular}{|c|c|c|}
\hline 
Label & Return type & Description\tabularnewline
\hline 
\hline 
x & vector & $x$-quadrature per channel\tabularnewline
\hline 
x2 & vector & $x$-quadrature squared per channel\tabularnewline
\hline 
p & vector & $p$-quadrature per channel\tabularnewline
\hline 
p2 & vector & $p$-quadrature squared per channel\tabularnewline
\hline 
pn & vector & Output photon number, $\hat{n}'_{j}$, per channel\tabularnewline
\hline 
nm & vector & Output photon number correlation in sequence $\hat{n}'_{1}\hat{n}'_{2}\dots$\tabularnewline
\hline 
k & vector & Clicks $\hat{\pi}_{j}(1)$ per channel\tabularnewline
\hline 
km & vector & Click correlation in sequence $\hat{\pi}_{1}(1)\hat{\pi}_{2}(1)\dots$\tabularnewline
\hline 
km2 & vector & Click correlation over two channels $\hat{\pi}_{j}(1)\hat{\pi}_{k}(1)$\tabularnewline
\hline 
km3 & vector & Click correlation over three channels $\hat{\pi}_{j}(1)\hat{\pi}_{k}(1)\hat{\pi}_{h}(1)$\tabularnewline
\hline 
kmsub & vector & Subset of click correlations per CO channels\tabularnewline
\hline 
k1 & vector & Binned click probability - single partition\tabularnewline
\hline 
n1 & vector & Binned photon number probability - single partition\tabularnewline
\hline 
kn & array & Binned click probability - $n$-fold partition\tabularnewline
\hline 
nn & array & Binned photon number probability - $n$-fold partition\tabularnewline
\hline 
\end{tabular}
\par\end{center}

\subsection{Compare functions}

The compare function is used for testing and for experimental data.
The comparisons may include error data. This generates comparison
plots, as well as error totals that are converted into a $\chi^{2}$-
error estimate when there are statistical variances. 

Standard compare functions given below are analogous to their observe
counterparts above, with compare functions denoted by the c at the
end of each label. Generally, such test cases are only applicable
when the output is unchanged from the input apart from a transmission
factor $p.tr$, and when the input is a thermalised or pure squeezed
state. For photon-number resolving (PNR) detector comparisons, uniform
squeezing parameters are required for the inputs. 

Functions labeled with an asterisk can be used with Gaussian states
having a coherent component $p.alpha$:\\

\begin{center}
\begin{tabular}{|c|c|c|}
\hline 
Label & Return type & Description\tabularnewline
\hline 
\hline 
xc & vector & $x$-quadrature per channel\tabularnewline
\hline 
x2c & vector & $x$-quadrature squared per channel\tabularnewline
\hline 
pc & vector & $p$-quadrature per channel\tabularnewline
\hline 
p2c & vector & $p$-quadrature squared per channel\tabularnewline
\hline 
$^{*}$pnc & vector & Output photon number, $\hat{n}'_{j}$, per channel\tabularnewline
\hline 
nmc & vector & Output photon number correlation in sequence $\hat{n}'_{1}\hat{n}'_{2}\dots$\tabularnewline
\hline 
kc & vector & Clicks $\hat{\pi}_{j}(1)$ per channel\tabularnewline
\hline 
kmc & vector & Click correlation in sequence $\hat{\pi}_{1}(1)\hat{\pi}_{2}(1)\dots$\tabularnewline
\hline 
km2c & vector & Click correlation over two channels $\hat{\pi}_{j}(1)\hat{\pi}_{k}(1)$\tabularnewline
\hline 
km3c & vector & Click correlation over three channels $\hat{\pi}_{j}(1)\hat{\pi}_{k}(1)\hat{\pi}_{h}(1)$\tabularnewline
\hline 
kmsubc & vector & Subset of click correlations per CO channels\tabularnewline
\hline 
k1c & vector & Binned click probability - single partition\tabularnewline
\hline 
n1thc & vector & Thermal input photon number probability - single partition\tabularnewline
\hline 
n1lsc & vector & Pure squeezed input photon number probability - single partition\tabularnewline
\hline 
knc & array & Binned click probability - $n$-fold partition\tabularnewline
\hline 
nnthc & array & Thermal input photon number probability - $n$-fold partition\tabularnewline
\hline 
\end{tabular}
\par\end{center}

\section{Sampling methods in phase-space\label{sec:Numerical-methods}}

Generating initial samples of a phase-space distribution is the first
step of any phase-space simulation, be that dynamical or input-output,
such as for GBS. In most cases, analytical forms of the initial samples
exist, which are straightforard to implement numerically. 

The section outlines the sampling methods used to simulate GBS input
and output distributions in phase-space, and perform comparisons with
either experimental data, or exact tests. 

\subsection{Input-output samples}

Simulating dynamic and static quantum systems in phase-space follows
the same, general procedure, at least initially. In dynamical applications,
initial stochastic samples generated from some phase-space distribution.
These can be from a default gaussian description of coherent, squeezed
or thermal states, given by the \emph{alpha}, \emph{sqz}, \emph{thermal
}and \emph{tr} parameters, or else from an \emph{initial} function,
which takes precedence over the gaussian parmeters.

To simulate quantum networks in phase-space, one must first generate
initial stochastic samples. This is achieved using the $\sigma$-ordering
scheme (see Subsection \ref{subsec:sigma-ordering} for theory) as
stochastic samples for any Gaussian input state in any representation
are generated following:

\begin{align}
\alpha_{j} & =\frac{1}{2}\left(\Delta_{\sigma x_{j}}w_{j}+i\Delta_{\sigma y_{j}}w_{j+M}\right)\nonumber \\
\beta_{j} & =\frac{1}{2}\left(\Delta_{\sigma x_{j}}w_{j}-i\Delta_{\sigma y_{j}}w_{j+M}\right),
\end{align}
where $\left\langle w_{j}w_{k}\right\rangle =\delta_{jk}$ are real
Gaussian noises and

\begin{align}
\Delta_{\sigma x_{j}}^{2} & =2(n_{j}+\sigma+\tilde{m}_{j})\nonumber \\
\Delta_{\sigma y_{j}}^{2} & =2(n_{j}+\sigma-\tilde{m}_{j}),\label{eq:quad_var_altered-1}
\end{align}
are thermal squeezed state quadrature variances which are altered
from the pure squeezed state definitions, (\ref{eq:x-quad}) and (\ref{eq:y-quad}).

For normally ordering, the input amplitudes $\boldsymbol{\alpha},\boldsymbol{\beta}$
are converted to outputs as

\begin{align}
\boldsymbol{\alpha}' & =\boldsymbol{T\alpha}\nonumber \\
\boldsymbol{\boldsymbol{\beta}}' & =\boldsymbol{T}^{*}\boldsymbol{\beta},
\end{align}
which follows from Eq.(\ref{eq:linear_combo}). However for non-normally
ordered methods, additional vacuum noise arising from the reservoir
modes must be included.

This is achieved using a hermitian decoherence matrix

\begin{equation}
\boldsymbol{D}=\boldsymbol{I}-\boldsymbol{T}^{\dagger}\boldsymbol{T},
\end{equation}
with decomposition $\boldsymbol{D}=\boldsymbol{U}\boldsymbol{\lambda}^{2}\boldsymbol{U}^{\dagger}$
where $\boldsymbol{B}=\boldsymbol{U}\boldsymbol{\lambda}\boldsymbol{U}^{\dagger}$
is the matrix square root and $\boldsymbol{\lambda}$ is a diagonal,
positive matrix. The output amplitudes when $\sigma>0$ are then obtained
as

\begin{equation}
\boldsymbol{\alpha}'=\boldsymbol{T\alpha}+\sqrt{\frac{\sigma}{2}}\boldsymbol{B}(\boldsymbol{u}+i\boldsymbol{v}),
\end{equation}
where $\boldsymbol{\beta}'=\boldsymbol{\alpha}^{\prime*}$ as these
are a classical phase-space.

\subsection{Grouped correlations computation: Threshold detectors}

Grouped correlations, or grouped count probabilities (GCPs), are binned
photon counting probability distributions that can be simulated in
xSPDE4 for photon-counting set-ups using both photon-number resolving
(PNR) or threshold detection. These are defined analytically in Section
\ref{sec:Grouped-correlations}, where all the nessecary background
theory is presented in Section \ref{sec:Linear-photonic-networks}. 

These observables are readily simulated in phase-space using the positive-P
representation. For threshold detectors, one replaces normally ordered
projection operator Eq.(\ref{eq:click proj}) with the positive-P
observable

\begin{equation}
\pi_{i}(c_{i})=:e^{-n'_{i}}\left(e^{n'_{i}}-1\right)^{c_{i}},
\end{equation}
where $n'_{i}=\alpha'_{i}\beta'_{i}$ is the output photon number.

The summation over exponentially many patterns implemented by GCPs
(see Eq.(\ref{eq:GCP}) is simulated using a multidimensional inverse
discrete Fourier transform

\begin{align}
\tilde{\mathcal{G}}_{\boldsymbol{S}}^{(n)}(\boldsymbol{k}) & =\left\langle \prod_{j=1}^{d}\bigotimes_{i\in S_{j}}\left(\pi_{i}(0)+\pi_{i}(1)e^{-ik_{j}\theta_{j}}\right)\right\rangle _{P},\nonumber \\
\mathcal{G}_{\boldsymbol{S}}^{(n)}(\boldsymbol{m}) & =\frac{1}{\prod_{j}(M_{j}+1)}\sum_{\boldsymbol{k}}\tilde{\mathcal{G}}_{\boldsymbol{S}}^{(n)}(\boldsymbol{k})e^{i\sum k_{j}\theta_{j}m_{j}},
\end{align}
where $\theta_{j}=2\pi/(M_{j}+1)$ and $k_{j}=0,\dots,M_{j}$.

The Fourier transform removes all patterns which don't contain $\boldsymbol{m}$
counts, in doing this the Fourier transform simulates all possible
correlations generated in a network. This reduces an otherwise computationally
complex task into a highly efficient and scalable one, allowing comparisons
to be performed on experimental correlations of any order.

\subsection{Grouped correlations computation: PNR detectors}

To simulate GCPs of PNR detectors, we are interested in the total
number of output photons contained in each subset $S_{j}$ defined
in operator form 

\begin{equation}
\hat{n}'_{S_{j}}=\sum_{i\in S_{j}}\hat{n}'_{i}.
\end{equation}
As was the case for threshold detectors, we replace this total number
operator with the phase-space observable 

\begin{equation}
n'_{S_{j}}=\sum_{i\in S_{j}}n'_{i},
\end{equation}
such that the actual GCP computation is performed as 

\begin{equation}
\mathcal{G}_{\boldsymbol{S}}^{(n)}(\boldsymbol{m})=\iint P(\boldsymbol{\alpha},\boldsymbol{\beta})\left[\prod_{j=1}^{d}\left[\frac{1}{m_{j}!}:(n'_{S_{j}})^{m_{j}}e^{-n'_{S_{j}}}:\right]\right]\text{d}^{2}\bm{\alpha}\text{d}^{2}\boldsymbol{\beta}.
\end{equation}

We note that this method allows one to avoid the multi-dimensional
Fourier transform needed for the threshold detector distribution.
The Fourier transform has the effect of introducing Fourier overheads
which can slow down simulations of multi-dimensional distributions,
especially for dimensions $d\ge4$. 

\chapter{Quantum phase-space theory \label{chap:Quantum-phase-space-theory}}

\textbf{\emph{This chapter describes the background of quantum phase-space
theory, including dynamical problems obtained by transforming the
master equation into a second-order partial differential equation
called the Fokker-Planck equation (FPE), as well as input-output problems
such as GBS. }}

\section{Phase-space representations}

Phase-space representations are an alternative description of quantum
mechanics where one maps operators and fields of various orderings
to classical probability distributions on a phase-space. This alternative
description was first introduced by Wigner \cite{Wigner1932Quantum},
and has since burgeoned into a vast field with applications to quantum
optics, atom-optics, quantum information, and many more. 

In this section, we introduce the phase-space representations implemented
in xSPDE4, although one is not restricted to these representations.
The outputs in phase-space are continuous real or complex variables
whose stochastic moments are equal to quantum expectation values,
including probability distributions obtained from experimental data
such as photon counting experiments. 

This assumes that the parameters are precisely known, and do not have
noise or fluctuations. Even then, some differences from sampling errors
due to finite numbers of experimental and theoretical counts. 

For clarity, throughout this chapter, hats like $\hat{a}$ are used
to indicate operators that do not commute with each other, as opposed
to stochastic variables like $\alpha$ that do commute. For any given
operator ordering, it is always possible to find a probability distribution
such that the expectation of an operator product equals the stochastic
variable correlations \cite{Dirac1945Analogy}.

\subsection{Glauber-Sudarshan P-representation}

The $M$-mode Glauber-Sudarshan P-representation expands the density
matrix as a sum of diagonal coherent state projectors

\begin{equation}
\hat{\rho}=\int P(\boldsymbol{\alpha})\left|\boldsymbol{\alpha}\right\rangle \left\langle \boldsymbol{\alpha}\right|\text{d}^{2M}\boldsymbol{\alpha},
\end{equation}
where $\left|\boldsymbol{\alpha}\right\rangle =\bigotimes_{j=1}^{M}\left|\alpha_{j}\right\rangle $
is a multimode coherent state eigenvector with corresponding eigenvalues
$\boldsymbol{\alpha}=[\alpha_{1},\dots,\alpha_{M}]$. Coherent states
\cite{Glauber1963Coherent} are the right eigenstate of the annihilation
operator, $\hat{a}\left|\alpha\right\rangle =\alpha\left|\alpha\right\rangle $,
and are the most commonly used basis state to define phase-space representations,
although other bases are possible, such as quadratures. 

For classical states such as coherent and thermal states, the P-distribution
$P(\boldsymbol{\alpha})$ satisfies the mathematical requirements
of a probability distribution: Real, positive, non-singular and normalizable
with 

\begin{equation}
\int P(\boldsymbol{\alpha})\text{d}^{2M}\boldsymbol{\alpha}=1.
\end{equation}
The P-representation breaks down for certain quantum states, giving
non-positive and singular distributions. This is due to the lack of
off-diagonal coherent state amplitudes required to represent nonclassical
superpositions.

\subsection{Positive P-representation}

Part of a family of generalized P-representations developed to extend
the Glauber-Sudarshan P-representation to quantum states \cite{Drummond1980Generalised},
the normally ordered positive P-representation always generates a
non-singular and positive distribution for any quantum state. The
trade-off is that it is non-unique, which can lead to growing sampling
errors for nonlinear Hamiltonians.

In the positive P-representation, the density matrix is defined as
an expansion over a multidimensional subspace of the complex plane:

\begin{equation}
\hat{\rho}=\iint P(\boldsymbol{\alpha},\boldsymbol{\beta})\hat{\Lambda}(\boldsymbol{\alpha},\boldsymbol{\beta})\text{d}^{2M}\boldsymbol{\alpha}\text{d}^{2M}\boldsymbol{\beta},
\end{equation}
where $P(\boldsymbol{\alpha},\boldsymbol{\beta})$ is the positive-P
distribution over coherent state amplitudes $\boldsymbol{\alpha},\boldsymbol{\beta}$
satisfying the normalization condition 

\begin{equation}
\iint P(\boldsymbol{\alpha},\boldsymbol{\beta})\text{d}^{2M}\boldsymbol{\alpha}\text{d}^{2M}\boldsymbol{\beta}=1.
\end{equation}

The projector

\begin{equation}
\hat{\Lambda}(\boldsymbol{\alpha},\boldsymbol{\beta})=\frac{\left|\boldsymbol{\alpha}\right\rangle \left\langle \boldsymbol{\beta}^{*}\right|}{\left\langle \boldsymbol{\beta}^{*}|\boldsymbol{\alpha}\right\rangle }=e^{-\boldsymbol{\alpha}\cdot\boldsymbol{\beta}+\frac{1}{2}\left|\boldsymbol{\alpha}\right|^{2}+\frac{1}{2}\left|\boldsymbol{\beta}\right|^{2}}\left|\boldsymbol{\alpha}\right\rangle \left\langle \boldsymbol{\beta}^{*}\right|,
\end{equation}
is responsible for the doubled classical phase-space dimension. This
allows off-diagonal amplitudes $\boldsymbol{\beta}\neq\boldsymbol{\alpha}^{*}$
to exist, such that $\boldsymbol{\alpha},\boldsymbol{\beta}$ are
now independent, generating two pairs of complex amplitudes.

One can restrict the distribution to a classical phase-space with
$\boldsymbol{\beta}=\boldsymbol{\alpha}^{*}$, in which case the diagonal
P-representation is obtained as a special case of the positive P-representation
via the substitution $P(\boldsymbol{\alpha},\boldsymbol{\beta})=P(\boldsymbol{\alpha})\delta(\boldsymbol{\alpha}^{*}-\boldsymbol{\beta})$.
As stated above, for non-classical states such as squeezed or Fock
states, this will lead to the well known singular behavior of the
distribution.

Moments of the positive-P distribution are equivalent to normally
ordered operator moments

\begin{align}
\left\langle \hat{a}_{j_{1}}^{\dagger},\dots,\hat{a}_{j_{n}}\right\rangle  & =\left\langle \beta_{j_{1}},\dots,\alpha_{j_{n}}\right\rangle _{P}\nonumber \\
 & =\iint P(\boldsymbol{\alpha},\boldsymbol{\beta})[\beta_{j_{1}},\dots,\alpha_{j_{n}}]\text{d}^{2M}\boldsymbol{\alpha}\text{d}^{2M}\boldsymbol{\beta},
\end{align}
where $\left\langle \dots\right\rangle $ denotes a quantum expectation
value and $\left\langle \dots\right\rangle _{P}=\left\langle \dots\right\rangle _{P,\infty}$
is the positive-P average in the limit of an infinite ensemble of
stochastic trajectories. 

\subsection{Wigner representation}

Although the diagonal P-representation is unsuitable for simulating
non-classical states, not all classical phase-space distributions
suffer from the same limitations. The symmetrically ordered Wigner
representation and anti-normally ordered Q-function both produce positive,
well defined distributions for Gaussian quantum states such as squeezed
states. Unfortunately, for non-Gaussian non-classical states such
as Fock and Schr\"{o}dinger cats states, the resulting Wigner distribution
is negative, and hence not a probability distribution. 

The negativity of the Wigner distribution for some states is why its
commonly referred to as a quasi-probability, and for an $M$-mode
system is defined as the Fourier transform of the symmetrically ordered
characteristic function such that

\begin{equation}
W(\boldsymbol{\alpha})=\frac{1}{\pi^{2M}}\int\text{d}^{2}\boldsymbol{z}\text{Tr}\left(\hat{\rho}e^{i\boldsymbol{z}(\hat{a}-\boldsymbol{\alpha})+i\boldsymbol{z}^{*}(\hat{a}^{\dagger}-\boldsymbol{\alpha}^{*})}\right),
\end{equation}
where $\text{Tr (\ensuremath{\dots})}$ is the matrix trace and $\boldsymbol{z}$
is a complex vector. For hermitian operators such as the density operator,
the Wigner distribution always exists as a real-valued function on
phase-space. This isn't the case for non-hermitian operators, in which
case Wigner distribution becomes complex. 

The Wigner representation is directly applicable to symmetrically
ordered operator products. Symmetric ordering, denoted $\{\dots\}_{sym}$,
is the average over all possible combinations of creation and annihilation
operators, for example:

\begin{align}
\{\hat{a}^{\dagger}\hat{a}\}_{sym} & =\frac{1}{2}(\hat{a}\hat{a}^{\dagger}+\hat{a}^{\dagger}\hat{a})\label{eq:sym_number_op}\\
\{\hat{a}^{\dagger}\hat{a}^{2}\}_{sym} & =\frac{1}{3}(\hat{a}^{2}\hat{a}^{\dagger}+\hat{a}\hat{a}^{\dagger}\hat{a}+\hat{a}^{\dagger}\hat{a}^{2}).
\end{align}
Therefore, moments of the Wigner distribution correspond to symmetrically
ordered operator moments 

\begin{align}
\left\langle \{\hat{a}_{j_{1}}^{\dagger},\dots,\hat{a}_{j_{n}}\}_{sym}\right\rangle  & =\left\langle \alpha_{j_{1}}^{*},\dots,\alpha_{j_{n}}\right\rangle _{W}\nonumber \\
 & =\int W(\boldsymbol{\alpha})[\alpha_{j_{1}}^{*},\dots,\alpha_{j_{n}}]\text{d}^{2}\boldsymbol{\alpha},
\end{align}
where $\left\langle \dots\right\rangle _{W}=\left\langle \dots\right\rangle _{W,\infty}$
is the Wigner ensemble average. 

For applications to photon-counting experiments, the Wigner representation
is suitable for simulating quadrature operators as measured by homodyne
detectors, which are symmetrically ordered. The ordering requirement
makes applications to normally-ordered detectors both cumbersome,
as one must reorder all operators to normal order, for example:

\begin{equation}
\left\langle \{\hat{a}^{\dagger}a\}_{sym}\right\rangle =|\alpha|^{2}+\frac{1}{2},
\end{equation}
as well as inaccurate. The additional term arises from the reordering,
as the Wigner representation adds half a quantum of vacuum noise per
mode, causing a rapid increase sampling errors. Hence, the Wigner
representation is unsuitable for simulations of normally-ordered photon
counting probabilities.

\subsection{Q-function}

The standard form of the anti-normally ordered, e.g. $\hat{a}\hat{a}^{\dagger}$,
$M$-mode Q-function is

\begin{equation}
Q(\boldsymbol{\alpha})=\frac{1}{\pi^{M}}\left\langle \boldsymbol{\alpha}|\hat{\rho}|\boldsymbol{\alpha}\right\rangle ,
\end{equation}
and, like the Wigner function, can be expressed as the Fourier transform
of the anti-normally ordered characteristic function.

Unlike the Wigner distribution, the Q-function distribution is always
positive for any classical or non-classical state but is only defined
for anti-normally ordered operator products with moments being obtained
as

\begin{align}
\left\langle \hat{a}_{j_{1}},\dots,\hat{a}_{j_{n}}^{\dagger}\right\rangle  & =\left\langle \alpha_{j_{1}},\dots,\alpha_{j_{n}}^{*}\right\rangle _{Q}\nonumber \\
 & =\int Q(\boldsymbol{\alpha})[\alpha_{j_{1}},\dots,\alpha_{j_{n}}^{*}]\text{d}^{2M}\boldsymbol{\alpha},
\end{align}
where $\left\langle \dots\right\rangle _{Q}=\left\langle \dots\right\rangle _{Q,\infty}$
denotes a Q-distribution ensemble average.

Like the Wigner function, this ordering requirment means that for
applications to normally-ordered photon counting experiments, or any
normally-ordered measurement, operators must be reordered. This is
readily illustrated using the standard bosonic commutation relations
Eqs.(\ref{eq:bose_comm_rel}), where anti-normal ordered expectation
value $\left\langle a\hat{a}^{\dagger}\right\rangle $ is reordered
to

\begin{equation}
\left\langle a\hat{a}^{\dagger}\right\rangle =|\alpha|^{2}+1.
\end{equation}
Therefore, the Q-function adds an entire quantum of vacuum noise per
mode, generating the largest increase in sampling errors of any phase-space
representation when used to simulate a normally ordered measurement.
This accumulation of vacuum noise for multimode linear photonic networks
rapidly causes Q-function simulations to become inaccurate.

\subsection{$\sigma$-ordering\label{subsec:sigma-ordering}}

The amount of vacuum noise added by each representation can be used
to define the operator ordering parameter $\sigma$, where $\sigma=0$
corresponds to normal ordering, $\sigma=1/2$ symmetric ordering and
$\sigma=1$ anti-normal ordering.

From the ability to define a common ordering scheme arises the ability
to define a $\sigma$-ordered phase-space distribution. Since the
Wigner and Q-function distributions can be defined as convolutions
of the positive P-representation, a $\sigma$-ordered representation
is defined as:

\begin{equation}
P_{\sigma}(\boldsymbol{\alpha})=\frac{1}{(\pi\sigma)^{M}}\int P(\boldsymbol{\alpha}_{0},\boldsymbol{\beta}_{0})e^{-(\boldsymbol{\alpha}-\boldsymbol{\alpha}_{0})(\boldsymbol{\alpha}^{*}-\boldsymbol{\beta}_{0})/\sigma}\text{d}^{2M}\boldsymbol{\alpha}\text{d}^{2M}\boldsymbol{\beta}.
\end{equation}
Here, $P_{\sigma}(\boldsymbol{\alpha})$ is a $\sigma$-ordered distribution,
$P(\boldsymbol{\alpha}_{0},\boldsymbol{\beta}_{0})$ is the positive-P
distribution and $\boldsymbol{\alpha}_{0},\boldsymbol{\beta}_{0}$
are used to denote the normal-ordered non-classical phase-space variables,
whilst $\boldsymbol{\alpha},\boldsymbol{\alpha}^{*}$ denote a classical
phase-space which is valid for $\sigma=1/2,1$.

Operator moments for any ordering can now be obtained via

\begin{align}
\left\langle \left\{ \hat{a}_{j_{1}}^{\dagger},\dots,\hat{a}_{j_{n}}\right\} _{\sigma}\right\rangle  & =\left\langle \alpha_{j_{1}}^{*},\dots,\alpha_{j_{n}}\right\rangle _{\sigma}\nonumber \\
 & =\int P_{\sigma}(\boldsymbol{\alpha})[\alpha_{j_{1}}^{*},\dots,\alpha_{j_{n}}]\text{d}^{2M}\boldsymbol{\alpha},
\end{align}
where, as above, $\left\langle \dots\right\rangle _{\sigma}=\left\langle \dots\right\rangle _{\sigma,\infty}$
is a $\sigma$-ordered ensemble average. 

\section{Dynamics in phase-space}

The density matrix master equation is basic to quantum theory, particularly
for open quantum systems, the theory of which will be treated in more
detail in Chapter \ref{chap:Quantum Theory}. The master equations
of most interest in this chapter have the standard Lindblad form:
\begin{align}
\dot{\hat{\rho}} & =\mathcal{L}_{J}\hat{\rho}\label{eq:Lindblad_Master_Equation-3}\\
 & =-i\left[\hat{H},\hat{\rho}\right]+\sum_{j=1}^{J}\gamma_{j}\left(2\hat{L}_{j}\rho\hat{L}_{j}^{\dagger}-\hat{L}_{j}^{\dagger}\hat{L}_{j}\rho-\rho\hat{L}_{j}^{\dagger}\hat{L}_{j}\right).\nonumber 
\end{align}
Here, $\mathcal{L}_{J}$ is the total super-operator for $J$ terms,
$\hat{H}$ is the reversible system Hamiltonian, $\hat{L}_{j}$ are
$J$ operators that couple the system to the dissipative reservoir,
and $\gamma_{j}$ is the decay rate. The dissipative operators can
be further classified by type $n$ and mode index $k$, including
vector indices if needed.

Solving the master equation using orthogonal basis methods is impractical,
as the master equation has a memory requirement that scales as $e^{2\lambda M}$
for $M$ modes, where $\lambda=\ln(N_{max})$, and $N_{max}$ is the
dimension of the Hilbert space of a single mode. As $M\rightarrow\infty$,
the Hilbert space dimension grows exponentially, limiting orthogonal
basis methods to small mode numbers as the memory and CPU time grows
rapidly. Therefore, large quantum systems like linear photonic networks
or quantum fields are inaccessible, apart from using various approximations
like mean-field or tensor network methods. Although even tensor networks
eventually succumb to computational limitations. 

One possible way to treat such large quantum systems is via quantum
phase-space expansions. These methods convert the master equation
into a stochastic differential equation, which are often more scalable
than other methods \cite{Opanchuk2019Mesoscopic,Opanchuk2019Robustness}.
There are trade-offs, and this often may require further approximations.

The advantage arises from the phase-space distribution being sampled
using random sampling, where each sample in phase-space requires a
polynomial amount of storage, typically growing linearly with the
number of modes. From this sampling procedure, it is also usually
relatively straightforward to estimate sampling errors.

This approach started when Schrödinger \cite{Schrodinger1926Constant}
pointed out that quantum oscillators can have classical equations.
This was extended to other systems \cite{Wigner1932Quantum,Moyal1949Quantum,Glauber1963Coherent},
especially including lasers and quantum optics \cite{Louisell1973Quantum,gardiner2004quantum,Carmichael2002Statistical,Drummond2014Quantum}.

\subsection{Operator mappings}

To perform dynamical simulations of a quantum system requires first
mapping the master equation into a second-order partial differential
equation called a Fokker-Planck equation (FPE) (see Chapter \ref{chap:SDE Theory}
for a theoretical review of FPEs). To do this, one requires a mapping
between operator products, such as $\hat{a}_{j}^{\dagger}\hat{\rho}$,
and partial derivatives of phase-space distributions, i.e. $\frac{\partial}{\partial\alpha_{j}}W(\boldsymbol{\alpha})$. 

Using the $\sigma$-ordered notation, unified operator identities
can be defined to perform this differential equation mapping as:

\begin{eqnarray}
\hat{a}_{j}^{\dagger}\hat{\rho} & \rightarrow & \left[\beta_{j}+\left(\sigma-1\right)\frac{\partial}{\partial\alpha_{j}}\right]P_{\sigma}\nonumber \\
\hat{a}_{j}\hat{\rho} & \rightarrow & \left[\alpha_{j}+\sigma\frac{\partial}{\partial\beta_{j}}\right]P_{\sigma}\nonumber \\
\hat{\rho}\hat{a}_{j} & \rightarrow & \left[\alpha_{j}+\left(\sigma-1\right)\frac{\partial}{\partial\beta_{j}}\right]P_{\sigma}\nonumber \\
\hat{\rho}\hat{a}_{j}^{\dagger} & \rightarrow & \left[\beta_{j}+\sigma\frac{\partial}{\partial\alpha_{j}}\right]P_{\sigma}\,.\label{eq:drummond_identities-1-1-1}
\end{eqnarray}

If the resulting differential equation obtained from this mapping
has a second-order positive-definite form it is an FPE, which is equivalent
to an SDE (see Chapter \ref{chap:SDE Theory} for theoretical background
on SDEs), or an SPDE for quantum fields \cite{Carter1987Squeezing}.
The noise can be additive or multiplicative, depending on the problem.
Although we have defined a unified mapping for any ordered phase-space
representation, not all methods give stable FPE equations \cite{Deuar2002Gauge},
such as the positive P-representation, which can suffer from boundary-term
corrections \cite{Deuar2002Gauge,Gilchrist_Gardiner_PD_PPR_Application_Validity}.
The Wigner representation meanwhile requires a truncation of larger
than second-order derivatives if the Hamiltonian is nonlinear \cite{Steel1998Dynamical},
hence simulations only approximate the system dynamics, although this
can be accurate in some cases \cite{Steel1998Dynamical}. The FPE
obtained from the Q-function on the other hand is no longer positive-definite,
which is a requirement for FPEs. 

The total noise includes internal quantum noise generated from the
Hamiltonian term $\hat{H}_{sys}$, as well as reservoir noise terms
generated from the coupling to the reservoir operators, which is proportional
to the damping rate $\Gamma_{j}$. There is a similar behavior in
classical systems, except that these correspond to a high-temperature
limit, and in most cases only have external reservoir noise from thermal
fluctuations.

\section{Damped harmonic oscillator\label{sec:Damped-harmonic-oscillator-1}}

As an example, take the driven quantum harmonic oscillator. This has
the Hamiltonian 
\begin{equation}
\hat{H}/\hbar=i\mathcal{E}\left(\hat{a}^{\dagger}-\hat{a}\right)+\omega_{0}\hat{a}^{\dagger}\hat{a},
\end{equation}
where $\mathcal{E}$ is the driving amplitude, and $\omega_{0}$ the
harmonic oscillator frequency. If damping is added, it obeys the master
equation 
\begin{align}
\frac{d\hat{\rho}}{dt} & =-i[i\left(\mathcal{E}\hat{a}^{\dagger}-\mathcal{E}^{*}\hat{a}\right)+\omega_{0}\hat{a}^{\dagger}\hat{a},\rho]+\gamma\left(1+\bar{n}\right)(2\hat{a}\rho\hat{a}^{\dagger}-\hat{a}^{\dagger}\hat{a}\rho-\rho\hat{a}^{\dagger}\hat{a})\nonumber \\
 & +\gamma\bar{n}(2\hat{a}^{\dagger}\rho\hat{a}-\hat{a}\hat{a}^{\dagger}\rho-\rho\hat{a}\hat{a}^{\dagger}),
\end{align}
where $\bar{n}$ is the temperature reservoir occupation (see Chapter
\ref{chap:Quantum Theory}). 

This leads to a random walk in a complex space \cite{Gardiner2009Stochastic,Drummond2014Quantum}:
\begin{align}
\frac{d\alpha}{dt} & =\mathcal{E}-\left(\gamma+i\omega_{0}\right)\alpha+\sqrt{2\gamma\left(\sigma+\bar{n}\right)}\zeta(t)\nonumber \\
\frac{d\beta}{dt} & =\mathcal{E}^{*}-\left(\gamma-i\omega_{0}\right)\beta+\sqrt{2\gamma\left(\sigma+\bar{n}\right)}\zeta^{*}(t),
\end{align}
where the noise is complex and $\zeta(t)=\left(w_{1}(t)+iw_{2}(t)\right)/\sqrt{2}$.
The correlations are 
\begin{align}
\left\langle \zeta(t)\left(\zeta(t')\right)^{*}\right\rangle  & =\delta\left(t-t'\right)\nonumber \\
\left\langle \zeta(\omega)\left(\zeta\left(\omega'\right)\right)^{*}\right\rangle  & =\delta\left(\omega-\omega'\right).
\end{align}

\subsection{Wigner dynamics}

In the undriven, zero temperature Wigner case with $\gamma=$1, $\sigma=1/2$,
and in a rotating frame so that $\omega_{0}=0$, using the mappings
Eq.(\ref{eq:drummond_identities-1-1-1}), the probability follows
the Fokker-Planck equation: 
\begin{equation}
\frac{\partial P_{1/2}}{\partial t}=\left[\frac{\partial}{\partial\alpha_{x}}\alpha_{x}\,+\frac{\partial}{\partial\alpha_{y}}\alpha_{y}+\frac{1}{4}\left(\frac{\partial^{2}}{\partial\alpha_{x}^{2}}\,+\frac{\partial^{2}}{\partial\alpha_{y}^{2}}\right)\,\right]P_{1/2}\,,\label{eq:FPE-1-1-1}
\end{equation}
which is an example of Eq.(\eqref{eq:FPE}). Ignoring terms that vanish
or can be obtained from symmetry, the first corresponding moment equations
in each of the real and imaginary directions are 
\begin{align}
\frac{\partial}{\partial t}\left\langle \alpha_{x}\right\rangle = & \left\langle -\alpha_{x}\frac{\partial}{\partial\alpha_{x}}\alpha_{x}\,\right\rangle =-\left\langle \alpha_{x}\,\right\rangle \nonumber \\
\frac{\partial}{\partial t}\left\langle \alpha_{x}\alpha_{y}\right\rangle = & \left\langle -\left(\alpha_{x}\frac{\partial}{\partial\alpha_{x}}+\alpha_{y}\frac{\partial}{\partial\alpha_{y}}\,\right)\alpha_{x}\alpha_{y}\,\right\rangle =-\left\langle \alpha_{x}\alpha_{y}\,\right\rangle \nonumber \\
\frac{\partial}{\partial t}\left\langle \alpha_{x}^{2}\right\rangle = & \left\langle \left(-\alpha_{x}\frac{\partial}{\partial\alpha_{x}}+\frac{1}{4}\frac{\partial^{2}}{\partial\alpha_{x}^{2}}\,\right)\alpha_{x}^{2}\,\right\rangle =\frac{1}{2}-2\left\langle \alpha_{x}^{2}\,\right\rangle .
\end{align}

The steady-state is therefore a Gaussian distribution with $\left\langle \alpha_{x,y}\,\right\rangle =0$,
$\left\langle \alpha_{x}\alpha_{y}\,\right\rangle =0$ and $\left\langle \alpha_{x,y}^{2}\,\right\rangle =1/4$.
One can use an initial condition of $\alpha=(v_{1}+iv_{2})/2$, with
$\left\langle v_{i}^{2}\right\rangle =1/2$, in order to replicate
the steady state, which is a Gaussian with$\left\langle \alpha_{x}\,\right\rangle =\left\langle \alpha_{y}\,\right\rangle =0$
and$\left\langle \alpha_{x}^{2}\,\right\rangle =\left\langle \alpha_{y}^{2}\,\right\rangle =1/4$.

\subsection{Internal spectrum}

Neglecting any boundary terms, the equation in frequency space is:

\begin{equation}
-i\omega\tilde{\alpha}(\omega)=-\tilde{\alpha}(\omega)+\tilde{\zeta}(\omega).
\end{equation}

For sufficiently long times, the solution in frequency space - where
$\omega=2\pi f$ is the angular frequency - is therefore given by:
\begin{equation}
\tilde{\alpha}\left(\omega\right)=\frac{\tilde{\zeta}(\omega)}{1-i\omega}.
\end{equation}
The expectation value of the noise spectrum, $\left\langle \left|\tilde{\alpha}(\omega)\right|^{2}\right\rangle $
in the long time limit, is: 
\begin{eqnarray}
\left\langle \left|\tilde{\alpha}(\omega)\right|^{2}\right\rangle  & = & \frac{1}{2\pi\left(1+\omega^{2}\right)}\int\int e^{-i\omega(t-t')}\left\langle \zeta(t)\zeta^{*}(t')\right\rangle dtdt'\,.\nonumber \\
 & = & \frac{T}{2\pi\left(1+\omega^{2}\right)}.\label{eq:spectra_SHO-1}
\end{eqnarray}

This equation can also be used for some classical problems, which
correspond to the high-temperature limit of $\bar{n}\gg1$.

\section{Stochastic gauge expansion}

In this approach, the density matrix is expanded as a weighted integral
over coherent state projection operators: 
\begin{equation}
\rho\left(t\right)=\int d\bm{\phi}P\left(t,\bm{\phi}\right)\Lambda\left(\bm{\phi}\right).
\end{equation}
Here, in the stochastic gauge method \cite{Deuar2002Gauge}, $\bm{\phi}\equiv\left[\Omega,\bm{\alpha},\bm{\beta}\right]$
, where $\bm{\alpha},\bm{\beta}$ are each $M$-dimensional complex
numbers, and $\Omega$ is a real or complex weight. The operator basis
$\Lambda$ is defined using un-normalized coherent states $\left\Vert \bm{\alpha}\right\rangle =\exp\left(\boldsymbol{\alpha}\cdot\hat{\boldsymbol{a}}^{\dagger}\right)\left|0\right\rangle $,
so that: 
\begin{align}
\Lambda\left(\bm{\phi}\right) & =\Omega\left\Vert \bm{\alpha}\right\rangle \left\langle \bm{\beta}^{*}\right\Vert e^{-\bm{\alpha}\cdot\bm{\beta}}.
\end{align}

There are standard identities available, namely:

\begin{align}
\hat{a}_{j}\Lambda & =\alpha_{j}\Lambda\nonumber \\
\hat{a}_{j}^{\dagger}\Lambda & =\left[\partial/\partial\alpha_{j}+\beta_{j}\right]\Lambda\nonumber \\
\Lambda\hat{a}_{j}^{\dagger} & =\beta_{j}\Lambda\nonumber \\
\Lambda\hat{a}_{j} & =\left[\partial/\partial\beta_{j}+\alpha_{j}\right]\Lambda\nonumber \\
0 & =\left[\Omega\partial/\partial\Omega-1\right]\Lambda\nonumber \\
0 & =\partial^{2}/\partial\Omega^{2}\Lambda
\end{align}
The hermiticity of $\rho$ means that every $\bm{\phi}$ has a conjugate
$\bm{\phi}^{*}$ of equal weight, so the integral is sampled in pairs
$\bm{\phi}_{s}$ and $\bm{\phi}_{s}^{*}$, corresponding to a sum
over $\mathcal{S}$ samples of the real part of $\Lambda$: 
\begin{equation}
\rho_{c}\left(t\right)=\lim_{\mathcal{S}\rightarrow\infty}\frac{1}{\mathcal{S}}\sum_{s}\Re\Lambda\left(\bm{\phi}_{s}\left(t\right)\right).
\end{equation}
Operator averages are obtained through defining a \emph{weighted}
average as the infinite ensemble limit of a sum of trajectories: 
\begin{equation}
\left\langle f\left(\bm{\phi}\right)\right\rangle \equiv\lim_{\mathcal{S}\rightarrow\infty}\left\langle f\left(\bm{\phi}\right)\right\rangle _{\mathcal{S}}.
\end{equation}

Here, for hermitian operators, 
\begin{equation}
\left\langle f\left(\bm{\phi}\right)\right\rangle _{\mathcal{S}}\equiv\frac{1}{\mathcal{S}}\sum_{s}\Re\left[\Omega_{s}f\left(\bm{\phi}_{s}\right)\right],
\end{equation}
with the approximation of taking only a finite number of samples $\mathcal{S}$.
For example, the quantum average particle number $\left\langle \hat{n}_{j}\right\rangle _{Q}$
is obtained on taking a weighted average of $n_{js}\equiv\alpha_{js}\beta_{js}$:
\begin{equation}
\left\langle \hat{n}_{j}\right\rangle _{Q}=\left\langle n_{j}\right\rangle .
\end{equation}
Individual trajectory photon numbers $\Re\left(\Omega_{s}n_{js}\right)$
can be negative, although their large-$\mathcal{S}$ average is non-negative.
These trajectories correspond to Schrodinger cat superpositions, causing
mixtures of positive and negative 'effective' photon numbers. Such
behavior is impossible in the diagonal Glauber-Sudarshan representation,
where for a probabilistic distribution, only classical photon statistics
occur \cite{Titulauer1965Correlation,Reid1986}.

\section{Input-output spectra}

The spectrum of an internal field variable is not the one that is
usually measured. An important application of stochastic equations
is therefore in calculating output, measured spectra of lasers, quantum
optics, opto-mechanics and quantum circuits \cite{gardiner2004quantum,Kiesewetter2014Scalable}.
These have the feature that the measured output spectrum may also
include noise from reflected fields at the input/output ports. If
the quantum noise term in the Heisenberg equations for a cavity operator
$\hat{a}_{c}$ is given by: $\dot{\hat{a}}_{c}\sim..+\sqrt{2\gamma}\hat{a}_{in}(t),$
then the corresponding operator input-output relations are $\hat{a}_{out}(t)+\hat{a}_{in}(t)=\sqrt{2\gamma}\hat{a}_{c}$.

In quantum phase-space for the case of the harmonic oscillator or
similar systems, $\alpha_{in}=\sqrt{\sigma+\bar{n}}\zeta$ is the
noise term in the Langevin equation. The output fields $\alpha_{out}$
that are measured are given by: 
\begin{align}
\alpha_{out} & =\sqrt{2\gamma}\alpha-\alpha_{in}.
\end{align}

Hence one must include in the spectrum both the internal mode variables
and the noise terms themselves. Solving for the spectra, one obtains
auxiliary fields with 
\begin{align}
\tilde{\alpha}_{in}(\omega) & =\sqrt{\sigma+\bar{n}}\tilde{\zeta}(\omega)\\
\tilde{\alpha}_{out}(\omega) & =\sqrt{2\gamma}\tilde{a}(\omega)-\sqrt{\sigma+\bar{n}}\tilde{\zeta}(\omega).\nonumber 
\end{align}

In summary, it is the output fields that are amplified and measured.
Hence one must be able to compute the spectra of the output fields
for experimental comparisons. These have the additional feature that
they include the reservoir noise $\tilde{\zeta}(\omega)$, evaluated
at the same time as the field is evaluated, since the reservoir noise
is the input here. In xSPDE these are called \emph{auxfields}.

\subsection{Steady-state result}

Consider the example of the damped quantum harmonic oscillator in
the Wigner representation case with $\gamma=1$, $\sigma=1/2$ and
$\bar{n}=0$. Over long time-scales, so that one is in the steady
state, the solution for $\tilde{a}_{out}$ is that: 
\begin{align}
\tilde{\alpha}_{out}(\omega) & =\sqrt{2}\left[\frac{1}{1-i\omega}-\frac{1}{2}\right]\tilde{\zeta}(\omega)\nonumber \\
 & =\frac{1}{\sqrt{2}}\left[\frac{1+i\omega}{1-i\omega}\right]\tilde{\zeta}(\omega).
\end{align}

This gives the following expectation values: 
\begin{align}
\left\langle \tilde{\alpha}_{out}(\omega)\left(\tilde{\alpha}_{out}(\omega)\left(\omega'\right)\right)^{*}\right\rangle  & =\frac{1}{2}\delta\left(\omega-\omega'\right)\nonumber \\
\left\langle \tilde{\alpha}_{in}(\omega)\left(\tilde{\alpha}_{in}(\omega)\left(\omega'\right)\right)^{*}\right\rangle  & =\frac{1}{2}\delta\left(\omega-\omega'\right).
\end{align}
These are the expectation values of the zero temperature quantum fluctuations
in the input and output channels. This means that the harmonic oscillator
in its ground state is in equilibrium with an external vacuum field
reservoir, also in its ground state. However, the internal spectral
correlations of the harmonic oscillator are modified by the coupling.

While this is a simple result, exactly the same general type of behavior
occurs in more sophisticated cases. These may include many coupled
modes with nonlinearities. Additional or auxiliary fields that depend
both on noise terms and internal stochastic variables are required.
The soluble case given above is a useful test case, and it is treated
numerically later in the manual.

\section{Linear photonic network theory\label{sec:Linear-photonic-networks}}

In some quantum systems, output observables are obtained not via dynamical
processes, but after a simple linear transformation of a multi-mode
density operator. In the case of linear photonic quantum computing
networks such as GBS, the mode transformation is traditionally generated
by a network of beam-splitters, phase shifters, and mirrors \cite{zhong2020quantum,deng2023gaussian},
although other set-ups include fibre delay lines \cite{deshpandeQuantumComputationalAdvantage2022a,madsenQuantumComputationalAdvantage2022}. 

These networks act as $M$-mode interferometers which interfere input
photon, generating large amounts of entanglement due to the exponential
number of interference pathways available to photons. In the ideal
lossless regime, the network itself is defined by an $M\times M$
Haar random unitary matrix $\boldsymbol{U}$, such that output modes
are linear combinations of each input mode: 

\begin{equation}
\hat{a}_{i}^{(\text{out})}=\sum_{j=1}^{M}U_{ij}\hat{a}_{j}^{(\text{in})},\label{eq:linear_combo}
\end{equation}
where $\hat{a}_{i}^{(\text{in})}$ and $\hat{a}_{j}^{(\text{out})}$
are the input and output annihilation operators for modes $i$, $j$
respectively.

Practically, photon loss in the network is commonplace, thus causing
the matrix to be non-unitary. Therefore, lossy networks are denoted
by the transmission matrix $\boldsymbol{T}$. These give a different
transformation law, where:

\begin{equation}
\hat{a}_{i}^{(\text{out})}=\sum_{j=1}^{M}T_{ij}\hat{a}_{j}^{(\text{in})}+\sum_{j=1}^{M}B_{ij}\hat{b}_{j}^{(\text{in})},\label{eq:linear_combo-1}
\end{equation}

Here, the $M$ operators $\hat{b}_{i}^{(\text{in})}$are noise operators
which are necessary to conserve the operator commutation relations 

\begin{align}
\left[\hat{a}_{i},\hat{a}_{j}\right] & =0\nonumber \\
\left[\hat{a}_{i},\hat{a}_{j}^{\dagger}\right] & =\delta_{ij}.\label{eq:bose_comm_rel}
\end{align}
The noise operators are independent, commuting operators, who comprise
inputs from the reservoirs that cause losses, where the reservoirs
are all in a vacuum state.

The inclusion of the loss matrix conserves the unitarity of the network.
Substituting Eq.(\ref{eq:linear_combo-1}) into Eq.(\ref{eq:bose_comm_rel})
and and taking expectation values for a vacuum state input gives 
\begin{align}
\delta_{ij} & =\left\langle \left[\hat{a}_{i}^{(\text{out})},\hat{a}_{j}^{\dagger(\text{out})}\right]\right\rangle \nonumber \\
 & =\sum_{k}\left(T_{ik}T_{jk}^{*}+B_{ik}B_{jk}^{*}\right).
\end{align}
Next, we can define a new $M\times M$ matrix 
\begin{equation}
\bm{D}=\bm{B}\bm{B}^{\dagger}=\bm{I}-\bm{T}\bm{T}^{\dagger}.
\end{equation}
This is hermitian, since $\bm{D}^{\dagger}=\bm{D}$, and so has a
diagonal representation as $D=\tilde{U}\lambda^{2}\tilde{U}^{\dagger}$,
for some unitary matrix $\tilde{U}$. We assume that the transmission
matrix $\bm{T}$ is lossy, so that $\bm{D}$ is positive definite
and $\lambda$ is real, representing absorption rather than gain.

\section{Quantum input states\label{subsec:Quantum-input-states}}

The density operator $\hat{\rho}^{(\text{in})}$ transformed by linear
network has $N$ input modes, where one can have $N=M$ or $N\subset M$,
in which case the remaining $M-N$ modes are vacuum inputs at unused
ports. If each input mode is independent, $\hat{\rho}^{(\text{in})}$
is a product of input states.

Currently, xSPDE can only generate input squeezed states and thermal
states as outlined below. Other inputs are possible, since the positive
P-representation and Q-representation are complete, positive representations,
and can be added through user customization.

For pure squeezed vacuum states, the input is defined as

\begin{equation}
\hat{\rho}^{(\text{in})}=\prod_{j=1}^{M}\left|r_{j}\right\rangle \left\langle r_{j}\right|,
\end{equation}
where $\boldsymbol{r}=[r_{1},\dots,r_{M}]$ is a vector of squeezing
parameters $r_{j}$ and

\begin{align}
\left|r_{j}\right\rangle  & =\hat{S}(r_{j})\left|0\right\rangle \nonumber \\
 & =\exp\left(r_{j}\frac{(\hat{a}_{j}^{\dagger(\text{in})})^{2}}{2}-r_{j}\frac{(\hat{a}_{j}^{(\text{in})})^{2}}{2}\right)\left|0\right\rangle ,\label{eq:squeezed_state_exp}
\end{align}
is the squeezed vacuum state with squeezing operator $\hat{S}(r_{j})$,
which satisfies the unitarity condition $\hat{S}\hat{S}^{\dagger}=\hat{S}^{\dagger}\hat{S}=1$.
Here, we have assumed the squeezed state phase is zero.

\subsection{Pure squeezed states}

Pure squeezed states are the default type of squeezed state generated
in xSPDE. Given each input mode is independent, one can use the well
known single-mode squeezed state theory \cite{loudonQuantumTheoryLight1983,walls2008quantum},
to derive the basic properties of pure squeezed states. 

Using the relations 

\begin{align}
\hat{S}^{\dagger}(r_{j})\hat{a}_{j}^{(\text{in})}\hat{S}(r_{j}) & =\hat{a}_{j}^{(\text{in})}\cosh(r_{j})-\hat{a}_{j}^{\dagger(\text{in})}\sinh(r_{j})\nonumber \\
\hat{S}^{\dagger}(r_{j})\hat{a}_{j}^{\dagger(\text{in})}\hat{S}(r_{j}) & =\hat{a}_{j}^{\dagger(\text{in})}\cosh(r_{j})-\hat{a}_{j}^{(\text{in})}\sinh(r_{j}),
\end{align}
the mean input photon number per mode is defined as 

\begin{align}
\bar{n}_{j} & =\left\langle \hat{a}_{j}^{\dagger(\text{in})}\hat{a}_{j}^{(\text{in})}\right\rangle \nonumber \\
 & =\left\langle 0\right|\hat{S}^{\dagger}(r_{j})\hat{a}_{j}^{\dagger(\text{in})}\hat{S}(r_{j})\hat{S}^{\dagger}(r_{j})\hat{a}_{j}^{(\text{in})}\hat{S}(r_{j})\left|0\right\rangle \nonumber \\
 & =\sinh^{2}(r_{j}),
\end{align}
while the mean input coherence per mode is

\begin{align}
m_{j} & =\left\langle \left(\hat{a}_{j}^{(\text{in})}\right)^{2}\right\rangle \nonumber \\
 & =\left\langle 0\right|\hat{S}^{\dagger}(r_{j})\hat{a}_{j}^{(\text{in})}\hat{S}(r_{j})\hat{S}^{\dagger}(r_{j})\hat{a}_{j}^{(\text{in})}\hat{S}(r_{j})\left|0\right\rangle \nonumber \\
 & =\sinh(r_{j})\cosh(r_{j}).
\end{align}
For pure squeezed states, the coherence and photon number are related
via $m_{j}^{2}-\bar{n}_{j}=\bar{n}_{j}^{2}$.

Squeezed states are minimum uncertainty states and are therefore defined
entirely by their quadrature variances. From the quadrature operators

\begin{align}
\hat{x}_{j}^{(\text{in})} & =\hat{a}_{j}^{(\text{in})}+\hat{a}_{j}^{\dagger(\text{in})}\nonumber \\
\hat{y}_{j}^{(\text{in})} & =-i\left(\hat{a}_{j}^{(\text{in})}-\hat{a}_{j}^{\dagger(\text{in})}\right),
\end{align}
which obey the commutation relation $\left[\hat{x}_{j}^{(\text{in})},\hat{y}_{k}^{(\text{in})}\right]=2i\delta_{jk}$,
the normally ordered $x_{j}^{(\text{in})}$-quadrature variance is
defined as

\begin{align}
\left\langle :(\Delta\hat{x}_{j}^{(\text{in})})^{2}:\right\rangle  & =\left\langle (\hat{x}_{j}^{(\text{in})})^{2}\right\rangle \nonumber \\
 & =2(\bar{n}_{j}+m_{j})\nonumber \\
 & =e^{2r_{j}}-1,\label{eq:x-quad}
\end{align}
while the normally ordered $y_{j}^{(\text{in})}$-quadrature variance
is

\begin{align}
\left\langle :(\Delta\hat{y}_{j}^{(\text{in})})^{2}:\right\rangle  & =\left\langle (\hat{y}_{j}^{(\text{in})})^{2}\right\rangle \nonumber \\
 & =2(\bar{n}_{j}-m_{j})\nonumber \\
 & =e^{-2r_{j}}-1.\label{eq:y-quad}
\end{align}

\subsection{Thermal squeezed states}

As stated above, xSPDE4 can currently simulate pure and thermalized
squeezed states, as well as classical thermal state inputs into a
photonic network. This is achieved using a model for thermal squeezed
states which alters the multi-mode input coherence as $\tilde{m}_{j}=(1-\epsilon)m_{j}$,
with $\epsilon$ being the thermalization component which is input
to xSPDE as \emph{p.thermal}, whilst keeping the input photon number
unchanged. This allows one to interpolate between pure thermal, $\epsilon=1$,
and pure squeezed, $\epsilon=0$, states.

Thermal states are classical states with fluctuations larger than
the vacuum limit such that their quadrature variances are $\left\langle :(\Delta\hat{x}_{j}^{(\text{in})})^{2}:\right\rangle =\left\langle :(\Delta\hat{y}_{j}^{(\text{in})})^{2}:\right\rangle >1$.
In terms of Fock states, the thermal state density operator for the
$j$-th mode is the single-mode state

\begin{equation}
\hat{\rho}_{j}^{(\text{in})}=\frac{1}{1+\bar{n}_{j}}\sum_{n_{j}=0}^{\infty}\left(\frac{\bar{n}_{j}}{1+\bar{n}_{j}}\right)^{n_{j}}\left|n_{j}\right\rangle \left\langle n_{j}\right|,
\end{equation}
which gives the well known single-mode photon number distribution

\begin{equation}
P\left(n_{j}\right)=\frac{\bar{n}_{j}^{n_{j}}}{\left(\bar{n}_{j}+1\right)^{n_{j}+1}}.\label{eq:thermal state dis}
\end{equation}

Thermal states can be used to generate thermal squeezed states with
initial occupation $n_{j}^{\text{th}}$, which gives \cite{marian1992higher}:
\begin{align}
\bar{n}_{j}= & n_{j}^{\text{th}}+\left(2n_{j}^{\text{th}}+1\right)\sinh^{2}\left(r_{j}\right)\nonumber \\
\tilde{m}_{j}= & \left(2n_{j}^{\text{th}}+1\right)\sinh\left(r_{j}\right)\cosh\left(r_{j}\right).\label{eq:thermal_squeezed_state_n_m}
\end{align}
In the thermalized case, the relationship between coherence and photon
number is modified, since to eliminate $r_{j}$ one must use the relationship
that 
\begin{align}
\frac{\tilde{m}_{j}^{2}}{\left(2n_{j}^{\text{th}}+1\right)^{2}} & =\sinh^{2}\left(r_{j}\right)\left(1+\sinh^{2}\left(r_{j}\right)\right)\nonumber \\
 & =\frac{\bar{n}_{j}-n_{j}^{\text{th}}}{\left(2n_{j}^{\text{th}}+1\right)}\left(1+\frac{\bar{n}_{j}-n_{j}^{\text{th}}}{\left(2n_{j}^{\text{th}}+1\right)}\right).
\end{align}
Therefore: 
\begin{align}
\tilde{m}_{j}^{2} & =\left(\bar{n}_{j}-n_{j}^{\text{th}}\right)\left(1+\bar{n}_{j}+n_{j}^{\text{th}}\right)\nonumber \\
 & =\bar{n}_{j}+\bar{n}_{j}^{2}-\left((n_{j}^{\text{th}})^{2}+n_{j}^{\text{th}}\right).
\end{align}

Using the above theory, thermalized squeezed states can also be used
as a test for numerical simulations of photon counting observables,
as one can define the threshold detector projection operators in terms
of the photon number and coherence as explained below. 

\section{Photon counting}

Although linear photonic networks are conceptually very simple, when
employed as quantum computers they generate samples from an output
distribution which corresponds to the $\#P$-hard matrix permanent,
Hafnian or Torontonian functions. Which type of matrix function is
evalutated depends on the input states to the network, with Fock states
corresponding to the permanent and squeezed states corresponding to
either the Hafnian or Torontonian functions, where the difference
between these distributions comes from the type of detector used. 

Currently in xSPDE4, only Gaussian states are generated natively,
hence we restrict our discussion here to Gaussian state photonic quantum
computing networks such as GBS. When photon-number resolving (PNR)
detectors are used, photon count patterns, which are our samples of
the output distribution, correspond ot the Hafnian function, while
the Torontonian requires threshold detectors that ``click'' for
a photon detection event. 

From standard photon counting theory, the projection operator for
observing $c_{j}=0,1,2,\dots,c_{j}^{(\text{max})}$ counts is denoted
by \cite{walls2008quantum}

\begin{equation}
\hat{p}_{j}(c_{j})=\frac{1}{c_{j}!}:(\hat{n}'_{j})^{c_{j}}e^{-\hat{n}'_{j}}:,\label{eq:photon_counting_proj}
\end{equation}
where $:\dots:$ denotes normal ordering and $\hat{n}'_{j}=a_{j}^{\dagger(\text{out})}a_{j}^{(\text{out})}$
is the output photon number and $c_{j}^{(\text{max})}$ is the maximum
observable count. 

For PNR detectors, which can discriminate between photon numbers,
each detector is defined by the above projector, with $c_{j}^{(\text{max})}$
varying depending on experimental implementation. Output photon count
patterns are denoted by the count vector $\boldsymbol{c}=[c_{1},c_{2}\dots,c_{M}]$,
and the projection operator for a specific output pattern given as 

\begin{equation}
\hat{P}(\boldsymbol{c})=\bigotimes_{j=i}^{M}\hat{p}_{j}(c_{j}).
\end{equation}

The expectation value of this pattern projector corresponds to the
Hafnian \cite{Hamilton2017PhysRevLett.119.170501}

\begin{equation}
\left\langle \hat{P}(\boldsymbol{c})\right\rangle =\frac{1}{\sqrt{\det(\boldsymbol{Q})}}\frac{\left|\text{Haf}(\boldsymbol{B}_{S})\right|^{2}}{\prod_{j=1}c_{j}!},
\end{equation}
which is $\#P$-hard to compute at large $M$. Here, $\boldsymbol{B}_{S}$
is the sub-matrix of $\boldsymbol{B}=\boldsymbol{U}\left(\bigoplus_{j=1}^{M}\tanh(r_{j})\right)\boldsymbol{U}^{T}$
formed from modes with detected counts and $\boldsymbol{Q}$ is a
$2M\times2M$ covariance matrix. 

Threshold detectors saturate for more than one count at a detector.
Therefore, outputs are binary with $c_{j}=c_{j}^{(\text{max})}=1$
denoting a detection event, or click, even if multiple photons hit
the same detector, and $c_{j}=0$ is no detection event. From Eq.(\ref{eq:photon_counting_proj}),
the click projection operator is obtained by summing over all $c_{j}>0$
counts such that 

\begin{align}
\hat{\pi}(1) & =:\sum_{c_{j}>0}\frac{(\hat{n}'_{j})^{c_{j}}}{c_{j}!}e^{-\hat{n}'_{j}}:\nonumber \\
 & =1-e^{-\hat{n}'_{j}},
\end{align}
which gives the standard threshold detector projection operator 

\begin{equation}
\hat{\pi}_{j}(c_{j})=:e^{-\hat{n}'_{j}}\left(e^{\hat{n}'_{j}}-1\right)^{c_{j}}:.\label{eq:click proj}
\end{equation}

The projection operator for a count pattern output is then similarly
defined as 

\begin{equation}
\hat{\Pi}(\boldsymbol{c})=\bigotimes_{j=i}^{M}\hat{\pi}_{j}(c_{j}),
\end{equation}
where the expectation value corresponds to the Torontonian function
\cite{quesada2018gaussian}

\begin{equation}
\left\langle \hat{\Pi}(\boldsymbol{c})\right\rangle =\frac{\text{Tor}\left(\boldsymbol{O}_{S}\right)}{\sqrt{\det\left(\boldsymbol{\Sigma}\right)}},
\end{equation}
where $\boldsymbol{O}_{S}$ is the sub-matrix of $\boldsymbol{O}=\boldsymbol{I}-\boldsymbol{\Sigma}^{-1}$
with covariance matrix $\boldsymbol{\Sigma}$. 

\subsection{Exact output examples: Threshold detectors\label{subsec:Exact-click}}

For photon counting probabilities obtained through threshold detection,
a variety of probabilities can be computed exactly. 

Initially, we are interested in computing the probability of detecting
no photons at the threshold detector output, i.e. $\left\langle \hat{\pi}\left(0\right)\right\rangle $.
From Marian \cite{marian1992higher}, the $n=0$ single-mode photon
number probability is known exactly from the full distribution of
a thermalized squeezed state, $P_{\text{sqth}}(n)$, such that 

\[
P_{\text{sqth}}(0)=\left\langle \hat{\pi}\left(0\right)\right\rangle =\frac{1}{n^{\text{th}}+1}\left(1+\frac{2n^{\text{th}}+1}{\left(n^{\text{th}}+1\right)^{2}}\sinh^{2}\left(r\right)\right)^{-(1/2)}
\]

Substituting the thermal squeezed state modified photon number Eq.(\ref{eq:thermal_squeezed_state_n_m})
the vacuum state detection event probability can be derived as 

\begin{align*}
\left\langle \hat{\pi}\left(0\right)\right\rangle  & =\left(\bar{n}-n_{th}+\left(n_{th}+1\right)^{2}\right)^{-(1/2)}\\
 & =\left(1+\bar{n}+n_{th}+n_{th}^{2}\right)^{-(1/2)}\\
 & =\left(\left(1+\bar{n}\right)^{2}-\tilde{m}^{2}\right)^{-(1/2)}.
\end{align*}
Hence, the click and no-click probabilities for thermalized squezed
states can be computed exactly in the limit of an identity transmission
matrix or in the case of a fully thermal state $(\epsilon=1)$ input
with Haar random unitary matrix as: 
\begin{align}
\left\langle \hat{\pi}\left(0\right)\right\rangle  & =\frac{1}{\sqrt{\left(1+\bar{n}\right)^{2}-\tilde{m}^{2}}}\nonumber \\
\left\langle \hat{\pi}\left(1\right)\right\rangle  & =1-\left\langle \hat{\pi}\left(0\right)\right\rangle =1-\frac{1}{\sqrt{\left(1+\bar{n}\right)^{2}-\tilde{m}^{2}}}.
\end{align}

\subsection{Exact output examples: PNR detectors}

For PNR detectors, the multi-mode photon counting distributions are
known exactly for both pure squeezed states and thermal states. These
exact distributions are only valid if one assumes each input state
has equal squeezing parameters $r=r_{1}=\dots=r_{N}$, which in turn
causes the input mean photon number to be equal for each mode $\bar{n}=\bar{n}_{1}=\dots=\bar{n}_{N}$. 

For thermal states, the single-mode photon counting distribution Eq.(\ref{eq:thermal state dis})
is a geometric distribution, which becomes clear by defining the success
probability as 

\begin{equation}
p=\frac{1}{1+\bar{n}},\label{eq:PNR_success_probability}
\end{equation}
such that 

\begin{equation}
P(n)=p(1-p)^{n}.
\end{equation}

This single-mode theory can be extended for multiple thermal state
inputs as the sum of a geometrically distributed random variable $X$
with success probability $p$ is the random variable

\begin{equation}
Y=\sum_{j}^{M}X_{j}
\end{equation}
which is negative binomially distributed with $M$ and $p$. The total
photon number distribution is then defined as

\begin{equation}
P(Y=m)=\left(\begin{array}{c}
m+M-1\\
m
\end{array}\right)p^{M}(1-p)^{m},\label{eq:Multi_mode_thermal_dist}
\end{equation}
where $m=0,1,2,\dots$ is the total, or binned, photon number. At
large $M$, the distribution is a Gaussian with mean and variance

\begin{align}
\mu_{m} & =\bar{n}M,\nonumber \\
\sigma_{m}^{2} & =\frac{\mu_{m}}{p}=\bar{n}M\left(\bar{n}+1\right).
\end{align}

For pure squeezed states transformed by a lossless unitary matrix,
the probability of observing $m=0,1,2,\dots$ total photon counts
from $M$ modes is \cite{zhuPhotocountDistributionsContinuouswave1990,huangPhotoncountingStatisticsMultimode1989}:

\begin{align}
P(2m) & =\binom{\frac{M}{2}+m-1}{m}\text{sech}^{M}(r)\tanh^{2m}(r)\nonumber \\
P(2m+1) & =0.
\end{align}
This distribution is well known due to the distinct oscillations between
even and odd photon count bins. Such oscillations arises due to the
generation of squeezed photons in highly correlated pairs in a parametric
down-conversion process. Therefore, only even numbers of photons are
ever generated, as in clear from the squeezing operator in Eq.(\ref{eq:squeezed_state_exp}). 

In the limit $M\rightarrow\infty$, the total photon counting distribution
$P(m)$ reduces to a Poisson distribution for the even counts \cite{zhuPhotocountDistributionsContinuouswave1990}

\begin{align*}
P(2m) & =\frac{1}{m!}e^{-Mn/2}\left(\frac{Mn}{2}\right)^{m}\\
P(2m+1) & =0,
\end{align*}
although the full distribution is considered super-Poissonian, i.e.
the variance is now larger than the mean. The lossless distribution
can also be written in terms of success probabilities as 

\begin{align}
P(2m) & =\binom{\frac{M}{2}+m-1}{m}p^{M/2}(1-p)^{m}\nonumber \\
P(2m+1) & =0,\label{eq:PNR_exact_squeezed_distribuion}
\end{align}
where $p$ is defined in Eq.(\ref{eq:PNR_success_probability}). 

For lossless linear networks, the total photon counting distribution
is also known exactly and is defined as \cite{deshpandeQuantumComputationalAdvantage2022a}

\begin{align}
P(2m) & =t^{4m}\binom{\frac{M}{2}+m-1}{m}p^{M/2}(1-p)^{m}{}_{2}F_{1}\left(m+\frac{1}{2},\frac{M}{2}+m;\frac{1}{2};(1-t^{2})^{2}(1-p)\right)\nonumber \\
P(2m+1) & =(1-t^{2})\left(\frac{m+1}{2}\right)t^{m-1}\binom{\frac{M}{2}+\frac{m+1}{4}-1}{\frac{m+1}{4}}p^{M/2}(1-p)^{(m+1)/4}\nonumber \\
 & \times{}_{2}F_{1}\left(\frac{m+3}{4},\frac{1}{4}(2M+m+1);\frac{3}{2};(1-t^{2})^{2}(1-p)\right),\label{eq:Multi_mode_squeezed_dist}
\end{align}
where $t$ is a uniform amplitude loss coffieicent, which is applied
as $t\boldsymbol{U}$ and is related to intensity loss via $t=\sqrt{\eta}$,
and $_{2}F_{1}(a,b;c;z)$ is the Gauss hypergeometric function. The
lossy distribution converges to the lossless distribution Eq.(\ref{eq:PNR_exact_squeezed_distribuion})
when $t=1$ as $_{2}F_{1}(a,b;c;0)=1$. 

When identical thermal or pure squeezed states are input into a optical
linear, one can use the above exact distributions to compare phase-space
simulated grouped count probabilities, which are explained below. 

\section{Intensity correlations}

We now explain two types of measurable correlations: Glauber intensity
correlations \cite{Glauber1963Coherent} and grouped correlations
\cite{drummondSimulatingComplexNetworks2022,delliosGBS_PNR2024},
also referred to as a grouped count probabilities (GCPs).

Intensity correlation simulations can only be performed on photon
number operator observables. Therefore, although they are valid for
determining photon number probabilities in click experiments, they
correspond directly to PNR detector outputs. Meanwhile GCPs are valid
for both threshold and PNR detectors. 

Glauber's $n$-th order intensity correlation is defined as \cite{Glauber1963Coherent}

\begin{equation}
G^{(n)}(c_{j})=\left\langle :(\hat{n}'_{j})^{c_{j}}\dots(\hat{n}'_{M})^{c_{M}}:\right\rangle ,
\end{equation}
where $n=\sum c_{j}$ is the correlation order. Multi-mode Glauber
correlations determine the probability of detecting $n$ photons at
$M$ modes.

The normal ordering requirement causes all creation operators to the
right and all annihilation operators to the left. For example, the
second-order correlation

\begin{equation}
G^{(2)}=\left\langle a_{1}^{\dagger(\text{out})}a_{2}^{\dagger(\text{out})}a_{2}^{(\text{out})}a_{1}^{(\text{out})}\right\rangle ,
\end{equation}
corresponds to detecting one photon at $M=1$, and one at $M=2$.

Upon reordering, one obtains

\begin{equation}
G^{(2)}=\left\langle a_{1}^{\dagger(\text{out})}a_{2}^{(\text{out})}\right\rangle \left\langle a_{2}^{\dagger(\text{out})}a_{1}^{(\text{out})}\right\rangle +\left\langle a_{1}^{\dagger(\text{out})}a_{1}^{(\text{out})}\right\rangle \left\langle a_{2}^{\dagger(\text{out})}a_{2}^{(\text{out})}\right\rangle .
\end{equation}
The first term describes non-local correlations, which is the interference
of photons between modes (or practically, detectors), while the second
term describes the photon intensity at each mode (or detector), which
are termed local correlations.

If the mean number of photons is small, such that a detector will
only ever observe one photon, the intensity correlation becomes a
coincidence count

\begin{equation}
P_{N}=\left\langle \prod_{j}\hat{n}'_{j}\right\rangle ,
\end{equation}
as we assume photons do not interfere at detectors, removing non-local
correlations.

\section{Grouped correlations\label{sec:Grouped-correlations}}

Grouped count probabilities (GCPs) are another observable correlation
implemented in the Quantum phase-space toolbox in xSPDE4. 

For threshold detectors, GCPs are defined as \cite{drummondSimulatingComplexNetworks2022}

\begin{equation}
\mathcal{G}_{\boldsymbol{S}}^{(n)}(\boldsymbol{m})=\left\langle \prod_{j=1}^{d}\left[\sum_{\sum c_{i}=m_{j}}\hat{\Pi}_{S_{j}}(\boldsymbol{c})\right]\right\rangle ,\label{eq:GCP}
\end{equation}
while for PNR detectors, GCPs are similiarly defined as \cite{delliosGBS_PNR2024}

\begin{equation}
\mathcal{G}_{\boldsymbol{S}}^{(n)}(\boldsymbol{m})=\left\langle \prod_{j=1}^{d}\left[\sum_{\sum c_{i}=m_{j}}\hat{P}_{S_{j}}(\boldsymbol{c})\right]\right\rangle .
\end{equation}

For both detector type, $\boldsymbol{m}=(m_{1},\dots,m_{d})$ is the
observed $d$-dimensional grouped count and $\boldsymbol{S}=(S_{1,}S_{2},\dots)$
is a vector of disjoint subsets of $\boldsymbol{M}=(M_{1},M_{2},\dots)$
modes. Each grouped count is obtained by summing over binary patterns
$m_{j}=\sum_{i}^{M}c_{i}$. Therefore, grouped counts contain $k$
bins, with each bin corresponding to the total number of clicks in
each pattern. In one-dimension, GCPs are the probability of observing
$m$ counts in any pattern with $n=M$ and $S=\{1,\dots,M\}$. This
observable is called total counts. 

Descriptions on how these GCPs are simulated in phase-space are given
in Section \ref{sec:Numerical-methods}. 

\subsection{Multi-dimensional binning}

For larger dimensions, each grouped count sums over detector outputs
for a subset of modes only such that $m_{j}=\sum_{i}^{M/d}c_{i}$.
The modes in each subset are denoted in the vector $\boldsymbol{S}$.
For example, in two-dimensions one has subsets $\boldsymbol{S}=(S_{1},S_{2})$
which contain modes 

\begin{align}
S_{1} & =\left\{ 1,\dots,\frac{M}{2}\right\} \nonumber \\
S_{2} & =\left\{ \frac{M+2}{2},\dots,M\right\} .
\end{align}
The output GCP is then a joint probability of observing $m_{1}=\sum_{i=1}^{M/2}c_{i}$
and $m_{2}=\sum_{i=M/2+1}^{M}c_{i}$ grouped counts with $k=(M/2+1)^{2}$
total bins. 

The implied segregation of output modes in the two-dimensional example
above is that $S_{1}$ will always contain the first $M/d$ modes,
$S_{2}$ the next $M/d+1\rightarrow2M/d$ modes, and so on for larger
dimensions. However, there is no practical restriction on the output
modes each subset can contain. 

Therefore, by randomly permuting each binary pattern we can change
the output modes that are contained in each subset giving 

\begin{equation}
\frac{\binom{M}{M/d}}{d}=\frac{M!}{d(M/d)!(M-M/d)!},
\end{equation}
possible ways of generating $m_{1},\dots,m_{d}$ grouped counts without
repeating a specific permutation. 

For example, when $M=4$ and $d=2$, including the standard division,
there are $3$ different orderings of outputs modes with subsets 

\begin{align}
\boldsymbol{S} & =(S_{1},S_{2})=(\{1,2\},\{3,4\}),\nonumber \\
\boldsymbol{S} & =(S_{1},S_{2})=(\{1,3\},\{2,4\}),\nonumber \\
\boldsymbol{S} & =(S_{1},S_{2})=(\{1,4\},\{2,3\}).
\end{align}
Each permutation generates a different correlation, where we assume
the commutation of GCP probabilities with subsets $(\{1,3\},\{2,4\})=(\{2,4\},\{1,3\})$. 

This permutation only changes the multidimensional GCP simulations,
as in the total count case all modes are contained in the same subset
$S=\{1,\dots,M\}$. This is also the case when simulating marginal
probabilities, which are obtained by setting $n<M$ such that $M-n$
inputs are ignored. 

\chapter{Phase-space examples\label{chap:Phase-space-examples}}

These examples show two ways to use xSPDE in quantum phase-space.
One way is to define all the noises and ordering methods in the input
files. Another way is to use the p.phase parameter, in which case
the default initialization will set up a Gaussian state network input
according to the parameters described in this chapter. Either method
works equivalently, depending on the preference of the user.

Note that choosing \emph{p.dimension=0 }corresponds to a network initialization,
followed by a quantum measurement. For this case the output axes which
correspond to different counting dimensions are treated as space dimensions.
This allows better graphics for the outputs of large networks, where
it is not practical to treat each different type of output as a separate
line on a single graph. 

\section{Saturated laser noise}

Consider the case where the laser saturates to a steady state:

\begin{equation}
\dot{a}=\left(1-\left|a\right|^{2}\right)a+bw(t)
\end{equation}

\begin{figure}
\centering{}\includegraphics[width=0.75\textwidth]{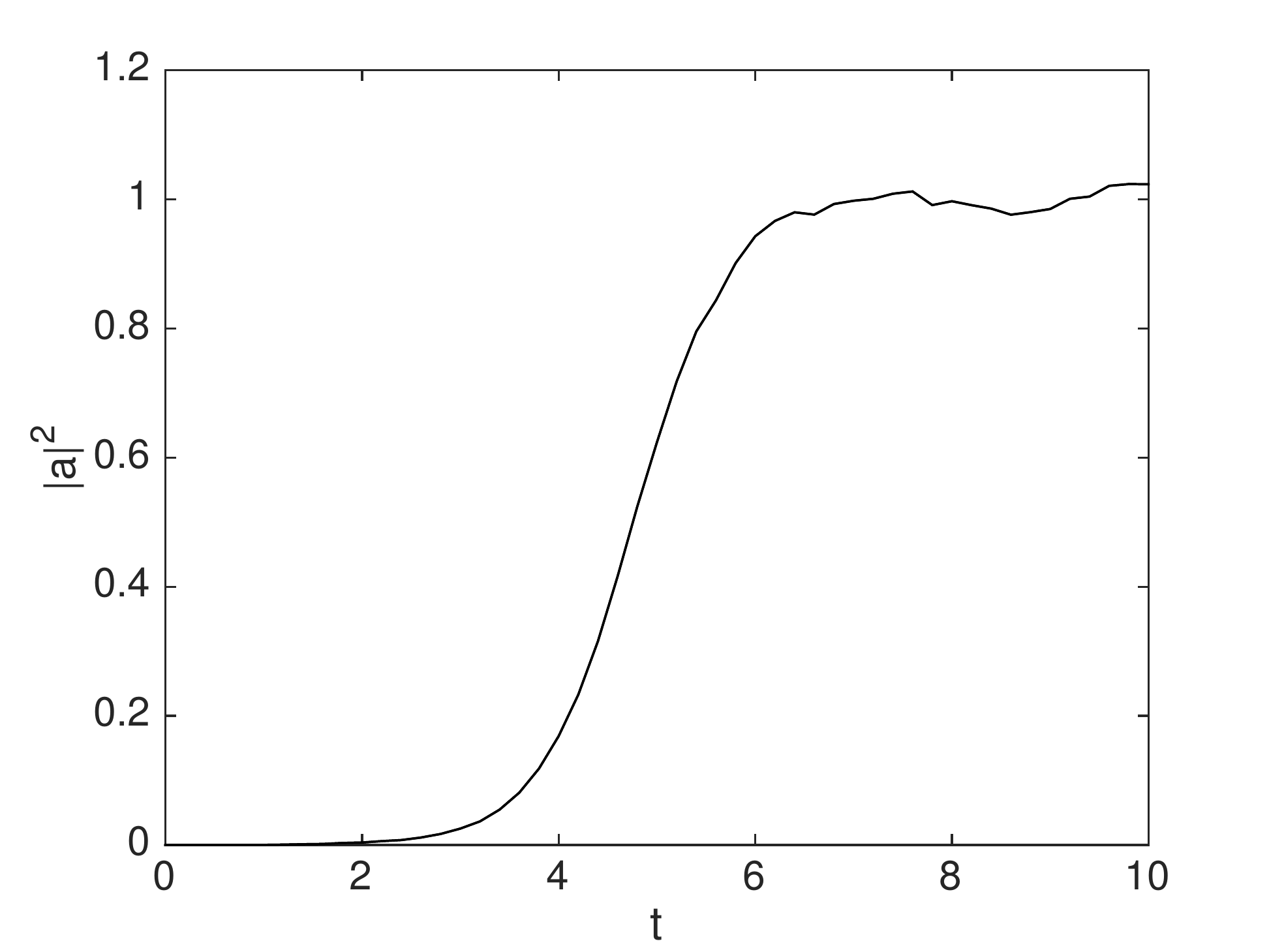}

\caption{\label{fig:The-laser-2}\emph{Simulation of the stochastic equation
describing a laser turning on.}}
\vspace{10pt}
\end{figure}

To learn how to use the function inputs, try the following: 
\begin{itemize}
\item \textbf{Solve for the saturated laser case} 
\end{itemize}
You should get the output graph in Fig (\ref{fig:The-laser-2}). 
\begin{center}
\doublebox{\begin{minipage}[t]{0.75\columnwidth}%
\texttt{clear}

\texttt{p.noises = 2;}

\texttt{p.observe = @(a,p) abs(a).\textasciicircum 2;}

\texttt{p.olabels = '\textbar a\textbar\textasciicircum 2';}

\texttt{p.deriv = @(a,w,p) (1-abs(a)\textasciicircum 2){*}a+0.01{*}(w(1)+1i{*}w(2));}

\texttt{xspde(p);}%
\end{minipage}} 
\par\end{center}

\section{Nonlinear quantum simulation}

This example involves a full nonlinear quantum phase-space simulation
using the positive-P representation described in Sec (\ref{chap:Quantum-phase-space-theory}),
in which the two variables are only conjugate in the mean. This allows
quantum superpositions of coherent states to be represented, or in
fact any state, including squeezed or entangled states in more general
cases.

A simple example is the nonlinear driven quantum subharmonic generator
- for example, an opto-mechanical, superconducting or nonlinear optical
medium in a driven cavity \cite{Drummond1981Non-equilibrium,Sun2019Schrodinger,sun2019discrete,Lugiato1988Bistability}.
This is derived from the Hamiltonian for a resonant, coupled two-mode
nonlinear interferometer, with $\hat{a}_{2}$ driven externally at
twice the frequency of $\hat{a}_{1}$: 
\begin{equation}
\hat{H}=i\hbar\left[\frac{\kappa}{2}\hat{a}_{2}\hat{a}_{1}^{\dagger2}+\mathcal{E}_{2}\hat{a}_{2}^{\dagger}-h.c.\right]
\end{equation}

After including losses in both modes in the positive P-representation,
assuming zero temperature reservoirs, and adiabatically eliminating
$\alpha_{2}$ with $\gamma_{2}\gg\gamma_{1}$, one has the following
Ito equation: 
\begin{align}
\frac{d\alpha_{1}}{dt} & =-\gamma_{1}\alpha_{1}+\alpha_{1}^{\dagger}\frac{\kappa\epsilon_{2}}{\gamma_{2}}\left[1-\frac{\kappa}{2\epsilon_{2}}\alpha_{1}^{2}\right]+\sqrt{\frac{\kappa\epsilon_{2}}{\gamma_{2}}-\frac{\kappa^{2}}{2\gamma_{2}}\alpha_{1}^{2}}w_{1}\left(t\right)\nonumber \\
\frac{d\alpha_{1}^{\dagger}}{dt} & =-\gamma_{1}\alpha_{1}^{\dagger}+\alpha_{1}\frac{\kappa\epsilon_{2}}{\gamma_{2}}\left[1-\frac{\kappa}{2\epsilon_{2}}\alpha_{1}^{\dagger2}\right]+\sqrt{\frac{\kappa\epsilon_{2}}{\gamma_{2}}-\frac{\kappa^{2}}{2\gamma_{2}}\alpha_{1}^{\dagger2}}w_{1}\left(t\right)
\end{align}
Rescaling the fields so that $\alpha_{1}=a_{1}\sqrt{n_{c}}$, $\alpha_{1}^{\dagger}=a_{2}\sqrt{n_{c}}$,
where $n_{c}=\frac{2\epsilon_{2}}{\kappa}$, then rescaling time by
letting $\tau=\frac{\kappa\epsilon_{2}}{\gamma_{2}}t$, defining $c=\frac{\gamma_{1}\gamma_{2}}{\kappa\epsilon_{2}}$,
and using Eq (\ref{eq:ItovsStratonovich}) to transform from an Ito
to a Stratonovich equation gives: 
\begin{align}
\frac{da_{1}}{d\tau} & =-(c-\frac{1}{2n_{c}})a_{1}+a_{2}\left[1-a_{1}^{2}\right]+\frac{1}{\sqrt{n_{c}}}\sqrt{1-a_{1}^{2}}w_{1}\left(\tau\right)\nonumber \\
\frac{da_{2}}{d\tau} & =-(c-\frac{1}{2n_{c}})a_{2}+a_{1}\left[1-a_{2}^{2}\right]+\frac{1}{\sqrt{n_{c}}}\sqrt{1-a_{2}^{2}}w_{2}\left(\tau\right)\,,
\end{align}
where $w_{1},w_{2}$ are delta-correlated real Gaussian noises.

There is a bistable region, which leads to a discrete time symmetry
breaking. The solution in the steady-state is 
\begin{equation}
P=\left(1-a_{1}^{2}\right)^{cn_{c}-1}\left(1-a_{2}^{2}\right)^{cn_{c}-1}e^{2n_{c}a_{1}a_{2}}
\end{equation}

The integration manifold is the region of real $a_{1}$, $a_{2}$,
such that $a_{1}^{2}\le1$ , $a_{2}^{2}\le1$. There are two physically
possible metastable values of the amplitudes. The physically observed
quantity is the amplitude and number: 
\begin{align}
\left\langle \hat{a}\right\rangle  & =\left\langle a_{1}+a_{2}\right\rangle \sqrt{\frac{n_{c}}{2}}\nonumber \\
\left\langle \hat{n}\right\rangle  & =n_{c}\left\langle a_{1}a_{2}\right\rangle .
\end{align}

Parameters that show bistable behavior on reasonable time-scales of
$T=100$ are $c=0.6$, $n_{c}=4$. To learn more, try the following: 
\begin{itemize}
\item \textbf{Simulate the nonlinear oscillator by creating a file, say,
$NonlinearQ.m$} 
\item \textbf{Can you observe quantum tunneling in the bistable regime?} 
\item \textbf{Do you see transient Schrödinger `cat states' with a negative
$n=\alpha_{1}\alpha_{2}$ value?} 
\end{itemize}
A negative value of $\alpha_{1}\alpha_{2}$ is evidence for a quantum
superposition! For experimental comparisons, one would measure correlation
functions and spectra. These calculations require long time scales,
$\mathtt{p.ranges}$, to observe tunneling, and of order $100$ time
steps per plotted time point, $\mathtt{p.steps}$, to maintain good
accuracy in the quantum simulations.

For lower damping and large nonlinearity, other methods should be
used, as the stochastic equations can become unstable in this limit.

The model is a simplified version of more recent quantum technologies
used to investigate Schrödinger cat formation in superconducting quantum
circuits \cite{leghtas2015confining}, and the CIM machine used to
solve NP-hard optimization problems with photonic circuits \cite{Marandi2014network,McMahon2019fully,Inagaki2019coherent},
although there are greater complexities in both these cases.

Similar methods can also be used to investigate quantum and chemical
non-equilibrium phase transitions \cite{drummond1981quasiprobability},
tunneling in open systems \cite{Kinsler1995Critical}, quantum entanglement
\cite{Kiesewetter2017Pulsed}, Einstein-Podolsky-Rosen paradoxes \cite{Reid1989Demonstration,Reid1989Correlations},
Bell violations \cite{RosalesZarate2014Probabilistic,Reid2014Quantum},
and many other problems treated in the literature \cite{gardiner2004quantum,Drummond2014Quantum}.

\section{Quantum linear oscillator}

This solves an SDE for a damped quantum harmonic oscillator in the
(truncated) Wigner phase-space calculus. It is initialized as a vacuum
state, corresponding to a complex Gaussian initial condition having
$\left\langle \left|a\left(0\right)\right|^{2}\right\rangle =1$.
It is subject to vacuum noise, here realized by the auxiliary field
$a_{in}$. An output field is given through the input-output relations
and is realized by the auxiliary field $a_{out}$.

\begin{align}
\frac{\partial a}{\partial t} & =-a+\sqrt{2}a_{in}.\nonumber \\
a_{in} & =\frac{1}{2}\left(w_{1}(t)+iw_{2}(t)\,\right)\nonumber \\
a_{out} & =\sqrt{2}a-a_{in}
\end{align}

The computed spectral variances are compared with exact solutions
and graphed, where: 
\begin{align}
\frac{2\pi}{T}\left\langle \left|a\left(\omega\right)\right|^{2}\right\rangle  & =\frac{1}{\left(1+\omega^{2}\right)}.\nonumber \\
\left\langle \left|a_{in}\left(\omega\right)\right|^{2}\right\rangle  & =\frac{1}{2}\nonumber \\
\left\langle \left|a_{out}\left(\omega\right)\right|^{2}\right\rangle  & =\frac{1}{2}.
\end{align}

\paragraph{Notes}
\begin{itemize}
\item Demonstrates how to include defined fields 
\item There are $4$ steps per point, to give better accuracy due to finite
steps 
\item The observe functions are all transformed, and include defined fields. 
\end{itemize}
\begin{center}
\doublebox{\begin{minipage}[t]{0.9\columnwidth}%
\texttt{function e = Quantum()}

\texttt{p.name = 'Quantum harmonic oscillator spectrum';}

\texttt{p.points = 160;}

\texttt{p.steps = 4;}

\texttt{p.ranges = 120;}

\texttt{p.fields = 1;}

\texttt{p.auxfields = 2;}

\texttt{p.noises = 2;}

\texttt{p.ensembles = {[}400,1,12{]};}

\texttt{p.initial = @(w,\textasciitilde ) (w(1,:)+1i{*}w(2,:))/(2);}

\texttt{p.a1 = @(w) (w(1,:)+1i{*}w(2,:))/2;}

\texttt{p.deriv = @(a,w,\textasciitilde ) -a(1,:)+sqrt(2){*}p.a1(w);}

\texttt{p.define = @(a,w,p) {[}p.a1(w);sqrt(2){*}a(1,:)-p.a1(w){]};}

\texttt{T = @(p) p.ranges(1);}

\texttt{p.observe\{1\} = @(a,x,p) (2.{*}pi/T(p)){*}a(1,:).{*}conj(a(1,:));}

\texttt{p.observe\{2\} = @(a,x,p) (2.{*}pi/T(p)){*}x(1,:).{*}conj(x(1,:));}

\texttt{p.observe\{3\} = @(a,x,p) (2.{*}pi/T(p)){*}x(2,:).{*}conj(x(2,:));}

\texttt{p.transforms = \{1,1,1\};}

\texttt{p.olabels\{1\} = '\textbar a(\textbackslash omega)\textbar\textasciicircum 2';}

\texttt{p.olabels\{2\} = '\textbar a\_\{in\}(\textbackslash omega)\textbar\textasciicircum 2';}

\texttt{p.olabels\{3\} = '\textbar a\_\{out\}(\textbackslash omega)\textbar\textasciicircum 2';}

\texttt{p.compare\{1\} = @(p) 1./(1+p.w.\textasciicircum 2);}

\texttt{p.compare\{2\} = @(p) 0.5;}

\texttt{p.compare\{3\} = @(p) 0.5;}

\texttt{e = xspde(p);}

\texttt{end}%
\end{minipage}} 
\par\end{center}

\begin{figure}[H]
\begin{centering}
\includegraphics[width=0.75\textwidth]{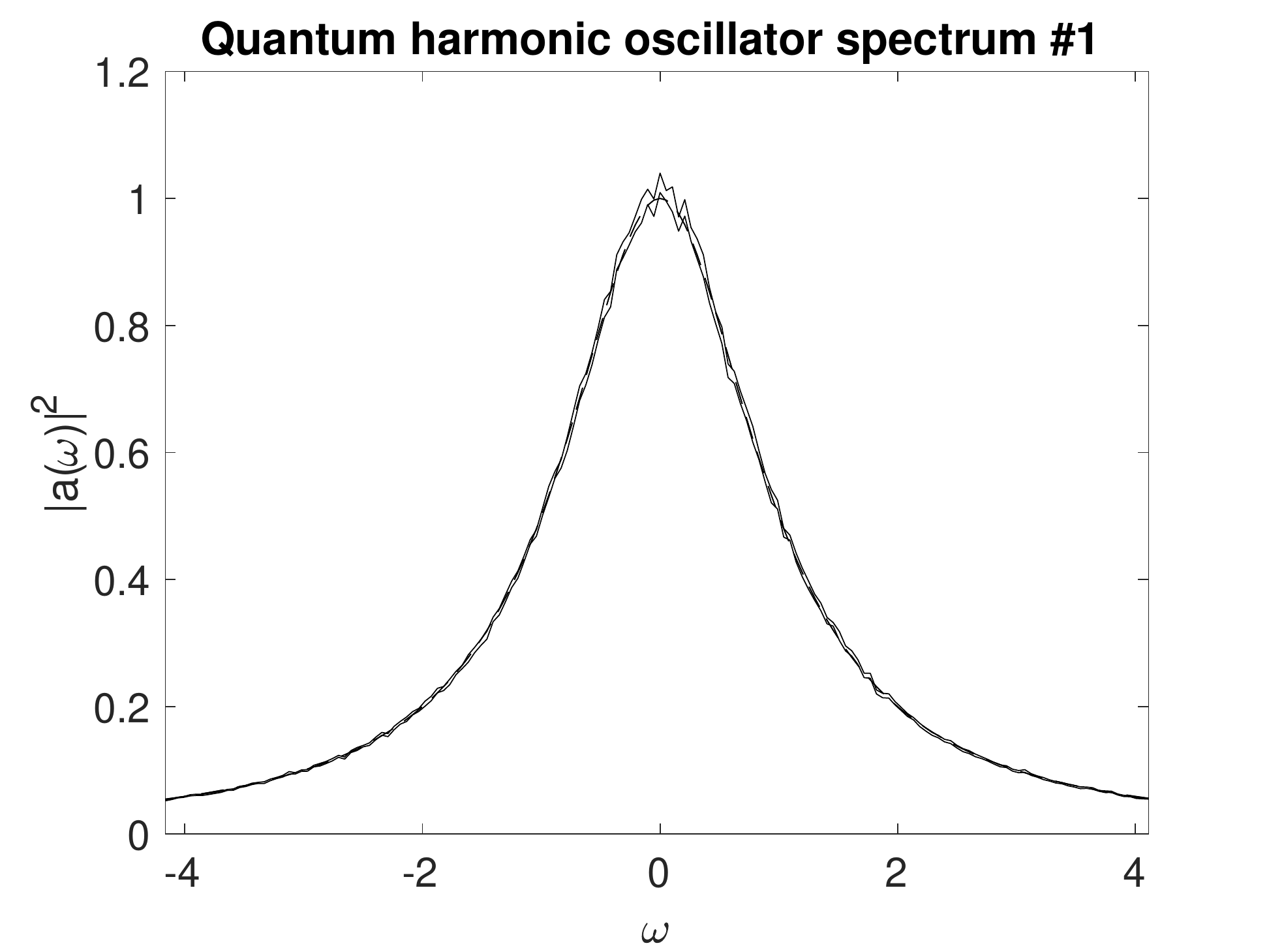}
\par\end{centering}
\centering{}\includegraphics[width=0.75\textwidth]{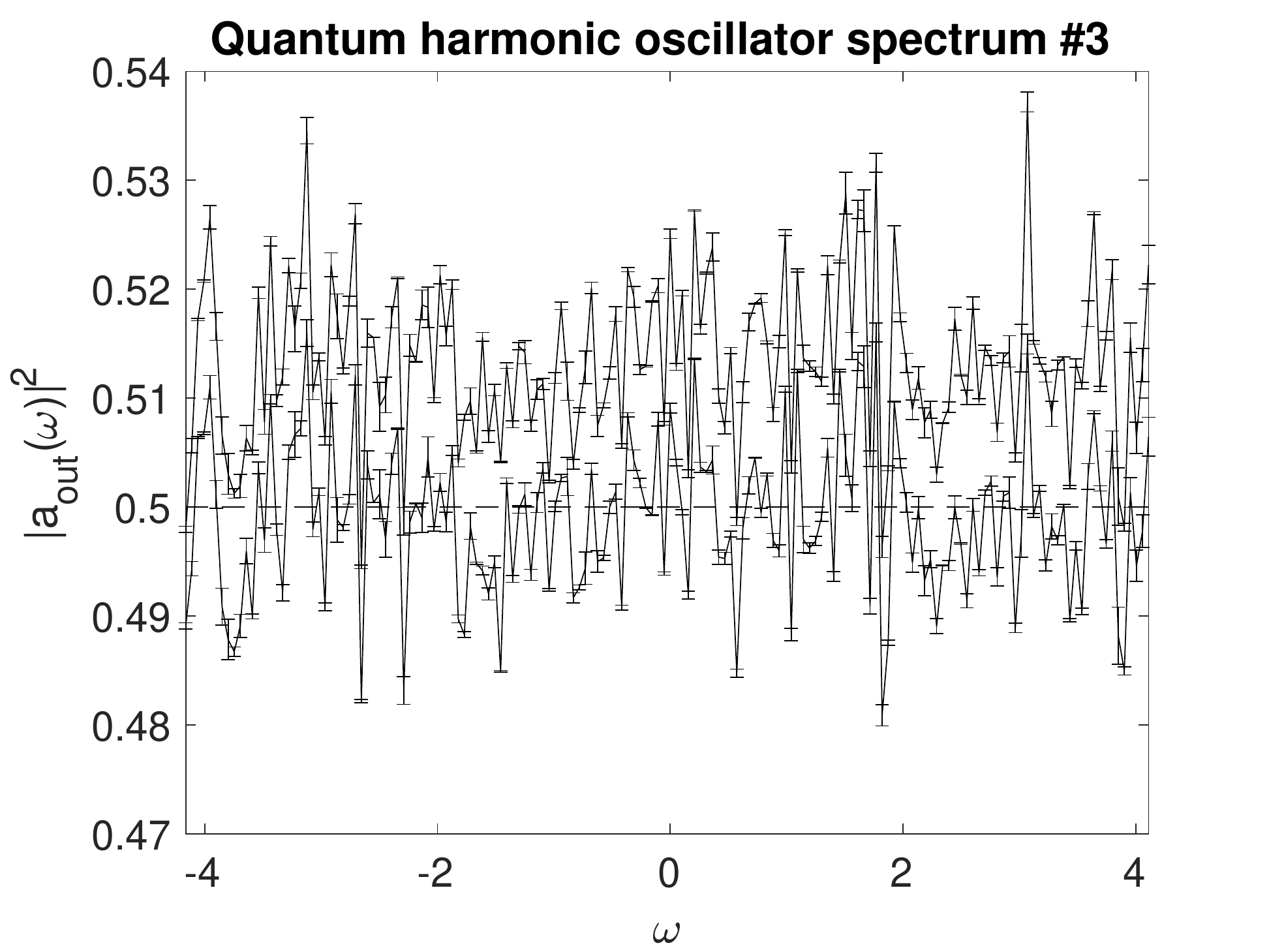}

\caption{\emph{Top figure: Spectral density of the quantum state. Bottom figure:
Spectral density of the output field. The solid lines indicate upper
and lower sampling error bounds $(\pm\sigma)$, from sampling the
stochastic equations. The dashed lines are exact results, the error-bars
indicate step-size errors. Error bars are less than the minimum size
for display in the top figure.}}
\vspace{10pt}
\end{figure}

\section{Quantum network}

This solves for a quantum network in the positive-P phase-space calculus.
It is initialized with a thermalized multi-mode squeezed state together
with a coherent component, giving Gaussian initial conditions. An
output field is obtained after an identity network transformation
for testing purposes. 

The output is the mean per-channel photon count, compared to an exact
prediction.
\begin{center}
\doublebox{\begin{minipage}[t]{0.9\columnwidth}%
function e1 = phase\_alphaGBS( )

p.dimensions = 0;

p.phase = 1; \%+P phase-space

p.modes = 50; \%matrix size m

p.name = sprintf('+P coherent, M=\%d',p.modes);

p.tr = .5{*}ones(1,p.modes); \%transmission

I = ones(1,p.modes/5); \%identity vector

p.sqz = {[}I/2,I,1.5{*}I,2{*}I,0{*}I{]}; \%nonuniform squeezing

p.alpha = {[}I/4,2{*}I,4{*}I,I/2,I{]}; \%nonuniform coherence

p.thermal = 0.5{*}ones(1,p.modes); \%thermal decoherence

p.ensembles = {[}1000,10,1{]}; \%ensmbles for averaging

p.observe = @pn;

p.compare = @nc;

p.glabels = \{\{' ','Mode j'\}\};

p.olabels = \{'\textless n\textgreater '\};

p.diffplot = \{1\};

e1 = xspde(p);

end%
\end{minipage}}
\par\end{center}

\begin{figure}
\centering{}\includegraphics[width=0.75\textwidth]{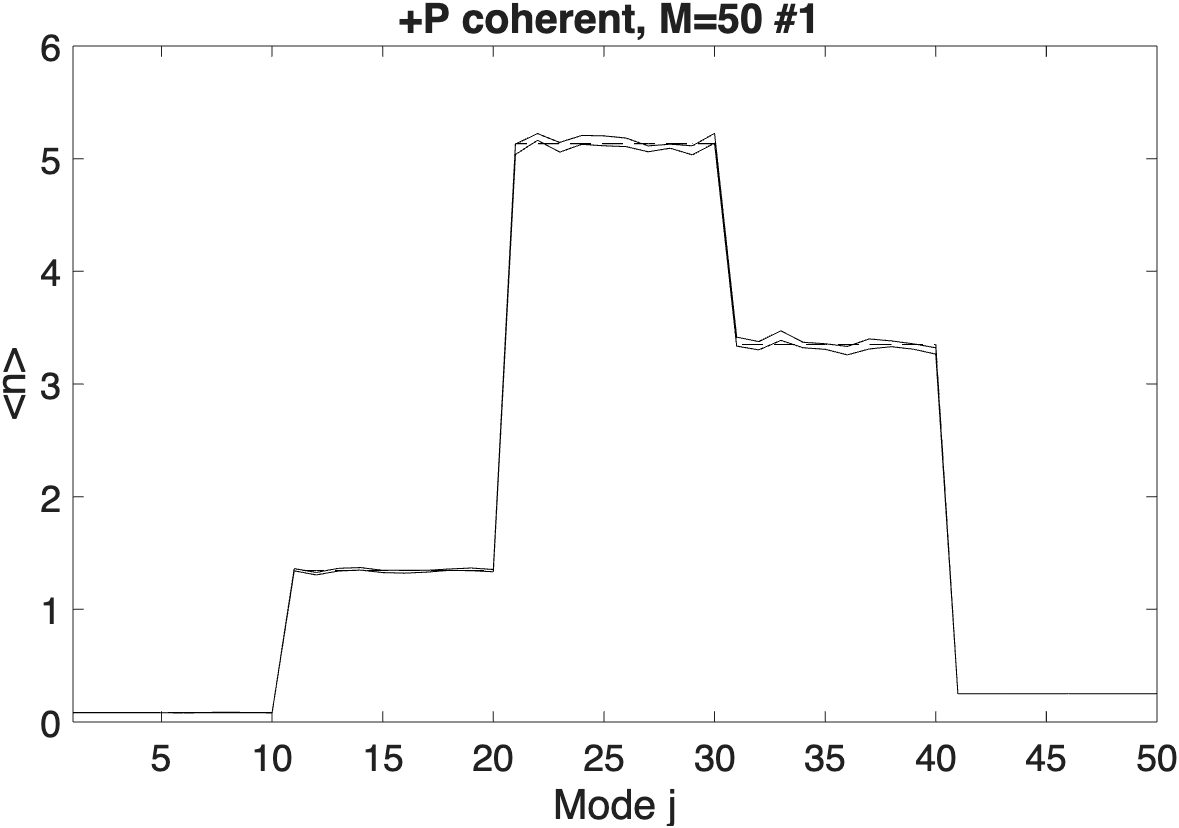}

\caption{\label{fig:The-laser-1-1}\emph{Simulation of the Gaussian boson sampling
channel count for a thermalized, squeezed input with a coherent part,
compared to an exact solution.}}
\vspace{10pt}
\end{figure}

\newpage{}

\part{Quantum Monte-Carlo}

\chapter{Monte-Carlo toolbox \label{chap:Monte-Carlo-toolbox}}

\textbf{This chapter describes how to use the xSPDE numerical toolbox
to solve network and quantum dynamical problems in phase-space. For
theoretical background, see Chapter (\ref{chap:Quantum Theory}).
For extended examples, see (\ref{chap:Quantum-examples}).}

\section{Wave-functions and density matrices}

The basic master equation treated here has the Markovian form: 
\begin{align}
\dot{\rho} & =-i\left[\hat{H},\rho\right]+\sum_{j}\gamma_{j}\left(2\hat{L}_{j}\rho\hat{L}_{j}^{\dagger}-\hat{L}_{j}^{\dagger}\hat{L}_{j}\rho-\rho\hat{L}_{j}^{\dagger}\hat{L}_{j}\right),
\end{align}
where: $j=\left[j_{1},j_{2}(,j_{3})\right]$. Here $j_{1}$ is for
the type of damping operator, $j_{2}$ is a mode index, and $j_{3}$
is an optional second mode index. 

The quantum toolbox in xSPDE has three methods for representing open
quantum systems, which allow the treatment of Hilbert spaces of increasing
dimensionality: 
\begin{enumerate}
\item Density matrices with sparse operators: p.quantum~=~2,~p.sparse~=~1. 
\item Stochastic wave-functions with sparse operators: p.quantum~=~1,~p.sparse~=~1. 
\item Stochastic wave-functions with functional operators: p.quantum~=~1,~p.sparse~=~0. 
\end{enumerate}
There is a speed/memory tradeoff here. The lowest numbered methods
are typically faster, but use more memory. In the second two cases,
one can use either a method using functions for operators, or else
a sparse matrix method, which requires the operators to be stored
in memory. The wave-function equations describe decoherence through
stochastic methods, so each of these two approaches can treat coupling
to reservoirs, up to the limits of time and memory constraints.

When using sparse methods, the multimode index $\bm{n}$ is packed
into the first single index $n$. This is automatic for density matrices,
but it is optional for stochastic wave-function calculations, which
can use either sparse or full vectors. While sparse methods are useful
for storing operators, these require memory, which must be allocated
when the matrices are generated. This can be minimized by only generated
the operators that are needed, rather than all possible ones.

Less memory is required if the operators effect on the wave-function
are calculated only when needed. This is a function call strategy,
It is currently available for stochastic wave-function calculations
only. It is slower than using sparse matrices, but it is more scalable.
Currently, this approach is not available for density matrix equations.

\section{Sparse matrix methods}

The different approaches have areas of applicability that depend on
the Hilbert space dimension. Suppose we use a stochastic method to
solve a Lindblad master equation for linear decay with initial condition
$\psi_{j}=\delta_{(N+1)j}$ and $L=a$, $\hat{H}=\hat{a}^{\dagger}\hat{a}$,
for $N=6$ , $\gamma=0.25$. The script below uses a sparse operator
method, and compares the solution with an exact result
\[
\left\langle \hat{n}\right\rangle =\left\langle \hat{n}\left(0\right)\right\rangle e^{-2\gamma t}.
\]

The alternative functional operator method inputs are explained in
Section \ref{sec:Lineardecaycomplex}.
\begin{center}
\doublebox{\begin{minipage}[t]{0.9\columnwidth}%
clear;

p.name = 'SSE linear decay, N=6';

p.ranges = 2;

p.nmax = 7;

p.sparse = 1;

p.quantum = 1;

p.a = mkbose(p);

p.ensembles = {[}100,1,10{]};

p.gamma\{1\} = @(p) 0.25;

p.compare\{1\} = @(p) 6{*}exp(-0.5{*}p.t);

p.L\{1\} = @(\textasciitilde ,p) p.a\{1\};

p.H = @(p) p.a\{1\}'{*}p.a\{1\};

p.diffplot = \{1,1\};

p.initial = @(\textasciitilde ,p) {[}0,0,0,0,0,0,1{]}';

p.expect\{1\} = @(p) p.a\{1\}'{*}p.a\{1\};

p.olabels = \{'\textbackslash langle N \textbackslash rangle'\};

xspde(p);%
\end{minipage}} 
\par\end{center}

\begin{flushleft}
With the sparse method, the function 'mkbose' is used to create the
operator matrix cell array 'p.a', before it is used. These are only
generated as needed. For large numbers of modes they can use a large
amount of storage, even though they are sparse matrices. The use of
p.L\{1\} indicates the first decay type is a linear loss, but there
could be other dissipative processes as well. 
\par\end{flushleft}

The use of p.quantum=1 shows that it is a stochastic wave-function
problem, while p.sparse=1 indicates the use of sparse matrices. Here,
$p.a\{1\}$ is the matrix version of the operator $\hat{a}_{1}$ ,
and $p.a\{1\}'$ is the matrix version of the operator $\hat{a}_{1}^{\dagger}$
. The number operator $\hat{n}_{1}$ is $p.a\{1\}'*p.a\{1\}$. To
use the master equation method, set p.quantum =2 and remove the p.ensembles
input. 

\begin{figure}[H]
\begin{centering}
\includegraphics[width=0.6\textwidth]{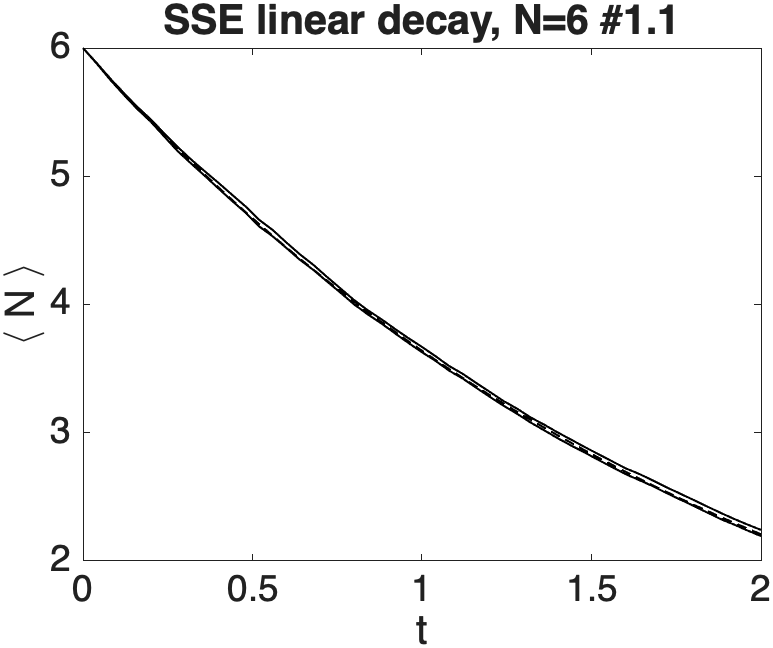}\\
 
\par\end{centering}
\caption{\emph{Example: Linear decay, including a comparison with the exact
result, using sparse methods.}}
\vspace{10pt}
\end{figure}

\subsection{Method versus memory requirements}

Algorithms for either wave-functions or density matrices, are selected
by choosing two parameters, quantum and sparse. If sparse is omitted,
the default is $sparse=0$. 
\begin{description}
\item [{p.quantum~=~1,~p.sparse~=~0}] - this is used to treat a full
wave-function, $psi(\bm{n},e)$, Here, $\bm{n}=n_{1},\ldots n_{m}$
is a wave-function index index, while $e$ is an ensemble index for
random ensembles, if used. Operators are treated as functions, so
there are no operator matrices stored. This minimizes the overall
memory requirement. 
\item [{p.quantum~=~1,~p.sparse~=~1}] - this is used to treat a packed
wave-function, $psi(n,e)$, Here, $n$ is a wave-function index index,
which is a packed version of the vector index $\bm{n}$, while $e$
is an ensemble index for random ensembles. Operators are treated as
sparse matrices, so these must be stored. This increases the overall
memory usage but is somewhat faster. 
\item [{p.quantum~=~2,~p.sparse~=~1}] - this is used to treat a packed
density matrix, $rho(n,\ell)$, Here, $n$ and $\ell$ are density
matrix indices, which are packed version of the vector index $\bm{n}$.
Operators are treated as sparse matrices. Due to the storage requirements
of a density matrix, this uses the most memory, and is the fastest.
There is no vector ensemble here. 
\end{description}

\section{Input parameters}

Input parameters are stored in a structure which is input to the xSPDE
program. This is a superset of the parameters already defined. In
the definitions below, the structure name is omitted. but we normally
use $p$ in the examples. For example, to specify a quantum wave-function
method, one would use $quantum=1$, as explained already. The input
parameters can be chosen not just in terms of the problem itself,
but also to suit the computational hardware that is available.

Not that while the $quantum$ toolbox and $phase$ toolbox share common
parameters listed below, but they are distinct toolboxes, and one
must choose to use either one or the other by setting $quantum>0$
or $phase>0$.

\subsection{Common parameters}
\begin{description}
\item [{modes}] gives the number of modes, hence $modes=3$ defines a 3
mode quantum system,. This can be given implicitly through nmax. 
\item [{ensembles(1)}] gives a vector of trajectories, $e=1,\ldots$ensembles(1).
This is fast, but increases memory use. It is not used for density
matrices. 
\item [{ensembles(2)}] gives the number of series repeats for stochastic
ensembles. It is always available, but slower. 
\item [{ensembles(3)}] gives the number of parallel repeats for stochastic
ensembles. It is useful for multicore processors with fast memory. 
\item [{jump~}] selects either a stochastic differential equation (jump
= 0), the default, or a stochastic jump equation (jump = 1). 
\item [{noises}] noise dimensions, set automatically for the built-in quantum
methods. 
\item [{points}] The number of integration points in time for data outputs.
The default setting is $51$. 
\item [{steps}] The integration steps used per time-step, used to reduce
time-step errors. The default is $1$. 
\item [{ranges}] The total integration range in time. The default setting
is$10$. 
\item [{initial}] The initial state is given by a function\texttt{ $initial$}.
This returns a column vector of size $fields\times1$ or $fields\times ensembles(1)$,
for wave-functions, or else of size $fields\times fields$ for density
matrix calculations. The default is the state with the first level
occupied. 
\item [{inrandoms}] are initial random number dimensions. They specify
the first argument of the function\texttt{ $initial(v,p)$} as a real
Gaussian noise vector $v$ with unit variance and length inrandoms.
These are used for an initially decoherent, randomized wave-function. 
\end{description}
The internal variable fields is used to specify the dimension of the
integrated variables, and is automatically set.

\subsection{Quantum parameters }
\begin{description}
\item [{quantum}] is the type of problem: $quantum=1$ for a wave-function,
$quantum=2$ for a density matrix. 
\item [{sparse}] indicates sparseness: if $sparse=1$, sparse matrices
are used to store operators. The default is $sparse=0$. 
\item [{nmax}] is the Hilbert dimension per mode. If this is a vector,
the dimension can be varied. 
\item [{mk...}] is a make function to generate sparse operators where required,
eg,\textbf{ mkbose}. 
\item [{operator:}] If sparse = 0, an operator is a function with inputs
of the mode index (or indices), and the wave-function $psi$. Operators
acting on multiple modes may have two or more indices. The function
$O_{k}$ returns a wave-function $\hat{O}_{k}\left|\psi\right\rangle $. 
\item [{sparse\_operator:}] When sparse = 1, operators are sparse matrices.
This is faster, but uses more memory. 
\item [{Hamiltonian:}] the function $H(psi,p)$ returns a wave-function
$\hat{H}\left|\psi\right\rangle $, if sparse = 0. Otherwise, if sparse
= 1, it is an operator function $H(p)$ that returns a sparse matrix. 
\end{description}
The following defaults are used to simplify input: 
\begin{itemize}
\item If modes is not specified, it is equal to the length of nmax. 
\item If modes and nmax are not specified the default is a single qubit:
modes=1 , nmax=2. 
\item If the nmax vector is shorter than modes, the last value of nmax is
repeated as necessary. 
\end{itemize}

\section{Dissipative parameters}

In order to explain the terminology for dissipative input, the following
list is useful. There are some differences that depend on whether
one uses sparse matrices or functional operators. In the list below,
$n$ is the channel index for the dissipative operators. One channel
index can generate any number of mode operators of the same type.
\begin{description}
\item [{L\{n\}:}] this is a cell array of dissipative functions. The first
argument is the mode index, $k$, or a vector of two indices, $[k_{1},k_{2}]$,
and the last argument p is the parameter structure. 
\begin{description}
\item [{@(k,p)}] is used for sparse operators, and returns a sparse matrix
$\hat{L}_{k}$. 
\item [{@(k,psi)}] is used for operator functions, returning $\hat{L}_{k}\left|\psi\right\rangle $. 
\end{description}
\item [{Conjugate~operators:}] for functions, a conjugate is returned
if the index or indices is negative.
\item [{gamma\{n\}(p):}] This is a cell array of functions for every type
of damping process. Cell array indices are used to distinguish different
dissipative processes. These return a vector or matrix of damping
rates for each type of Lindblad operator. The resulting vector indices
give the mode (or modes, if the operator acts on more than one mode).
\item [{alpha\{n\}(k):}] This is a cell array of noise amplitude vectors
or matrices for each type of damping process with real noises. If
\textbf{alpha} is zero, which is the default, a complex noise is used. 
\item [{measure:}] This gives the number of measured channel operators.
\end{description}
Note that: 
\begin{itemize}
\item Operators may have one or two mode indices. 
\item Functional operators are slower than sparse operators for small Hilbert
spaces. 
\item Only the required index combinations are accessed by the Lindblad
functions, reducing storage. 
\end{itemize}

\section{Functional operators}

These are linear functions that act on the quantum wave-function.
New ones can readily be added. They reduce memory requirements, which
is an advantage for large Hilbert spaces, where storing even sparse
operators can require large quantities of memory.

The xSPDE code includes internal functions for bosonic and spin operators.
The predefined operators also return auxiliary quantities used in
dissipative equations if required, as they have variable input and
output lists.

\subsection{Bosonic operators}
\begin{center}
\begin{tabular}{|c|c|c|}
\hline 
Label & Inputs & Output(s)\tabularnewline
\hline 
\hline 
a & $(m,psi)$ & $\hat{a}_{m}\left|\psi\right\rangle $\tabularnewline
\hline 
a2 & $\left(m,psi\right)$ & $\hat{a}_{m}^{2}\left|\psi\right\rangle $\tabularnewline
\hline 
n & $\left([m_{1}\left(,m_{2}\right)],psi\right)$ & $\hat{a}_{m_{1}}^{\dagger}\hat{a}_{m_{2}}\left|\psi\right\rangle $\tabularnewline
\hline 
\end{tabular}
\par\end{center}

Operators have scalar or vector indices. For a complete description,
see (\ref{subsec:Bosonic-operator-table}).

\subsection{Qubit and Pauli spin operators}

The following set of operators are used for spin chain evolution. 
\begin{center}
\begin{tabular}{|c|c|c|}
\hline 
Label & Inputs & Output(s)\tabularnewline
\hline 
\hline 
sx & $(m,psi)$ & $\hat{\sigma}_{m}^{x}\left|\psi\right\rangle $\tabularnewline
\hline 
sy & $(m,psi)$ & $\hat{\sigma}_{m}^{y}\left|\psi\right\rangle $\tabularnewline
\hline 
sz & $(m,psi)$ & $\hat{\sigma}_{m}^{z}\left|\psi\right\rangle $\tabularnewline
\hline 
sx2 & $(\left[m_{1},m_{2}\right],psi)$ & $\hat{\sigma}_{m_{1}}^{x}\hat{\sigma}_{m_{2}}^{x}\left|\psi\right\rangle $\tabularnewline
\hline 
sy2 & $(\left[m_{1},m_{2}\right],psi)$ & $\hat{\sigma}_{m_{1}}^{y}\hat{\sigma}_{m_{2}}^{y}\left|\psi\right\rangle $\tabularnewline
\hline 
sz2 & $(\left[m_{1},m_{2}\right],psi)$ & $\hat{\sigma}_{m_{1}}^{z}\hat{\sigma}_{m_{2}}^{z}\left|\psi\right\rangle $\tabularnewline
\hline 
\end{tabular}
\par\end{center}

\subsection{Qubit gate operators}

The following operators can be used to implement quantum logic gates,
in addition to the standard Pauli operators. These assume qubit or
two-state qubit logic in each mode. 
\begin{center}
\begin{tabular}{|c|c|c|}
\hline 
Label & Inputs & Output(s)\tabularnewline
\hline 
\hline 
cx & $(\left[m_{1},m_{2}\right],psi)$ & Controlled Not\tabularnewline
\hline 
ha & $(m,psi)$ & Hadamard\tabularnewline
\hline 
p8 & $(m,psi)$ & $\pi/8$\tabularnewline
\hline 
ph & $(m,psi)$ & Phase\tabularnewline
\hline 
\end{tabular}
\par\end{center}

\section{Sparse operators }

The xSPDE code includes internal functions to generate operators.
These are either sparse or full. Sparse operators are generated if
needed requiring a mk function call to create the required index combinations,
before they are used.

\subsection{Sparse bosonic operators: mkbose}

These are a cell array of annihilation operators, generated using
mkbose. 
\begin{center}
\begin{tabular}{|c|c|c|}
\hline 
Label & Indices & Meaning\tabularnewline
\hline 
\hline 
a & $\{m\}$ & $\hat{a}_{m}$\tabularnewline
\hline 
a' & $\{m\}$ & $\hat{a}_{m}^{\dagger}$\tabularnewline
\hline 
\end{tabular}
\par\end{center}
\begin{description}
\item [{p.a~=~mkbose((list,)~p)}] Returns a cell array of annihilation
operators defined either at all modes, if there is no list, or at
the listed mode locations. Here list is a vector of integers, p is
the parameter structure. 
\end{description}

\section{Observe, expect, output and compare}

There are four types of possible outputs. The observe and compare
functions are computed during the time-evolution, so that the entire
wave-function doesn't need to be stored in time, reducing the storage
needs. Additional functional transformations for either can be used
as well, called output functions. Finally, a compare function allows
comparison plots. 
\begin{description}
\item [{observe}] is a cell array of any stochastic function. xSPDE expects
a (named or anonymous) function that takes two parameters, namely
the wave-function $psi$ or density matrix $\rho$, and the input
structure $p$. The function return a real or complex matrix of dimension
$(\ell,ensembles(1))$, where $\ell$ indexes a vector observable.
xSPDE then averages over the second index, to calculate the observable.
This allows an average of any type. 
\item [{expect}] is a cell array of operators defining a quantum expectation
value. For full matrices, expect is a (named or anonymous) function
that takes two inputs, the wave-function $psi$ and the structure
$p$. For sparse matrices, the expect function returns a matrix. xSPDE
internally averages over both the quantum and stochastic degrees of
freedom to calculate the observable.\\
 To plot the mean number in mode $m=1$, using:\\
 a) the number operator $\hat{n}=\hat{a}^{\dagger}\hat{a}$ with sparse
operators:\\
 ~~%
\doublebox{\begin{minipage}[t]{0.75\columnwidth}%
\texttt{p.expect\{1\} = @(p) p.a\{1\}'{*}p.a\{1\};}%
\end{minipage}}\\
 ~\\
 b) the number operator $\hat{n}=\hat{a}^{\dagger}\hat{a}$, with
function calls: \\
\doublebox{\begin{minipage}[t]{0.75\columnwidth}%
\texttt{p.expect\{1\} = @(psi,p) n(1,psi);}%
\end{minipage}} 
\item [{output}] All observe and expect results are stored. Transformations
of both can be introduced. These may include multiple averages and/or
different times. These are called the output functions. Default outputs
pass through observe and expect results with no change. Defined outputs,
like p.output\{1\}, replace the defaults, or add new outputs. Graphed
data uses the outputs, which include sampling errors if ensembles
are used, and step-size errors if checks is turned on. 
\item [{compare}] Comparison functions can be used to obtain comparison
graphs and differences. 
\end{description}
The output numbering that is used is the \textbf{same} for all four
types of function. This can lead to overwriting, with the precedence
that output\textgreater expect\textgreater observe. To prevent overwriting,
use different cell-indices. Compare functions are plotted independently,
as an extra line on an existing graph, so they don't overwrite, and
can be compared with any of output, expect, observe.

\section{SSE derivative}

The SSE derivative terms are calculated from $SSE(a,w,p)$ for solving
Eq \ref{eq:nonlinear SSE}. The equation can be solved by any $method$
for a Stratonovich SDE. Projective normalization of wavefunction equations
is automatic for the standard methods of xSPDE. Mathematical details
are given in Section (\ref{sec:Stratonovich-SSE}).

\section{Solving with the MCWF method \label{sec:Solving-with-MCWF-1}}

At each time step in the numerical simulation with the MCWF method,
the jump probability $\Delta P$ is first calculated. This is carried
out by computing the jump probability per unit time of each jump operator
$L_{m}$ in the master equation, which is given by 
\begin{equation}
\Delta P_{m}=2\gamma_{m}\langle\psi(t)|L_{m}^{\dagger}L_{m}|\psi(t)\rangle\,.\label{eq:individual_jump_prob-1}
\end{equation}

The calculated jump rate $\Delta P_{m}$ is then compared with a uniform,
randomly generated number $r_{m}$ between zero and $1/\Delta t$.
If $\Delta P_{m}>r_{m}$, the state then undergoes the given jump.
As $\Delta t\rightarrow0$, a jump in any given step become increasingly
rare.

After this, the state vector evolves according to the non-hermitian
Hamiltonian $H_{eff}$ in Eq. (\ref{eq:non-Hermitian_Hamiltonian})
as follows: 
\begin{equation}
\frac{d}{dt}|\psi(t)\rangle=-iH_{eff}|\psi(t)\rangle\label{eq:ODE-1}
\end{equation}
This differential equation is solved by a midpoint or Runge-Kutta
algorithm, or others available. These steps are repeated till the
final time step, and they constitute a single trajectory. Many trajectories
are taken to compute the expectation values for the observables of
interest.

For error-checking, fine step results are checked against a coarse
step with a noise given by $r_{c}=\min\left(r_{1},r_{2}\right)$,
so that the coarse jump occurs if a jump takes place in either fine
step. This allows errors due to step-size to be accurately estimated
by comparing the fine and coarse step-size results, just as with continuous
noise.

The MCWF algorithm described above is carried out simply by setting
$p.jump=1$. No further inputs from the user are required. All other
parameters are input in exactly the same way as in the SSE numerical
simulation. Projective normalization of wavefunction equations is
automatic for the standard methods of xSPDE. Mathematical details
are given in Section (\ref{sec:MCWF Theory}).

\pagebreak{}

\chapter{Quantum Monte-Carlo theory\label{chap:Quantum Theory}}

\textbf{This chapter describes the quantum theory used in xSPDE, to
explain the background to the open system methods available.}

\section{Master equations}

The master equation \cite{Carmichael2002Statistical,Drummond2014Quantum}
is a standard tool for solving Markovian open quantum system dynamics:
\begin{align}
\dot{\rho} & =\mathcal{L}_{J}\rho\label{eq:Lindblad_Master_Equation}\\
 & =-i\left[\hat{H},\rho\right]+\sum_{j=1}^{J}\gamma_{j}\left(2\hat{L}_{j}\rho\hat{L}_{j}^{\dagger}-\hat{L}_{j}^{\dagger}\hat{L}_{j}\rho-\rho\hat{L}_{j}^{\dagger}\hat{L}_{j}\right)
\end{align}
Here, $\mathcal{L}_{J}$ is the total super-operator for $J$ terms,$H$
is the reversible system Hamiltonian, $L_{j}$ are $J$ operators
that couple the system to the dissipative reservoir, and $\gamma_{j}$
is the decay rate. The dissipative operators can be further classified
by type $n$ and mode index $k$, including vector indices if needed.

Provide that $\rho$ is normalized, the expectation values of observables
$\hat{O}$ are given by: 
\begin{equation}
\left\langle \hat{O}\right\rangle =Tr\left(\rho\hat{O}\right)
\end{equation}
The definitions used here mean that for the case of linear damping
with $\hat{L}=\hat{a}$, the rate $\gamma$ is the amplitude decay
rate. This abstract notation does not include the effects of finite
temperatures, which are explained below.

The rate is written explicitly here. This is useful for xSPDE inputs,
which use standard dimensionless operators. Alternative approaches
include combining the rate with the operator \cite{gisin1992quantum},
implying $\gamma=1.$ Others use a rate constant $\kappa=2\gamma$,
i.e., the number decay rate. Some combine this with the operator,
defining $\hat{c}_{j}=\sqrt{2\gamma_{j}}\hat{L}_{j}$, giving a fourth
operator convention.

Including finite temperature reservoir occupation numbers $\bar{n}_{j}$
explicitly, the quantum master equation with damping rates $\Gamma_{j}$
is 
\begin{align}
{\frac{\partial\hat{\rho}}{\partial t}} & =-i\left[\hat{H},\hat{\rho}\right]+\sum_{j}\Gamma_{j}\left(\bar{n}_{j}+1\right)(2\hat{A}_{j}\hat{\rho}\hat{A}_{j}^{\dagger}-\hat{A}_{j}^{\dagger}\hat{A}_{j}\hat{\rho}-\hat{\rho}\hat{A}_{j}^{\dagger}\hat{A}_{j})\,\,\nonumber \\
 & +\sum_{j}\Gamma_{j}\bar{n}_{j}(2\hat{A}_{j}^{\dagger}\hat{\rho}\hat{A}_{j}-\hat{A}_{j}\hat{A}_{j}^{\dagger}\hat{\rho}-\hat{\rho}\hat{A}_{j}\hat{A}_{j}^{\dagger})\,\,,\label{eq:master-equation-1}
\end{align}
where $\Gamma_{j}$ is a zero temperature damping rate for reservoir
couplings to the operator $\hat{A}_{j}$, $\bar{n}_{j}$ is the finite
temperature reservoir occupation. In the numerical toolbox, the finite-temperature
reservoirs are included explicitly as a separate Lindblad term.

\subsection{Bosonic Hilbert spaces}

The operators available in xSPDE include multimode bosonic operators
$\hat{a}_{j}$. For these, typical damping operators are:

\begin{table}
\begin{centering}
\begin{tabular}{|c|c|c|}
\hline 
Damping operator ($\hat{A}_{j}$)  & $\Gamma_{j}$  & Physical interpretation\tabularnewline
\hline 
$\hat{a}_{j}$  & $\gamma_{j}$  & Linear amplitude loss (units $s^{-1}$)\tabularnewline
\hline 
$\hat{a}_{j}^{\dagger}$  & $g_{j}$  & Linear amplitude gain (units $s^{-1}$)\tabularnewline
\hline 
$\hat{a}_{j}^{\dagger}\hat{a}_{j}$  & $\gamma_{j}^{p}$  & Phase decay rate gain (units $s^{-1}$)\tabularnewline
\hline 
$\hat{a}_{j}^{2}$  & $\kappa_{j}/2$  & nonlinear amplitude loss (units $s^{-1}$).\tabularnewline
\hline 
\end{tabular}
\par\end{centering}
\caption{Typical types of quantum decoherence term. Note that when the damping
operator is a number operator, it conserves particle number but causes
phase coherence decay.}
\end{table}

\subsection{Qubit Hilbert spaces}

Additionally, xSPDE includes finite Hilbert spaces, focusing on $SU(2)$
or qubit cases. Operators available are: $\hat{\sigma}_{m}^{x}$,$\hat{\sigma}_{m}^{y},$$\hat{\sigma}_{m}^{z},$
together with quantum logic gates: hadamard, controlled-not, phase
and $\pi/8$. These can be combined to give a complete set of logic
gates, allowing a simulation of quantum computers.

Using the master-equation toolbox, one can also include decoherence
and loss. As usual, this is limited by exponential growth in the Hilbert
space dimension, but the stochastic Schrödinger equation and related
methods improve memory efficiency compared to the full density matrix.

\section{Stochastic Schrödinger equation (SSE)}

While master equations can be solved directly, they grow in size quadratically
with Hilbert space dimension. An alternative to reduce memory size
is to use quantum Monte Carlo methods, which although still restricted
to small mode numbers, can reduce memory requirements substantially.

A stochastic Schrödinger equation (SSE) is an equation with noise
terms used to solve a dissipative master equation by random sampling,
and was originally developed for applications in quantum foundations
\cite{pearle1979toward,Ghirardi1986unified,gisin1992quantum}. It
has the advantage over a master equation that for large numbers of
modes it uses less storage. This requires $e^{\lambda M}$complex
numbers for the storage of an $M-$mode quantum system, compared to
$\sim e^{2\lambda M}$ for the master equation, where $\lambda=\log_{e}(N)$
for an $N$-level local Hilbert space. This is exponentially large,
but the memory required is less than with a density matrix equation,
doubling the number of modes that are accessible. The drawback is
that many parallel trajectories must be averaged in order to give
a low final sampling error.

An SSE can also be regarded in certain cases as providing a direct
simulation of the measurement process, which means that the information
recorded in a simulation is similar to the information measured in
a quantum experiment. This requires a suitable choice of the method,
sometimes called an ``unraveling'' and the stochastic integration
algorithm. Different unravellings mean different measurements and
different convergence rates.

There are many versions of the SSE, which use different normalization,
different random noises or different types of stochastic calculus.
Noises can be real or complex, and either continuous or with discrete
jumps. These are different ``unravellings'', and correspond physically
to distinct measurement devices and outcomes. They also have different
sampling errors.

Compared to phase-space expansions, the SSE method has the problem
that storage requirements are exponential in the system size, although
it may have lower sampling errors for high nonlinearities. This limits
mode numbers to $10-50$, depending on the size of the Hilbert space
per mode and the computational resources. The approach is most useful
for small mode numbers, especially for large nonlinearities.

\subsection{Normalized Ito SSE}

A widely used form of the continuous noise, normalized SSE is as follows
\cite{gisin1992quantum}, in the Ito calculus:

\begin{align}
d\left|\Psi\right\rangle  & =\left\{ -i\hat{H}+\sum_{j}\gamma_{j}\left(2\left\langle \hat{L}_{j}^{\dagger}\right\rangle _{\Psi}\hat{L}_{j}-\left\langle \hat{L}_{j}^{\dagger}\right\rangle _{\Psi}\left\langle \hat{L}_{j}\right\rangle _{\Psi}-\hat{L}_{j}^{\dagger}\hat{L}_{j}\right)\right\} \left|\Psi\right\rangle dt\nonumber \\
 & +\sum_{j}\sqrt{2\gamma_{j}}\Delta\hat{L}_{j}\left|\Psi\right\rangle d\xi_{j}.\label{eq:nonlinear SSE-1-2}
\end{align}
where $\Delta\hat{L}_{j}=\hat{L}_{j}-\left\langle \hat{L}_{j}\right\rangle _{\Psi}$,
and 
\begin{align}
\left\langle d\xi_{j}^{*}d\xi_{k}\right\rangle  & =\delta_{jk}dt.\nonumber \\
\left\langle d\xi_{j}d\xi_{k}\right\rangle  & =0.\label{eq:Complex noise}
\end{align}

A more general form of the normalized SSE \cite{rigo1997continuous}
in the Ito calculus is:

\begin{align}
d\left|\Psi\right\rangle  & =\left\{ -i\hat{H}+\sum_{j}\gamma_{j}\left(2\left\langle \hat{L}_{j}^{\dagger}\right\rangle _{\Psi}\hat{L}_{j}-\left\langle \hat{L}_{j}^{\dagger}\right\rangle _{\Psi}\left\langle \hat{L}_{j}\right\rangle _{\Psi}-\hat{L}_{j}^{\dagger}\hat{L}_{j}\right)\right\} \left|\Psi\right\rangle dt\nonumber \\
 & +\sum_{jn}\sqrt{2\gamma_{j}}\Delta\hat{L}_{j}\left|\Psi\right\rangle \alpha_{jn}d\zeta_{jn}.\label{eq:nonlinear SSE-1}
\end{align}
where we require that $\sum_{n}\left|\alpha_{jn}\right|^{2}=1$, and
\begin{align}
\left\langle d\zeta_{jn}\left(t\right)d\zeta_{km}\left(t'\right)\right\rangle  & =\delta_{jk}\delta_{nm}dt.
\end{align}

One can also include a unitary transformation, which we set to a delta
function for simplicity. When there is one noise per decay channel,
then $\alpha_{j}\left(t\right)=e^{i\phi_{j}}$, where $\phi_{j}$
is arbitrary. If there are two noises, then one can choose $\alpha_{j1}\left(t\right)=1/\sqrt{2}$,
and $\alpha_{j2}\left(t\right)=i/\sqrt{2}$, giving a complex noise
SDE, with $\xi_{j}=\left(\zeta_{1}+i\zeta_{2}\right)/\sqrt{2}$, as
above.

This can be written for $\left|\Psi\right\rangle \rightarrow\Psi_{\nu}$,
and $d\zeta_{jn}\rightarrow dw_{\sigma}$, as: 
\[
d\psi_{\mu}=A_{\mu}dt+B_{\mu\sigma}dw_{\sigma}
\]

\section{Stratonovich SSE\label{sec:Stratonovich-SSE}}

To obtain standard calculus for an SSE, one must transform to the
Stratonovich equation. This form of stochastic calculus allows integration
algorithms that often give lower errors \cite{Drummond1991Computer}.
There are also higher order methods for Ito equations, but these have
greatly increased complexity. Here we derive the Stratonovich correction
\cite{stratonovich1960theory,Gardiner2009Stochastic}, which is obtained
with $\hat{L}_{j}\left|\Psi\right\rangle \rightarrow L_{j\mu\nu}\Psi_{\nu}$
so that $\left\langle \hat{L}_{j}^{\dagger}\right\rangle _{\Psi}=\sum_{\sigma\rho}\Psi_{\sigma}^{*}L_{j\rho\sigma}^{*}\Psi_{\rho}$.

For complex noise as in Eq (\ref{eq:Complex noise}), the Stratonovich
drift is given by:

\begin{equation}
A_{\mu}=A_{\mu}^{(I)}-\frac{1}{2}\sum_{j\nu}B_{\nu j}^{*}\partial_{\nu}^{*}B_{\mu j}.
\end{equation}
On taking matrix elements in an orthogonal basis, and defining: 
\begin{equation}
\Delta L_{j\mu\nu}=L_{j\mu\nu}-\delta_{\mu\nu}\sum_{\sigma\rho}\Psi_{\rho}^{*}L_{j\rho\sigma}\Psi_{\sigma},
\end{equation}
one has:

\begin{align}
B_{\mu j} & =\sqrt{2\gamma_{j}}\Delta L_{j\mu\beta}\Psi_{\beta}\\
B_{\mu j}^{*} & =\sqrt{2\gamma_{j}}\Psi_{\beta}^{*}\Delta L_{j\mu\beta}^{*}.\nonumber 
\end{align}
On differentiating one therefore obtains: 
\begin{align}
\partial_{\nu}^{*}B_{\mu j} & =\sqrt{2\gamma_{j}}\sum_{\beta}\left[-\Psi_{\beta}\partial_{\nu}^{*}\left(\delta_{\mu\beta}\sum_{\sigma\rho}\Psi_{\rho}^{*}L_{j\rho\sigma}\Psi_{\sigma}\right)\right]\nonumber \\
 & =\sqrt{2\gamma_{j}}\left[-\Psi_{\mu}\sum_{\sigma}L_{j\nu\sigma}\Psi_{\sigma}\right].
\end{align}
The Stratonovich correction is given by: 
\begin{align}
-\frac{1}{2}\sum_{\nu,j}B_{\nu j}^{*}\partial_{\nu}^{*}B_{\mu j} & =-\sum_{j}\gamma_{j}\sum_{\nu,\alpha}\Psi_{\alpha}^{*}\Delta L_{j\nu\alpha}^{*}\left[-\Psi_{\mu}\sum_{\sigma}L_{j\nu\sigma}\Psi_{\sigma}\right]\nonumber \\
 & =\sum_{j}\gamma_{j}\left[\sum_{\nu\sigma\alpha}\Psi_{\alpha}^{*}\left(L_{j\nu\alpha}^{*}-\delta_{\nu\alpha}\left\langle \hat{L}_{j}^{\dagger}\right\rangle _{\Psi}\right)L_{j\nu\sigma}\Psi_{\sigma}\right]\Psi_{\mu}\nonumber \\
 & =\sum_{j}\gamma_{j}\left[\left\langle \hat{L}_{j}^{\dagger}\hat{L}_{j}\right\rangle _{\Psi}-\left\langle \hat{L}_{j}^{\dagger}\right\rangle _{\Psi}\left\langle \hat{L}_{j}\right\rangle _{\Psi}\right]\Psi_{\mu}.
\end{align}

In summary, the complex Ito SSE can be transformed to a nonlinear
Stratonovich stochastic differential equation which locally preserves
normalization for zero step-size \cite{diosi1989models}. This is
called the quantum state diffusion model:

\begin{align}
\frac{d\left|\Psi\right\rangle }{dt} & =\left(-i\hat{H}+\sum_{j}\gamma_{j}\left(2\Delta\hat{L}_{j}\left\langle \hat{L}_{j}^{\dagger}\right\rangle _{\Psi}-\Delta\left[\hat{L}_{j}^{\dagger}\hat{L}_{j}\right]\right)\right)\left|\Psi\right\rangle \nonumber \\
 & +\sum_{j}\sqrt{2\gamma_{j}}\xi_{j}\Delta\hat{L}_{j}\left|\Psi\right\rangle \label{eq:nonlinear SSE}
\end{align}
where: 
\begin{equation}
\left\langle \xi_{k}\left(t\right)\xi_{j}^{*}\left(t'\right)\right\rangle =\delta_{kj}\delta\left(t-t'\right).
\end{equation}

Here, $\Delta\hat{L}_{j}\equiv\hat{L}_{j}-\left\langle \hat{L}_{j}\right\rangle _{\Psi}$
and the equation uses Stratonovich calculus. This preserves the norm
of the wave-function. Suppose the Stratonovich form has a dissipative
term $\Delta\hat{\mathcal{L}}_{s}$, where

\begin{equation}
\frac{d\left|\Psi\right\rangle }{dt}=\left\{ \Delta\hat{\mathcal{L}}_{s}-i\hat{H}\right\} \left|\Psi\right\rangle 
\end{equation}
Since it is a Stratonovich equation, one can use ordinary calculus
rules. Only dissipative terms can change the norm, and:

\begin{align}
\frac{d}{dt}\left\langle \Psi\right.\left|\Psi\right\rangle  & =\left\langle \Delta\left(\hat{\mathcal{L}}_{s}+\hat{\mathcal{L}}_{s}^{\dagger}\right)\right\rangle _{\psi}\nonumber \\
 & =\left\langle \hat{\mathcal{L}}_{s}+\hat{\mathcal{L}}_{s}^{\dagger}\right\rangle _{\psi}-\left\langle \hat{\mathcal{L}}_{s}+\hat{\mathcal{L}}_{s}^{\dagger}\right\rangle _{\psi}=0
\end{align}

For Ito equations, the trajectories have a norm error that grows with
time. While there are projective methods to prevent this, the result
has higher step-size errors \cite{Joseph2023midpoint}.To obtain observables,
one must use the ``double'' expectation indicating a quantum and
stochastic mean, where the wave-functions $\left|\Psi\right\rangle $
are normalized, and have all the same weight: 
\begin{equation}
\left\langle O\right\rangle \equiv\left\langle \left\langle \Psi\right|O\left|\Psi\right\rangle \right\rangle _{\xi}.
\end{equation}

Integrating this equation is best carried out with a projection at
each time-step to prevent the normalization changing, as derived elsewhere
\cite{Joseph2023midpoint}. This is implemented automatically within
xSPDE.

\subsection{Real noise Stratonovich equation}

For the real noise case, the correction term is: 
\begin{equation}
A_{\mu}=A_{\mu}^{(I)}-\frac{1}{2}\sum_{\sigma\nu}\left(B_{\nu\sigma}\partial_{\nu}+B_{\nu\sigma}^{*}\partial_{\nu}^{*}\right)B_{\mu\sigma},
\end{equation}
where $B_{\mu j}=\sqrt{2\gamma_{j}}\Delta L_{j\mu\beta}\Psi_{\beta}$.
Taking $n=1$ and $j=\sigma,$ the conjugate correction is given above
and is independent of $\alpha_{j}$. The first term is obtained from
differentiation of the noise matrix:

\begin{equation}
B_{\mu j}=\alpha_{j}\sqrt{2\gamma_{j}}\Delta L_{j\mu\beta}\Psi_{\beta}.
\end{equation}
hence one obtains that: 
\begin{align}
\partial_{\nu}B_{\mu j} & =\sqrt{2\gamma_{j}}\alpha_{j}\left[\Delta L_{j\mu\nu}-\sum_{\beta}\delta_{\mu\beta}\left[\Psi_{\rho}^{*}L_{j\rho\nu}\right]\Psi_{\beta}\right].
\end{align}
The additional correction is as follows: 
\begin{align}
-\frac{1}{2}\sum_{\nu j}B_{\nu j}\partial_{\nu}B_{\mu j} & =-\sum_{\nu j}\gamma_{j}\alpha_{j}^{2}\left[\Delta L_{j\mu\nu}-\Psi_{\rho}^{*}L_{j\rho\nu}\Psi_{\mu}\right]\Delta L_{j\nu\sigma}\Psi_{\sigma}\nonumber \\
 & =-\sum_{\nu j}\gamma_{j}\alpha_{j}^{2}\left(\Delta L_{j\mu\nu}\Delta L_{j\nu\sigma}\Psi_{\sigma}-\left[\Psi_{\rho}^{*}L_{j\rho\nu}\Delta L_{j\nu\sigma}\Psi_{\sigma}\right]\Psi_{\mu}\right).
\end{align}

Written in operator/wave-function terminology, the real correction
$\left|\delta A^{r}\right\rangle $is 
\begin{align}
\left|\delta A^{r}\right\rangle  & =-\sum_{\nu j}\gamma_{j}\alpha_{j}^{2}\left(\left[\Delta\hat{L}_{j}\Delta\hat{L}_{j}\right]-\left\langle \hat{L}_{j}\Delta\hat{L}_{j}\right\rangle _{\Psi}\right)\left|\Psi\right\rangle \nonumber \\
 & =-\sum_{\nu j}\gamma_{j}\alpha_{j}^{2}\left(\left[\hat{L}_{j}^{2}-2\hat{L}_{j}\left\langle \hat{L}_{j}\right\rangle _{\Psi}+\left\langle \hat{L}_{j}\right\rangle _{\Psi}^{2}\right]-\left\langle \hat{L}_{j}^{2}\right\rangle _{\Psi}+\left\langle \hat{L}_{j}\right\rangle _{\Psi}^{2}\right)\left|\Psi\right\rangle \nonumber \\
 & =\sum_{\nu j}\gamma_{j}\left(2\Delta\hat{L}_{j}\left\langle \alpha_{j}^{2}\hat{L}_{j}\right\rangle _{\Psi}-\Delta\left[\alpha_{j}^{2}\hat{L}_{j}^{2}\right]\right)\left|\Psi\right\rangle .
\end{align}
Combining both terms, and defining $\hat{X}_{j}=\hat{L}_{j}^{\dagger}+\alpha_{j}^{2}\hat{L}_{j}$,
one obtains a result known in the literature \cite{Gambetta2002PRA}
for the case $\alpha=1$; 
\begin{align}
\frac{d\left|\Psi\right\rangle }{dt} & =\left\{ -i\hat{H}+\sum_{j}\gamma_{j}\left(2\left\langle \hat{X}_{j}\right\rangle _{\Psi}\Delta\hat{L}_{j}-\Delta\left[\hat{X}_{j}\hat{L}_{j}\right]\right)\right\} \left|\Psi\right\rangle \nonumber \\
 & +\sum_{j}\sqrt{2\gamma_{j}}\alpha_{j}\zeta_{j}\left(t\right)\Delta\hat{L}_{j}\left|\Psi\right\rangle 
\end{align}

As with the complex noise case, this is explicitly norm-preserving
since the dissipative terms have zero quantum mean values for every
noise realization. This generic result reduces to the complex case
if one sets$\hat{X}_{j}=\hat{L}_{j}^{\dagger}$ and $\alpha_{j}\zeta_{j}\rightarrow\xi_{j}$.

\section{Monte Carlo wave-function method\label{sec:MCWF Theory}}

The Monte Carlo or quantum jump method is another approach to solve
a master equation. The master equation treated here has the form given
in Eq (\ref{eq:Lindblad_Master_Equation}).

\subsection{Integer noise}

A jump SSE is obtained by using an Ito stochastic differential equation
with real noise, in the form:

\begin{align}
d\left|\phi\right\rangle  & =\left\{ -i\hat{H}-\sum_{j}\gamma_{j}\left[\hat{L}_{j}^{\dagger}\hat{L}_{j}-\left\langle \hat{L}_{j}^{\dagger}\hat{L}_{j}\right\rangle \right]\right\} \left|\phi\right\rangle dt\nonumber \\
 & +\sum_{j}\left(\hat{L}_{j}/\sqrt{\left\langle \hat{L}_{j}^{\dagger}\hat{L}_{j}\right\rangle }-1\right)\left|\phi\right\rangle dN_{j},
\end{align}
where the real integer noise $dN=[0,1]$ has correlations of: 
\begin{align}
\left\langle dN_{j}\left(t\right)\right\rangle  & =2\gamma_{j}\left\langle \hat{L}_{j}^{\dagger}\hat{L}_{j}\right\rangle dt.
\end{align}
In any interval $dt$, $dN$ is unity with probability $p=2\gamma_{j}\left\langle \hat{L}_{j}^{\dagger}\hat{L}_{j}\right\rangle dt$
, and zero otherwise.

To generate integer noise, one first obtains a random real number
$r$ where $0<r<1/dt$. From this, one can choose $dN=1$ if $r<2\gamma_{j}\left\langle \hat{L}_{j}^{\dagger}\hat{L}_{j}\right\rangle $.

\subsection{MCWF method}

In the MCWF method, state vectors evolve according to an effective
Hamiltonian, 
\begin{equation}
H_{e}=H-i\sum_{m}\gamma_{m}L_{m}^{\dagger}L_{m}\,,\label{eq:non-Hermitian_Hamiltonian}
\end{equation}
punctuated by quantum jumps 
\begin{equation}
|\psi\rangle\rightarrow L_{m}|\psi\rangle\,,\label{eq:jump}
\end{equation}
where $L_{m}$ is one of the possible operators in the master equation.
At each step in time, the system will either evolve according to the
non-Hermitian Hamiltonian Eq. (\ref{eq:non-Hermitian_Hamiltonian})
or undergo a jump operation, depending on the jump probability $\Delta P$.

A sequence of quantum jumps or photo-counts giving total counts $\bm{c}=c_{1},\ldots c_{M}$
is obtained. For times when there is no jump, 
\begin{equation}
\dot{\psi}_{c}=-i\hat{H}_{e}\psi_{c}.\label{eq:Lindblad_Master_Equation-1}
\end{equation}

Jumps occur at random times given by choosing random numbers $r_{m}$
such that $0<r_{m}<1/\Delta t$, where $r_{m}$ determines the jump
probability for the $m-th$ process.

The jump changes counts so that $c_{j}\rightarrow c_{j}+1$ . Afterwards,
one resets $\rho_{c}$ after an infinitesimal time $\epsilon$ so
that 
\begin{equation}
|\psi\left(t_{c}+\epsilon\right)\rangle=\frac{|\psi_{j}\rangle}{\sqrt{\left\langle \psi_{j}\right|\left.\psi_{j}\right\rangle }}.
\end{equation}
The MCWF algorithm is presented in the numerical section.

\subsection{Monte-Carlo master equations}

Monte Carlo master equation theory \cite{mollow1975pure,Zoller1987,molmer1993monte,gardiner1992wavefunction,Dalibard1992Wave,Carmichael1993Quantum}
implements the Copenhagen model for measurement as a sequential wave-function
projections. It treats dissipative evolution whose average behavior
is given by a master equation, where if there is one decay channel
$\hat{L}_{j}$ per mode $M$: 
\begin{equation}
\frac{d\rho}{dt}=-i\left[\hat{H},\rho\right]+\sum_{j=1}^{M}\gamma_{j}\left(2\hat{L}_{j}\rho\hat{L}_{j}^{\dagger}-\left[\hat{L}_{j}^{\dagger}\hat{L}_{j},\rho\right]_{+}\right).\label{eq:Lindblad_Master_Equation-2}
\end{equation}

An equivalent sequence of quantum jumps or photo-counts giving total
counts $\bm{c}=c_{1},\ldots c_{M}$ is described by a conditional
density matrix equation, which is a nonlinear Ito discrete SDE in
the form: 
\begin{align}
d\rho & =-i\left[\hat{H}_{e}\rho-\rho\hat{H}_{e}^{\dagger}\right]dt+\sum_{j}\left(\frac{\hat{L}_{j}\rho\hat{L}_{j}^{\dagger}}{\left\langle \hat{L}_{j}^{\dagger}\hat{L}_{j}\right\rangle }-\rho\right)dN_{j}\nonumber \\
 & \,\,,
\end{align}
where the effective Hamiltonian $\hat{H}_{e}$ is non-hermitian: 
\[
\hat{H}_{e}=H-i\sum_{j=1}^{M}\gamma_{j}\left[\hat{L}_{j}^{\dagger}\hat{L}_{j}-\left\langle \hat{L}_{j}^{\dagger}\hat{L}_{j}\right\rangle \right]
\]
and the real integer noise $dN_{j}=[0,1]$ has correlations of: 
\begin{align}
\left\langle dN_{j}\left(t\right)\right\rangle  & =2\gamma_{j}\left\langle \hat{L}_{j}^{\dagger}\hat{L}_{j}\right\rangle dt.
\end{align}
In any interval $dt$, $dN$ is unity with probability $p=2\gamma_{j}\left\langle \hat{L}_{j}^{\dagger}\hat{L}_{j}\right\rangle dt$
, and zero otherwise. This is not a standard Lindblad form due to
the nonlinear terms, but it conserves probabilities, and has an average
behavior that corresponds to the full master equation.

Jumps occur at times given as above by choosing random numbers $r_{j}$
in $[0,1/dt]$ such that $dN=1$ if 
\begin{equation}
r_{j}<2\gamma_{j}\left\langle \hat{L}_{j}^{\dagger}\hat{L}_{j}\right\rangle .
\end{equation}

The Ito density matrix equation can be integrated by integrating the
deterministic part over a small time interval, then deciding whether
or not to jump. A jump changes detector counts so $c_{j}\rightarrow c_{j}+1$.
One must correspondingly project $\rho$ after an infinitesimal time
to give the new density matrix, given by the discontinuous jump $d\rho_{N}$,
where 
\begin{equation}
d\rho_{N}=\sum_{j}\left(\frac{\hat{L}_{j}\rho\hat{L}_{j}^{\dagger}}{\left\langle \hat{L}_{j}^{\dagger}\hat{L}_{j}\right\rangle }-\rho\right)dN_{j}.\label{eq:projected density}
\end{equation}

\section{Examples}

We now consider examples of linear and nonlinear dissipative operators. 

\subsection{Linear master equation}

The standard case of linear losses in quantum optics, gives:

\begin{align}
L & =a
\end{align}
The corresponding master equation is;

\begin{equation}
\dot{\rho}=2a\rho a^{\dagger}-a^{\dagger}a\rho-\rho a^{\dagger}a.
\end{equation}
This leads to a linear decay in amplitude and occupation number:

\begin{align}
\left\langle \dot{n}\right\rangle  & =Tr\left[\left(2a\rho a^{\dagger}a^{\dagger}a-n^{2}\rho-\rho n^{2}\right)\right]\nonumber \\
 & =2Tr\left[\rho a^{\dagger2}a^{2}-n^{2}\rho\right]\nonumber \\
 & =2Tr\left[\rho\left(n^{2}-n\right)-n^{2}\rho\right]\nonumber \\
 & =-2\left\langle n\right\rangle .
\end{align}

The effect of the operator on the state expansion is 
\begin{align}
\left|\phi\right\rangle  & =\sum_{n}\phi_{n}\left|n\right\rangle \nonumber \\
a\left|n\right\rangle  & =\sqrt{n}\left|n-1\right\rangle \nonumber \\
a^{\dagger}\left|n\right\rangle  & =\sqrt{n+1}\left|n+1\right\rangle \nonumber \\
a^{\dagger}a\left|n\right\rangle  & =n\left|n\right\rangle .
\end{align}

Therefore for a number state expansion of the density operator: 
\begin{align}
\left\langle a\right\rangle  & =\sum_{nm}\phi_{j}^{*}\left\langle m\right|\phi_{n}a\left|n\right\rangle \nonumber \\
 & =\sum_{nm}\phi_{j}^{*}\left\langle m\right|\phi_{n}\sqrt{n}\left|n-1\right\rangle \nonumber \\
 & =\sum_{n=0}^{\infty}\phi_{n}^{*}\phi_{n+1}\sqrt{n+1}.
\end{align}
also, for the conjugate, 
\begin{equation}
\left\langle a^{\dagger}\right\rangle =\sum_{n=1}^{\infty}\phi_{n}^{*}\phi_{n-1}\sqrt{n}.
\end{equation}

\subsection{Linear stochastic equation}

This is the simplest case:

\begin{align}
\frac{d\left|\phi\right\rangle }{dt} & =\sum_{n}\left(-a^{\dagger}a+a\xi\right)\phi_{n}\left|n\right\rangle \nonumber \\
 & =\sum_{n}\left(-n\left|n\right\rangle +\sqrt{n}\left|n-1\right\rangle \xi\right)\phi_{n}.
\end{align}

Taking matrix elements, one obtains: 
\begin{align*}
\frac{d\phi_{j}}{dt} & =\sqrt{m+1}\phi_{m+1}\xi-m\phi_{j}
\end{align*}

\subsection{Normalized, nonlinear stochastic equation}

\begin{align*}
\frac{d\left|\phi\right\rangle }{dt} & =\sum_{n}\left(\left[\left\langle a^{\dagger}a\right\rangle -a^{\dagger}a\right]+\left[a-\left\langle a\right\rangle \right]\left[\xi+2\left\langle a^{\dagger}\right\rangle \right]\right)\phi_{n}\left|n\right\rangle \\
 & =\sum_{n}\left(\left[\left\langle n\right\rangle -n\right]\left|n\right\rangle +\left[\xi+2\left\langle a^{\dagger}\right\rangle \right]\left[\sqrt{n}\left|n-1\right\rangle -\left\langle a\right\rangle \left|n\right\rangle \right]\right)\phi_{n}
\end{align*}
Taking matrix elements, 
\begin{align}
\frac{d\phi_{j}}{dt} & =\left[\xi+2\left\langle a^{\dagger}\right\rangle \right]\left[\sqrt{m+1}\phi_{m+1}-\left\langle a\right\rangle \phi_{j}\right]+\left[\left\langle n\right\rangle -m\right]\phi_{j}.
\end{align}

\subsection{Nonlinear absorber}

The next case of nonlinear two-photon losses in quantum optics, gives:

\begin{align*}
L & =a^{2}
\end{align*}
where we recall that: 
\begin{align*}
\left|\phi\right\rangle  & =\sum_{n}\phi_{n}\left|n\right\rangle \\
a^{2}\left|n\right\rangle  & =\sqrt{n\left(n-1\right)}\left|n-2\right\rangle \\
a^{\dagger2}\left|n\right\rangle  & =\sqrt{\left(n+1\right)\left(n+2\right)}\left|n+2\right\rangle \\
a^{\dagger2}a^{2}\left|n\right\rangle  & =n\left(n-1\right)\left|n\right\rangle 
\end{align*}

\subsection{Master equation}

The quantum expectations in a pure state are given by: 
\begin{align*}
\left\langle a^{2}\right\rangle  & =\sum_{nm}\phi_{j}^{*}\left\langle m\right|\phi_{n}a^{2}\left|n\right\rangle \\
 & =\sum_{nm}\phi_{j}^{*}\left\langle m\right|\phi_{n}\sqrt{n(n-1)}\left|n-2\right\rangle \\
 & =\sum_{n=0}^{\infty}\phi_{n}^{*}\phi_{n+2}\sqrt{(n+2)(n+1)}
\end{align*}

The diagonal master equation in a number state basis is therefore:
\[
\dot{\rho}_{n}=-2n(n-1)\rho_{n}+2(n+1)(n+2)\rho_{n+2}.
\]
This equation is generated automatically using the master equation
quantum method in the numerical toolbox.

\newpage{}

\chapter{Quantum examples\label{chap:Quantum-examples}}

\section{Linear decay, complex SSE\label{sec:Lineardecaycomplex}}

This solves a standard Lindblad master equation for linear decay with
initial condition $\psi_{j}=\delta_{Nj}$ and $L=\sqrt{\gamma}a$,
$\hat{H}=\hat{a}^{\dagger}\hat{a}$; for $N=6$ , $\gamma=0.25$:
\[
\dot{\rho}=-i[\hat{H},\rho]+2L\rho L^{\dagger}-L^{\dagger}L\rho-\rho L^{\dagger}L
\]

\paragraph{Function operator method}
\begin{flushleft}
\doublebox{\begin{minipage}[t]{0.9\columnwidth}%
\texttt{function {[}e{]} = SSElin}

\texttt{\%Uses an SSE to solve for a linear decay}

\texttt{p.name = 'SSE linear decay, N=6 initial photons';}

\texttt{p.N = 6;}

\texttt{p.nmax = p.N+1;}

\texttt{p.ranges = 2;}

\texttt{p.quantum = 1;}

\texttt{p.ensembles = {[}100, 10{]};}

\texttt{p.gamma\{1\} = @(p) 0.25;}

\texttt{p.H = @(psi,p) n(1,psi);}

\texttt{p.compare = @(p) p.N{*}exp(-0.5{*}p.t);}

\texttt{p.L\{1\} = @a;}

\texttt{p.diffplot = 1;}

\texttt{p.initial = @(w,psi) {[}0,0,0,0,0,0,1{]}';}

\texttt{p.expect = @(psi,p) n(1,psi);}

\texttt{p.olabels = \{'\textbackslash langle N \textbackslash rangle'\};}

\texttt{e = xspde(p);}

\texttt{end}%
\end{minipage}} 
\par\end{flushleft}

\begin{flushleft}
With the function method, the function 'mkbose' is not required. Instead,
the effect of the operators is obtained through a function call to
the handle '@a' . For large numbers of modes this method uses a reduced
amount of memory as there is no stored matrix involved in this case. 
\par\end{flushleft}

One cannot simply write p.H=@n here, because the Hamiltonian is a
function of the wave-function $psi$ and the parameters p, while the
number operator is a function of the mode number and the wave-function.
For Lindblad operators, these arguments are inserted automatically.

The flag p.diffplot =1 is used by the graphics code to create a plot
of the difference between the comparison solution and the simulation.

Note that one can determine the relative size of the sampling errors
and step-size errors from the difference plot, although these are
also printed out.

\begin{figure}[H]
\begin{centering}
\includegraphics[width=0.6\textwidth]{SSElin1}\\
 \includegraphics[width=0.6\textwidth]{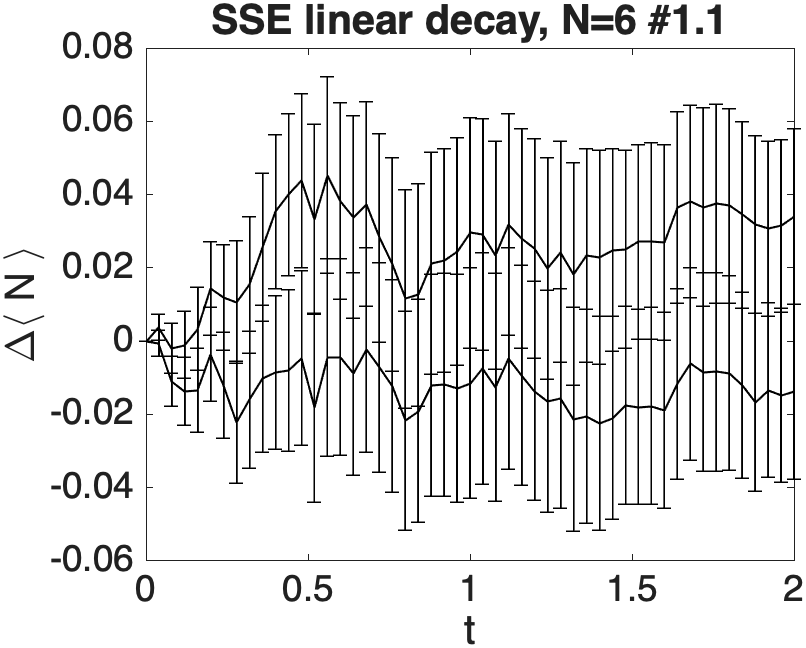}
\par\end{centering}
\caption{\emph{Example: Linear decay, including a comparison with the exact
result, below. The graph shows the sampling error-bars as two parallel
lines. The discretization error-bars are less than the minimum, and
are not shown.}}
\vspace{10pt}
\end{figure}

\pagebreak{}

\section{Time-dependent decay, real SSE}

This solves a Lindblad master equation for linear time-dependent decay
with two modes. and real noises, corresponding to homodyne detection.
The initial condition is $\psi_{j}=\delta_{Nj}$ and $L_{1}=a$, for
$\bm{N}=[3,6]$ .

The decay rates are: 
\begin{align*}
\gamma_{1} & =[0.5,1]*t
\end{align*}
As above, the sparse and functional methods give identical results,
but the sparse method is faster. For comparison purposes, the following
results are expected: 
\[
\bm{N}=[3e^{-t^{2}/2},6e^{-t^{2}}]
\]

\paragraph{Sparse operator method}
\begin{center}
\doublebox{\begin{minipage}[t]{0.9\columnwidth}%
\texttt{function e = SSElin2spr}

\texttt{\%Uses a sparse SSE to solve for a linear two-mode decay}

\texttt{p.name = 'SSE sparse real, N = 3,6';}

\texttt{p.N = 3;}

\texttt{p.Om = 1;}

\texttt{p.noises = 4;}

\texttt{p.ranges = 2;}

\texttt{p.nmax = {[}p.N+1,2{*}p.N+1{]};}

\texttt{p.a = mkbose(p);}

\texttt{p.quantum = 1;}

\texttt{p.sparse = 1;}

\texttt{p.ensembles = {[}100,1,10{]};}

\texttt{p.theta\{1\} = {[}1,1{]};}

\texttt{p.gamma\{1\} = @(p) {[}0.5,1{]}{*}p.t;}

\texttt{p.L\{1\} = @(m,p) p.a\{m\};}

\texttt{p.H = @(p) p.Om{*}(p.a\{1\}'{*}p.a\{1\}+p.a\{2\}'{*}p.a\{2\});}

\texttt{p.initial = @(\textasciitilde ,p) kron({[}0,0,0,1{]},{[}0,0,0,0,0,0,1{]})';}

\texttt{p.expect\{1\} = @(p) p.a\{1\}'{*}p.a\{1\};}

\texttt{p.expect\{2\} = @(p) p.a\{2\}'{*}p.a\{2\};}

\texttt{p.compare\{1\} = @(p) p.N{*}exp(-p.t.\textasciicircum 2/2);}

\texttt{p.compare\{2\} = @(p) 2{*}p.N{*}exp(-p.t.\textasciicircum 2);}

\texttt{p.diffplot = \{1,1\};}

\texttt{p.olabels = \{' \textless n\_1 \textgreater{} ','\textless{}
n\_2 \textgreater{} '\};}

\texttt{e = xspde(p);}

\texttt{end}%
\end{minipage}} 
\par\end{center}

The use of p.quantum=1 shows that it is a stochastic wave-function
problem, while p.sparse=1 indicates sparse matrices, and p.theta =
\{{[}1,1{]}\} specifies that all channels have real noises. To use
the master equation method, set p.quantum =2 and remove the p.ensembles
and p.theta inputs.

\paragraph{Function operator method}
\begin{flushleft}
\doublebox{\begin{minipage}[t]{0.9\columnwidth}%
\texttt{function e = SSElin2r}

\texttt{\%Uses a non-sparse SSE to solve for a linear two-mode decay}

\texttt{p.name = 'SSE, N = 3,6';}

\texttt{p.N = 3;}

\texttt{p.Om = 1;}

\texttt{p.ranges = 2;}

\texttt{p.nmax = {[}p.N+1,2{*}p.N+1{]};}

\texttt{p.quantum = 1;}

\texttt{p.ensembles = {[}100, 10{]};}

\texttt{p.gamma\{1\} = @(p){[}0.5,1{]}{*}p.t;}

\texttt{p.theta\{1\} = {[}1,1{]};}

\texttt{p.L\{1\} = @a;}

\texttt{p.H = @(psi,p) p.Om{*}(n(1,psi)+n(2,psi));}

\texttt{p.initial = @(\textasciitilde ,p) kron({[}0,0,0,1{]}',{[}0,0,0,0,0,0,1{]});}

\texttt{p.expect\{1\} = @(psi,p) n(1,psi);}

\texttt{p.expect\{2\} = @(psi,p) n(2,psi);}

\texttt{p.compare\{1\} = @(p) p.N{*}exp(-p.t.\textasciicircum 2/2);}

\texttt{p.compare\{2\} = @(p) 2{*}p.N{*}exp(-p.t.\textasciicircum 2);}

\texttt{p.olabels = \{'n\_1','n\_2'\};}

\texttt{e = xspde(p);}

\texttt{end}%
\end{minipage}} 
\par\end{flushleft}

\begin{flushleft}
With the function method, the function 'mkbose' is not required. Instead,
the effect of the operators is obtained through a function call to
the handles '@a' and '@a2' . For large numbers of modes this method
uses a reduced amount of memory as there is no stored matrix involved
in this case. 
\par\end{flushleft}

\begin{figure}[H]
\begin{centering}
\includegraphics[width=0.75\textwidth]{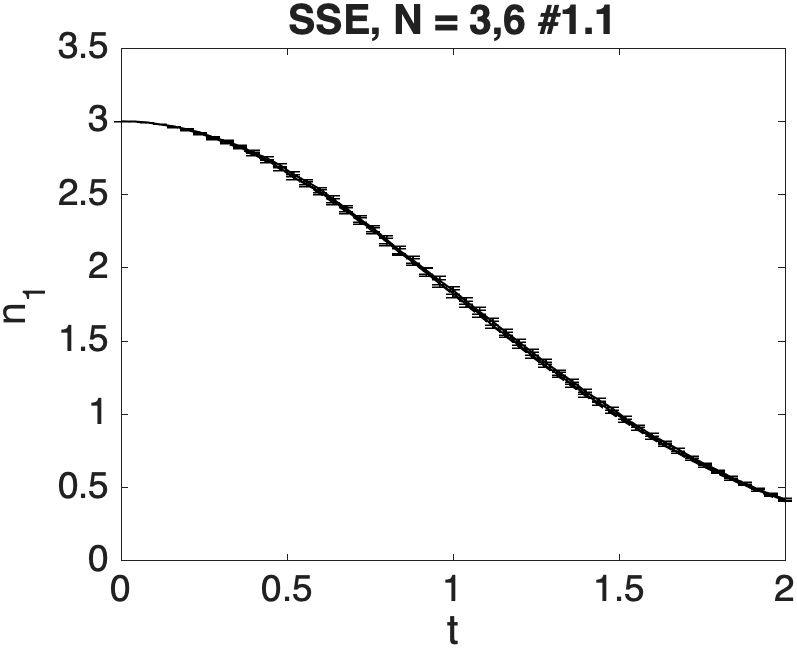}\\
 \includegraphics[width=0.75\textwidth]{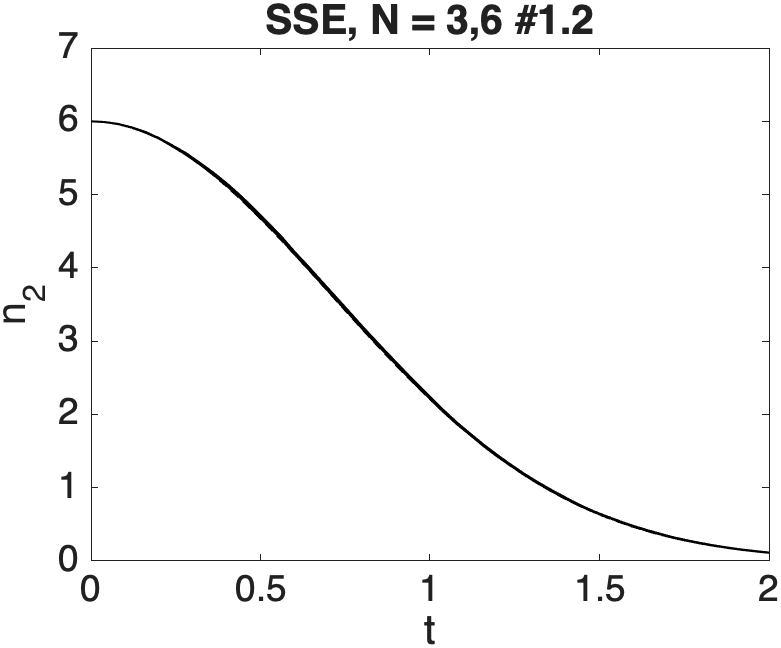}
\par\end{centering}
\caption{\emph{Example: SSE linear decay, with a time-dependent decay rate..
Top graph has $N=3$ , lower graph has $N=6.$}}
\vspace{10pt}
\end{figure}

\pagebreak{}

\section{Nonlinear decay, real SSE}

This solves a Lindblad master equation for nonlinear decay with two
modes and two decay channels. The initial condition is $\psi_{j}=\delta_{Nj}$
and $L_{1}=a$, $L_{2}=a^{2}$, for $N=[3,6]$ .

The decay rates are: 
\begin{align*}
\gamma_{1} & =[0.01,0.01]\\
\gamma_{2} & =[.5,.25],
\end{align*}
The function uses the midpoint algorithm with the SSE derivative,
and has real noise terms. The sparse and functional methods give identical
results, but the sparse method is faster.

\paragraph{Sparse operator method}
\begin{center}
\doublebox{\begin{minipage}[t]{0.9\columnwidth}%
\texttt{function e = SSEnonlin2spr}

\texttt{\%Uses a real sparse SSE to solve for nonlinear two-mode decay}

\texttt{p.name = 'Real sparse SSE, M=2, N=3,6';}

\texttt{p.nmax = {[}4,7{]};}

\texttt{p.steps = 8;}

\texttt{p.a = mkbose(p);}

\texttt{p.a2 = mkbose(1:2,2,p);}

\texttt{p.ensembles = {[}10,10,10{]};}

\texttt{p.quantum = 1;}

\texttt{p.sparse = 1;}

\texttt{p.gamma = \{@(p){[}0.01,0.01{]},@(p){[}.5,.1{]}\};}

\texttt{p.theta = \{{[}1,1{]},{[}1,1{]}\};}

\texttt{p.L = \{@(m,p) p.a\{m\},@(m,p) p.a2\{m\}\};}

\texttt{p.initial = @(\textasciitilde ,p) kron({[}0,0,0,1{]},{[}0,0,0,0,0,0,1{]})';}

\texttt{p.expect\{1\} = @(p) p.a\{1\}'{*}p.a\{1\};}

\texttt{p.expect\{2\} = @(p) p.a\{2\}'{*}p.a\{2\};}

\texttt{p.olabels = \{'n\_1','n\_2'\};}

\texttt{e = xspde(p);}

\texttt{end}%
\end{minipage}} 
\par\end{center}

\begin{flushleft}
With the sparse method, the function 'mkbose' is used twice to create
the operator matrix cell array 'p.a', and 'p.a2' before they are used. 
\par\end{flushleft}

The use of p.quantum=1 shows that it is a stochastic wavefunction
problem, p.sparse=1 indicates sparse matrices, and p.theta = \{{[}1,1{]},{[}1,1{]}\}
specifies that all channels have real noises.

\paragraph{Function operator method}
\begin{flushleft}
\doublebox{\begin{minipage}[t]{0.9\columnwidth}%
\texttt{function e = SSEnonlin2r}

\texttt{\%Uses an SSE to solve for a linear two-mode decay}

\texttt{p.name = 'Real nonlinear SSE, 2-modes, N = 3,6';}

\texttt{p.nmax = {[}4,7{]};}

\texttt{p.steps = 8;}

\texttt{p.ensembles = {[}10,10,10{]};}

\texttt{p.quantum = 1;}

\texttt{p.gamma = \{@(p){[}0.01,0.01{]},@(p) {[}.5,.1{]}\};}

\texttt{p.L = \{@a,@a2\};}

\texttt{p.theta = \{{[}1,1{]},{[}1,1{]}\};}

p.H = @(psi,p) (n(1,psi)+n(2,psi));

\texttt{p.initial = @(\textasciitilde ,p) kron({[}0,0,0,1{]}',{[}0,0,0,0,0,0,1{]});}

\texttt{p.expect\{1\} = @(psi,p) n(1,psi);}

\texttt{p.expect\{2\} = @(psi,p) n(2,psi);}

\texttt{p.olabels = \{'n\_1','n\_2'\};}

\texttt{e = xspde(p);}

\texttt{end}%
\end{minipage}} 
\par\end{flushleft}

\begin{flushleft}
With the function method, the function 'mkbose' is not required. Instead,
the effect of the operators is obtained through a function call to
the handles '@a' and '@a2' . For large numbers of modes this method
uses a reduced amount of memory as there is no stored matrix involved
in this case. 
\par\end{flushleft}

\begin{figure}[H]
\begin{centering}
\includegraphics[width=0.75\textwidth]{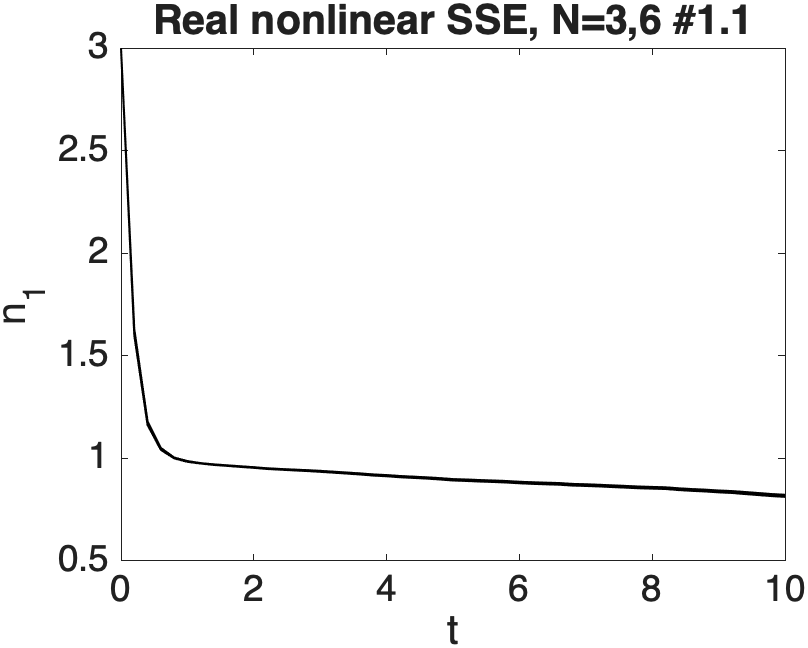}\\
 \includegraphics[width=0.75\textwidth]{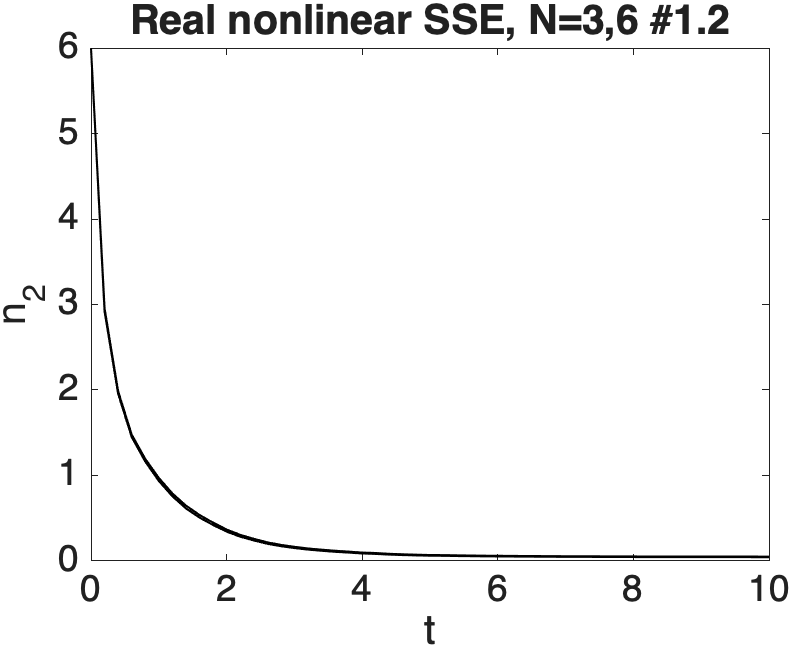}
\par\end{centering}
\caption{\emph{Example: SSE nonlinear decay, with a small linear decay., real
noise and either odd or even number starting points.Top graph has
$N=3$ , lower graph has $N=6.$}}
\vspace{10pt}
\end{figure}

\newpage{}

\part{Methods, API and Examples}

\chapter{Stochastic methods \label{sec:Algorithms}}

\textbf{This chapter describes the integration methods available,
and how to add custom algorithms.}

\section{Introduction to algorithms}

Stochastic, partial and ordinary differential equations are central
to numerical mathematics. Ordinary differential equations have been
known in some form ever since calculus was invented. There are an
extraordinary number of algorithms used to solve these equations.
This chapter provides an overview of the included algorithms.\\
 xSPDE has six built-in choices of algorithm, with defaults. All built-in
methods have an interaction picture and can be used with any space
dimension, including $dimensions=1$, which is an ordinary stochastic
equation. All can be used with stochastic or with non-stochastic equations,
and with order extrapolation. \\
 For stochastic equations, the Euler method requires an Ito form of
stochastic equation, the implicit Euler method requires an implicit
Ito form, while the others should be used with the Stratonovich form
of calculus. Each is chosen to be able to use an interaction picture
to take care of exactly soluble linear terms.

The default methods will solve most DE, SDE, PDE and SPDE problems
reliably, but other ones can be included if needed.

\subsection{Standard methods}

The standard xSIM algorithms given below are available for ODEs, PDEs,
SDEs and SPDEs. More advanced algorithms for specialized cases are
described in section \ref{sec:Algorithms}.

For stochastic differential equations, which are non-differentiable,
the usual rules of calculus do not apply because stochastic noise
is non-differentiable. It has fluctuations proportional to $1/\sqrt{dtdV}$,
for noise defined on a lattice with temporal cell-size $dt$ and spatial
cell-size $dV$. Hence, the usual differentiability and smoothness
properties required to give high-order convergence for standard Runge-Kutta
methods are simply not present. Instead, xSPDE has a built-in extrapolation
to zero step-size for high-order stochastic convergence.

Many more complex higher order algorithms for stochastic integration
exist but are not included in the current xSPDE distribution, and
users are encouraged to contribute their favorite methods.

We note here that there are multiple error sources possible. SDE/SPDE
errors are often dominated by the sampling error, not discretization.
In addition, all convergence theorems only apply to the limit of zero
step-size. One may be very far from this regime in a given practical
calculation. Analytic error estimates also have pre-factors which
are hard to calculate. However, xSPDE can numerically estimate both
the discretization and sampling error for any given average observable.

\subsection{Advanced methods}

Three more advanced method libraries are included here, namely weighted,
projected and forward-backward stochastic differential equations.
If you have a favorite algorithm that is not included, user-defined
algorithms and libraries can be added. The existing methods are listed
below, and the corresponding {.m}-files can be used as a model.

Define the routine, for example {"myalgorithm.m"}, set \textit{$p.method=@myalgorithm$},
then adjust the input value of ipsteps and order if these need be
changed to a new value. The interaction-picture transform, prop, can
also be changed if the built-in choice is not sufficient.The xSPDE
algorithms available currently treat 
\begin{itemize}
\item ordinary (and partial) differential equations 
\item stochastic differential equations 
\item stochastic partial differential equations 
\item weighted stochastic differential equations 
\item projected stochastic differential equations, 
\item forward-backward stochastic differential equations 
\end{itemize}
Some of the more advanced features of the libraries require additional
input parameters. In particular: 
\begin{description}
\item [{backfields}] is used for forward-backward stochastic equations,
describing backward time components. These are described in the Forward-backward
section. Note that \textbf{fields} is still used, and it gives the
total number of forward+backward fields. 
\item [{auxfields}] gives the number of auxiliary fields. These have a
functional definition (defines) that includes both a field and noise
variable, as needed for spectral observables. Cell index numbers $i$
greater than the maximum field cells access the auxiliary fields in
the observe function. 
\end{description}

\section{General differential form}

The general equation treated is given in differential form as 
\begin{equation}
\begin{split}\frac{\partial\boldsymbol{a}}{\partial t}=\boldsymbol{A}\left[\boldsymbol{\nabla},\boldsymbol{a},t\right]+\underline{\mathbf{B}}\left[\boldsymbol{\nabla},\boldsymbol{a},t\right]\cdot\boldsymbol{\zeta}(t)+\underline{\mathbf{L}}\left[\boldsymbol{\nabla}\right]\cdot\boldsymbol{a}.\end{split}
\label{eq:Standard xspde SDE}
\end{equation}
It is convenient for the purposes of describing interaction picture
methods, to introduce an abbreviated notation as: 
\begin{equation}
\begin{split}\begin{aligned}\mathcal{D}\left[\mathbf{a},t\right]=\boldsymbol{A}\left[\boldsymbol{a},t\right]+\underline{\mathbf{B}}\left[\boldsymbol{a},t\right]\cdot\boldsymbol{\zeta}(t).\end{aligned}
\end{split}
\end{equation}
Hence, we can rewrite the differential equation in the form: 
\begin{equation}
\begin{split}\frac{\partial\boldsymbol{a}}{\partial t}=\mathcal{D}\left[\mathbf{a},t\right]+\underline{\mathbf{L}}\left[\boldsymbol{\nabla}\right]\cdot\boldsymbol{a}.\end{split}
\end{equation}

\subsection{Linear propagator}

Next, we define a linear propagator. This is given formally by: 
\begin{equation}
\begin{split}\mathcal{P}\left(\Delta t\right)=\exp\left(\Delta t\underline{\mathbf{L}}\left[\boldsymbol{\nabla}\right]\right)\end{split}
.
\end{equation}
Typically, but not necessarily, this is evaluated in Fourier space,
where it is a diagonal term in the momentum vector conjugate to the
transverse space coordinate. It involves a Fourier transform, multiplication
by a function of momentum, and an inverse Fourier transform. For simplicity,
the stochastic noise is assumed constant throughout the interval $dt$.
The reader is referred to the literature for more details.

It is simple to add your own algorithm if you prefer a different one.
Note that if they use an interaction picture, then ipsteps must be
given explicitly to specify the interaction picture duration, where
ipsteps gives the number of sequential propagator steps in time required
for the method.

\section{Standard methods}

The standard methods are listed below. All of these can be used with
any equation: ODE, SDE, PDE or SPDE, either with or without a linear
interaction picture term. The basic equation used here is:
\[
\begin{split}\frac{\partial\tilde{\mathbf{a}}}{\partial t}=\mathcal{D}\left[\mathcal{P}\left(t,\tilde{t}\right)\tilde{\mathbf{a}},t\right]=\tilde{\mathcal{D}}\left[\tilde{\mathbf{a}},t\right].\end{split}
\]

\subsection{Euler: Ito-Euler}

\label{algorithms:euler} This is an explicit Ito-Euler method using
an interaction picture. While traditional, it is not generally recommended.
If it is used, very small step-sizes will generally be necessary to
reduce errors to a usable level. This is because it is is only convergent
to first order deterministically and tends to have large errors.

It is designed for use with an Ito form of stochastic equation. It
requires one IP transform per step (\textit{$p.ipsteps=1$}). Choosing
the origin of the interaction picture at $\tilde{t}=t_{n}$, one has
$\mathbf{a}_{n}\equiv\tilde{\mathbf{a}}_{n}$, so:

\[
\Delta\tilde{\mathbf{a}}_{n+1}=\tilde{\mathbf{a}}_{n+1}-\tilde{\mathbf{a}}_{n}=\Delta t\mathcal{D}\left[\mathbf{a}_{n},t_{n}\right]
\]

To get the next time point at $t=t_{n+1}=t_{n}+\Delta t$, one calculates:
\begin{equation}
\begin{split}\begin{aligned}\Delta\tilde{\mathbf{a}}_{n+1} & =\Delta t\mathcal{D}\left[\mathbf{a}_{n},t_{n}\right]\\
\mathbf{a}_{n+1} & =\mathcal{P}\left(\Delta t\right)\cdot\left[\mathbf{a}_{n}+\Delta\tilde{\mathbf{a}}_{n+1}\right]
\end{aligned}
\end{split}
\end{equation}

\subsection{Implicit: implicit Ito-Euler}

\label{algorithms:implicit} This is a fully implicit Ito-Euler method
using an interaction picture. It is more robust, though slower, than
the explicit form. If it is used, very small step-sizes will generally
be necessary to reduce errors to a usable level.

This is because it is is only convergent to first order, and therefore
tends to have large errors. It is designed for use with an implicit
Ito form of stochastic equation. Note that this implies double the
usual Stratonovich correction!

It requires one IP transform per step ($p.ipsteps=1$). Choosing the
origin of the interaction picture at $\tilde{t}=t_{n+1}$, one has
$\mathbf{a}_{n+1}\equiv\tilde{\mathbf{a}}_{n+1}$, so:

\[
\Delta\tilde{\mathbf{a}}_{n+1}=\tilde{\mathbf{a}}_{n+1}-\tilde{\mathbf{a}}_{n}=\Delta t\mathcal{D}\left[\mathbf{a}_{n+1},t_{n}\right]
\]
Starting from time $t=t_{n}$, to get the next time point at $t=t_{n+1}=t_{n}+\Delta t$,
one calculates, using iteration to get the implicit result of the
next time-point:

\begin{equation}
\begin{split}\begin{aligned}\bar{\mathbf{a}}^{(0)} & =\mathcal{P}\left(t_{n+1},t_{n}\right)\cdot\left[\mathbf{a}_{n}\right]\\
\bar{\mathbf{a}}^{(i)} & =\bar{\mathbf{a}}^{(0)}+\Delta t\mathcal{D}\left[\bar{\mathbf{a}}^{(i-1)},t_{n+1}\right]\\
\mathbf{a}_{n+1} & =\tilde{\mathbf{a}}_{n+1}=\bar{\mathbf{a}}^{(iter)}
\end{aligned}
\end{split}
\end{equation}

Here the result of $\bar{\mathbf{a}}^{(iter)}$ is obtained after
a fixed number of iterations of $\bar{\mathbf{a}}^{(i)}$.

\subsection{MP: Midpoint}

\label{algorithms:midpoint} This is a semi-implicit midpoint method
using an interaction picture. It gives good results for stochastic
and stochastic partial differential equations. It is convergent to
second order in time for deterministic equations and for stochastic
equations with commuting noise. It is strongly convergent and robust.
It requires two half-length IP transforms per step ($p.ipsteps=2$).

To get the next time point, one calculates a midpoint derivative iteratively
at time to get the next time point at $t=t_{n+1/2}=t_{n}+\Delta t/2$,
to give an estimated midpoint field $\bar{\mathbf{a}}^{(i)}$, usually
with four iterations. The number of iterations can be changed: 
\begin{equation}
\begin{split}\begin{aligned}\bar{\mathbf{a}}^{(0)} & =\mathcal{P}\left(t_{n+1/2},t_{n}\right)\cdot\left[\mathbf{a}_{n}\right]\\
\bar{\mathbf{a}}^{(i)} & =\bar{\mathbf{a}}^{(0)}+\frac{\Delta t}{2}\mathcal{D}\left[\bar{\mathbf{a}}^{(i-1)},t_{n+1/2}\right]\\
\mathbf{a}_{n+1} & =\mathcal{P}\left(t_{n+1},t_{n+1/2}\right)\cdot\left[2\bar{\mathbf{a}}^{(iter)}-\bar{\mathbf{a}}^{(0)}\right]
\end{aligned}
\end{split}
\end{equation}

This is the default method for stochastic cases.

\subsection{MPadapt: adaptive midpoint}

\label{algorithms:midpoint-1} This is an implicit midpoint method
using an interaction picture, together with an adaptive technique
for integrating highly nonlinear equations. At low amplitudes it is
identical to the standard midpoint method. For amplitudes $|a_{i}|^{2}$
above a critical value, p.adapt, the amplitude is inverted and propagated
using the differential equation for its inverse.

Initially a switch $p$ is set to $1$ for low amplitudes, and $-1$
for high amplitudes. To get the next time point, one calculates a
midpoint derivative iteratively at time to get the next time point
at $t=t_{n+1/2}=t_{n}+\Delta t/2$, to give an estimated midpoint
field $\bar{\mathbf{a}}^{(i)}$, as above, but with the derivative
modified to give the derivative of $a_{i}^{p}$: 
\begin{equation}
\begin{split}\begin{aligned}\bar{\mathbf{a}}^{(0)} & =\mathcal{P}\left(t_{n+1/2},t_{n}\right)\cdot\left[\mathbf{a}_{n}\right]\\
\tilde{\mathbf{a}}^{(0)} & =\mathbf{a}_{n}^{p}\\
\tilde{\mathbf{a}}^{(i)} & =\tilde{\mathbf{a}}^{(0)}+\frac{\Delta t}{2}p\left[\tilde{\mathbf{a}}^{(i-1)}\right]{}^{1-p}\left(\mathcal{D}\left[[\tilde{\mathbf{a}}^{(i-1)}]^{p},t_{n+1/2}\right]\right)\\
\mathbf{a}_{n+1} & =\mathcal{P}\left(t_{n+1},t_{n+1/2}\right)\cdot\left[2\tilde{\mathbf{a}}^{(iter)}-\tilde{\mathbf{a}}^{(0)}\right]^{p}
\end{aligned}
\end{split}
\end{equation}

\subsection{RK2: second order Runge-Kutta}

\label{algorithms:second-order-runge-kutta} This is a second order
Runge-Kutta method using an interaction picture. It is convergent
to second order in time for non-stochastic equations, and for stochastic
equations with additive noise, but otherwise it is first order. It
often has higher errors than midpoint methods. It requires two IP
transforms per step, but each is a full time-step long ($p.ipsteps=1$).
The basic RK2 method is defined by:

\begin{align*}
\tilde{\mathbf{a}}^{(1)} & =\tilde{\mathbf{a}}_{n}+\Delta t\tilde{\mathcal{D}}\left[\tilde{\mathbf{a}}_{n},t_{n}\right]\\
\tilde{\mathbf{a}}^{(2)} & =\tilde{\mathbf{a}}_{n}+\Delta t\tilde{\mathcal{D}}\left[\tilde{\mathbf{a}}^{(1)},t_{n+1}\right]\\
\tilde{\mathbf{a}}_{n+1} & =\left(\tilde{\mathbf{a}}^{(1)}+\tilde{\mathbf{a}}^{(2)}\right)/2
\end{align*}

Including the interaction picture transforms, based at $t_{n}$, one
calculates: 
\begin{equation}
\begin{split}\begin{aligned}\bar{\mathbf{a}} & =\mathcal{P}\left(t_{n+1},t_{n}\right)\cdot\mathbf{a}_{n}\\
\mathbf{a}^{(1)} & =\mathcal{P}\left(t_{n+1},t_{n}\right)\left(\mathbf{a}_{n}+\Delta t\cdot\mathcal{D}\left[\mathbf{a}_{n},t_{n}\right]\right)\\
\mathbf{a}^{(2)} & =\bar{\mathbf{a}}+\Delta t\mathcal{D}\left[\mathbf{a}^{(1)},t_{n+1}\right]\\
\mathbf{a}_{n+1} & =\left(\mathbf{a}^{(1)}+\mathbf{a}^{(2)}\right)/2.
\end{aligned}
\end{split}
\end{equation}

\subsection{RK4: fourth order Runge-Kutta}

\label{algorithms:fourth-order-runge-kutta} This is a fourth order
Runge-Kutta method using an interaction picture. It is convergent
to fourth order in time for non-stochastic equations, but for stochastic
equations it can be more slowly convergent than the midpoint method.
It requires four half-length IP transforms per step (ipsteps = 2).
To get the next time point, one calculates four derivatives sequentially:
\begin{equation}
\begin{split}\begin{aligned}\bar{\mathbf{a}} & =\mathcal{P}\left(\frac{\Delta t}{2}\right)\cdot\left[\mathbf{a}_{n}\right]\\
\mathbf{d}^{(1)} & =\frac{\Delta t}{2}\mathcal{P}\left(\frac{\Delta t}{2}\right)\cdot\mathcal{D}\left[\mathbf{a}_{n},t_{n}\right]\\
\mathbf{d}^{(2)} & =\frac{\Delta t}{2}\mathcal{D}\left[\bar{\mathbf{a}}+\mathbf{d}^{(1)},t_{n+1/2}\right]\\
\mathbf{d}^{(3)} & =\frac{\Delta t}{2}\mathcal{D}\left[\bar{\mathbf{a}}+\mathbf{d}^{(2)},t_{n+1/2}\right]\\
\mathbf{d}^{(4)} & =\frac{\Delta t}{2}\mathcal{D}\left[\mathcal{P}\left(\frac{\Delta t}{2}\right)\left[\bar{\mathbf{a}}+2\mathbf{d}^{(3)},t_{n+1}\right]\right]\\
\mathbf{a}_{n+1} & =\mathcal{P}\left(\frac{\Delta t}{2}\right)\cdot\left[\bar{\mathbf{a}}+\left(\mathbf{d}^{(1)}+2\left(\mathbf{d}^{(2)}+\mathbf{d}^{(3)}\right)\right)/3\right]+\mathbf{d}^{(4)}/3
\end{aligned}
\end{split}
\end{equation}
This might seem the obvious choice, having the highest order. However,
it can converge at a range of apparent rates, depending on the relative
importance of stochastic and non-stochastic terms. Due to its use
of differentiability, it may converge more slowly than the midpoint
method with stochastic terms present. It is the default for ODE and
PDE cases.

\section{Weighted library}

In some types of stochastic equation, there is a weight associated
with each trajectory, which is used to weight the probability of the
trajectory \cite{kiesewetter2022coherent}. This type of equation
is sometimes found when dealing with quantum trajectories \cite{Dalibard1992Wave,Carmichael1993Quantum}
and feedback \cite{Hush2013Controlling}.

The equations still have the standard form of Eq \eqref{eq:SDE},
with an extra weight equation, Eq \eqref{eq:SDE-2}. However, the
results for mean values are weighted by a term $\exp\left(\Omega\left(t\right)\right)$,
so that: 
\begin{equation}
\left\langle \mathbf{O}\right\rangle _{\Omega}=\frac{\sum_{n}\mathbf{O}\left(\mathbf{a}^{\left(n\right)}\right)\exp\left(\Omega^{\left(n\right)}\left(t\right)\right)}{\sum_{n}\exp\left(\Omega^{\left(n\right)}\left(t\right)\right)}.\label{eq:Weighted-averages}
\end{equation}

This reduces to the standard expression of Eq \eqref{eq:averages-1}
in the case that $\Omega\left(t\right)=0$. To simulate these equations
automatically, the weight exponent $\Omega$ is integrated as the
last field in the vector $\mathbf{a}$, which must have at least two
components. A nonzero threshold weight, $thresholdw$, must be entered
to allow calculation of breeding.

With these changes, averages in each vector ensemble are calculated
using Eq \eqref{eq:Weighted-averages}. Before each plotted step in
the calculation, a breeding calculation is carried out. There are
$p.steps(1)-1$ of these in total. During breeding, any weight such
that $\exp\left(\Omega^{(n)}\right)<thresholdw/\left\langle \exp\left(\Omega\right)\right\rangle $
is removed.

The most probable trajectory is then duplicated to replace the low-weight
trajectory. Both exponential weights are halved, so the total weight
of the remaining trajectories is unchanged. If they are complex, weights
such that $\exp\left(Re\left(\Omega^{(n)}\right)\right)<thresholdw/\left\langle \exp\left(Re\left(\Omega\right)\right)\right\rangle $
are removed, and the real weight of the bred trajectory is reduced,
which removes any low-weight trajectories that don't contribute. When
used, the internal variable p.breedw is set to allow the fraction
of trajectories that are bred per step to be monitored. For weighted
SPDEs, the spatial weights $\Omega(x_{j})$ are summed over space
points to obtain $\Omega$.

\subsection{Example}

The following example shows how weights are implemented. 
\begin{center}
\doublebox{\begin{minipage}[t]{0.75\columnwidth}%
\texttt{function {[}e{]} = Weightcheck()}

\texttt{p.name = 'Weightcheck';}

\texttt{p.ensembles = {[}10000,10,1{]};}

\texttt{p.fields = 2;}

\texttt{p.points = 6;}

\texttt{p.order = 2;}

\texttt{p.thresholdw = 0.1;}

\texttt{p.diffplot = 1;}

\texttt{p.initial = @(w,p) {[}1+w(1,:);0{*}w(2,:){]};}

\texttt{p.deriv = @(a,z,p) {[}-a(1,:)+ z(1,:);-a(2,:)+...}

\texttt{z(2,:){]};}

\texttt{p.observe\{1\} = @(a,p) a(1,:);}

\texttt{p.observe\{2\} = @(a,p) p.breedw;}

\texttt{p.compare\{1\} = @(p) exp(-p.t);}

\texttt{p.olabels\{1\} = '\textless a\textgreater ';}

\texttt{p.olabels\{2\} = '\textless fractional breeds per step\textgreater
';}

\texttt{e = xcheck(2,p);}

\texttt{end}%
\end{minipage}} 
\par\end{center}

This algorithm converges with second-order accuracy for this exercise,
due to the structure of the equation. The example also demonstrates
how to use the xcheck function instead of xspde, to check convergence.

\section{Projection library}

When numerically integrating projected SDEs or SPDEs, it is also useful
to have a normal projection $\mathcal{P}^{\perp}$available. This
is used to normally project to the nearest point on the manifold,
to eliminate constraint errors. These are solved using functions collected
in a projection library, to provide the specialized methods that are
needed for this purpose.

The projection library has three predefined algorithms, 
\begin{itemize}
\item \textbf{Enproj},  
\item \textbf{MPproj}, 
\item \textbf{MPnproj}. 
\end{itemize}
Here the capital E stands for Euler, MP for midpoint. All use tangential
projection. The letter n=normal indicates if an additional normal
projection is used. In all cases, if it is present, a normal projection
is used last. The recommended type is \textbf{MPnproj}, due to its
much lower errors.

Tangential and normal projections are needed to define the geometry
of any sub-manifold. These are input by setting the variable project
equal to a function handle that defines the projection. These can
be user provided if required. There are three different predefined
manifold geometry types, which need different inputs, given below.

\subsection{Calling the project function}

The calling arguments for the project function are: (d,a,n,(c,)p),
where d is a vector to be tangentially projected at location a, a
is the current (near)-manifold field or cell array, n is an option
switch, and p is the parameter structure. This can be used for a field
a, without the cell index c, or for a cell array of fields, including
the cell index c.

The options available in any project implementation are defined as: 
\begin{itemize}
\item \textit{n = 0} returns the tangent vector for testing 
\item \textit{n = 1} returns the tangential projection of d at a 
\item \textit{n = 2} returns the normal projection of a, where d is not
used 
\item \textit{n = 4} returns the constraint function at a for testing 
\end{itemize}
The projections defined in an xSPDE project function can be of any
type. Arbitrary dimension reduction and manifold geometry is possible.
Currently in the examples, dimensionality is reduced by 1, and normal
projections use fixed point iterations, defined by iterations.

\subsection{The predefined manifold geometries}

The current manifolds, by setting p.project = @Quadproj ..., are as
follows: 
\begin{enumerate}
\item Quadratic - Quadproj - needs: qcproj defined by $f=\sum q{}_{ij}x^{i}x^{j}-1=0$ 
\item Polynomial - Polproj - needs: vcproj defined by $f=\sum v_{i}(x^{i})^{p}-1=0$ 
\item Catenoid - Catproj - uses fixed coefficients defined by $f=(x_{1})^{2}+(x_{2})^{2}-(sinh(x_{3}))^{2}-1=0$ 
\end{enumerate}
Any other manifold can be used by replacing these predefined manifolds
with an appropriate project function.

\newpage{}

\chapter{Errors\label{chap:Errors}}

\textbf{This chapter describes the estimation and control of integration
errors.}

\section{Time-step discretization errors}

To check convergence, xSPDE default settings will repeat the calculations
twice for checking time-steps, and many times more in stochastic cases
to estimate sampling errors. Since the checks make xSPDE slower, they
can be turned off, but then there are no error-estimate. Whatever
the application, error-estimates useful, and generally should be used.

If the errors are too large relative to the application, you should
decrease the time-steps or increase the number of samples. Which is
needed depends on the type of error.

Errors caused by the finite time-domain step-size are checked automatically,
since $p.checks(1)=1$ is the default option. If $p.checks=0$ is
used, there is no time-domain error check.

Errors due to a finite step-size are estimated by running a check
simulation with half the initial step-size and the same random sequence,
extrapolating to zero step-size if $order>0$ is specified. The program
returns an error bound as the difference of the two most accurate
results. Any 2D output graph plots error-bars if $checks=1$ was specified,
provided they are large enough to plot. 

RMS output error summaries are also reported in the text outputs.
Even more error information is available if p.verbose=1 is specified.
Individual time-step error bounds, $e\left(o\right)$ are given in
the output data, and the plots give $\bar{o}\pm e\left(o\right)$.

Error-bars below a minimum relative size compared to the vertical
range of the plot, specified by the graphics variable $minbar$, are
not plotted. The default for this is $minbar=0.01$. All error bars
are calculated individually for each type of data average. Minbar
is a cell array that can can be set for each type of average or graph.
If the cell argument is omitted, it applies globally. Error estimates
are also given for functional transforms of averages.

If the errors are too large, one can either increase the points, which
gives more plotted points and lower errors, or increase the steps,
which reduces the step size without changing the data resolution.
The default algorithm and extrapolation order can also be changed.
Error bars on graphs can be removed by setting $checks=0$ or increasing
$minbar$.

Discretization errors caused by the finite spatial lattice are not
currently checked in the xSIM code. They must be checked by comparing
results with different transverse lattice ranges and step-size. Similarly,
errors from discrete probability bin sizes are not checked.

If computed, the discretization error is included in the graphical
data outputs for all observables. It is accessed by setting the last
index for the output data equal 2. The raw discretization error is
generally a very cautious estimate, and may overestimate the errors.
This estimate can be improved using extrapolation, explained next.

\subsection{Extrapolation}

\label{algorithms:convergence-checks}

xSPDE can use extrapolation to improve convergence, which requires
input of the method order. If this is non-zero, and checks are set
to 1 to allow successive integration with different step-sizes, the
output of all data graphed will be extrapolated by assuming the method
has the specified order. To implement extrapolation and obtain a less
conservative mean and error-bar result, set $p.order>0$. Note that
this value is user-defined.

Although convergence rates are somewhat problem-dependent, all xSPDE
methods will return their theoretical convergence order for deterministic
and stochastic calculations respectively. To extrapolate using these
theoretical orders, specify $p.order=-1$, which gives the method
order. The deterministic order is used if there is one ensemble.

\label{algorithms:extrapolation-order-and-error-bars}

Extrapolation is valuable for improving the accuracy of a differential
equation solver. It is valid for small time-steps. Suppose an algorithm
has a correct solution $R_{0}$, but returns a numerical result $R$
with an error order $n$. For small step-size, integration results
$R\left(dt\right)$ with step-size $dt$ have an error of order $dt^{n}$,
that is: 
\begin{equation}
\begin{split}R\left(dt\right)=R_{0}+e\left(R\right)=R_{0}+k.dt^{n}.\end{split}
\end{equation}
Hence, from two results at different values of $dt,$ differing by
a factor of $2$, one would obtain 
\begin{equation}
\begin{split}\begin{aligned}R_{1} & =R\left(dt\right)=R_{0}+k.dt^{n}\\
R_{2} & =R\left(2dt\right)=R_{0}+2^{n}k.dt^{n}.
\end{aligned}
\end{split}
\end{equation}
The true result, extrapolated to the small-step size limit, is obtained
by giving more weight to the fine step-size result, while subtracting
from this a correction due to the coarse step-size calculation, to
cancel the leading error term: 
\begin{equation}
\begin{split}R_{0}=\frac{\left[R_{1}-R_{2}2^{-n}\right]}{\left[1-2^{-n}\right]}.\end{split}
\end{equation}
Thus, if we define a factor $\epsilon$ as 
\begin{equation}
\begin{split}\epsilon\left(n\right)=\frac{1}{\left[2^{n}-1\right]}=\left(1,\frac{1}{3},\frac{1}{7}\ldots\right),\end{split}
\end{equation}
the true results are obtained from extrapolation to zero step-size
as: 
\begin{equation}
\begin{split}R_{0}=\left(1+\epsilon\right)R_{1}-\epsilon R_{2}.\end{split}
\end{equation}
The built-in algorithms have an order as ordinary differential equation
integrators of 1, 1, 2, 2, 2, 4 respectively and will converge to
this order at small step-sizes. Weak first order convergence is always
obtainable for these single noise-step SDE methods \cite{burrage2006comment}.
Second order weak convergence is obtained in some cases with RK4 algorithms.

Higher order convergence for the raw data is not guaranteed for the
built-in SDE algorithms. The algorithms used do\textbf{ not }always
converge to the standard ODE order when used for stochastic equations.
Hence extrapolation to higher than first order should be used with
caution in stochastic calculations, unless more complex methods are
used \cite{Kloeden1992numerical}.

\subsection{Extrapolated error-bars}

If extrapolation is used, the error bar half-size is the difference
of the best raw estimate and the extrapolation. Extrapolated results
are usually inside those given by the error-bars, however, note that: 
\begin{itemize}
\item \textbf{extrapolation with too high an order may under-estimate error
bars} 
\item \textbf{extrapolation with too low an order reduces the accuracy} 
\end{itemize}
A conservative order estimate of order = 1 can be used for all SDE
and SPDE cases, although there are higher order methods available.
This gives an extrapolated weak order of $2$ for stochastic cases.
One can set order = 0 to remove the default, or use a higher order
if preferred, although, as explained above, it requires some caution.
For an ODE or PDE the usual deterministic order should be used. For
the default RK4 deterministic method, order = 4. All orders are improved
by one with extrapolation.

High-order convergence without extrapolation can also be obtained,
either in special cases using the xSPDE methods, or by adding user-specified
techniques. The xSPDE libraries can be readily extended by the user
to include these, through defining a modified method function appropriately.

\section{Statistical errors}

Sampling error estimation in xSIM uses three different techniques. 
\begin{itemize}
\item xSIM uses sub-ensemble averaging, requiring high-level ensembles. 
\item For probability estimates, a Poissonian sampling error is used, based
on counts. 
\item If there is a comparison probability, this is used for sampling error
estimates. 
\end{itemize}
This procedure leads to reliable sampling error estimates, and makes
efficient use of the vector instruction sets used by Matlab. Ensembles
are specified in three levels. The first, ensembles(1), is called
the number of samples for brevity. All computed quantities returned
by the \textbf{observe} functions are first averaged over the samples,
which are calculated efficiently using a parallel vector of trajectories.
By the central limit theorem, these low-level sample averages are
distributed as a normal distribution at large sample number.

Next, the sample averages are averaged again over the two higher level
ensembles, if specified. This time, the variance is accumulated. The
variance of these distributions is used to estimate a standard deviation
in the mean, since each computed quantity is now a normally distributed
result. This method is applied to all the observables. The two lines
generated represent $\bar{o}\pm\sigma\left(o\right)$, where $o$
is the observe function output, and $\sigma$ is the standard deviation
in the mean.

Here, ensembles(2) specifies ensembles computed in series. The highest
level ensemble, ensembles(3), is used for parallel simulations. This
is faster for a multiple core CPU or when the codes are run in a supercomputing
environment, which requires the Matlab parallel toolbox. Either type
of high-level ensemble, or both together, can be used to calculate
sampling errors.

If $ensembles(2)>1$ or $ensembles(3)>1$, which allows xSPDE to calculate
sampling errors, it will plot upper and lower limits of one standard
deviation. If the sampling errors are too large, try increasing $ensembles(1)$,
which increases the trajectories in a single thread. An alternative
is to increase $ensembles(2)$, which is slower, but is only limited
by the compute time, or else to increase $ensembles(3)$, which gives
higher level parallelization.

Each is limited in different ways: the first by memory, the second
by time, the third by the number of cores. Sampling error control
helps ensures accuracy.

\subsection{Sampling error}

Quantitative sampling error estimation in xSPDE uses sub-ensemble
averaging. Ensembles are specified in three levels, using vector,
serial and parallel methods, respectively. The vector ensemble length,
\textit{p.ensembles(1)}, is called the number of samples for brevity.
All quantities returned by the observe functions are averaged over
the samples, which are calculated efficiently using a vector of trajectories.

By the central limit theorem, the sample averages are distributed
as a normal distribution at large sample number. Next, the sample
averages are averaged over the two higher level ensembles, if specified.
The variance of this data is used to estimate a standard deviation
in the mean, since each is normally distributed.

The next level, p.ensembles(2), is for serial calculations of ensembles.
The highest level ensemble, p.ensembles(3), is used for parallel simulations.
This requires the Matlab parallel toolbox. Either type of high-level
ensemble, or both together, can be used to calculate sampling errors.

Note that one standard deviation is not a strong bound; errors are
expected to exceed this value in $32\%$ of observed measurements.
Another point to remember is that stochastic errors are often correlated,
so that a group of points may all have similar errors due to statistical
sampling.

The statistical error due to finite samples of trajectories is called
the sampling error. The RMS value of the relative sampling error for
each computed function, normalized by the maximum modulus of the observable,
is printed out after each xSPDE simulation. If the expected comparison
value is zero, the absolute value is given.

Averages over stochastic ensembles are the specialty of xSPDE, which
requires specification of the ensemble size. A hierarchy of ensemble
specifications in three levels allows maximum resource utilization,
so that: 
\[
p.ensembles=[ensembles(1),ensembles(2),ensembles(3)]\,.
\]
The local ensemble, $ensembles\left(1\right)$, gives within-thread
parallelism, allowing vector instruction use for single-core efficiency.
The serial ensemble, $ensembles\left(2\right)$, gives the number
of independent sub-ensembles of trajectories calculated serially.

The parallel ensemble, $ensembles\left(3\right)$, gives multi-core
parallelism, and requires the Matlab parallel toolbox. This improves
speed when there are multiple cores. One should optimally put $ensembles\left(3\right)$
equal to the available number of CPU cores.

The total number of stochastic trajectories or samples is 
\[
ensembles(1)\times ensembles(2)\times ensembles(3)\,.
\]

Either $ensembles(2)$ or $ensembles(3)$ are required if sampling
error-bars are to be calculated, owing to the sub-ensemble averaging
method used in xSPDE to calculate sampling errors accurately.

Two lines are graphed for an upper and lower standard deviation departure
from the mean. This is only plotted if the total number of serial
or parallel ensembles is greater than one, preferably at least 10--20
to give reliable estimates. The sampling error is reasonably accurate,
but may underestimate errors for observe function results that have
highly non-Gaussian trajectory distributions, especially with asymmetries.
These estimates are available for all observables in any dimension.
The two lines generated in the graphs represent $\bar{o}\pm\sigma$,
where $o$ is the mean output, and $\sigma$ is the computed standard
deviation in the mean.

\label{algorithms:sampling-errors}

\subsection{Comparisons: compare}

Every observe function can be accompanied by a comparison function,
with a function handle $compare\{n\}$. This generates a vector of
analytic solutions or experimental data-points which is compared to
the average of the stochastic results. Results are plotted as additional
lines on the two-dimensional graphical outputs, and a summary of comparison
differences is printed.

A cell array of functions is used to obtain comparison results. These
are calculated from the user-specified \textbf{compare\{n\}(p)} handle
where the function argument is the parameter structure p, giving a
extra dashed line on the two-dimensional graphs. Other graphics options
are available as well. These optional comparisons can be input in
all dimensions. When there are error estimates, a chi-squared test
is carried out to determine if the difference is within the expected
step-size and sampling error bars. If the comparison has errors, for
example from experimental data, the chi-squared test will include
the experimental errors.

\subsection{Convergence: xcheck}

The convergence checker, xcheck(checks,p), is designed for use where
there are analytic results available for comparisons. This will automatically
run xSIM a total of checks times, increasing the initial steps by
2 after each run, to reduce the step-size by 2. It then runs xGRAPH
to display the most accurate result. It prints the time-step, the
maximum difference with an input compare and the estimated errors
found at the relevant point.

\subsection*{Exercise}
\begin{itemize}
\item \textbf{Simulate the Kubo oscillator using the file, $Kubocheck.m$,
with xcheck.} 
\end{itemize}
\begin{center}
\doublebox{\begin{minipage}[t]{0.75\columnwidth}%
\texttt{function {[}e{]} = Kubocheck()}

\texttt{p.name = 'Kubo with convergence checks';}

\texttt{p.ensembles = {[}1000,10{]};}

\texttt{p.initial = @(w,p) 1;}

\texttt{p.range = 2;}

\texttt{p.deriv = @(a,xi,p) 1i{*}xi.{*}a;}

\texttt{p.observe\{1\} = @(a,p) real(a(1,:));}

\texttt{p.observe\{2\} = @(a,p) a(1,:).{*}conj(a(1,:));}

\texttt{p.olabels = \{'\textless a\textgreater{} ','\textless{}
a\textasciicircum 2\textgreater{} '\};}

\texttt{p.xlabels = \{'\textbackslash tau'\};}

\texttt{p.compare\{1\} = @(p) exp(-p.t/2);}

\texttt{p.compare\{2\} = @(p) 1;}

\texttt{e = xcheck(2,p);}

\texttt{end}%
\end{minipage}} 
\par\end{center}

\section{Chi-squared estimates}

Chi-squared error estimates are reported in cases that have statistical
sampling errors and comparison functions. These allow estimates of
goodness of fit for probabilities. For $N_{p}$ independent points
graphed or measured, if $O_{i}$ is an observable with measured mean
$\bar{O}_{i}$ and statistical fluctuations $\Delta O_{i},$ one has
that: 
\begin{equation}
\chi^{2}/N_{p}=\frac{1}{N_{p}}\sum_{i}\frac{\left\langle \left[\left(\bar{O}_{i}+\Delta O_{i}\right)-O_{i}^{a}\right]^{2}\right\rangle }{\sigma_{i}^{2}}
\end{equation}

Here $\sigma_{i}^{2}$ is an estimated variance. Provided that $\left\langle \Delta O_{i}^{2}\right\rangle =\sigma_{i}^{2}$
and $\bar{O}_{i}=O_{i}^{a}$, one should obtain the expected result
of $\chi^{2}/N_{p}\approx1$. The exact distribution is known in special
cases, but this requires that all data is independent and has a Gaussian
distribution, which is not the case for stochastic trajectories.

Because of the lack of independence from point to point, these error
sums are not identical to Pearson's original definition of $\chi^{2}$,
and therefore should be used with caution. Nevertheless, the definition
provides a way of evaluating goodness of fit that is useful.

The value of $\sigma_{i}^{2}$ is obtained by including all known
statistical error sources, so 
\begin{equation}
\sigma_{i}^{2}=\sum_{n=1}^{2}\left(\sigma_{i}^{(n)}\right)^{2}.
\end{equation}
where: 
\begin{enumerate}
\item If higher ensembles are used, the estimated $\sigma_{i}^{2}$ includes
numerical sampling errors. 
\item If comparisons have known statistical errors, these are included as
well. 
\end{enumerate}

\subsection{Probability comparisons}

Comparisons of trajectory probabilities and analytic probabilities
do not always result in perfect agreement. This is because the limitations
of memory and simulation time mean that trajectories have to be binned,
which leads to an additional discretization error. Note that xSPDE
approximates the comparison analytic probability of a bin by the central
bin value of the probability, which is the simplest procedure.

To explain this, comparisons of probabilities ought to use the average
probability density over the bin, which is different from the central
value. Suppose one has a comparison distribution $p^{a}\left(x\right)$.
Using Simpson's rule, the average analytic probability density integrated
over a bin size $\Delta x$ is approximately: 
\begin{align}
p_{o}^{a} & =\frac{1}{\Delta x}\int_{x_{0}-\Delta x/2}^{x_{0}+\Delta x/2}p^{a}(x)dx\\
 & \approx\frac{1}{6}\left[4p^{a}(x_{0})+p^{a}\left(x_{0}+\frac{\Delta x}{2}\right)+p^{a}\left(x_{0}-\frac{\Delta x}{2}\right).\right]\nonumber 
\end{align}

This is equivalent to a cubic polynomial fit. It can be used to improve
the analytic binning comparisons. It is especially important for multi-dimensional
comparisons. It results in $9$ distinct terms for two dimensions.
This correction should be inserted manually in the comparison functions.

\subsection{Scaling of \textrm{\textmd{\normalsize{}$\chi^{2}$}}{\normalsize{}
errors}}

Because chi-squared probability tests are sensitive, it helps to understand
how they scale with bin-size. With $N_{s}$ total samples, the estimated
probability $P_{i}$ in a bin with probability density $p\left(\mathbf{a}\right)$
and sampled counts of $N_{i}$ is given by $P_{i}=N_{i}/N_{s}=p_{i}A$
for a bin $b_{i}$ with area $A$, where: 
\begin{equation}
p_{i}=\frac{1}{A}\int_{b_{i}}p\left(\mathbf{a}\right)dA
\end{equation}
The Poissonian variance of the counts in the bin is $\left\langle \Delta N_{i}\right\rangle =\left\langle N_{i}\right\rangle $.
The expected probability variance is therefore 
\begin{equation}
\left\langle \Delta P^{2}\right\rangle =\left\langle \Delta N_{i}^{2}/N_{s}^{2}\right\rangle =\left\langle N_{i}\right\rangle /N_{s}^{2}.
\end{equation}

Let $\left\langle N_{i}\right\rangle =N_{i}^{a}$, the analytic or
expected count number. The expected probability density variance at
a point is therefore 
\begin{equation}
\left\langle \Delta p_{i}^{2}\right\rangle =\left\langle \Delta N_{i}^{2}/A^{2}N_{s}^{2}\right\rangle =N_{i}^{a}/A^{2}N_{s}^{2}=p_{i}^{a}/AN_{s}.
\end{equation}
Here $p_{i}^{a}$ is the analytic or comparison probability density,
and $\left\langle \Delta p_{i}^{2}\right\rangle ^{a}=p_{i}^{a}/AN_{s}$
is the expected analytic variance. The $\chi^{2}$ variable, that
follows the Pearson $\chi^{2}$ distribution, is defined as follows:

\begin{equation}
\chi^{2}/N_{p}=\frac{1}{N_{p}}\sum_{i}\frac{\left\langle \left[p_{i}-p_{i}^{a}\right]^{2}\right\rangle }{\left\langle \Delta p_{i}^{2}\right\rangle }
\end{equation}

Here, $p_{i}^{a}$ is obtained by integrating over the $i$-th probability
bin. It can be estimated by using the central value, $p_{i}^{a}\approx p\left(\mathbf{a}_{i}\right)$,
although cubic interpolation is more precise.

This could lead to a fixed error in the analytic probability density
$p_{i}^{a}$, so $p_{i}^{a}\rightarrow p_{i}^{a}+\epsilon_{i}$, possibly
localized to some fraction of bins $f$ which may change with the
bin size. Suppose, for simplicity, that $\epsilon$ is due to an integration
error in integrating the exact distribution or any other error in
the 'exact' distribution, and it does not change with changes to the
bin area $A$.

From the definition of $\chi^{2}$, if the generated samples have
negligible step-size errors:

\begin{equation}
\chi^{2}/N_{p}=\frac{1}{N_{p}}\sum_{i}\frac{\left\langle \left[\left(p_{i}^{a}+\Delta p_{i}\right)-p_{i}^{a}-\epsilon_{i}\right]^{2}\right\rangle }{\left\langle \Delta p_{i}^{2}\right\rangle }
\end{equation}
For simplicity, if we consider the large sample limit with uniform
probabilities, 
\begin{align}
\chi^{2}/N_{p} & =1+\frac{f\epsilon^{2}}{\left\langle \Delta p^{2}\right\rangle }=1+\frac{f\epsilon^{2}AN_{s}}{p^{a}}
\end{align}

Increasing the bin area $A$ will increase $\chi^{2}/N_{p}$ above
its usual value of 1 by an amount proportional to $A$. This is simply
because smaller bins have less intrinsic accuracy, due to a larger
sampling error. As a result, it is often preferable to use more accurate
probability estimates with larger bins having more counts, since these
are much more sensitive to effects like this.

Often, simulated and comparison graphs may appear identical visually,
but even if they have small errors they may still be very significant.
Such comparison binning errors can be reduced by using cubic spline
interpolations, as explained above.

\section{\label{sec:Error-outputs}Error outputs}

There are six types of data outputs: data, step errors, sampling errors,
comparisons, comparison systematic errors, and comparison random errors.
Summaries of this will appear in the printed outputs, with greater
details if p.verbose\textgreater 0 is chosen. Step errors and sampling
errors, as well as comparison data are stored in the output data arrays.

\subsection{Numerical error outputs}

The last data index $c$ is used to obtain errors and comparisons
in data outputs. To obtain comparison data, a comparison function
is defined for each output function. This can include, for example,
experimental data, experimental errors or exact analytic comparisons
where they are available. 
\begin{enumerate}
\item Means are in $c=1$ data, except if scatters\textgreater 1, which
gives individual trajectories. 
\item If checks\textgreater 0, all the step errors are in $c=2$ data. 
\item If $ensembles(2,3)>1,$ the sampling errors are in $c=3$ data. 
\item Comparison values from compare functions are in $c=4$ data. 
\item Comparison systematic errors can be included in $c=5$ data. 
\item Comparison statistical errors can be included in $c=6$ data. 
\end{enumerate}

\subsection{Graphical error outputs}

These are explained in detail in the xGRAPH reference section. 
\begin{enumerate}
\item Mean values or trajectories are graphed as separate data lines. 
\item Step errors generate graph error bars 
\item Sampling errors are graphed as parallel solid lines 
\item Dashed lines indicate comparison values from compare functions. 
\item Comparison systematic errors give additional error bars 
\item Comparison statistical errors can be included as parallel lines 
\end{enumerate}
Because multiple errors can generate very complex graphs, there is
additional control of error bar generation, explained in the xGRAPH
reference section. One can also obtain difference graphs with comparisons,
which allow errors to be examined more closely, and error bars can
be combined in different ways.

Graphics data is only available for two-dimensional graphs, and is
subject to selection using the axes inputs.

\subsection{Printed error outputs}

Printed error summaries are generated for each data output if p.verbose
\textgreater{} 0. The defaults are root mean square (RMS) and maximum
errors, all normalized. Normalization is by the modulus of the largest
data value in a given output data set, including all lines. If available,
the largest comparison values is used. If it is zero or p.relerr =
0, then no normalization is carried out.

After computing RMS values for each output dataset, again including
all lines and grid-points, a second RMS average is taken over all
the outputs, weighting each total equally, and including all functions
and sequence datasets where there are nonzero errors. Data with no
errors, below a tolerance of $10^{-10}$, are not included in the
mean RMS total errors for each category. 

There is a final RMS average taken over the step, sampling and comparison
totals. This ignores categories with no errors. This printout occurs
even with verbose = 0, to allow a rapid comparison in case there are
unexpected errors, which might require a new simulation with more
steps or random trajectories.

Printed errors are summarized in three main categories 
\begin{enumerate}
\item Discretization or step errors 
\item Sampling errors 
\item Comparison or difference errors 
\end{enumerate}
Comparison data may not be available over an entire lattice. If this
is the case, the axes point selections can be used to restrict the
relevant datas points used for these comparisons. This also applies
to the goodness of fit and error-vector outputs, since they make use
of comparison data where it is available.

\subsection{Goodness of fit ($\chi^{2}$) }

The $\chi^{2}$ statistics are obtained by normalizing the comparison
squared differences by the sum of squares of all the data and comparison
errors at that point. These are summed over every data point with
relevant data, and the number of relevant data points, $k$, is stored.
The ratio of $\chi^{2}/k$ should be order 1 for statistical errors.

These are summarized for each functional data output type, as well
as giving rise to an error total.

\subsection{Error vector output}

When used as a function call in batch mode, the first type of data
returned by xSIM is a six-component error vector. This can be used
for summarizing error data in a batch job, to determine if a specified
error-threshold is reached, to allow an iterative increase in the
number of time-steps or trajectories.

The error-vector components are all RMS averages: 
\begin{enumerate}
\item Total error overall, including step, discretization and comparisons 
\item Total step-size error 
\item Total sampling error 
\item Total comparison error 
\item Total $\chi^{2}/k$ goodness of fit 
\item Simulation elapsed time 
\end{enumerate}

\subsection{Error summaries}

There are six types of data outputs: data, errors, comparisons and
comparison errors. Summaries will appear in the printed outputs, depending
on the verbosity setting. Step errors and sampling errors, as well
as comparison data are stored in output data arrays. These are also
available graphically in two-dimensional graphs.

\newpage{}

\chapter{Simulation parameters and extensibility \label{chap:API-reference}}

\textbf{This chapter gives a reference guide to the xSPDE simulation
parameters and functions.}

\section{Overview}

Simulations carried out by xSPDE are performed by xSIM, then graphed
by xGRAPH. Input parameters come from an \textbf{input} sequence of
parameter structures, while output is saved in a \textbf{data} array,
and optionally in data files. During the simulation, global averages
are calculated for time-step and sampling errors, together with comparisons.
When completed, timing and errors are printed.

\subsection{Output data storage and batch jobs}

An xSPDE session can either run simulations interactively, described
in section \ref{chap:Simulating-an-SDE}, or else using a function
file called a project file. In either case, the Matlab path must include
the xSPDE folder. For generating graphs automatically, the script
input or project function should end with the combined function \textbf{xspde}.

Alternatively, it can be useful to divide xSPDE into its simulation
function, xSIM, and its graphics function, xGRAPH, to allow graphs
to be made at a later time from the simulation. In this case the function\textbf{
$xsim$} runs the simulation, and $xgraph$ makes the graphs. The
two-stage option is better for running batch jobs which you can graph
at a later time.

\subsection{Batch input template}

To create a data file, you must enter the filename when running the
simulation, using the $p.file=filename$ input. A typical xSPDE project
function of this type, where all the data is stored is as follows: 
\begin{center}
\doublebox{\begin{minipage}[t]{0.75\columnwidth}%
\texttt{function e = project.m}

\texttt{p.{[}label1{]} = {[}parameter1{]};}

\texttt{p.{[}label2{]} = ...;}

\texttt{p.file = '{[}myfile{]}.mat'}

\texttt{{[}e,\textasciitilde ,p{]} = xsim(p);}

\texttt{xgraph(p.file);}

\texttt{end}%
\end{minipage}} 
\par\end{center}

Alternatively, for an interactive session one can use the commands:

\doublebox{\begin{minipage}[t]{0.75\columnwidth}%
\texttt{...}

\texttt{{[}e,data,p{]} = xsim(p);}

\texttt{xgraph(data,p);}

\texttt{...}%
\end{minipage}}\texttt{ }

This is specially useful if one wishes to have direct access to the
data and graphics options, with possible multiple trials. When preparing
a project file using the editor, click on the Run arrow above the
editor window to run the job.

A batch job workflow is as follows: 
\begin{itemize}
\item Create the metadata $p$, including a file name, eg, p.file='myfile.mat'. 
\item Change the Matlab directory path to your preferred directory. 
\item Run the simulation with\textbf{ }{[}e,data, p{]} = xsim(p), or just
xsim(p). 
\item Run xgraph(p.file), and the data will be graphed. 
\item Alternatively, xgraph(p.file,p) allows you to change the inputs in
the structure $p$. 
\item Graph outputs can be stored using the p.saveeps=1 and/or p.savefig=1
options. 
\end{itemize}
You can use either Matlab (.mat) or standard HDF5 (.h5) file-types
for data storage. If raw data is generated it will be stored too,
but the files can be large. For stored graphics files the options
are encapsulated postscript (.eps) files or Matlab graphics (.fig)
files, obtained using the graphics input switches p.saveeps and/or
p.savefig.

\section{Input, output and logic}

To explain xSPDE in full detail, 
\begin{itemize}
\item Simulation parameters are stored in the \textbf{input} list. 
\item This describes a sequence of parameter structures, so that \textbf{input=p1,p2,...}. 
\item Each structure \textbf{p1,p2,...} generates an output which is the
input of the next. 
\item The main simulation function is called using \textbf{xsim(input).} 
\item The RMS errors and integration time are returned in the \textbf{error
}vector 
\item Parameters including defaults are returned in the \textbf{output}
cell array. 
\item Averages are recorded sequentially in the \textbf{data} cell array. 
\item Raw trajectory data is optionally stored in the \textbf{raw} cell
array. 
\end{itemize}
The sequence input defines a sequence of individual simulations, with
parameters that specify the simulation functions and give the equations
and observables. If there is only one simulation, just one data structure
is needed, without a cell array. In addition, xSPDE can generates
graphs with its own graphics program, xGRAPH.

\subsection{Applications}

The parameters that xSIM uses are divided into applications for ease
of use. Almost all parameters have default values. The SDE parameters
are common to all applications, but the default values may be changed
in more specialized cases. Defaults are defined through preference
functions that are included in each application folder. Parameters
are shared between the applications where this is meaningful.

Current application folders are as follows: 
\begin{description}
\item [{SDE}] Stochastic differential equation data and methods 
\item [{SPDE}] Partial differential equation extensions and grids. 
\item [{PROJECTIONS}] This is the projective library, used to solve projected
SDE/SPDEs 
\item [{QUANTUM}] Stochastic Schrödinger and master equations, including
logic gates 
\end{description}
All applications use a common definition of cell arrays of integration
variables, and cell arrays of output averages. In all cases, a single
variable, vector or array can be used instead of a multicomponent
cell array. All data outputs are xGRAPH compatible, except for raw
trajectory outputs that need to be further processed if graphs are
needed.

\subsection{User functions}

The xSIM input objects include parameters and functions, with an extensible
architecture. All xSIM functions are modular and replaceable. This
is as easy as just defining a new function handle to replace the default
value.

There are two types of functions: 
\begin{itemize}
\item User functions define equations, and have default values. The defaults
are usually obtained by adding 'x' in front of the name. In the case
of method, the default depends on the problem. 
\item Helper functions usually start with 'x'. In some cases these are defaults
for user functions. They are like the reserved functions in C, Python,
Matlab or Julia. 
\item All arguments in square brackets are optional, but may be needed only
in specific cases. 
\item The last argument, p, is the parameter structure. 
\end{itemize}
For example, to define your own integration function, include in the
xSPDE/xSIM input the line: 
\begin{center}
\doublebox{\begin{minipage}[t]{0.75\columnwidth}%
\texttt{p.method = @Mystep;}%
\end{minipage}} 
\par\end{center}

\begin{center}
Next, include anywhere on your Matlab path the function definition,
for example: 
\par\end{center}

\begin{center}
\doublebox{\begin{minipage}[t]{0.75\columnwidth}%
\texttt{function a = Mystep(a,w,p)}

\texttt{\% a = Mystep(a,w,p) propagates a step my way.}

\texttt{..}

\texttt{a = ...;}

\texttt{end}%
\end{minipage}} 
\par\end{center}

\section{xSIM Parameters \label{sec:xSIM-parameters}}

Simulation parameters are stored in a parameter structure which is
passed to the $xsim$ program. Constants can be included, but must
not be reserved names. Names starting with a capital letter like 'A...'
- except the reserved 'D' for derivatives - are always available.
Globals are incompatible with the Matlab parallel toolbox. Graphics
data is stored for the graphics program to use.

Standard inputs have default values, which are user-modifiable through
the xpreferences function. Defaults can be checked by including the
input $verbose=2$. All the inputs are part of a structure passed
to xSPDE. If a cell array of multiple structures are input, these
are executed in sequence, with the output of the first simulation
passed to the second, then the third, and so on.

Library functions inputs do not have defaults, as these are subject
to change.

\begin{tabular}{|c|c|c|}
\hline 
Label  & Default value  & Description\tabularnewline
\hline 
\hline 
version  & 'xSIM4.xx'  & Current version number\tabularnewline
\hline 
name  & ''  & Simulation name\tabularnewline
\hline 
dimensions  & $1$  & Space-time dimensions\tabularnewline
\hline 
fields  & $1$  & Stochastic field dimensions (or cell)\tabularnewline
\hline 
backfields  & $0$  & Number of backward fields\tabularnewline
\hline 
auxfields  & $0$  & Auxiliary field dimensions\tabularnewline
\hline 
ranges  & $[10,..]$  & Range of coordinates in {[}t,x,y,z,..{]}\tabularnewline
\hline 
origins  & {[}0,..{]}  & Origin of coordinates in {[}t,x,y,z,..{]}\tabularnewline
\hline 
points  & {[}51,...  & Output lattice points in {[}t,x,y,z,..{]}\tabularnewline
\hline 
noises  & fields & Number of noise fields (or cell)\tabularnewline
\hline 
knoises  & 0 & Filtered noise fields (or cell)\tabularnewline
\hline 
unoises  & 0 & Uniform noise fields (or cell)\tabularnewline
\hline 
inrandoms  & noises & Initial random fields (or cell)\tabularnewline
\hline 
krandoms  & noises & Filtered initial randoms (or cell)\tabularnewline
\hline 
urandoms  & noises & Uniform initial randoms (or cell)\tabularnewline
\hline 
ensembles  & {[}1, 1, 1{]}  & Size of {[}vector, serial, parallel{]} ensembles\tabularnewline
\hline 
steps  & 1  & Integration steps per output point\tabularnewline
\hline 
iterations  & 4  & Maximum implicit or midpoint iterations\tabularnewline
\hline 
order  & 1  & Extrapolation order: depends on the method\tabularnewline
\hline 
checks  & {[}1,0,0..{]}  & Check errors for time and space grids: 0 or 1\tabularnewline
\hline 
seed  & 0  & Seed for random number generator\tabularnewline
\hline 
file  & ''  & File-name: 'f.mat' = Matlab, 'f.h5' = HDF5\tabularnewline
\hline 
boundaries\{n\}  & $[0,0;0,0]$  & Boundary: '-1,0,1'=Neum, periodic, Dirichlet boundary.\tabularnewline
\hline 
binranges\{n\}  & \{0,0,...\}  & Observable binning ranges for probabilities\tabularnewline
\hline 
cutoff  & $10^{-12}$  & Global lower data cutoff for chi-squared estimates\tabularnewline
\hline 
cutoffs\{n\}  & $cutoff$  & Lower data cutoff for chi-squared estimates\tabularnewline
\hline 
mincount  & 0  & Lower count cutoff for chi-squared estimates\tabularnewline
\hline 
averages & 1:max(observe) & Optional list of computed observe functions\tabularnewline
\hline 
ipsteps  & 2  & IP transforms per time-step: depends on the method\tabularnewline
\hline 
numberaxis  & 0  & If 1, forces use of numerical axis labels\tabularnewline
\hline 
verbose  & 0  & 0 for brief, 1 for informative, 2 for full output\tabularnewline
\hline 
$A,B,C,\ldots$  & -  & User specified static parameters\tabularnewline
\hline 
transforms  & \{{[}0 0 0 0{]},..\}  & Fourier transforms in {[}t,x,y,z,..{]} per observable\tabularnewline
\hline 
ftransforms  & \{{[}0 0 0 0{]},..\}  & Fourier transforms in {[}t,x,y,z,..{]} per function\tabularnewline
\hline 
rawdata  & 0  & Raw data switch: 1 for raw output\tabularnewline
\hline 
scatters  & \{0,..\}  & Specify to obtain scatter plots, not averages\tabularnewline
\hline 
octave  & 0  & Force octave syntax: 1 for octave\tabularnewline
\hline 
thresholdw  & $0$  & Threshold for weighted simulation breeding\tabularnewline
\hline 
qcproj  & -  & Quadratic projection coefficients\tabularnewline
\hline 
vcproj  & -  & Vector projection coefficients\tabularnewline
\hline 
\end{tabular}

~~

Detailed descriptions are as follows:

\subsection{auxfields\{c\}}
\begin{description}
\item [{Default:}] 0 
\end{description}
These are real or complex auxiliary fields stored at each lattice
point, specified using define. They are useful for input/output spectral
calculations, and can be functions of the noise. Like fields, there
can be several of these, defined in a cell array.
\begin{description}
\item [{Examples:}] p.auxfields = 2 ,p.auxfields = \{1,2\} 
\end{description}

\subsection{averages}
\begin{description}
\item [{Default:}] 1:maximum\_observable 
\end{description}
This optional input gives a vector of average indices to calculate.
Default value is all the observe functions. This is used to suppress
unwanted outputs, for example, while testing an input script.
\begin{description}
\item [{Examples:}] p.averages = {[}2,3,4{]}; 
\end{description}

\subsection{axes\{n\}}
\begin{description}
\item [{Default:}] \{0,0,0,..\} 
\end{description}
Gives the axis points used for comparisons in the $n$-th output function,
in each dimension. For each function, the axes can be individually
specified in each dimension. Each entry value is a vector range for
a particular dimension, for $d$=1,...p.dimensions. Thus, 5 gives
the fifth point only in that dimension, and an input 1:4:41 plots
every fourth point. Zero or negative values are shorthand: -1 generates
a default point at the midpoint, -2 the endpoint, and 0 is the default
value that gives the vector for the every axis point. This data is
also used to control graphics outputs. It can be input separately
for each graph if required. If there are extra space points using
the p.steps input, then the spatial points are expanded internally,
and axes is used to select the output points. If desired, this can
be changed, but the larger number of space points should be taken
into account. 
\begin{description}
\item [{Example:}] p.axes\{4\} = \{1:2:10,0,0,-1\} 
\end{description}

\subsection{backfields\{c\}}
\begin{description}
\item [{Default:}] 0 
\end{description}
The optional input \textbf{\textit{backfields}} is the number of backward-time
stochastic fields that are integrated, as part of the overall vector
of integrated fields components. Requires a forward-backward method
like MPfb. 
\begin{description}
\item [{Example:}] p.backfields = 2 
\end{description}

\subsection{binranges\{n\}}
\begin{description}
\item [{Default:}] \{\} 
\end{description}
Nested cell array, $binranges\{n\}\{m\}$, that defines the probability
plotted for observable $n$. If null or zero, the mean of the observable
is calculated as usual. The second cell index, $m=1,\ldots M$, corresponds
to the line index returned by the corresponding $n$-th observe function.
When nonzero, the probability of the $n$-th observable is calculated
and plotted according to the specified vector of axis points. This
sets extra dimensions in the data, depending on the range of $m$
values, with $[o_{1},o_{2},\ldots o_{K}]$, being the start and end
of each of the bins used to accumulate probabilities. The $k-th$
bin is centered at $(o_{k}+o_{k+1})/2$. In this version of xSPDE,
each bin must have the same width for an observable and line number.
The output is the average probability density versus the (vector)
value of the observable. Hence $M$ extra output dimensions are added
to the generated probability data. 
\begin{description}
\item [{Example:}] p.binranges\{n\}\{1\} = \{-5:0.1:5,-2:0.1:2\} 
\end{description}

\subsection{boundaries\{c,d\}}
\begin{description}
\item [{Default:}] {[}0, 0{]} 
\end{description}
Cell array for type of spatial boundary conditions used, set for each
dimension and field component independently, and used in the equation
solutions. The cell index is $dir=2,3,..$, indicating the dimension.
The boundary conditions are defined as a matrix. The first index is
the field index i and the second index the boundary j, with $j=1$
for the lower and $j=2$ for the upper boundary. The options are $b=-1,0,1$. 
\begin{itemize}
\item The default option, or 0, is periodic. 
\item If -1, Robin/Neumann boundaries are used, with derivatives set to
prescribed values. 
\item If 1, Dirichlet boundaries are used, with fields set to prescribed
values. 
\end{itemize}
In the current code, only default boundaries are available using spectral
(linear) methods. Using arbitrary non-periodic boundaries requires
the use of finite difference derivatives, without the option of an
interaction picture derivative. In such general cases, arbitrary boundary
values are set by boundfun(a,d,p). If the cell index $c$ is omitted,
the first cell is used.
\begin{description}
\item [{Example:}] p.boundaries\{d\} = {[}-1,1;0,0;1,-1{]} 
\end{description}

\subsection{C...}

The starting letter C is reserved to store user-specified constants
and parameters. It is passed to user functions and can be any data.
All inputs --- including C data --- are copied into the stored data
files via the lattice structure p, to give a permanent record of simulation
parameter values along with the output data. 

Most other capital letters are available, unless reserved for specific
method functions.
\begin{description}
\item [{Example:}] p.Constant = 2{*}pi 
\end{description}

\subsection{checks}
\begin{description}
\item [{Default:}] 1 
\end{description}
This defines if a repeat integration is carried out for error-checking
purposes. If p.checks = 0, there is one integration, with no checking
at smaller time-steps. For error checking, set p.checks = 1, which
repeats the calculation at half the time-step --- but with identical
noise --- to obtain error bars. This is the default value, taking
three times longer overall, but with increased accuracy and error-estimates.

Also see the order parameter, below. 
\begin{description}
\item [{Example:}] p.checks = 0 
\end{description}

\subsection{dimensions}
\begin{description}
\item [{Default:}] 1 
\end{description}
This is the space-time dimension for an SPDE. If omitted, dimensions=1,
giving an SDE. It is arbitrary apart from the obvious memory requirements
at large dimensionality. 
\begin{description}
\item [{Example:}] p.dimensions = 4 
\end{description}

\subsection{ensembles}
\begin{description}
\item [{Default:}] {[}1, 1, 1{]} 
\end{description}
Number of independent stochastic trajectories simulated. This has
three levels to maximize efficiency. The first is within-thread parallelism,
allowing vector instructions. The second gives a number of independent
trajectories calculated serially. The third gives multi-core parallelism
and requires the Matlab parallel toolbox. Either p.ensembles(2) or
p.ensembles(3) are required to obtain sampling error-bars. The total
number of stochastic trajectories or samples is $ensembles(1)\times ensembles(2)\times ensembles(3)$.
The second and third ensembles cannot be changed during a sequence
of simulations. 
\begin{description}
\item [{Example:}] p.ensembles = {[}1000,100,10{]} 
\end{description}

\subsection{fields}
\begin{description}
\item [{Default:}] 1 
\end{description}
These are real or complex variables stored at each lattice point that
are the independent variables for integration. The fields are vectors
or arrays that can have any number of components or dimensions. The
fields input is the number of real or complex components initialized
by the initial function and integrated using the deriv derivative.
One array can be used, or cell arrays of multiple named fields. See
the specific method for details.
\begin{description}
\item [{Example:}] p.fields = \{2,{[}3,3{]}\} 
\end{description}

\subsection{file}
\begin{description}
\item [{Default:}] ’ ’ 
\end{description}
Matlab or HDF5 file name for output data. Includes all data and parameter
values, including raw trajectories if $p.rawdata=1$. If not needed
just omit this. A Matlab filename should end in .mat, while an HDF5
file requires the filename to end in .h5. For a sequence of inputs,
the filename should be given in the first structure of the sequence,
and the entire sequence is stored. This cannot be changed for successive
parts of the overall sequence. 
\begin{description}
\item [{Example:}] p.file = 'file-name' 
\end{description}

\subsection{ftransforms\{n\}}
\begin{description}
\item [{Default:}] transforms\{n\} 
\end{description}
Cell array defining the Fourier transform switches for output n. There
is one ftransform vector per output function. The n-th flag indicates
a Fourier transform if it is set to one, and none if set to zero.
The default value is the observe transform switch. If there are more
functions than observe handles, the additional transform switches
default to zero. 

This is used to identify which outputs are from an initially transformed
observe average, so they can be graphed with the correct axis labels.
This is only needed if there are multiple outputs generated from one
transformed observe average. Otherwise, the default is completely
adequate.
\begin{description}
\item [{Example:}] p.ftransforms\{n\} = {[}1,0,0,1{]} 
\end{description}

\subsection{inrandoms\{n\}}
\begin{description}
\item [{Default:}] noises 
\end{description}
This defines the initial random Gaussian fields generated per lattice
point in coordinate and momentum space. Set to zero ($p.inrandoms=0$)
for no random fields. Random fields are delta-correlated in x-space.
This can be a scalar, vector or an array. It can optionally be a cell
array of multiple vectors. The maximum number of inrandom cells equals
the number of field cells plus auxiliary field cells.
\begin{description}
\item [{Example:}] p.inrandoms = 2 
\end{description}

\subsection{ipsteps}
\begin{description}
\item [{Default:}] 1 for Euler, Implicit and RK2; 2 for MP, MPadapt and
RK4; 0 otherwise 
\end{description}
This specifies the number of interaction picture time-steps needed
in an integration time-step. Default values are specified in method.
Can always be changed for custom integration methods. This must be
initialized if a non-standard integration method is used that requires
an interaction picture, and the relevant data isn't returned by method. 
\begin{description}
\item [{Example:}] p.ipsteps = 1 
\end{description}

\subsection{iterations}
\begin{description}
\item [{Default:}] 4 
\end{description}
For iterative algorithms like the implicit midpoint method, the iteration
count is set here, typically around 3-4. Will increase the integration
accuracy if set higher, but it may be better to increase steps if
this is needed. With non-iterated algorithms, this input is not used.
Also used to specify the iterations in projection methods.
\begin{description}
\item [{Example:}] p.iterations = 3 
\end{description}

\subsection{knoises}
\begin{description}
\item [{Default:}] {[} {]} 
\end{description}
This gives the number of Gaussian noises generated per lattice point
in momentum space. This allows use of finite correlation lengths,
by including a frequency filter function that is used to modify the
noise in Fourier-space. The Fourier-space random variance is defined
by the filter function. This takes the noises in Fourier space and
returns a filtered version, which is inverse Fourier transformed before
use. Filtered noises have a finite correlation length. This can be
a scalar, vector or an array. It can optionally be a cell array of
multiple vectors. The maximum number of cells is the number of field
cells plus auxiliary cells. Omitted if it is not input, or null.
\begin{description}
\item [{Example:}] p.knoises = {[}2,4{]}. 
\end{description}

\subsection{krandoms}
\begin{description}
\item [{Default:}] {[} {]} 
\end{description}
This gives the number of initial random Gaussian fields generated
per lattice point in momentum space. The fields are delta-correlated
in momentum space, with a variance modified by the filter function.
This takes initial random fields in Fourier space and returns a filtered
version, which is inverse Fourier transformed before use. This can
be a scalar, vector or an array. It can optionally be a cell array
of multiple vectors. The maximum number of cells equals the number
of field cells plus auxiliary field cells. The filtered random inputs
have a finite correlation length. 
\begin{description}
\item [{Example:}] p.krandoms = 2 
\end{description}

\subsection{name}
\begin{description}
\item [{Default:}] ’ ’ 
\end{description}
Name used to label simulation, usually corresponding to the equation
or problem solved. This can be removed from graphs using headers equal
to a single blank space when running xgraph. 
\begin{description}
\item [{Example:}] p.name = 'your project name' 
\end{description}

\subsection{noises}
\begin{description}
\item [{Default:}] {[} {]}, or p.fields if no other noises are specified
\end{description}
This gives the number of Gaussian noises generated per lattice point,
in coordinate and momentum space, respectively. Set to zero ($p.noises=0$)
for no noises. Noises are delta-correlated in x-space. This can be
a scalar, vector or an array. It can optionally be a cell array of
multiple vectors. The maximum number of cells equals the number of
field cells plus auxiliary field cells. 
\begin{description}
\item [{Example:}] p.noises = \{2,4\}. 
\end{description}

\subsection{order}
\begin{description}
\item [{Default:}] 0 
\end{description}
This is the extrapolation order, which is only used if $p.checks=1$.
The program uses the estimated convergence order to extrapolate to
zero step-size, with reduced errors. If p.order = 0, no extrapolation
is used, which is the most conservative input. The specific default
order returned by the method can be used if one specifies p.order
= -1.

The extrapolation order cannot be changed during a sequence. The default
deterministic orders of the six preset methods used without stochastic
ensembles are: 
\begin{description}
\item [{1}] for Euler and Implicit; 
\item [{2}] for RK2, MP and MPadapt; 
\item [{4}] for RK4. 
\item [{Example:}] p.order = 0 
\end{description}

\subsection{origins}
\begin{description}
\item [{Default:}] {[}0, -p.ranges/2{]} 
\end{description}
This displaces the graph origin for each simulation to a user-defined
value. If omitted, all initial times in a sequence are zero, and the
space origin is set to -p.ranges/2 to give results that are symmetric
about the origin. As an example, for the x-dimension, the problem
is solved on an interval of $x=[O_{2},O_{2}+R_{2}]$, with a default
origin of $-R_{2}/2$, so that $x=[-R_{2}/2,R_{2}/2].$ There is no
cell index used.
\begin{description}
\item [{Example:}] p.origins = {[}0,-20,-20{]} 
\end{description}

\subsection{points\{n\}}
\begin{description}
\item [{Default:}] {[}51, 35, ..., 35{]} 
\end{description}
The rectangular lattice of points plotted for each dimension and field
cell $n$, are defined by a vector giving the number of points in
each dimension. The default values are given as a guide for initial
calculations. Large, high dimensional lattices take more time to integrate.
Increasing points improves graphics resolution and gives better accuracy
in each relevant dimension as well, but requires more memory. 

Cells for $n>$1 can be reduced to singleton dimensions to treat boundary
fields, but the smallest space-steps used in the integrations are
defined relative to points\{1\}. Speed when using spectral methods
is improved when the lattice points are a product of small prime factors.
In order to discretize the problem, the $p_{i}$ lattice points are
fitted into the range $R_{i}$ so that $dx_{i}=R_{i}/(p_{i}-1)$,
ie: 
\begin{equation}
x_{i}=O_{i}+(i-1)dx_{i}\,.
\end{equation}

\begin{description}
\item [{Example:}] p.points = {[}30,40,40{]} 
\end{description}

\subsection{ranges}
\begin{description}
\item [{Default:}] {[}10, 10, ...{]} 
\end{description}
Each lattice dimension has a coordinate range. The default value is
10 in each dimension. In the temporal graphs, the first coordinate
is plotted over $0:p.ranges(1)$. All other coordinates are plotted
over $-p.ranges(n)/2:p.ranges(n)/2$. The starting value in any dimension
can be changed using the origins variable. This is not a cell array,
since the ranges are the same for all field cells (see: points). 
\begin{description}
\item [{Example:}] p.ranges = {[}1, 10{]} 
\end{description}

\subsection{rawdata}
\begin{description}
\item [{Default:}] 0 
\end{description}
Flag for storing raw trajectory data. If this flag is turned on, raw
trajectories are stored in memory. The raw data is returned in function
calls and also written to a file on completion, if a file-name is
included. 
\begin{description}
\item [{Example:}] p.rawdata = 1 
\end{description}

\subsection{relerr}
\begin{description}
\item [{Default:}] 1 
\end{description}
Flag for normalizing the error data. If p.relerr = 1 then all errors
are normalized either by the maximum output value, or else by the
maximum comparison value, if there is one. If p.relerr = 0 then the
absolute error values are output, without normalization.
\begin{description}
\item [{Example:}] p.relerr = 0 
\end{description}

\subsection{rmserr}
\begin{description}
\item [{Default:}] 1 
\end{description}
Flag for averaging the error data. If p.rmserr = 1 then all errors
are calculate as RMS averages over the space and time grid of the
output values. If p.rmserr = 0 , then the error outputs are the maximum
values, not the space-time averages.
\begin{description}
\item [{Example:}] p.rmserr = 0 
\end{description}

\subsection{scatters\{n\}}
\begin{description}
\item [{Default:}] 0 
\end{description}
Cell array that defines the number of scatter trajectories plotted
for observable $n$. If absent or zero, the mean of the observable
is calculated as usual. If nonzero, a set of $s$ observables that
correspond to independent stochastic fields are accumulated, with
no averaging. This cannot be combined with probabilities or with parallel
ensembles. There must be at least s trajectories in ensembles(1),
otherwise the number of stored trajectories is reduced. 
\begin{description}
\item [{Example:}] p.scatters\{n\} = 20 
\end{description}

\subsection{seed}
\begin{description}
\item [{Default:}] 0 
\end{description}
Random noise generation seed, for obtaining reproducible noise sequences.
Set to unique and distinct values for the different parallel ensembles.
Used if $p.noises>0$ or $p.inrandoms>0$. 
\begin{description}
\item [{Example:}] p.seed = 42 
\end{description}

\subsection{steps}
\begin{description}
\item [{Default:}] 1 
\end{description}
Number of internal steps per plotted point. The total number of integration
time-steps in a simulation is therefore p.steps$\times$(p.points(1)-1).
Thus, steps can be increased to improve the accuracy, but gives no
change in graphics resolution. Increasing the steps will give a lower
time-discretization error. If this is a vector, then the number of
internal space points is also increased, with each dimension changed
independently, otherwise only the time-step is changed.
\begin{description}
\item [{Example:}] p.steps = {[}1, 2, ...{]} 
\end{description}

\subsection{transforms\{n\}}
\begin{description}
\item [{Default:}] {[}0,0,..{]} 
\end{description}
Cell array defining the Fourier transforms used for an observable
n. There is one transform vector per observable. The $n$-th flag
indicates a Fourier transform on the $n$-th axis if set to one, starting
with the time axis. The default value is zero, indicating no transform.
The normalization of the Fourier transform is such that the $k=0$
value in momentum space corresponds to the integral over space with
a factor of $1/\sqrt{2\pi}$ in each transformed dimension. The Fourier
transform that is graphed has $k=0$ as the central value. The default
is no Fourier transform. Must be set for any functional transform
of a Fourier observable, to give the correct graph axes. 
\begin{description}
\item [{Example:}] p.transforms\{n\} = {[}1,0,0,1{]} 
\end{description}

\subsection{unoises}
\begin{description}
\item [{Default:}] {[} {]}
\end{description}
This gives the number of uniform noises generated per lattice point,
in coordinate space. This can be a scalar, vector or an array. It
can optionally be a cell array of multiple vectors. The maximum number
of cells equals the number of field cells plus auxiliary field cells.
Omitted if it is not input, or null.
\begin{description}
\item [{Example:}] p.unoises = {[}2,4{]}. 
\end{description}

\subsection{urandoms\{n\}}
\begin{description}
\item [{Default:}] {[} {]} 
\end{description}
This gives the number of initial uniform random fields generated per
lattice point. The fields are uncorrelated in ordinary space. This
can be a scalar, vector or an array. It can also be a cell array of
multiple vectors. The maximum number of cells equals the number of
field cells plus auxiliary field cells. Omitted if it is not input,
or null.
\begin{description}
\item [{Example:}] p.urandoms = \{1,2\}
\end{description}

\subsection{verbose}
\begin{description}
\item [{Default:}] 0 
\end{description}
Print flag for output information while running xSIM. Print options
are: 
\begin{itemize}
\item Brief if verbose = 0: Additionally prints the final, total integration
errors 
\item Informative if verbose = 1: Also prints the individual function RMS
errors and progress indicators 
\item Full if verbose = 2: Prints everything, including the internal parameter
structure data. 
\end{itemize}
In summary, if verbose = 0, most output is suppressed except the final
data, while verbose = 1 displays a progress report, and verbose =
2 additionally generates a readable summary of the parameter input
as a record. 
\begin{description}
\item [{Example:}] p.verbose = 2 
\end{description}

\subsection{version}
\begin{description}
\item [{Default:}] ’xSIM4’ 
\end{description}
Sets the current version number of the simulation program. There is
no need to input this except for project documentation for a customized
version. 
\begin{description}
\item [{Example:}] p.version = 'current version name' 
\end{description}

\subsection{User-defined functions.}

These functions define the stochastic problem. The three most important
ones are given in boldface. These are generated automatically by the
quantum application, to simplify the user inputs and interface. Their
calling arguments, and purpose, are:\\

\begin{tabular}{|c|c|c|}
\hline 
Label  & Arguments  & Purpose\tabularnewline
\hline 
\hline 
\textbf{initial\{n\}}  & \textbf{$(z,p)$}  & \textbf{Functions to initialize fields}\tabularnewline
\hline 
\textbf{deriv\{n\}}  & \textbf{$(a,..w,..p)$}  & \textbf{Total stochastic derivatives}\tabularnewline
\hline 
\textbf{observe\{n\}}  & \textbf{$(a,p)$}  & \textbf{Observable functions}\tabularnewline
\hline 
derivA\{n\}  & $(a,p)$  & Drift derivative term\tabularnewline
\hline 
derivB\{n\}  & $(a,p)$  & Noise derivative term\tabularnewline
\hline 
linear \{n\}  & $(p)$  & Linear derivative function\tabularnewline
\hline 
transfer \{n\}  & $(a0,z,p)$  & Transfer inside a sequence\tabularnewline
\hline 
method  & $(a,w,p)$  & Algorithm defining a time-step{*}\tabularnewline
\hline 
output\{n\}  & $(o,p)$  & Output function\tabularnewline
\hline 
compare\{n\}  & $(p)$  & Function for differences and $\chi^{2}$\tabularnewline
\hline 
define \{n\}  & $(a,w,p)$  & Defines an auxiliary field value\tabularnewline
\hline 
boundfun  & $(a,c,d,p)$  & Boundary function\tabularnewline
\hline 
project  & $(d,a,n,(c,)p)$  & Defines projections\tabularnewline
\hline 
\end{tabular}\\

$^{*}$In all cases except for method, the calling variables are a
list of field and noise arrays. The method function inputs cell arrays
of fields and noises, and a parameter structure. It outputs a field
cell array.

If cell arrays have more than one member, then $(a,w,p)\rightarrow(a,b,c,..w,x,y,..p)$,
where $a,b,c$ are the fields and $w,x,y,$ are the noises. The cell
array of deriv, initial, or transfer functions must be as large as
the cell array of integrated fields.

\subsection{Integrals and derivatives}

For details of the internal integration and differentiation functions
that can be used in deriv, observe and define see section \ref{sec:xSIM-internal-functions}
and sections \ref{sec:Algorithms} and \ref{sec:Algorithms}. All
xSPDE internal functions are capitalized. Note that D1, D2 use finite
differences, DS uses spectral methods. These require a cell index
$c$ to specify the boundary conditions. The functions are:\\

\begin{tabular}{|c|c|c|}
\hline 
Label  & Arguments  & Purpose\tabularnewline
\hline 
\hline 
Ave  & $(a,[av,]p)$  & Averages over a spatial lattice\tabularnewline
\hline 
D1  & $(a,[d,c,ind,]p)$  & First derivative\tabularnewline
\hline 
D2  & $(a,[d,c,ind,]p)$  & Second derivative\tabularnewline
\hline 
DS  & $(a,[n,d,c,ind,]p)$  & Spectral derivative, n-th order\tabularnewline
\hline 
Int  & $(a,[dx\,or\,dk,bounds],p)$  & Integrates over space or momentum\tabularnewline
\hline 
\end{tabular}\\

\begin{itemize}
\item For derivatives, $d$ is the dimension, $c$ the cell, $ind$ the
first index values. 
\item Defaults are $d=1$ , $c=1$, and all indices.
\item For $Int$, one can integrate either with respect to $dx$ or $dk$,
in either ordinary space or momentum space, by changing the second
argument passed to $xint$. 
\item For integration in momentum space, fields that are passed to $Int$
are only transformed if the observe function is used with Fourier
transforms selected using transforms. 
\item For integrating functions like function\{n\} with transforms, the
transform flags transforms\{n\} should be used both for the function
and any observe averages used, to ensure correct graphical output.
Data is always transformed before averaging. 
\end{itemize}

\subsection{Extensible functions}

Extensible functions define the numerical methods used. They use a
similar pattern of (fields..,noises.., parameters). For generality,
these all pass and return cell arrays of fields and noises. 

They all have defaults, and needn't be input in user code when the
default available is used. Any compatible user function can be employed
instead. If required, use the following syntax:
\begin{center}
\inputencoding{latin9}\begin{lstlisting}
p.method = @My_extended_method;
\end{lstlisting}
\inputencoding{utf8}.
\par\end{center}

\begin{center}
The system default values don't usually have to be changed unless
required. %
\begin{tabular}{|c|c|c|c|}
\hline 
Label  & Standard Value  & Arguments  & Purpose\tabularnewline
\hline 
\hline 
method  & $@MP,RK4$  & $(a,w,p)$  & Algorithm defining a time-step\tabularnewline
\hline 
grid  & @xgrid  & $(p)$  & Grid calculator for the lattice\tabularnewline
\hline 
prop  & @xprop  & $(a,p)$  & Interaction picture propagator\tabularnewline
\hline 
propfactor  & @xpropfactor  & $(nc,p)$  & Propagator array calculation\tabularnewline
\hline 
randomgen  & @xrandom  & $(p)$  & Initial random generator\tabularnewline
\hline 
noisegen  & @xnoise  & $(p)$  & Noise generator\tabularnewline
\hline 
\end{tabular}
\par\end{center}

\subsection{SDE methods table}

For details of the internal methods available, see section \ref{sec:xSIM-internal-functions}
and sections \ref{sec:Algorithms} and \ref{sec:Algorithms}. All
xSIM internal method functions are capitalized. Currently only the
MP method is available for jump processes as well as SDEs.

They are:\\

\begin{tabular}{|c|c|c|}
\hline 
Label  & Arguments  & Purpose\tabularnewline
\hline 
\hline 
{*}Euler  & $(a,w,p)$  & Euler algorithm\tabularnewline
\hline 
MP  & $(a,w,p)$  & Midpoint algorithm\tabularnewline
\hline 
MPadapt  & $(a,w,p)$  & Midpoint adaptive algorithm\tabularnewline
\hline 
RK2  & $(a,w,p)$  & Runge-Kutta (2) algorithm\tabularnewline
\hline 
RK4  & $(a,w,p)$  & Runge-Kutta (4) algorithm\tabularnewline
\hline 
{*}{*}Implicit  & $(a,w,p)$  & Implicit or time-reversed\tabularnewline
\hline 
\end{tabular}\\

All standard methods can use the xprop interaction picture propagator,
which also projects onto boundaries. They can all normalize quantum
wave-functions and density matrices if $p.quantum>1$, and can treat
vectors, arrays and cells. The MP and RK methods are intended for
Stratonovich equations.The '{*}' methods are for Ito equations, while
'{*}{*}' methods are for time-reversed Ito equations.

In general, the input and output,'a' is a cell array of fields. The
fields themselves can be scalars, vectors or tensors. These can have
definitions that include spatial indices. However, there also are
special cases:
\begin{itemize}
\item Field tensors can't be integrated in space as well as time, although
this will change in future.
\item A weak, second order Ito method, RKWP21 is available for scalar ODEs,
and extensions are planned.
\item Currently, quantum fields and non-quantum fields can't be mixed, but
this will be extended. 
\end{itemize}

\subsection{Projection methods }

More advanced methods are also available:\\

\begin{tabular}{|c|c|c|}
\hline 
Label  & Arguments  & Purpose\tabularnewline
\hline 
\hline 
Catproj  & $(d,a,n,(c,)p)$  & Catenoid projector\tabularnewline
\hline 
Quadproj  & $(d,a,n,(c,)p)$  & General quadratic projector\tabularnewline
\hline 
Polproj  & $(d,a,n,(c,)p)$  & Diagonal polynomial projector\tabularnewline
\hline 
Enproj  & $(a,w,p)$  & Euler normal projection method\tabularnewline
\hline 
MPproj  & $(a,w,p)$  & Midpoint projection method\tabularnewline
\hline 
MPnproj  & $(a,w,p)$  & Midpoint normal projection method\tabularnewline
\hline 
\end{tabular}\\

\begin{itemize}
\item Projection algorithms with a 'proj' suffix require a project function. 
\item The method functions take cell inputs and outputs.
\item The projectors with 4 arguments take individual arrays as inputs and
outputs.
\item If the cell index c is added, the projector field input a must be
a cell array
\end{itemize}

\section{xSIM exported data }

The following table show how xSPDE output data is stored, which helps
customize and extend the code. There are several different types of
arrays used. Averages are generated from the observe functions, p.observe.
These are modified, if required, by user functions p.output, and exported
as graphics data. The exported data has additional sequence and check
indices.

The internal averages and the exported graphics data are as follows: 
\begin{center}
\begin{tabular}{|c|c|c|}
\hline 
Label & Indices & Description\tabularnewline
\hline 
\hline 
av & $\{n\}(\ell,\mathbf{j})$ & Internal averages\tabularnewline
\hline 
d & $\{s\}\{n\}(\ell,\mathbf{j},c)$ & Graph data\tabularnewline
\hline 
\end{tabular}
\par\end{center}

Here: 
\begin{itemize}
\item $s$ is the sequence index 
\item $n$ is the graph index 
\item $\ell$ is the graphics line index 
\item $j_{1}$ is the time index 
\item $\mathbf{j}=j_{1},j_{2},\dots j_{d}$ is the space-time index 
\item $c$ is the check index 
\end{itemize}

\subsection{Check index uses}

There are multiple uses for the last index, c. It can be omitted if
needed. If present, it stores data for errors and comparisons. This
is indicated by the input parameter field $p.errors>0$, which is
the index of the largest error field. If there are no parameters,
or $p.errors=0,$ there is no error or comparison index. The standard
value that xSIM outputs is $p.errors=3.$

When the check index present, the index values are defined as follows: 
\begin{description}
\item [{$c=1$}] for the average of the n-th output function 
\item [{$c=2$}] for the time-step error, 
\item [{$c=3$}] for the sampling error. 
\item [{$c=4$}] for (optional) comparisons 
\item [{$c=5$}] for (optional) systematic comparison errors 
\item [{$c=6$}] for (optional) statistical comparison errors 
\end{description}
If xGRAPH is used with data from an other source, with no simulation
error fields, but with comparisons, then one simply puts $p.errors=1$,
or if there is just one input error field $p.errors=2$.

\subsection{Comparisons}

For every type of observation in xSIM, the observe function can be
accompanied by a comparison function, compare(p). This generates a
vector of analytic solutions or experimental data which is compared
to the stochastic results. Results are plotted as additional lines
on the two-dimensional graphical outputs, and comparison differences
can be graphed in any dimension.

Comparisons are possible for either moments or probabilities, and
can be input in any number of dimensions. When there are error estimates,
a chi-squared test is carried out to determine if the difference is
within the expected step-size and sampling error bars. If the comparison
has errors, for example from experimental data, the chi-squared test
will include the experimental errors.

Comparison data can be added to the graphics files from any source.
It must match the corresponding space-time lattice or probability
bins that are in the graphed data. Note that the compare functions
are specified during the simulation. The graphics code does not generate
comparison data, as it is dedicated to graphics, not to generating
data.

\section{QUANTUM and PHASE parameters \label{sec:Simulation-parameters-1}}

\subsection{Quantum parameters}

The QUANTUM parameters are identical to the xSIM parameters, with
additional functions and methods. Currently only one cell-array index
is available.

There are three switchable options that can be chosen:

\begin{tabular}{|c|c|c|}
\hline 
Label  & Value  & Purpose\tabularnewline
\hline 
\hline 
quantum  & $0,1,2$  & Use wave-functions (1), or density matrices (2)\tabularnewline
\hline 
sparse  & $0,1$  & Use functional (0) or sparse (1) operators\tabularnewline
\hline 
jump  & $0,1$  & Use continuous (0) or jump (1) methods\tabularnewline
\hline 
\end{tabular}

Wave-functions are stored in a packed, one-dimensional form with sparse
operators, in a packed, two-dimensional form with density matrices,
and in a multidimensional array with functional operators. Sparse
operators usually give faster results, but require greater overall
memory storage for larger Hilbert spaces.

For operators with two mode indices, the second mode index can be
omitted if identical to the first. Hermitian conjugate operators are
returned if the mode index is negative.

\subsection{Bosonic operator table\label{subsec:Bosonic-operator-table}}
\begin{center}
\begin{tabular}{|c|c|c|}
\hline 
Label  & Inputs  & Output(s)\tabularnewline
\hline 
\hline 
a  & $(m,psi)$  & $\hat{a}_{m}\left|\psi\right\rangle $\tabularnewline
\hline 
a2  & $\left(m,psi\right)$  & $\hat{a}_{m}^{2}\left|\psi\right\rangle $\tabularnewline
\hline 
n  & $\left([m_{1}\left(,m_{2}\right)],psi\right)$  & $\hat{a}_{m_{1}}^{\dagger}\hat{a}_{m_{2}}\left|\psi\right\rangle $\tabularnewline
\hline 
\end{tabular}
\par\end{center}

\subsection{Qubit and Pauli spin operators}
\begin{center}
\begin{tabular}{|c|c|c|}
\hline 
Label  & Inputs  & Output(s)\tabularnewline
\hline 
\hline 
sx  & $(m,psi)$  & $\hat{\sigma}_{m}^{x}\left|\psi\right\rangle $\tabularnewline
\hline 
sy  & $(m,psi)$  & $\hat{\sigma}_{m}^{y}\left|\psi\right\rangle $\tabularnewline
\hline 
sz  & $(m,psi)$  & $\hat{\sigma}_{m}^{z}\left|\psi\right\rangle $\tabularnewline
\hline 
sx2  & $\left([m_{1}\left(,m_{2}\right)],psi,\right)$  & $\hat{\sigma}_{m_{1}}^{x}\hat{\sigma}_{m_{2}}^{x}\left|\psi\right\rangle $\tabularnewline
\hline 
sy2  & $\left([m_{1}\left(,m_{2}\right)],psi\right)$  & $\hat{\sigma}_{m_{1}}^{y}\hat{\sigma}_{m_{2}}^{y}\left|\psi\right\rangle $\tabularnewline
\hline 
sz2  & $\left([m_{1}\left(,m_{2}\right)],psi\right)$  & $\hat{\sigma}_{m_{1}}^{z}\hat{\sigma}_{m_{2}}^{z}\left|\psi\right\rangle $\tabularnewline
\hline 
\end{tabular}
\par\end{center}

\subsection{Quantum logic gate operators}
\begin{center}
\begin{tabular}{|c|c|c|}
\hline 
Label & Inputs & Output(s)\tabularnewline
\hline 
\hline 
ha & $(m,psi)$ & $h\left|\psi\right\rangle $\tabularnewline
\hline 
ph & $(m,psi)$ & $p\left|\psi\right\rangle $\tabularnewline
\hline 
p8 & $(m,psi)$ & $t\left|\psi\right\rangle $\tabularnewline
\hline 
cx & $\left([m_{1}\left(,m_{2}\right)],psi,\right)$ & $cx\left|\psi\right\rangle $\tabularnewline
\hline 
\end{tabular}
\par\end{center}

\subsection{PHASE parameters \label{sec:Simulation-parameters-1-1}}

The PHASE parameters are identical to the xSIM parameters, with additional
functions and methods. Currently only one cell-array index is available.

There are three switchable phase-space options that can be chosen.
The other parameters define types of Gaussian inputs, and measurements.

\begin{tabular}{|c|c|c|}
\hline 
Label & Value & Purpose\tabularnewline
\hline 
\hline 
phase & $1,2,3$ & Use +P, Wigner or Q ordering\tabularnewline
\hline 
sqz & $real$vector & Squeezing vector\tabularnewline
\hline 
alpha & complex vector & Coherent input amplitude\tabularnewline
\hline 
matrix & complex matrix function & Returns transmission matrix\tabularnewline
\hline 
tr & $real$vector & Amplitude transmission vector\tabularnewline
\hline 
thermal & $real$vector & Thermal fraction of input\tabularnewline
\hline 
\end{tabular}

\section{User function reference \label{sec:xSIM-functions}}

The following user-defined function inputs define the differential
equation that is solved. They are specified in an xSPDE/xSIM input
file using p.(fun) = @(Myfun). They can be inline or externally defined
functions. Externally defined functions must be in the same file as
the input parameters, or on the execution path.

\subsection{boundfun(a, c, d, p)}
\begin{description}
\item [{Default:}] xboundfun() 
\end{description}
The boundary function boundfun(a,c,d,p) is called for specified boundary
conditions for field cell $c$ in the $d$-th dimension. This returns
the boundary values used for the fields or their first derivatives
in space dimension $d>1$, as an array indexed as $b(f,\mathbf{i},e)$
in the standard way. Here $f$ is the field index, $\mathbf{i}\equiv\left[j_{2},\dots j_{d}\right]$
are the space indices, and $e$ is the ensemble index.

Only two values are needed for $j_{d}$, which is the index of the
dimension whose boundary values are specified. These are $j_{d}=1,2$,
for the lower and upper boundary values, which are either field values
or derivatives. Boundary values may be constant or a function of the
fields $a$ and space-time $t,\mathbf{x}$.

If boundary values have stochastic values which are calculated only
once, they must be initialized. To allow for this, boundfun(a,c,d,p)
is initially called with time $t=origins(1)-1$, and with the input
field a set to random values from randomgen, which are independent
of those that initialize the field at $t=origins(1)$.

They are reproducible for different $check$ cycles, to allow noise-independent
error-checking. The initial results for the boundaries are stored
in an array boundval\{c,d\} for later use by boundfun.

The default boundary value is zero, or equal to boundval if it is
specified initially. It is automatically set by the default boundary
function xboundfun(a,c,d,p).

\subsection{compare\{n\}(p)}
\begin{description}
\item [{Default:}] compare\{n\}= {[}{]} 
\end{description}
This is for comparisons to experimental or analytic data. The output
is an array with $d+2$ dimensions. The first dimension is the line
index, the next $d$ dimensions are time and space, while the last
index is an error index. This can have up to two additional entries
for systematic and/or statistical error bars in the comparison data,
from analytic or experimental results. Error-bars are optional if
not available.

\subsection{define(a\{:\},w\{:\}, p)}
\begin{description}
\item [{Default:}] xdefine() 
\end{description}
Calculates a list of auxiliary fields, which are combinations of fields
and noises. They can be accessed in observe functions as part of the
input cell list, after the propagating fields. These are used in spectral
calculations to access the noise fields, which are needed in quantum
input-output calculations. The default, xdefine(), sets the auxiliary
fields to zero.

\subsection{deriv(a\{:\},w\{:\},p)}
\begin{description}
\item [{Default:}] deriv()= 0 
\end{description}
This defines the stochastic time derivative, given the current field
cells $a$, delta-correlated noise terms w, and parameters $p$. It
is defined explicitly in (\ref{eq:deriv_without_linear_term}). This
is the right-hand-side of (\ref{eq:SDE}) or (\ref{eq:spde}), without
the linear term if it is specified separately. In the case of multiple
cell calculations, this user defined function must return a full list
of all propagating derivative terms in the form of {[}da\{1\}, da\{2\},..{]}.

\subsection{firstfb(a0,nc,p)}
\begin{description}
\item [{Default:}] xfirstfb() 
\end{description}
Returns the zero-th order field estimates in a forward-backward iteration.
Here $nc$ is the time-step check index. This is needed because the
number of time-points to be initialized depends on $nc$. The default
function is xfirstfb, which sets each field in either direction equal
to its initial value at the time boundaries, given by $a0$. Other
estimates may give faster convergence.

\subsection{grid(p)}
\begin{description}
\item [{default}] xgrid 
\end{description}
Calculates the spatial grid for specialized purposes like non-uniform
grids. The default, xgrid, returns a homogeneous rectangular grid
in both ordinary and momentum space, as part of the parameter structure
p. Grids are removed from stored data.

\subsection{initial\{c\}(rv, p)}
\begin{description}
\item [{Default:}] xinitial() 
\end{description}
This is used to initialize each field cell integration in time. It
is a user-defined function which can involve random numbers for an
initial probability distribution. This creates a stochastic field
on the spatial lattice\textbf{.} The returned first dimension is p.fields(1).
The initial Gaussian random field variable, rv, has unit variance
if dimension is 1 or else is delta-correlated in space, with variance
$1/p.dv=1/(dx_{2}...dx_{d}))$ for $d$ space-time dimensions. If
inrandoms is given in the input parameter structure, rv has a first
dimension of inrandoms(1) + inrandoms(2). If not specified, the default
for inrandoms is noises. The default function is xinitial, which sets
fields to zero. The function can be either a cell array of initial
functions, or a single function if there is just one cell.

\subsection{linear\{c\}(p)}
\begin{description}
\item [{Default:}] xlinear() 
\end{description}
A cell array of user-defined linear response functions. It is a vector
for an SDE or ODE. For an SPDE or PDE, it includes transverse derivatives
in space, returning linear coefficients $L$ in FFT/DST/DCT space,
which are assumed diagonal in the field index. These are functions
of differential terms Dx, Dy, Dz, which correspond to $\partial/\partial x$,
$\partial/\partial y$, $\partial/\partial z$, respectively. Each
component has a dimension the same as the coordinate lattice. For
axes that are numbered, use D\{2\}, D\{3\} etc. The default, xlinear,
sets L to zero. The function can be either a cell array of linear
functions, or a single function if there is just one cell.

\subsection{method(a, w, p)}
\begin{description}
\item [{Default:}] @MP (stochastic); @RK4 (deterministic) 
\end{description}
Gives the integration method for the field cell array a, noise cell
array w, parameters p. It returns the new field cell array. It uses
the current reduced step in time p.dtr and current time p.t. This
function can be set to any of the predefined stochastic integration
routines provided with xSPDE, described in the Algorithms section.
User-written functions can also be used. The default deterministic
method, RK4, is a fourth-order interaction picture Runge-Kutta. The
default stochastic method, MP, is an interaction picture midpoint
integrator which is used if $ensembles$ is not {[}1,1,1{]}.

\subsection{nfilter\{n\} (w,p)}
\begin{description}
\item [{Default:}] xnfilter() 
\end{description}
Returns the $n-th$momentum-space filter function for the propagation
noise terms in momentum-space. Each component has an array dimension
the same as the random noises in momentum space, that is, the return
dimension is {[}knoises\{n\}, d.lattice{]}.

\subsection{noisegen(p)}
\begin{description}
\item [{Default:}] xnoisegen(p) 
\end{description}
Generates arrays of noise terms for each point in time. The default,
xnoisegen() returns noises Gaussian real noises that are delta-correlated
in time and space, and knoises Gaussian real noises that are delta-correlated
in time and momentum space, unless nfilter is used to modify momentum
space correlations.

\subsection{observe\{n\}(a..., p)}
\begin{description}
\item [{Default:}] xobserve\{1\}=@(a,p) a 
\end{description}
Cell array of function handles that take the current field(s) and
returns an observable o. Note the braces for cell arrays! One can
input these as p.observe\{n\} = @(a,p) o(a,p). An omitted function
less than the maximum index is replaced by the default. This is all
the real field amplitudes in the first cell, with tensor fields and
field arrays reshaped into vectors.

\subsection{output\{n\}(o,p)}
\begin{description}
\item [{Default:}] @(o,p) o\{n\} 
\end{description}
This is a user-defined cell array of output functions of the observe
results after averaging over ensembles(1), possibly involving combinations
of several observed averages. The input to the n-th output function
is the cell array of all averages, and the output is the data for
the n-th graph. This function is compatible with all error estimates.
The default values generate all the observe averages that are in the
data.

The output data format of the output functions is an array with $d+1$
dimensions. The first dimension is the line index, the next $d$ dimensions
are time and space.

The xSIM program augments the outputs with columns of errors and comparison
data, if available, before graphing.

\subsection{prop(a, p)}
\begin{description}
\item [{Default:}] xprop() 
\end{description}
Returns the fields propagated for one step in the interaction picture,
given an initial field a, using the propagator array. The time-step
used in propagator depends on the input time-step, the error-checking
and the algorithm. The default, xprop, takes a Fourier transform of
$\mathbf{a}$, multiplies by propfactor to propagate in time, then
takes an inverse Fourier transform.

\subsection{propfactor(nc, p)}
\begin{description}
\item [{Default:}] xpropfactor() 
\end{description}
Returns the interaction picture propagator used by the prop function.
The time propagated is a fraction of the current integration time-step,
dt. It is equal to $1/ipsteps$ of the integration time-step. It uses
data from the $\mathbf{linear}$ function to calculate this.

\subsection{randomgen(p)}
\begin{description}
\item [{Default:}] xrandom() 
\end{description}
Generates a cell array of initial random fields v to initialize the
fields simulated. The default, $\mathbf{xrandomgen}$, returns Gaussian
real fields that have inrandoms\{n\}(1) components delta-correlated
in space, with inrandoms\{n\}(2) delta-correlated in momentum space.
The default uses a user-defined cell array, rfilter\{n\}, of filter
functions, to modify correlations in momentum space, if specified.

\subsection{rfilter\{n\}(w, p)}
\begin{description}
\item [{Default:}] xrfilter() 
\end{description}
Returns the momentum-space filter function for the momentum-space
random terms. Each component has an array dimension the same as the
input random fields in momentum space, that is, the return dimension
of cell $n$ is {[}inrandoms\{n\}(2),... nlattice{]}.

\subsection{transfer\{c\}(a0\{:\},v\{:\},p)}
\begin{description}
\item [{Default:}] xtransfer() 
\end{description}
This function initializes sequential simulations,\textbf{ }where the
previous field a0 can be used as an input to the next stage in the
integration sequence. The default, xtransfer(), takes each output,
a0\{c\} of the previous simulation to initialize the field $a\{c\}$.
Otherwise, this function is identical to initial(). 

\section{System function reference\label{sec:xSIM-internal-functions}}

The following xSIM predefined system functions are available to define
the differential equations and averages. They all start with a capital
letter. Algorithms are documented in section \ref{sec:Algorithms}.
Fields can be differentiated or integrated only in space, observables
in space or time.

\subsection{Ave(o{[}, av {]}, p)}

This function takes a field or observable and returns an average over
one or more dimensions. The input includes an optional averaging switch
\textit{av}. If $av(j)>0$, an average is taken over dimension j.
If the av vector is omitted, the average is taken over all space directions.

\subsection{Bin(o{[}, dx {]}, p)}

The $Bin$ function takes a field o and returns probabilities on space
axes that are defined by a vector dx. This allows binning of position
probabilities if the observable is a mean position that is plotted
on an axis. If j is the first index with $dx(j)>0$, the binning is
taken over dimension j. The results returned are the probability of
o in the bin, normalized by $1/dx\left(j\right)$. If the input array
is Fourier transformed, by using the transforms attribute in the observe
function, then one must set $dx(j)=p.dk(j)$ for transformed dimensions
j. If the dx vector is omitted, or a scalar dx is used, the binning
is over the first space direction.

\subsection{D1(a{[}, d,c,ind{]}, p)}

Takes a scalar or vector field a and returns a derivative in dimension
d using central finite differences. Set \textit{d = 2} for an x-derivative,
\textit{d = 3} for a y-derivative, etc. The default value is $d=2$.
The cell index is the index of the cell that is differentiated, which
is needed when there is more than one cell with different boundary
types. An index list $ind$ can be included to take a derivative of
one component or a specified list. If omitted, derivatives of all
components are returned. 

Boundary values are stored in p.boundval, and are needed for Neumann/Robin
boundaries. Hence, in multi-cell computations, if $a$ is from a cell
index $c>1$, the cell index $c$ must be included to identify which
boundary value to use. If there are Neumann/Robin boundaries, the
entire field a must be input unless all boundary values are the same.

For other types of boundaries, the cell index is not needed, and D1
can differentiate a single field component without having to identify
the component. The method is of second order in the space step. It
is used in the deriv, observe and output functions, with automatic
compensation for the presence/absence of a time index.

For Dirichlet boundaries, the derivatives are ambiguous at the boundaries,
and a periodic derivative is returned. This is not needed for time-evolution,
as the boundary value overrides it.

\subsection{D2(a{[}, d,c,ind{]}, p)}

This takes a scalar or vector field a and returns the second derivative
in dimension d using central finite differences. Other properties
are the same as D1(). The method is of second order in the space step,
except at a Neumann/Robin boundary, where the boundary result is of
first order.

\subsection{DS(a{[},order,d,c,ind{]}, p)}

This takes a scalar or vector field a and returns the spectral derivative
of a given order in dimension d. Other properties are the same as
D1(). The method uses Fourier, sine or cosine transforms to compute
derivatives on an equally spaced grid, depending on the boundary conditions
used. For Fourier transforms, with periodic boundaries, any integer
order can be used. With non-periodic boundaries, only even orders
are available, and use is restricted to the midpoint (MP) method.

\subsection{Int(o{[}, dx, bounds,c{]}, p)}

This function takes any vector or scalar field or observable and returns
a space integral over selected dimensions with vector measure dx.
If $dx(j)>0$, dimension j is integrated. Time integrals are only
possible for observables. Space dimensions are labelled from j = 2,3,...dimensions.
To integrate over the lattice, set dx = p.dx, otherwise set dx(j)
= p.dx(j) for integrated dimensions and dx(j) = 0 for non-integrated
dimensions.

If the input array is Fourier transformed by using the p.transforms
attribute, \textbf{one must set dx(j) = p.dk(j) for transformed dimensions
j, to get correct results}. If the dx vector is omitted, the integral
is over all available space dimensions, assuming no Fourier transforms. 

The optional input bounds is an array of size {[}p.dimensions,2{]},
which specifies lower and upper integration bounds in each direction.
This is only available if dx or dk is input. If omitted, integration
is over the whole domain. The optional input switch $c$is used to
indicate that the input should be reshaped to the implicit shape of
cell c.

\section{Internal parameters}

Knowing the details of array indexing inside xSPDE isn't usually necessary.
Yet it becomes important if you want to write your own functions to
extend xSPDE, interface xSPDE with other functions, or read and write
xSPDE data files with external programs. It also helps to understand
how the program works.

\subsection{Array tables}

There are two main internal xSPDE arrays:, fields labelled $a$ and
output data labelled $d$. The fields contain stochastic variables,
the data contains the averaged outputs and errors estimates.

Important array and index definitions are:\\

\begin{center}
\begin{tabular}{|c|c|c|}
\hline 
Label  & Indices  & Description\tabularnewline
\hline 
\hline 
$a$  & $\{n_{1}\}\left[f,\mathbf{i},e_{1}\right]$  & Stochastic field array\tabularnewline
\hline 
v  & $\{n_{1}\}\left[m_{1},\mathbf{i},e_{1}\right]$  & Initial random variable array\tabularnewline
\hline 
w  & $\{n_{1}\}\left[m_{2},\mathbf{i},e_{1}\right]$  & Noise field array\tabularnewline
\hline 
r\{2\},k\{2\}....  & $(1,\mathbf{i},1)$  & Numbered space/momentum coordinates\tabularnewline
\hline 
x,y,z,kx,ky,kz  & $(1,\mathbf{i},1)$  & Labelled space/momentum coordinates\tabularnewline
\hline 
$o$  & $\{n_{2}\}(\ell,\mathbf{j})$  & Cell array of all observed averages\tabularnewline
\hline 
$data$  & $\{s\}\{n_{2}\}(\ell,\mathbf{j},c)$  & Cell array of output data with checks\tabularnewline
\hline 
$raw$  & $\{s,c,h\}\{n_{1}\}(f,\mathbf{j},e_{1})$  & Raw trajectories\tabularnewline
\hline 
$points$  & $\left[pt_{1},pt_{2}\ldots pt_{d}\right]$  & Vector of lattice sizes\tabularnewline
\hline 
$ensembles$  & $\left[h_{1},h_{2},h_{3}\right]$  & Vector of ensemble sizes\tabularnewline
\hline 
\end{tabular}
\par\end{center}

Here: 
\begin{itemize}
\item $s$ is the sequence index 
\item $f$ is the field internal index 
\item $\mathbf{i}$ is the space index 
\item $e_{1}$ is the first ensemble index 
\item $c$ is the check index for errors and comparisons 
\item $\mathbf{m}$ is the random or noise index 
\item $\mathbf{j}=[j_{1},\mathbf{i}]$ is the space-time index 
\item $n_{1}$ is the cell index of a computational field or noise 
\item $n_{2}$ is the cell index of an observe and/or output function 
\item $\ell$ is the line index of an output 
\item $e$ is the high-level ensemble index (combines $e_{2},e_{3}$ indices) 
\end{itemize}
When fields are passed to observe or to raw outputs, the defined auxiliary
fields are included as well. Apart from the internal field dimension(s),
the common dimensionality for internal arrays used in computations
is $[d.space,ensembles(1)]$. The number of points in $d.space$ can
be changed depending on the cell index, for different integrated fields.

\subsection{Simulation data in xSIM}

In xSIM, the space-time dimension $d$ is unlimited. xGRAPH can plot
up to three chosen axes. All fields are stored in cell arrays that
contain real or complex numerical arrays. Average results are stored
are stored in cell arrays of real numerical arrays, usually of rank
$2+d$, although this can change in special cases like the plot of
a probability, which requires extra axes.

The array index ordering in xSPDE integrated fields is $\{n_{1}\}(\bm{f},\mathbf{i},e_{1})$,
where: 
\begin{itemize}
\item The internal cell index $n_{1}$ labels distinct integrated variables. 
\item The internal field index $\bm{f}$, is a field index or indices, not
including auxiliary fields 
\item The next $d-1$ indices are $\mathbf{i}$, which is a space index
with no time index. 
\item The last is an ensemble index $e_{1}$, to store low-level parallel
trajectories. 
\end{itemize}
The array index ordering in graphical averaged data is $(\ell,\mathbf{j},c)$where: 
\begin{itemize}
\item The first index is a line index $\ell$. 
\item The next $d$ indices are $\mathbf{j}=\left[j_{1},\ldots j_{d}\right]=\left[j,\mathbf{i}\right]$,
for time and space. 
\item The last is a check index $c$, for comparisons and errors. 
\end{itemize}
Stored data uses heterogenous cell arrays to package numerical arrays
with additional high level indices. The first cell index is the sequence
index, $s$. Inside each sequence, data cell arrays have a graph index
$n$. This distinguishes the different averages generated for output
graphs and data. Raw data has cell indices for the sequence, time-step
and high level ensembles.

In summary, the xSPDE internal arrays are as follows: 
\begin{itemize}
\item \textbf{Field} arrays $a\{n_{1}\}(\bm{f},\mathbf{i},e)$ - these have
a field index, a space index and low-level ensemble index $e$. 
\item \textbf{Auxiliary} arrays $a_{x}\{n_{1}\}(\bm{f},\mathbf{i},e)$ -
these are appended to the field cells for raw data and observables. 
\item \textbf{Random} and \textbf{noise} arrays $w\{n_{1}\}(m,\mathbf{i},e)$
- these are initial random fields or noise fields. The first index
may have a different range to the field index. 
\item \textbf{Coordinate} arrays $x(1,\mathbf{i})$ - these contain the
coordinates at grid-points, with labels $x,y,z$, and $j_{1}=1$.
Numeric labels $x\{l\}$ are used for $d>4$, where $l=2,\ldots d$.
The same sizes are used for: 
\begin{itemize}
\item momentum coordinates $kx,ky,kz$ (alternatively $k\{2\},k\{3\},\ldots$) 
\item spectral derivative arrays $Dx,Dy,Dz$ (alternatively $D\{2\},D\{3\},\ldots$)
. 
\end{itemize}
\item \textbf{Raw} \textbf{data} arrays $r\{s,c,e\}\{n_{1}\}(\bm{f},\mathbf{j},e_{1})$
- these are cell arrays of generated trajectories, including integrated
and defined field values. They are optional, as they use large amounts
of memory. These are saved in cell arrays with indices $s$ for the
sequence, $c$ for the time-step error-check and $h$ for high level
ensemble index. The cell indices are: 
\begin{itemize}
\item $s=1,\ldots S$ for the sequence number, 
\item $c=1,2,3..$ for the error-checking step used: first fine, then coarse
in each dimension checked. 
\item $e=1,\ldots ensembles(2)*ensembles(3)$ for a high level parallel
and serial ensemble index. 
\end{itemize}
\item \textbf{Observe} arrays $o\{n_{2}\}(\ell,\mathbf{j})$ - these are
generated in xSIM by the observe functions, then used to store generated
average data at all time points. The cell index $n$ is the observe
index, which indexes overs the observe functions.The internal index
$\ell$ is a line index generated by an observe function. 
\item \textbf{Data} arrays $d\{s\}\{n_{2}\}(\ell,\mathbf{j},c))$ - these
store the final results. The $\mathbf{j}$ indices may be Fourier
indices if transforms are specified, and may include extra axes for
probabilities. \\
 Check indices are used for error estimates and comparisons, where
$c=1$ for the average, $c=2$ for the total step error, and $c=3$
for the sampling error. The total step error is a composite of all
step errors that are checked.\\
 If there is comparison data, it uses $c=4$ up to $c=6$, to allow
for any error bars. The output data uses cell indices $\{s\}$ for
the sequence index, and $\{n_{2}\}$ for the data index. This has
a default of the index of the observe function. \\
 If this data is modified by an xSIM output function, the data index
equals the relevant output function index. 
\end{itemize}

\subsection{Internal parameter table\label{sec:Table-of-parameters}}

The internal parameter structures in xSPDE are available to the user
if required. Internally, all xSPDE parameters are stored in the parameter
structures passed to functions. This includes the data given above
from the input structures. In addition, it includes the computed parameters
given below, which includes internal array dimensions.

When accessing these in a function, prefix them by the structure label,
usually $p.$ in the examples, eg, $p.t$. Where the space points
change with the cell, the labels below refer to properties of the
first cell index. Fields, noises and random fields have a cell index
when stored internally, but are passed to user functions as arrays,
in order of the index. Spatial cell volumes are reduced if there are
multiple spatial steps for increased spatial resolution. 

Data in $k-$space is stored in two alternative lattices, each having
their own axis vectors. The propagation grid is used while propagating,
and is compatible with numerical FFT conventions where the first index
value is $k=0$. The graphics grid is centered around $k=0$, and
is used for graphics and data storage, following scientific conventions.

For more than four total dimensions, the spatial grid, momentum grid
and derivative grid notation of $t,x,y,z$, $\omega,kx,ky,kz$ and
$Dx,Dy,Dz$ is changed to use numerical labels that correspond to
the dimension numbers, i.e., $D\{2\},\dots D\{d\}$, $r\{1\},\dots r\{d\}$,
$k\{1\},\dots k\{d\}$.

Numeric dimension labeling can also be used even for lower dimensionality
if preferred.

\begin{tabular}{|c|c|c|c|}
\hline 
Label  & Type  & Typical value  & Description\tabularnewline
\hline 
\hline 
$t,x,y,z$  & array  & -  & Space-time grid of $t,x,y,z$\tabularnewline
\hline 
$\omega,kx,ky,kz$  & array  & -  & Frequency-momentum grid of $k_{x},k_{y},k_{z}$\tabularnewline
\hline 
$Dx,Dy,Dz$  & array  & -  & Derivative grid of $D_{x},D_{y},D_{z}$\tabularnewline
\hline 
$r\{1\},\ldots r\{d\}$  & array  & -  & Space-time grid of $r_{1},\ldots r_{d}$\tabularnewline
\hline 
$k\{1\},\ldots k\{d\}$  & array  & -  & Graphics momentum grid of $k_{1},\ldots k_{d}$\tabularnewline
\hline 
$D\{2\},\ldots D\{d\}$  & array  & -  & Derivative grid of $D_{2},\ldots D_{d}$\tabularnewline
\hline 
dx  & vector  & {[}0.2,..{]}  & Steps in $[t,x,y,z]$\tabularnewline
\hline 
dk  & vector  & {[}0.61,....{]}  & Steps in $[\omega,k_{x},k_{y},k_{z}]$\tabularnewline
\hline 
dt  & double  & 0.2000  & Output time-step\tabularnewline
\hline 
dtr  & double  & 0.1000  & Computational time-step\tabularnewline
\hline 
v  & real  & 1  & Spatial lattice volume\tabularnewline
\hline 
kv  & real  & 1  & Momentum lattice volume\tabularnewline
\hline 
dv  & real  & 1  & Spatial cell volume\tabularnewline
\hline 
dkv  & real  & 1  & Momentum cell volume\tabularnewline
\hline 
xc\{d\}  & cells of vectors  & {[}-5,... 5{]}  & Coordinate axes in $t,x,y,z$\tabularnewline
\hline 
kc\{d\}  & cells of vectors  & {[}-5,..5{]}  & Momentum axes in$[\omega,k_{x},k_{y},k_{z}]$\tabularnewline
\hline 
nspace  & integer  & 35  & Number of spatial lattice points\tabularnewline
\hline 
inrandoms  & vector  & \{2\}  & Initial random fields per cell\tabularnewline
\hline 
krandoms  & vector  & \{2\}  & Initial random fields in kspace per cell\tabularnewline
\hline 
noises  & vector  & \{2\}  & Number of noise fields per cell\tabularnewline
\hline 
knoises  & vector  & \{2\}  & Noise fields in kspace per cell\tabularnewline
\hline 
\end{tabular}

\section{Examples, testing and structure\label{sec:Examples}}

Additional examples are given in the Examples folder distributed with
xSPDE. These can all be run using Batchtest.m, which has a typical
runtime of $50-100s$, and runs $35$ different case studies. This
shows your distribution is intact. All the graphs produced are deleted.
It lists the different examples available, some of which are given
below.

The batch testing code will run each different example sequentially.
It prints the RMS relative errors for the step-size, sampling and
difference error, as well as the total RMS error combining all three,
the chi-square error normalized by the number of points, and the timing.
The geometric mean of the $35$ RMS total errors is computed as a
benchmark.

As Matlab random noise is reproducible with a fixed seed, this geometric
mean error is fixed. The total is printed to more than six decimals
for verification, and an error is indicated if it varies by a factor
of more than $\pm10^{-3}$. Due to different random noise algorithms
used in some Octave versions, the Octave error may vary by up to $\pm20\%$.

\subsection{xSPDE structure}

The control program, $xspde$, calls the xsim integration and xgraph
graphics functions successively

\[
\mathbf{xspde}\rightarrow\begin{cases}
\begin{array}{c}
\mathbf{xsim}\,\,(simulations)\\
\mathbf{xgraph}\,\,(graphics)
\end{array}\end{cases}
\]

For convergence checking, a useful alternative to xspde which repeats
the calculation checks times while halving the time-step each time,
and reports the resulting errors for averaged observables, is: 
\begin{itemize}
\item xcheck (checks,p) 
\end{itemize}

\subsection{xSIM}

The integration function, $xsim$, generates all data. It first carries
out elementary checks in xpreferences and constructs the grid of lattice
points in xlattice. Then it generates the nested ensembles in xensemble,
and integrates each subensemble using xpath. The output data is written
to files, if required, in xwrite.

\begin{align*}
\mathbf{xsim} & \rightarrow\mathbf{xpreferences}\rightarrow\mathbf{xlattice}\,\,(checks\,inputs)\\
 & \rightarrow\mathbf{xensemble}\leftrightarrow\mathbf{xpath}\leftrightarrow\mathbf{xdata}\,\,(simulates)\\
 & \rightarrow\mathbf{xwrite}\,\,\,\,(stores\,data)
\end{align*}

\newpage{}

\chapter{Graphics parameters \label{chap:API-reference-1}}

\textbf{This chapter gives a reference guide to the xGRAPH parameters
and functions.}

\section{xGRAPH overview}

The graphics function provided is a general purpose multidimensional
batch graphics code, xGRAPH, which is automatically called by xSPDE
when xSIM is finished. The results are graphed and output if required.
Alternatively, xGRAPH can be replaced by another graphics code, or
it can be used to process the data generated by the xSIM function
at a later time.

The xgraph function call syntax is: 
\begin{itemize}
\item \textbf{xgraph (data {[},input{]})} 
\end{itemize}
This takes simulation data and input cell arrays, then plots graphs.
The data should have as many cells as there are input cells, for sequences.

If data = 'filename.h5' or 'filename.mat', the specified file is read
both for input and data. Here .h5 indicates an HDF5 file, and .mat
indicates a Matlab file.

When the data input is a filename, parameters in the file can be replaced
by new input parameters that are specified. Any stored input in the
file is then overwritten when graphs are generated. This allows graphs
of data to be modified retrospectively, if the simulation takes too
long to be run again in a reasonable timeframe.

\subsection{Parameter and data structures}

This is a batch graphics function, intended to process quantities
of graphics data, input as a cell array of multi-dimensional data.
Theoretical and/or experimental data is passed to the graphics program,
including the complete data cell array and a cell array of graphics
parameters for plotting each graph.

To explain xGRAPH in full detail, 
\begin{itemize}
\item Data to be graphed are recorded sequentially in a cell array, with
data=\{d1,d2,...\}. 
\item Graphics parameters including defaults are given in the input cell
array. 
\item This describes a sequence of graph parameters, so that input=\{p1,p2,...\}. 
\item For a one member sequence, a dataset and parameter structure can be
used on its own. 
\item Each dataset and parameter structure describes a set of graphs. 
\end{itemize}
The data input to xGRAPH can either come from a file, or from data
generated directly with xSIM. The main graphics data is a nested cell
array. It contains several numerical graphics arrays. Each defines
one independent set of averaged data, the observed data averages,
stored in a cell array indexed as $data\{s\}\{n\}(\ell,\mathbf{j},c)$.
To graph these also requires a corresponding cell array of structures
of graphics parameters.

The output is unlimited, apart from memory limits. The program also
generates error comparisons and chi-squared values if required. The
data structure for input is as follows: 
\begin{enumerate}
\item The input data is a cell array of datasets, which can be collapsed
to a single dataset 
\item The parameters are also a cell array of parameter structures, which
can be collapsed to one structure 
\item The dataset is a cell array of multidimensional graphs, each with
arbitrary dimensionality. 
\item The first or line index of each graph array allows multiple lines,
with different line-styles 
\item The last or check index of each graph array is optionally used for
error and comparison fields. 
\item Each graph array can generate multiple graphic plots, as defined by
the parameters. 
\end{enumerate}

\section{Parameter table\label{sec:Graphics-parameter-table}}

The complete cell array of the simulation data is passed to the xGRAPH
program, along with graphics parameters for each observable, to create
an extended graphics data structure. Graphics parameters have default
values which are user-modifiable by editing the xgpreferences function.

Some input parameters are global parameters for all graphs. However,
most xGRAPH parameters are cell arrays indexed by graph index. These
graphics parameters are individually set for each output that is plotted,
using the cell index $\{n\}$ in a curly bracket. If present they
replace the global parameters like labels.

If a graph index is omitted, and the parameter is not a nested array,
the program will use the same value for all graphs. The axes, glabels,
legends, lines, logs, and xfunctions of each graph are nested cell
arrays, as there can be any number of lines and axis dimensions. In
the case of the logs switch, the observable axis is treated as an
extra dimension.

The plotted result can be an arbitrary function of the generated average
data, by using the optional input gfunction. If this is omitted, the
generated average data that is input is plotted.

Comparisons are plotted if present in the input data indexed by the
last or check index $c$, with $c>errors$, where $errors=3$ is the
usual maximum value.

A table of the graphics parameters is given below.\\

\begin{tabular}{|c|c|c|}
\hline 
Label  & Default value  & Description\tabularnewline
\hline 
\hline 
axes\{n\}  & \{0,..\}  & Points plotted for each axis\tabularnewline
\hline 
chisqplot\{n\}  & 0  & Chi-square plot options\tabularnewline
\hline 
cutoff  & 1.e-12  & Global lower cutoff for chi-squares\tabularnewline
\hline 
cutoffs\{n\}  & cutoff  & Probability cutoff for n-th graph\tabularnewline
\hline 
diffplot\{n\}  & 0  & Comparison difference plot options\tabularnewline
\hline 
errors  & 0  & Index of last error field in data\tabularnewline
\hline 
esample\{n\}  & 1  & Size and type of sampling error-bar\tabularnewline
\hline 
font\{n\}  & 18  & Font size for graph labels\tabularnewline
\hline 
gfunction\{n\}  & @(d,\textasciitilde )~d\{n\}  & Functions of graphics data\tabularnewline
\hline 
glabels\{n\}  & \{'t' ,'x' ,'y' ,'z'\}  & Graph-specific axis labels\tabularnewline
\hline 
graphs  & $[1:max]$  & Vector of all the required graphs\tabularnewline
\hline 
gsqplot\{n\}  & 0  & G-square (likelihood) plot options\tabularnewline
\hline 
headers\{n\}  & ''  & Graph headers\tabularnewline
\hline 
images\{n\}  & 0  & Number of movie images\tabularnewline
\hline 
imagetype\{n\}  & 0  & Type of 3D image\tabularnewline
\hline 
klabels  & \{'\textbackslash omega' ,'k\_x' ,'k\_y' ,'k\_z'...\}  & Global transformed axis labels\tabularnewline
\hline 
legends\{n\}  & \{'label1',..\}  & Legends for multi-line graphs\tabularnewline
\hline 
limits\{n\}  & \{{[}lc1,uc1{]},{[}lc2,uc2{]}\}  & Axis limits, first lower then upper\tabularnewline
\hline 
linestyle\{n\}  & \{'-',..\}  & Line styles for multiline 2D graphs\tabularnewline
\hline 
linewidth\{n\}  & 0.5  & Line width for 2D graphs (in points)\tabularnewline
\hline 
logs\{n\}  & \{0,..\}  & Axis logarithmic switch: $0$ linear, $1$ log\tabularnewline
\hline 
minbar\{n\}  & 0.01  & Minimum relative error-bar\tabularnewline
\hline 
mincount  & 10  & Global counts for chi-square cutoffs\tabularnewline
\hline 
name  & ''  & Global graph header\tabularnewline
\hline 
olabels\{n\}  & 'a\_1'  & Observable labels\tabularnewline
\hline 
pdimension\{n\}  & 3  & Maximum plot dimensions\tabularnewline
\hline 
saveeps  & 0  & Switch, set to 1 to save eps files\tabularnewline
\hline 
savefig  & 0  & Switch, set to 1 to save figure files\tabularnewline
\hline 
scale\{n\}  & 1  & Scaling: Counts/ probability density\tabularnewline
\hline 
transverse\{n\}  & 0  & Number of transverse plots\tabularnewline
\hline 
xfunctions\{n\}  & \{@(t,\textasciitilde ) t,@(x,\textasciitilde ) x,..\}  & Axis transformations\tabularnewline
\hline 
verbose   & 0  & 0 for brief, 1 for informative, 2 for full output\tabularnewline
\hline 
xlabels  & \{'t' ,'x' ,'y' ,'z'...\}  & Global axis labels\tabularnewline
\hline 
octave  & 0  & 0 for Matlab, 1 for octave environment\tabularnewline
\hline 
\end{tabular}
\begin{itemize}
\item Up to 6 types of input data can occur, including errors and comparisons,
indexed by the last index. The original mean data always has c =1.
If there are no errors or comparisons, one graph is plotted for each
dimensional reduction. 
\item The input data has up to two error bars (I and II), and optional comparisons
also with up to two error bars. 
\item Type I errors labeled $c=2$ have standard vertical error bars. Type
II errors labeled $c=3$, which are usually standard deviation errors
from sampling, have two solid lines. 
\item If esample = -1, both error bars are combined and the RMS errors are
plotted as a single error bar. 
\item If $diffplot>0$, differences are plotted as unnormalized ($diffplot=1$),
or normalized ($diffplot=2$) by the total RMS errors. If $diffplot=3$,
raw comparison data is plotted. 
\item When differences are plotted, the total comparison errors are treated
as type I error bars, while total simulation errors are treated as
type II errors with parallel lines in the graphs, in order to distinguish
them. 
\end{itemize}
A detailed description of each parameter is listed in Sec (\ref{sec:Parameter-reference}).

\subsection{Example}

A simple example of data and input parameters, but without errors
or comparisons is as follows 
\begin{center}
\doublebox{\begin{minipage}[t]{0.75\columnwidth}%
\texttt{p.name = 'Sine and cosine functions';}

\texttt{p.olabels = \{'sine(m\_1\textbackslash pi/100)','cosine(m\_1\textbackslash pi/100)'\};}

\texttt{data = \{sin({[}1:100{*}pi{]}/100),cos({[}1:100{*}pi{]}/100)\};}

\texttt{xgraph(data,p);}%
\end{minipage}} 
\par\end{center}

\begin{figure}
\includegraphics[width=0.5\textwidth]{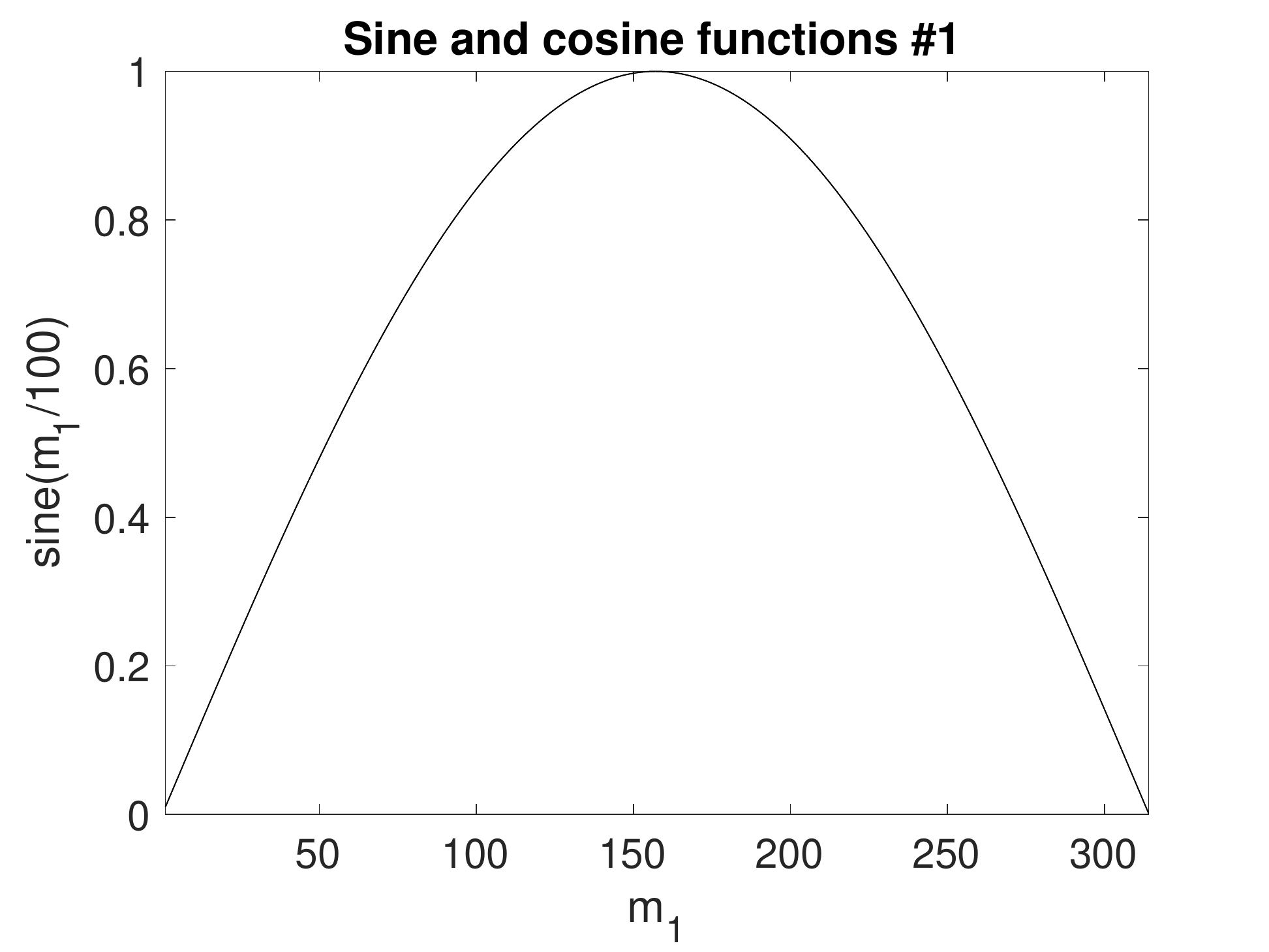}\includegraphics[width=0.5\textwidth]{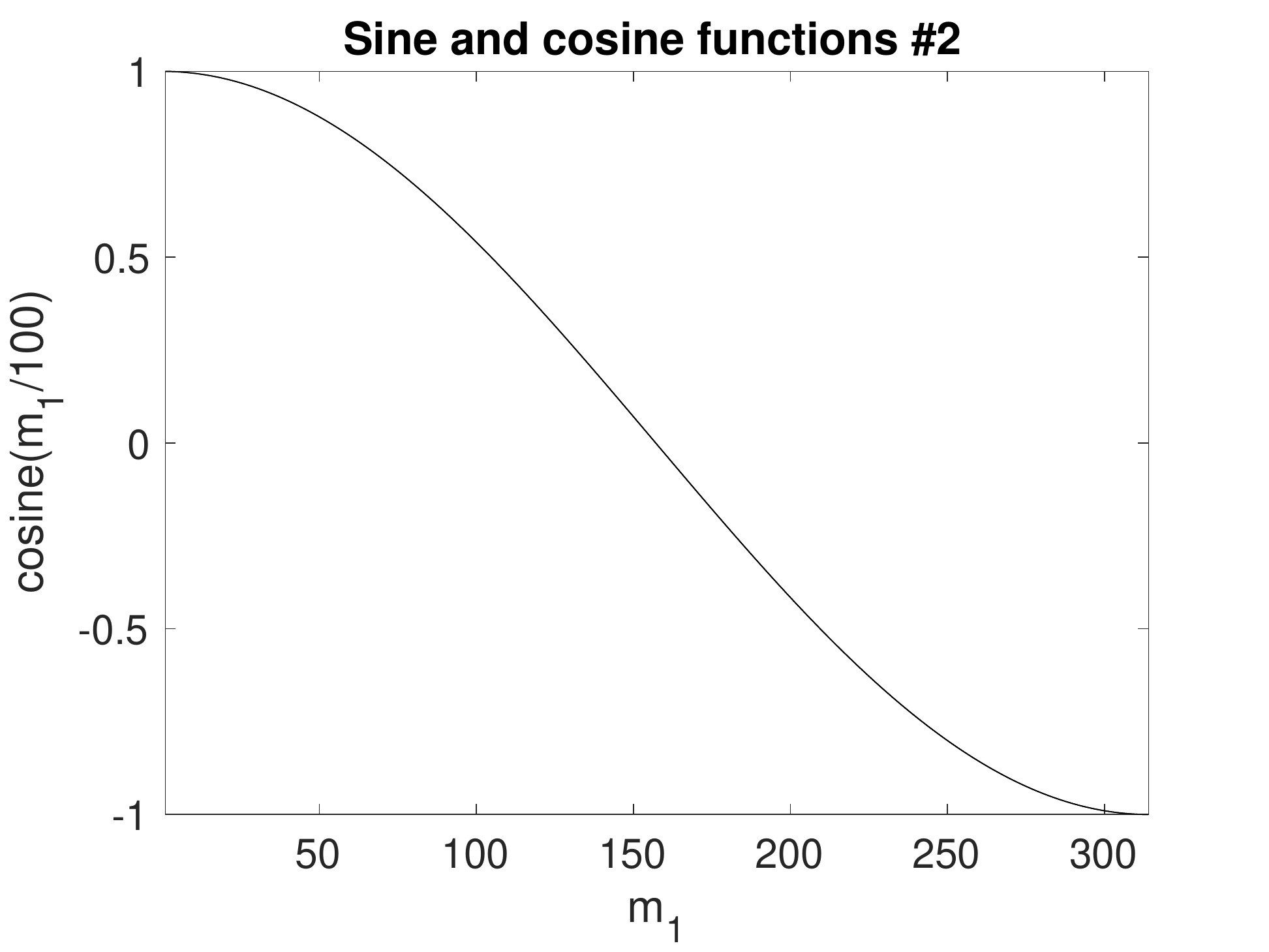}

\caption{\emph{Example: xgraph output of two plots.}}
\vspace{10pt}
 
\end{figure}

Note that in this case the default setting of p.errors=0 is used,
with no check index used in the data arrays, because these are simple
graphs without error-bars or comparisons.

\subsection{xGRAPH data arrays}

The data input to xGRAPH can come from a file, or from data generated
directly from any compatible program.

The data is stored in a cell array $data$ with structure: 
\[
data\{s\}\{n\}(\ell,\mathbf{j},c)
\]
Each member of the outer cell array data\{s\} defines a number of
related sets of graphical data, all described by common parameters
input\{s\}. Comparisons and errors are plotted if there are errors
and comparison data in the input, indexed by c. This generates comparison
plots, as well as error totals and $\chi$- squared error estimate
when there are statistical variances available.

An individual member of data\{s\}\{n\} is a multidimensional array,
called a graph in the xSPDE User's guide. For each graph, multiple
different plots with different dimensionality can be obtained from
the dataset data\{s\}\{n\}, either through projections and slices
or by generating additional data defined with graphics functions.
Either or both alternatives are available.

Note that: 
\begin{itemize}
\item If a sequence has one member, the outer cell array can be omitted. 
\item In this simplified case, if there is only one graph array, the inner
cell array can be omitted. 
\end{itemize}
The graphics data for a single dataset is held in a multidimensional
real array, where: 
\begin{itemize}
\item $\ell$ is the index for lines in the graph. Even for one line, the
first dimension is retained. 
\item $\mathbf{j}=j_{1},\ldots j_{d}$ is the array index in each dimension,
where $d\ge1$. 
\item Averages in momentum space have the momentum origin as the central
index. 
\item If integrals or spatial averages are used, the corresponding dimension
has one index $j_{d}=1$. 
\item With probabilities, extra dimensions are added to $\mathbf{j}$ to
store the bin indices. 
\item c indexes error-checks and comparisons. If not present, omit p.errors
and the last dimension. 
\item If $c>p.errors$, the extra fields are comparison inputs, where $p.errors$
is the largest data index. 
\end{itemize}
When the optional comparison fields are used, an input parameter $errors$
is required to indicate the maximum error index, to distinguish data
from comparisons. Parameter structures from xSIM have $errors=3$
set to allow for both sampling errors and discretization errors. If
this is omitted, the default is $errors=0$, which implies that there
is no error or comparison data

If $errors>0$, the last index can have larger values with $c>errors$,
for comparisons. The special case of $errors=1$ is used if the data
has no error bars, but there are comparisons in the data. Larger indices
are used to index the comparison data, which can also have two types
of errors. The largest usable last index is $errors+3$.

It is possible to directly plot the raw data using xGRAPH. One can
even combine the raw data with a graphics parameter input. But since
the raw data has no error estimates - it is raw data - one must set
$p.errors=0$, since the xsim output parameters have a normal setting
of $p.errors=3$. This will give a single trajectory.

However, the raw data from a simulation typically includes many trajectories
if $ensembles(1)>0$. One must select particular trajectory datasets
from the raw cell array, to plot just one.

\subsection{Input parameters and defaults}

A sequence of graph parameters is obtained from inputs in a cell array,
as input = \{in1, in2, ...\}. The input parameters of each simulation
in the sequence are specified in a Matlab structure. The inputs are
numbers, vectors, strings, functions and cell arrays. All metadata
has preferred values, so only changes from the preferences need to
be input. The resulting data is stored internally as a sequence of
structures in a cell array, to describe the simulation sequence.

The graphics parameters are also stored in the cell array input as
a sequence of structures p. This only need to be input when the graphs
are generated and can be changed at a later time to alter the graphics
output. A sequence of simulations is graphed from input specifications.

If there is one simulation, just one structure can be input, without
the sequence braces. The standard way to input each parameter value
is:

\[
p.label=parameter
\]

The standard way to input a function handle is:

\[
p.label=@function
\]

The inputs are scalar or vector parameters or function handles. Quantities
relating to graphed averages are cell arrays, indexed by the graph
number. The available inputs, with their default values in brackets,
are given below.

Simulation metadata, including default values that were used in a
particular simulation, can be included in the input data files. This
is done in both the .mat and the .h5 output files generated by xSIM,
so the entire graphics input can be reconstructed or changed.

Parameters can be numbers, vectors, strings or cell arrays. Conventions
that are used are that: 
\begin{itemize}
\item All input parameters have default values 
\item Vector inputs of numbers are enclosed in square brackets, {[}...{]}. 
\item Cell arrays of strings, functions or vectors are enclosed in curly
brackets. 
\item Vector or cell array inputs with only one member don’t require brackets. 
\item Incomplete parameter inputs are completed with the last used default
value. 
\item Function definitions can be handles pointing elsewhere, or defined
inline. 
\end{itemize}
If any inputs are omitted, there are default values which are set
by the internal function xgpreferences. The defaults can be changed
by editing xgpreferences.

In the following descriptions, graphs is the total number of graphed
variables of all types. The space coordinate, image, image-type and
transverse data can be omitted if there is no spatial lattice, that
is, if the dimension variable is set to one.

For uniformity, the graphics parameters that reference an individual
data object are cell arrays. These are indexed over the graph number
using braces \{\}. If a different type of input is used, like a scalar
or matrix, xSPDE will attempt to convert the type to a cell array.

Axis labels are cell arrays, indexed over dimension. The graph number
used to index these cell arrays refers to the data object. In each
case there can be multiple generated plots, depending on the graphics
input.

\subsection{Cascaded plots}

The xGRAPH function generates a default range of graphs, but this
can be modified to suit the user. In the simplest case of one dimension,
one graph dataset will generate a single plot. For higher dimensions,
a cascade of plots is generated to allow visualization, starting from
3D movies, then 3D static plots and finally 2D slices. These can also
be user modified.

Note that for all probabilities, the plot dimension is increased by
the bin range dimensionality.

\subsection{Plot dimensions}

The pdimension input sets the maximum plotted dimensions. For example,
$pdimension\{1\}=1$ means that only plots vs $r_{1}$ are output
for the first function plotted. Default values are used for the non-plotted
dimensions, unless there are axes specified, as indicated below.

The graphs cascade down from higher to lower dimensions, generating
different types of graphs. Each type of graph is generated once for
each function index.

\subsection{Plot axes}

The graphics axes that are used for plotting and the points plotted
are defined using the optional axes\textbf{ }input parameters, where
$axes\{n\}$ indicates the n-th specified graph or set of generated
graph data.

If there are no axes\textbf{ }inputs, or the axes inputs are zero
- for example, $axes\{1\}=\{0,0,0\}$ - only the lowest dimensions
are plotted, up to 3. If either the data or axes\textbf{ }inputs project
one point in a given dimension, - for example, $axes\{1\}=\{0,31,-1,0\}$,
this dimension is suppressed in the plots, which reduces the effective
dimension of the data - in this case to two dimensions.

Examples: 
\begin{itemize}
\item $axes\{1\}=\{0\}$ - For function 1, plot all the first dimensional
points; higher dimensions get defaults. 
\item $axes\{2\}=\{-2,0\}$ - For function 2, plot the maximum value of
$r_{1}$ (the default) and all higher-dimensional x-points. 
\item $axes\{3\}=\{1:4:51,32,64\}$ - For function 3, plot every 4-th $x_{1}$
point at $x_{2}$ point 32, $x_{3}$ point 64 
\item $axes\{4\}=\{0,2:4:48,0\}$ - For function 4, plot every $x_{1}$
point , every 4-th $x_{2}$ point, and all $x_{3}$-points. 
\end{itemize}
Points labelled $-1$ indicates a default `typical' point, which is
the midpoint. If one uses $-2$, this is the last point.

Lower dimensions are replaced by corresponding higher dimensions if
there are dimensions or axes that are suppressed. Slices can be taken
at any desired point, not just the midpoint. The notation of $axes\{1\}=\{6:3:81\}$,
is used to modify the starting, interval, and finishing points for
complete control on the plot points.

The graphics results depend on the resulting \textbf{effective} dimension,
which is equal to the actual input data dimension unless there is
an axes suppression, described above. Since the plot has to include
a data axis, the plot itself will usually have an extra data axis.

One can plot only three axes directly using standard graphics tools.
The strategy to deal with the higher effective dimensionality is as
follows. For simplicity, ``time'' is used to label the first effective
dimension, although in fact any first dimension is possible: 
\begin{description}
\item [{dimensions~=~1}] For one lattice dimension, a 2D plot of observable
vs t is plotted, with data at each lattice point in time. Exact results,
error bars and sampling error bounds are included if available. 
\item [{dimensions~=~2}] For two lattice dimensions, a 3D image of observable
vs x,t is plotted. A movie of distinct 2D graphic plots is also possible.
Otherwise, a slice through $x=0$ is used tp reduce the lattice dimension
to $1$. 
\item [{dimensions~=~3}] For three lattice dimensions, if $images>1$,
a movie of distinct 3D graphic images of observables are plotted as
$images$ slices versus the first plot dimension. Otherwise, a slice
through the chosen point, is used at the highest dimension to reduce
the lattice dimension to $2$. 
\item [{dimensions~=~4,5..}] For higher lattice dimensions, a slice through
a chosen point, or the default midpoint is used to reduce the lattice
dimension to $3$. 
\end{description}
As explained above, in addition to graphs versus $x_{1}$ the \textbf{xGRAPH}
function can generate images (3D) and transverse (2D) plots at specified
points, up to a maximum given by the number of points specified. The
number of these can be individually specified for each graph number.
The images available are specified as imagetype$=1,\ldots4$, giving: 
\begin{enumerate}
\item 3D perspective plots (Matlab surf - the default) 
\item 2D filled color plots (Matlab contourf ) 
\item contour plots (Matlab contour ) 
\item pseudo-color plots (Matlab pcolor ) 
\end{enumerate}
Error bars, sampling errors and multiple lines for comparisons are
only graphed for 2D plots. Error-bars are not plotted when they are
below a user-specified size, with a default of $1\%$ of the maximum
range, to improve graphics quality. Higher dimensional graphs do not
output error-bar data, but they are still recorded in the data files.

\subsection{Probabilities and parametric plots}

Probability data can be input and plotted like any other data. It
is typically generated from simulation programs using the $binranges$
data for binning. It is plotted like any other graph, with any dimension,
except that the total dimension is extended by the number of variables
or lines in the observe function.

\subsection{Chi-squared plots}

In addition the program can make a $\chi^{2}$ plot, which is a plot
of the $\chi^{2}$ comparison with a comparison probability density
against space and/or time. This allows a test of the simulated data
against a known target probability distribution, provided that the
following input data conditions are satisfied: 
\begin{itemize}
\item The input data dimension exceeds the p.dimensions parameter, 
\item The switch p.chisqplot is set to $1$or 2, and 
\item The input data includes comparison function data. 
\end{itemize}
The $\chi^{2}$ plots, depending on $p.chisqplot$ are: 
\begin{enumerate}
\item a plot of $\chi^{2}$ and $k$, where $k$ is the number of valid
data points, 
\item a plot of $\sqrt{2\chi^{2}}$ and $\sqrt{2k-1}$, which should have
a unit variance. 
\end{enumerate}
Here, for one point in space and time, with $m$ bins, $N_{j}$ counts
per bin and $E_{j}$ expected counts: 
\begin{equation}
\chi^{2}=\sum_{j=1}^{m}\frac{\left(N_{j}-E_{j}\right)^{2}}{E_{j}}.
\end{equation}

The number $k$ is the number of valid counts, with $N_{j},E_{j}>mincount$.
This is partly determined from the requirement that the probability
count data per bin is greater than the $p.mincount$ parameter. The
default is set to give a number of samples $>10$. The program prints
a summary that sums over of all the $\chi^{2}$ data.

The $p.scale\{n\}$ parameter gives the number of counts per bin at
unit probability density. This is needed to set the scale of the $\chi^{2}$
results, ie, $N_{j}=scale\{n\}\times p_{j}$, where $p_{j}$ is the
probability density that is compared and plotted in the simulation
data. Note that a uniform bin size is assumed here, to give a uniform
scaling.

\subsection{Comparisons with variances}

It can be useful to compare two probability distributions with different
variances. For one point in space and time, with $m$ bins, $p_{j}$
probability density and $e_{j}$ expected probability density, 
\begin{equation}
\chi^{2}=\sum_{j=1}^{m}\frac{\left(p_{j}-e_{j}\right)^{2}}{\sigma_{j}^{2}+\sigma_{e,j}^{2}}.
\end{equation}
In this case, $\sigma_{j}^{2}$ and $\sigma_{e,j}^{2}$ are the sampling
errors in the simulation data and comparison data, so that built-in
error fields in the data are used to work out the $\chi^{2}$ results.
This option is chosen if $p.scale\{n\}=0$, and the cutoff for the
data is then specified so that $p_{j},e_{j}>p.cutoffs\{n\}$. The
default value is the global cutoff, $p.cutoff$, which has a default
of $10^{-12}$. 

This output only has a $\chi^{2}$ distribution with $\chi^{2}\approx m$
if all the points are independent. The measured $\chi^{2}/m$ includes
all the space-time points above the cutoff value. This can be less
than unity when comparing an expected exact result with a computed
SDE solution where data is often correlated. At the other extreme,
if the cutoff is too low, the data may not be reliable, and one can
obtain too large a value.

\subsection{Maximum likelihood}

It is also possible to plot the $G^{2}$ or maximum likelihood plot
of the data, which is an alternative means to compare distributions,
where 
\begin{equation}
G^{2}=2\sum_{j=1}^{m}N_{j}\ln\left(N_{j}/E_{j}\right).
\end{equation}
The expected values $E_{j}$ are automatically scaled so that $\sum N_{j}=\sum E_{j},$with
the same minimum count cutoff that is used for the $\chi^{2}$ data.
The result is similar to the $\chi^{2}$ results. It is obtained if
p.gsqplot is set to $1$ or 2 and requires for the input that $p.scale\{n\}>0.$
It is sometimes regarded as a preferred method for comparisons.

\subsection{Parametric plots}

Any input dataset can be converted to a parametric plot, where a second
data input is plotted along the horizontal axis instead of the time
coordinate. It is also possible to substitute a second data input
for the x-axis data if a parametric plot in space is required instead.
This allows visualization of how one type of data changes as a function
of a second type of data input.

The two datasets that are plotted must have the same number of lines,
that is, the first index range should be the same, in order that multiple
lines can be compared. This is achieved where required using the p.scatters
input in the simulation code. The details of the parametric plot are
specified using the input:  
\begin{equation}
p.parametric\{n\}=[n1,p2]
\end{equation}

Here $n$ is the graph number which is plotted, and must correspond
to an input dataset. The number $n1$ is the graph number of the observable
that is plotted on the horizontal axis, ignoring functional transformations.
The second number is the axis number where the parametric value is
substituted, which can be the time (axis 1) or the x-coordinate (axis
2), if present.

In all cases the vertical axis is used to plot the original data.
The specified horizontal axis is used for the parametric variable.
Only vertical error-bars are available. An example is given in xAMPLES/SDE\_1/SHO,
which is a noise-driven harmonic oscillator, with several lines plotted
of $x$ vs y.

\section{xGRAPH Parameter reference\label{sec:Parameter-reference}}

\subsection{axes\{n\}}
\begin{description}
\item [{Default:}] \{0,0,0,..\} 
\end{description}
Gives the axis points plotted for the $n$-th plotted function, in
each dimension. Each entry value is a vector range for a particular
plot and dimension. Thus, p = 5 gives the fifth point only, and a
vector input p = 1:4:41 plots every fourth point. Single points generate
graphics projections, allowing the other dimensions to be plotted.
Zero or negative values are shorthand. For example, p = -1 generates
a default point at the midpoint, p = -2 the endpoint, and p = 0 is
the default value that gives the vector for the every axis point.
For each graph type, i.e. n=1,..graphs the axes can be individually
specified in each dimension, d=1,..dimensions. If more than three
axes are specified to be vectors, only the first three are used, and
others are set to default values in the plots. 
\begin{description}
\item [{Example:}] p.axes\{4\} = \{1:2:10,0,0,-1\} 
\end{description}

\subsection{diffplot\{n\}}
\begin{description}
\item [{Default:}] 0 
\end{description}
Differences are plotted as a comparison dashed line on $2D$ plots
as a default. Otherwise, a separate difference plot is obtained which
is unnormalized (diffplot = 1), or normalized (diffplot = 2) by the
total RMS errors. If diffplot = 3, the comparison data is plotted
directly as an additional graph. 
\begin{description}
\item [{Example:}] p.diffplot\{3\} = 2 
\end{description}

\subsection{errors}
\begin{description}
\item [{Default:}] 0 
\end{description}
Indicates if the last index in the graphics input data arrays is used
for error-bars and/or comparisons. Should be set to zero if there
is no error or comparison data. If non-zero, this will give the highest
last index used for errors. The standard xsim output sets $p.errors=3$
automatically. As a special case, $p.errors=1$ is used to indicate
that there is comparison data but no error data.

If $p.errors>0$ , the data indexed up to p.errors gives the data,
then a maximum of two types of error bars. Up to three further index
values, up to $p.errors+3$, are available to index all comparison
data and its error fields. The maximum last index value used is $6$. 
\begin{description}
\item [{Example:}] p.errors = 2 
\end{description}

\subsection{esample\{n\}}
\begin{description}
\item [{Default:}] 1 
\end{description}
This sets the type and size of sampling errors that are plotted. If
esample = 0, no sampling error lines are plotted, just the mean. If
$esample=-n$, $\pm n\sigma$ sampling errors are included in the
error-bars. If $esample=n$, separate upper and lower $\pm n\sigma$
sampling error lines are plotted. In both cases, the magnitude of
esample sets the number of standard deviations used. 
\begin{description}
\item [{Example:}] p.esample\{3\} = -1 
\end{description}

\subsection{font\{n\}}
\begin{description}
\item [{Default:}] 18 
\end{description}
This sets the default font sizes for the graph labels, indexed by
graph. This can be changed per graph. 
\begin{description}
\item [{Example:}] p.font\{4\}=18 
\end{description}

\subsection{functions}
\begin{description}
\item [{Default:}] number of functional transformations 
\end{description}
This gives the maximum number of output graph functions and is available
to restrict graphical output. The default is the length of the cell
array of input data. Normally, the default will be used. 
\begin{description}
\item [{Example:}] p.functions = 10 
\end{description}

\subsection{glabels\{n\}}
\begin{description}
\item [{Default:}] xlabels or klabels 
\end{description}
Graph-dependent labels for the independent variable labels. This is
a nested cell array with first dimension of graphs and second dimension
of dimensions. This is used to replace the global values of xlabels
or klabels if the axis labels change from graph to graph, for example,
if the coordinates have a functional transform. These can be set for
an individual coordinate on one graph if needed. 
\begin{description}
\item [{Example:}] p.glabels\{4\}\{2\} = 'x\textasciicircum 2' 
\end{description}

\subsection{graphs}
\begin{description}
\item [{Default:}] observables to plot 
\end{description}
This gives the observables to plot. The default is a vector of indices
from one to the length of the cell array of observe functions. Normally
not initialized, as the default is used. Mostly used to reduce graphical
output on a long file. 
\begin{description}
\item [{Example:}] p.graphs = 10 
\end{description}

\subsection{gtransforms\{n\}}
\begin{description}
\item [{Default:}] {[}0,0,...{]} 
\end{description}
This switch specifies the Fourier transformed graphs and axes for
graphics labeling. Automatically equal to ftransforms if from an earlier
xSIM input, but can be changed. If altered for a given graph, all
the axis Fourier switches should be reset. This is ignored if there
is no dimensions setting to indicate space dimensions. 
\begin{description}
\item [{Example:}] p.gtransforms\{1\} = {[}0,0,1{]} 
\end{description}

\subsection{headers\{n\}}
\begin{description}
\item [{Default:}] '' 
\end{description}
This is a string variable giving the graph headers for each type of
function plotted. The default value is an empty string. Otherwise,
the header string that is input is used. Either is combined with the
simulation name and a graph number to identify the graph. This is
used to include simulation headers to identify graphs in simulation
outputs. Graph headers may not be needed in a final published result.
For this, either edit the graph, or use a space to make plot headers
blank: p.headers\{n\} = ' ', or p.name = ' ' . 
\begin{description}
\item [{Example:}] p.headers\{n\} = 'my\_graph\_header' 
\end{description}

\subsection{images\{n\}}
\begin{description}
\item [{Default:}] 0 
\end{description}
This is the number of 3D, transverse o-x-y images plotted as discrete
time slices. Only valid if the input data dimension is greater than
2. If present, the coordinates not plotted are set to their central
value when plotting the transverse images. This input should have
a value from zero up to a maximum value of the number of plotted points.
It has a vector length equal to graphs. 
\begin{description}
\item [{Example:}] p.images\{4\} = 5 
\end{description}

\subsection{imagetype\{n\}}
\begin{description}
\item [{Default:}] 1 
\end{description}
This is the type of transverse o-x-y movie images plotted. It has
a vector length equal to graphs. 
\begin{itemize}
\item imagetype = 1 gives a perspective surface plot 
\item imagetype = 2, gives a 2D plot with colors 
\item imagetype = 3 gives a contour plot with 10 equally spaced contours 
\item imagetype = 4 gives a pseudo-color map 
\end{itemize}
\begin{description}
\item [{Example:}] p.imagetype\{n\} = 1, 2, 3, 4 
\end{description}

\subsection{klabels}
\begin{description}
\item [{Default:}] \{'\textbackslash omega', 'k\_x', 'k\_y', 'k\_z'\}``
or ``\{'k\_1', 'k\_2', 'k\_3', 'k\_4',...\} 
\end{description}
Labels for the graph axis Fourier transform labels, vector length
of dimensions. The numerical labeling default is used when the ``p.numberaxis``
option is set. Note, these are typeset in Latex mathematics mode!
When changing from the default values, all the required new labels
must be set. 
\begin{description}
\item [{Example:}] p.klabels= \{'\textbackslash Omega', 'K\_x', 'K\_y',\} 
\end{description}

\subsection{legends\{n\}}
\begin{description}
\item [{Default:}] \{'',''\} 
\end{description}
Graph-dependent legends, specified as a nested cell array of strings
for each line. 
\begin{description}
\item [{Example:}] p.legends\{n\} = \{labels(1), ..., labels(lines)\} 
\end{description}

\subsection{limits\{n\}}
\begin{description}
\item [{Default:}] \{0,0,0,0; ...\} 
\end{description}
Graph-dependent limits specified as a cell array with dimension graphs.
Each entry is a cell array of graph limits indexed by the dimension,
starting from $d=1$ for the time dimension. The limits are vectors,
indexed as 1,2 for the lower and upper plot limits. This is useful
if the limits required change from graph to graph. If an automatic
limit is required for either the upper or lower limit, it is set to
inf. 

An invalid, scalar or empty limit vector, like {[}0,0{]} or $0$ or
{[}{]} is ignored, and an automatic graph limit is used. 
\begin{description}
\item [{Example:}] p.limits\{n\} = \{{[}t1,t2{]},{[}x1,x2{]},{[}y1,y2{]}
...,\} 
\end{description}

\subsection{linestyle\{n\}}
\begin{description}
\item [{Default:}] \{'-k','-{}-k',':k','-.k','-ok','-{}-ok',':ok','-.ok','-+k','-{}-+k'\} 
\end{description}
Line types for each line in every two-dimensional graph plotted. If
a given line on a two-dimensional line is to be removed completely,
set the relevant line-style to zero. For example, to remove the first
line from graph 3, set p.linestyle\{3\} =\{0\}. This is useful when
generating and changing graphics output from a saved data file. The
linestyle uses Matlab terminology. It allows setting the line pattern,
marker symbols and color for every line. The default lines are black
('k'), but any other color can be used instead.

The specifiers must be chosen from the list below, eg, '-ok', although
the marker can be omitted if not required. 
\begin{itemize}
\item Line patterns: '-' (solid), '--' (dashed), ':' (dotted) ,'-.' (dash-dot) 
\item Marker symbols: '+','o','{*}','.','x','s','d','\textasciicircum ','v','\textgreater
','\textless ','p' 
\item Colors: 'r','g','b','c','m','y','k','w' 
\end{itemize}
\begin{description}
\item [{Example:}] p.linestyle\{4\} = \{'-k','-{}-ok',':g','-.b',\} 
\end{description}

\subsection{linewidth\{n\}}
\begin{description}
\item [{Default:}] 0.5 
\end{description}
Line width for plotted lines in two-dimensional graphs. For example,
to make the lines wider in graph 3, set p.linewidth\{3\} =1. This
is useful for changing graphics output appearance if the default lines
are too thin. 
\begin{description}
\item [{Example:}] p.linewidth\{n\} = 1 
\end{description}

\subsection{minbar\{n\}}
\begin{description}
\item [{Default:}] \{0.01, ...\} 
\end{description}
This is the minimum relative error-bar that is plotted. Set to a large
value to suppress unwanted error-bars, although its best not to ignore
the error-bar information! This can be changed per graph. 
\begin{description}
\item [{Example:}] p.minbar\{n\} = 0 
\end{description}

\subsection{name}
\begin{description}
\item [{Default:}] '' 
\end{description}
Name used to label simulation graphs, usually corresponding to the
equation or problem solved. This can be removed from individual graphs
by using headers\{n\} equal to a single blank space. The default is
a null string. To remove all headers globally, set name equal to a
single blank space: name = ' '. 
\begin{description}
\item [{Example:}] p.name = 'Wiener process simulation' 
\end{description}

\subsection{olabels\{n\}}
\begin{description}
\item [{Default:}] 'a' 
\end{description}
Cell array of labels for the graph axis observables and functions.
These are text labels that are used on the graph axes. The default
value is 'a\_1' if the default observable is used, otherwise it is
blank. This is overwritten by any subsequent label input when the
graphics program is run: 
\begin{description}
\item [{Example:}] p.olabels\{4\} = 'v' 
\end{description}

\subsection{parametric\{n\}}
\begin{description}
\item [{Default:}] {[}0,0{]} 
\end{description}
Cell array that defines parametric plots, for each graph number. The
first number is the graph number of the alternative observable plotted
on the horizontal axis. The second number is the axis number where
the parametric value is substituted, which can be the time (axis 1)
or the x-coordinate (axis 2), if present.

If both are zero, the plot against an independent space-time coordinate
is calculated as usual. If nonzero, a parametric plot is made for
two-dimensional plots. In all cases the vertical axis is used to plot
the original data. The specified horizontal axis is used for the parametric
variable. Only vertical error-bars are available. Can be usefully
combined with scatters\{n\} to plot individual trajectories, but the
number of scatters should be the same in each of the two graphs that
are parametrically plotted against each other. 
\begin{description}
\item [{Example:}] p.parametric\{n\} = {[}p1,p2{]} \textgreater = 0 
\end{description}

\subsection{\label{subsec:pdimension}pdimension\{n\}}
\begin{description}
\item [{Default:}] 3 
\end{description}
This is the maximum plotted space-time dimension for each plotted
quantity. The purpose is eliminate unwanted graphs. For example, it
is useful to reduce the maximum dimension when averaging in space.
Higher dimensional graphs are not needed, as the data is duplicated.
Averaging can be useful for checking conservation laws, or for averaging
over homogeneous data to reduce sampling errors. All graphs are suppressed
if it is set to zero. Any three dimensions can be chosen to be plotted,
using the axes parameter to suppress the unwanted data points in other
dimensions. 
\begin{description}
\item [{Example:}] p.pdimension\{4\} = 2 
\end{description}

\subsection{saveeps}
\begin{description}
\item [{Default:}] 0 
\end{description}
If set to $1$, all plots are saved to the current folder as .eps
files, numbered consecutively. It is best to use the close all command
first to remove unwanted displayed xFIGURES, before running xgraph
with this option. 
\begin{description}
\item [{Example:}] p.saveeps =1 
\end{description}

\subsection{savefig}
\begin{description}
\item [{Default:}] 0 
\end{description}
If set to $1$, all plots are saved to the current folder as .fig
files, numbered consecutively. It is best to use the close all command
first to remove unwanted displayed xFIGURES, before running xgraph
with this option. 
\begin{description}
\item [{Example:}] p.savefig =1 
\end{description}

\subsection{transverse\{n\}}
\begin{description}
\item [{Default:}] 0 
\end{description}
This is the number of 2D transverse images plotted as discrete time
slices. Only valid if dimensions is greater than 2. If present, the
$y,z$-coordinates are set to their central values when plotting transverse
images. Each element can be from 0 up to the number of plotted time-points.
The cell array has a vector length equal to graphs. 
\begin{description}
\item [{Example:}] p.transverse\{n\}= 6 
\end{description}

\subsection{verbose}
\begin{description}
\item [{Default:}] 0 
\end{description}
Print flag for output information while running xGRAPH. Print options
are: 
\begin{itemize}
\item Minimal if verbose = -1: Prints just the start-up time and hard error
messages 
\item Brief if verbose = 0: Additionally prints the final, total chi-squared
errors where present 
\item Informative if verbose = 1: Also prints the graph progress indicators 
\item Full if verbose = 2: Prints everything including the internal parameter
structure data. 
\end{itemize}
In summary, if verbose = 0, most output is suppressed except the final
data, verbose = 1 displays a progress report, and verbose = 2 additionally
generates a readable summary of the graphics parameter input. 
\begin{description}
\item [{Example:}] p.verbose = 0 
\end{description}

\subsection{xlabels}
\begin{description}
\item [{Default:}] \{'t', 'x', 'y', 'z'\} or \{'x\_1', 'x\_2', 'x\_3',
'x\_4',...\} 
\end{description}
Global labels for the independent variable labels, vector length equal
to dimensions. The numerical labeling default is used when the numberaxis
option is true. These are typeset in Latex mathematics mode. When
changing from the default values, all the required new labels must
be set. 
\begin{description}
\item [{Example:}] p.xlabels = \{'tau'\} 
\end{description}

\subsection{gfunction\{n\} (d,p)}

This is a cell array of graphics function handles. Use when a graph
is needed that is a functional transformation of the observed averages.
The default value generates the n-th graph data array directly from
the n-th input data. The input is the data cell array for all the
graphs in the current sequence number with their graph parameters
x, and the output is the n-th data array that is plotted.

An arbitrary number of functions of these observables can be plotted,
including vector observables. The input to graphics functions is the
observed data averages or functions of averages in a given sequence,
each stored in a cell array $d\{n\}(\ell,\mathbf{j},c)$. If there
are more graphics functions than input data cells, this generate additional
data for plotting.

\subsection{xfunctions\{n\} \{nd\} (ax,p)}

This is a nested cell array of axis transformations. Use when a graph
is needed with an axis that is a function of the original axes. The
input is the original axis coordinates, and the output is the new
coordinate set. The default value generates the input axes. Called
as xfunctions\{n\}\{nd\}(ax,p) for the n-th graph and axis direction
dir, where ax is a vector of coordinates for that axis.There is one
graphics function for each separate graph dimension or axis. The default
value is the coordinate vector $xk\{nd\}$ stored in the input parameter
structure p, or else the relevant index if xk\{nd\} is omitted.

\section{xGRAPH structure}

The graphics function, $xgraph$, plots the simulation data. The general
structure is:

\begin{align*}
\mathbf{xgraph} & \rightarrow\mathbf{xgpreferences}\,\,(checks\,inputs)\\
 & \rightarrow\mathbf{xmultigraph}\leftrightarrow\mathbf{xreduce\leftrightarrow\mathbf{xcompress}}\,\,(structures\,data\,arrays)\\
 & \rightarrow\mathbf{ximages}\rightarrow\mathbf{xtransverse}\rightarrow\mathbf{xplot3}\rightarrow\mathbf{xplot2}\,\,(graphs\,all\,data)
\end{align*}

Most graphics functions simply work, but two important functions are
listed below for reference.

\subsection{xgraph(data,input)}

The xgraph function graphs multidimensional data files. 
\begin{itemize}
\item Input: graphics data cells data, input parameter cells input. 
\item Output: graphs, displayed and/or stored as eps or fig files. 
\item If no numeric data present, reads data from a file named data. 
\item If data is present but without any input parameters it plots using
default parameters. 
\item First data dimension is the line index, last dimension are the error-bars
and comparisons 
\item Needs: xread, xmakecell, xgpreferences, xmultiplot 
\end{itemize}

\subsection{xgpreferences (input,oldinput)}

The xgpreferences function sets default values for graphics inputs. 
\begin{itemize}
\item Input: input cell array and optionally previous inputs from a datafile,
oldinput. 
\item Note that each cell array is a sequence of graphics parameter structures 
\item Output: the updated plus default graphics parameters 
\item Called by: xgraph 
\item Needs: xprefer, xcprefer 
\end{itemize}
\pagebreak{}

\section*{Acknowledgements}

We would like to thank the users whose feedback was invaluable, including
Rodney Polkinghorne, Simon Kiesewetter, Bogdan Opanchuk, King Ng,
Jesse van Rhijn and Thomas Rodriguez. This work was funded through
the Australian Research Council Discovery Project scheme under Grants
DP180102470 and DP190101480. The authors also wish to thank NTT Research
and the Templeton Foundation for their financial and technical support.

\bibliographystyle{SciPost_bibstyle}

\end{document}